\documentclass[11pt,a4paper]{article} 

\PassOptionsToPackage{bbgreekl}{mathbbol}

\usepackage[showcomments]{ajtex}

\usepackage{tikz}

\usetikzlibrary{shapes.misc}

\usetikzlibrary{decorations.markings,decorations.pathmorphing,arrows.meta}

\newcommand\tikzsetnextfilename[1]{}

\tikzset{aux/.style={decorate,decoration={snake,segment length=1.5mm,amplitude=0.4mm}}}
\tikzset{bold/.style={ultra thick}}
\tikzset{cdotted/.style={line width=1.2pt, dash pattern=on 0pt off 1.2mm, line cap=round}}
\tikzset{bcdotted/.style={line width=2pt, dash pattern=on 0pt off 1.5mm, line cap=round}}
\tikzset{aux2/.style={double,double distance=0.6pt,decorate,decoration={snake,segment length=1.5mm,amplitude=0.4mm}}}
\tikzset{sol2/.style={double,double distance=0.6pt}}

\tikzset{left/.style={arrows={Stealth[scale length=0.5, scale width=1.5][sep=2pt]-}}}
\tikzset{right/.style={arrows={-[sep=-2pt]Stealth[scale length=0.5, scale width=1.5]}}}
\tikzset{lleft/.style={arrows={Stealth[scale length=0.5, scale width=1.5][sep=2pt]-}}}
\tikzset{right/.style={arrows={-[sep=-2pt]Stealth[scale length=0.5, scale width=1.5]}}}

\usepackage{mdframed}
\newmdenv[
  backgroundcolor=gray!20,
  linecolor=gray!20,
  linewidth=0pt,
  skipabove=\baselineskip,
  skipbelow=\baselineskip,
  leftmargin=-.5cm,
  rightmargin=-.5cm,
  innerleftmargin=.5cm,
  innerrightmargin=.5cm,
  innertopmargin=10pt,
  innerbottommargin=15pt,
  splittopskip=20pt plus 0pt minus 20pt,
  splitbottomskip=10pt plus 0pt minus 10pt,
]{myshaded}

\DeclareMathOperator{\Sch}{Sch}
\DeclareMathOperator\PSL{PSL}

\def\bx{\mathbf x}
\def\by{\mathbf y}
\def\bk{\mathbf k}
\def\bq{\mathbf q}

\def \beq {\begin{equation}}
\def \eeq {\end{equation}}

\def\le{\left}
\def\ri{\right}
\newcommand\ov{\over}
\newcommand{\es}[2] {\begin{equation} \ifx#1\empty\else\label{#1}\fi \begin{split} #2 \end{split} \end{equation}}

\newcommand\lam{\lambda}

\setlist[itemize]{itemsep=0pt,topsep=5pt}
\setlist[enumerate]{itemsep=0pt,topsep=5pt}

\makeatletter
\AddToHook{cmd/appendix/before}{%
\def\cref@section@alias{appendix}%
\def\cref@subsection@alias{appendix}%
}
\makeatother

\newenvironment{calign}
{\renewcommand\nn{}%
\begin{equation}\begin{aligned}}
{\end{aligned}\end{equation}\ignorespacesafterend}

\newenvironment{cgather}
{\renewcommand\nn{}%
\begin{equation}\begin{gathered}}
{\end{gathered}\end{equation}\ignorespacesafterend}

\title{Out-of-time-order hydrodynamics and scrambling}

\author[a]{Marta Bucca}\email{marta.bucca@maths.ox.ac.uk}
\author[a]{Akash Jain}\email{akash.jain@maths.ox.ac.uk}
\author[a]{M\'ark Mezei}\email{mark.mezei@maths.ox.ac.uk}
\author[b]{Alexey Milekhin}\email{milekhin@uky.edu}         
 
 \affiliation[a]{Mathematical Institute, University of Oxford, Woodstock Road, Oxford, OX2 6GG, United Kingdom}  

\affiliation[b]{Department of Physics and Astronomy, University of Kentucky, 506 Library Drive, Lexington, KY 40506}  

\abstract{Hydrodynamics as an effective field theory (EFT) has been immensely successful for predicting the behaviour of time-ordered correlation functions in the long-time, large-distance regime. In this paper, we analyse the predictions of hydrodynamics for scrambling as probed by out-of-time order correlators (OTOCs). To do so, we consider an all-derivative order generalisation of the Schwinger-Keldysh EFT of hydrodynamics extended to a four-fold time contour required to capture OTOCs. A crucial ingredient of the construction is a local $\SL(2,\mathbb{R})$ symmetry, which has previously been suggested as a mechanism for scrambling. We also derive this exact EFT from the microscopic formulation of the maximally chaotic low-temperature limit of the SYK lattice, thereby determining all transport coefficients for this model. As a novel application, we compute the OTOC of energy currents, find that its exponential growth saturates the chaos bound, and match it to expectations based on a shockwave (or scramblon) effective description. Our work shows that the universality of hydrodynamics degrees of freedom extends beyond traditional time-ordered correlators to more complex observables such as OTOCs.
}

\begin{document}

\maketitle

\clearpage

\section{Introduction}

\paragraph*{Hydrodynamic effective field theories.} 

In a chaotic system, most excitations decay quickly at finite temperature. The exception are excitations associated with the local excess or deficit of a conserved quantity, like energy, momentum, or charge. These degrees of freedom can be described by the effective field theory (EFT) of hydrodynamics~\cite{Liu:2018kfw,Delacretaz:2026owo}. For the long-time, large-distance limit of time-ordered correlation functions,\footnote{Here we mean to include retarded and advanced correlation functions in the loosely defined term ``time-ordered''. Our aim is to  distinguish these correlators from OTOCs.} the EFT is formulated on the Schwinger-Keldysh (SK) closed-time contour requiring the doubling of degrees of freedom. 

We expect that hydrodynamic degrees of freedom should also play a role in the description of the long time-large distance limit of more complex observables that require multiple copies of the time evolution operator to be defined: out-of-time ordered correlators (OTOCs)~\cite{Khemani:2017nda}, the spectral form factor \cite{Friedman:2019gyi,Winer:2020gdp}, or R\'enyi entropies \cite{Rakovszky:2019oht,Huang:2019hts,Zhou:2019xjd,Lastres:2026nrt}. In this paper we examine the case of OTOCs.

\begin{figure}[t]
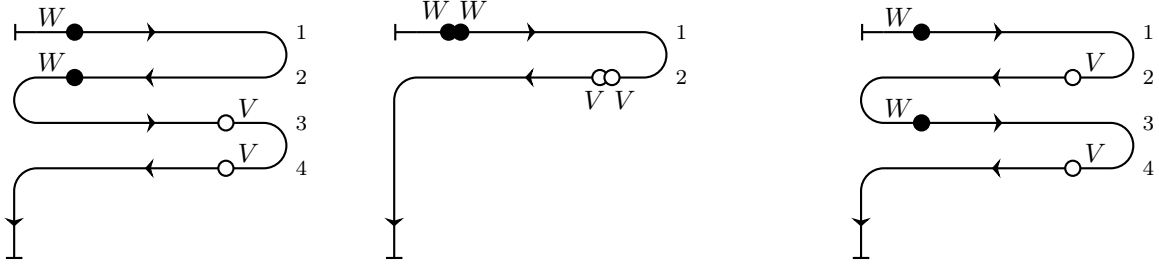

\centering
\tikzsetnextfilename{TOC-4-basic}
\raisebox{-16mm}{
\tikz[thick]{
    \draw[|-] (-0.3,0) -- (0,0);
    \draw[|-,right] (0,0) -- (1.5,0);
    \draw[right] (1.5,0) -- (3,0) arc (90:-90:.3) -- (1.5,-.6);
    \draw[right] (1.5,-.6) -- (0,-.6) arc (90:270:.3) -- (1.5,-1.2);
    \draw[right] (1.5,-1.2) -- (3,-1.2) arc (90:-90:.3) -- (1.5,-1.8);
    \draw[right] (1.5,-1.8) -- (0,-1.8) arc (90:180:.3) -- (-0.3,-2.5);
    \draw[-|] (-0.3,-2.5) -- (-0.3,-3);
    \draw [fill=white] (2.5,-1.2) circle (1mm);
    \draw [fill=white] (2.5,-1.8) circle (1mm);
    \draw [fill=black] (.5,0) circle (1mm);
    \draw [fill=black] (.5,-.6) circle (1mm);
    \node at (2.8,-1.2+.2) {\small $V$};
    \node at (2.8,-1.8+.2) {\small $V$};
    \node at (0.2,0+.2) {\small $W$};
    \node at (0.2,-.6+.2) {\small $W$};
    \node at (3.5,-1.8) {\scriptsize 4};
    \node at (3.5,-1.2) {\scriptsize 3};
    \node at (3.5,-.6) {\scriptsize 2};
    \node at (3.5,0) {\scriptsize 1};
}
}
\quad
\tikzsetnextfilename{TOC-2-basic}
\raisebox{-16mm}{
\tikz[thick]{
    \draw[|-] (-0.3,0) -- (0,0);
    \draw[|-,right] (0,0) -- (1.5,0);
    \draw[right] (1.5,0) -- (3,0) arc (90:-90:.3) -- (1.5,-.6);
    \draw[right] (1.5,-.6) -- (0,-.6) arc (90:180:.3) -- (-0.3,-2.5);
    \draw[-|] (-0.3,-2.5) -- (-0.3,-3);
    \draw [fill=white] (2.42,-.6) circle (1mm);
    \draw [fill=white] (2.58,-.6) circle (1mm);
    \draw [fill=black] (.42,-0) circle (1mm);
    \draw [fill=black] (.58,-0) circle (1mm);
    \node at (2.55,-1.2+.3) {\small $V~V$};
    \node at (0.5,-0.2+.5) {\small $W~W$};
    \node at (3.5,-.6) {\scriptsize 2};
    \node at (3.5,0) {\scriptsize 1};
}
}
\qquad\qquad  
\tikzsetnextfilename{OTOC-basic}
\raisebox{-16mm}{
\tikz[thick]{
    \draw[|-] (-0.3,0) -- (0,0);
    \draw[|-,right] (0,0) -- (1.5,0);
    \draw[right] (1.5,0) -- (3,0) arc (90:-90:.3) -- (1.5,-.6);
    \draw[right] (1.5,-.6) -- (0,-.6) arc (90:270:.3) -- (1.5,-1.2);
    \draw[right] (1.5,-1.2) -- (3,-1.2) arc (90:-90:.3) -- (1.5,-1.8);
    \draw[right] (1.5,-1.8) -- (0,-1.8) arc (90:180:.3) -- (-0.3,-2.5);
    \draw[-|] (-0.3,-2.5) -- (-0.3,-3);
    \draw [fill=white] (2.5,-0.6) circle (1mm);
    \draw [fill=white] (2.5,-1.8) circle (1mm);
    \draw [fill=black] (.5,0) circle (1mm);
    \draw [fill=black] (.5,-1.2) circle (1mm);
    \node at (2.8,-.6+.2) {\small $V$};
    \node at (2.8,-1.8+.2) {\small $V$};
    \node at (0.2,0+.2) {\small $W$};
    \node at (0.2,-1.2+.2) {\small $W$};
    \node at (3.5,-1.8) {\scriptsize 4};
    \node at (3.5,-1.2) {\scriptsize 3};
    \node at (3.5,-.6) {\scriptsize 2};
    \node at (3.5,0) {\scriptsize 1};
}
}
\caption{Schwinger-Keldysh closed-time contour representation of four-point TOCs (left) and OTOCs (right). White and black circles denote operator insertions. While generic four-point correlation functions require a four-fold closed time contour, the representation of TOCs can be collapsed onto a two-fold contour.}
\label{fig:TOCvsOTOC}
\end{figure}

The OTOC requires a four-fold closed-time contour for its definition; see Figure~\ref{fig:TOCvsOTOC}. Hence its hydrodynamical description constitutes a generalisation of the SK EFT to this four-fold contour and the quadrupling of degrees of freedom. Such a construction was carried out for charge diffusion at leading order in the derivative expansion recently in~\cite{Mishra:2025vlf}. Their construction is appropriate to determine the ultimate long-time behaviour of OTOCs, usually referred to as long-time tails~\cite{Alder:1970zza,Pomeau:1974hg,Kovtun:2003vj,Chen-Lin:2018kfl,Abbasi:2021fcz,Michailidis:2023mkd,Jain:2026obh}.

In this paper, however, we are also interested in the medium-time exponential Lyapunov behaviour that OTOCs in chaotic large-$N$ systems  are expected to exhibit~\cite{Larkin:1969abc,kitaev2014hidden,Shenker:2013pqa,Roberts:2014isa,Shenker:2014cwa,Maldacena:2015waa,KitaevTalks}. To wit, OTOCs of generic local operators $\hat W(0)$ and $\hat V(t)$ are expected to take the form:
\es{OTOCexpectation}{
\frac{\Big\langle \hat V(t)\, \hat W(0)\, \hat V(t)\, \hat W(0)\Big\rangle_\beta}
{\Big\langle \hat V(t) \hat V(t) \Big\rangle_\beta \,
\Big\langle \hat W(0)\hat W(0) \Big\rangle_\beta}
&= 1-\frac{c}{N}
 \E{\lambda\,t}
 +\dots
, \qquad t_\text{hyd}\ll t\ll \frac{1}{\lambda}\log N\,,
}
where $c=O(1)$ is a constant and $\lam$ is the quantum Lyapunov exponent.\footnote{We use hats to denote operators in the Hilbert space, while reserving the respective unhatted notation for the respective fields in the EFT. In general, the operator insertions can also depend on spatial points, which we have suppressed here for clarity.}
The Lyapunov coefficient $\lambda$ is bounded from above by $\lam\leq 2\pi/\beta$~\cite{Maldacena:2015waa}, where $\beta$ is the inverse temperature. The exponential growth holds in the time window with the hydrodynamic onset time $t_\text{hyd}$, which at strong coupling is expected \cite{qptbook,Zaanen2004,Hartnoll:2021ydi,Delacretaz:2023pxm,Qi:2026vht}  to be $O(\beta)$, and the scrambling time $t_\text{scr}=1/\lambda\,\log N$~\cite{Hayden:2007cs,Sekino:2008he,Lashkari:2011yi,Shenker:2013pqa}, which is $O(\beta\log N)$ at strong coupling. It was pointed out early on in~\cite{Blake:2017ris} that this exponential growth leads to the breakdown of the hydrodynamic derivative expansion.\footnote{In more detail, it would require $(\beta\partial_t \bullet)/\bullet\ll 1$ for the connected correlator in~\cref{OTOCexpectation} (i.e.~the exponential), which translates to $\lambda\ll 2\pi/\beta$ not obeyed in generic systems. However, the exponential spatial dependence expected from the Lieb--Robinson bound~\cite{Lieb:1972wy} (or its finite temperature version, the butterfly cone) can be less rapid, allowing to keep only the leading spatial derivatives in certain cases.} Hence, in order to possibly reproduce~\cref{OTOCexpectation}, we need to work with an all-order version of hydrodynamics~\cite{Bu:2014sia,Bu:2014ena,Grozdanov:2019kge}, which is still an EFT in the sense that it only has fields representing hydrodynamic degrees of freedom, but these now have a nonlocal effective action.

\paragraph*{\bf State of the art.} For  the computation of the OTOC for generic operators $V,W$,  to the order exhibited in~\cref{OTOCexpectation}, it suffices to know a simpler version of the theory, i.e. the all-order SK EFT that lives only on the two-fold closed-time contour. To wit, this can be understood as the SK EFT obtained by collapsing two consecutive folds of the four-fold SK EFT; see \cref{fig:TOCvsOTOC}.
Such a hydrodynamic EFT for energy transport was proposed in~\cite{Blake:2017ris,Blake:2021wqj}; see also~\cite{Haehl:2018izb} for the analogous
effective description of chaos in (1+1)d CFTs and~\cite{Grozdanov:2017ajz,Blake:2018leo} for
the connection between energy dynamics and chaos in holography. While the work of \cite{Blake:2017ris,Blake:2021wqj} proposed this two-fold energy-diffusion SK EFT based on a bottom-up prescription of symmetries, we recently derived this explicitly for the microscopic model of SYK lattices~\cite{Bucca:2026xmh}.  

The reason for this simplification to two-fold SK EFT can be understood through the simplest example of such an EFT: the Schwarzian quantum mechanics that governs the energy dynamics of the Sachdev-Ye-Kitaev (SYK) model~\cite{Sachdev_1993,KitaevTalks,Maldacena:2016hyu,Kitaev:2017awl} at low temperatures. Schematically, these low energy modes (``reparametrizations'' $F$) couple directly to bilocals  $WW$ and $VV$ and in the leading $1/N$ order the answer is given by the exchange of a reparametrisation mode $\delta F$~\cite{Maldacena:2016hyu,Maldacena:2016upp,Jensen:2016pah,Kitaev:2017awl}:
\begin{calign}\label{eq:bilocal-diagram}
\tikzsetnextfilename{bilocal-diagram}
\raisebox{-8mm}{
\tikz[thick,scale=0.7]{
    \draw [dashed] (20mm,0mm) arc (0:160:10mm);
    \draw [dashed] (20mm,0mm) arc (0:-160:10mm);
    \draw [ultra thick] (20mm,0) -- (40mm,0);
    \draw [dashed] (40mm,0mm) arc (180:20:10mm);
    \draw [dashed] (40mm,0mm) arc (-180:-20:10mm);
    \draw [fill=black] (0.6mm,3.4mm) circle (1.25mm);
    \draw [fill=black] (0.6mm,-3.4mm) circle (1.25mm);
    \draw [fill=white] (59.4mm,3.4mm) circle (1.25mm);
    \draw [fill=white] (59.4mm,-3.4mm) circle (1.25mm);
    \node at (-9mm,3.4mm) {\small $W(t_4)$};
    \node at (-9mm,-3.4mm) {\small $W(t_3)$};
    \node at (72mm,-3.4mm) {\small $V(t+t_2)$};
    \node at (72mm,3.4mm) {\small $V(t+t_1)$};
    \node at (30mm,4mm) {\small $\delta F$};
}
}
\end{calign}
Since all two-point functions can be captured by the two-fold SK EFT, it suffices for the purpose of this computation. 

While the above simplification applies to the OTOCs of generic ``few-body'' operators in the theory, it does not extend to collective hydrodynamic observables themselves, such as the energy-momentum tensor. This is because a single energy-momentum tensor insertion is represented as a local expression in the hydrodynamic fields such that the full four-fold EFT field content is activated because, even at the leading order, it is sensitive to interactions of low-energy degrees of freedom. Assessing scrambling through such hydrodynamic operators is particularly intriguing because these are universally defined in any physical system with the particular set of global symmetries, and thus likely carry more universal information compared to generic local operators.
In this paper we construct the all-order four-fold EFT for energy diffusion in an expansion in fields up to cubic vertices, and apply it to compute the OTOC of energy current. The resulting OTOC shows exponential growth as in~\cref{OTOCexpectation}.

\paragraph*{General SK EFT on the four-fold contour.}

In this paper, we construct the general, all-derivative order, SK EFT for energy diffusion on a four-fold contour up to cubic interactions starting from symmetry and unitarity principles. Our EFT will be formulated in terms of spatially-local diffeomorphisms $F(t,\bx)$ (reparametrizations) of the closed-time contour. The general formalism for the respective two-fold SK EFT has been proposed by \cite{Blake:2017ris}. In two-fold SK EFTs, it is known that the thermal KMS symmetry entirely fixes the effective action up to cubic order in fluctuations in terms of the constitutive relations; see e.g.~\cite{Jain:2020zhu}. The same applies to four-fold SK EFTs as well~\cite{Mishra:2025vlf}. In fact, we find that for a restricted class of physical systems, where the rate of entropy production is strictly odd under time-reversal, the four-fold SK effective action $S[F_1,F_2,F_3,F_4]$ takes a simple form in terms of the two-fold one $S[F_1,F_2]$ at this order; to wit\footnote{In principle, since we work to all orders in derivatives, the calculations could also be set up in Euclidean formalism and analytically continued at the end; this setup would include only one copy of the degrees of freedom, but would lead to complicated analytic continuations. (Note that the derivative expansion is intrinsically Lorentzian and truncating it as in conventional EFTs would prevent such flexible analytic continuations.) We chose not to pursue this direction, as the Lorentzian setup is conceptually cleaner.}
\es{eq:4fold-action}{
    S[F_1,F_2,F_3,F_4]
    &= S[F_1,F_2] 
    - S[F_1,F_3] 
    + S[F_1,F_4] 
    + S[F_2,F_3] 
    - S[F_2,F_4] 
    + S[F_3,F_4] \\
    &\qquad 
    + \text{(additional four-fold interactions)}\,.
}
Here the subscripts on the reparametrisation field $F$ denote the Lorentzian fold of the closed time contour on which they reside; see \cref{fig:TOCvsOTOC}. This simple form enables us to obtain universal results for how the OTOCs of general local operators in the EFT exhibit Lyapunov growth. In particular, we are able to probe scrambling through OTOCs of universal hydrodynamic operators, in this case the energy current. A crucial ingredient of our construction is a spatially-local $\SL(2,\bbR)$ symmetry, on which we will comment momentarily. This symmetry causes OTOCs to grow at maximal rate $\lambda=2\pi/\beta$, while shielding all time-ordered correlators (TOCs) from such growth, which would otherwise signal instabilities.

\paragraph*{Top-down constructions and checks.} One of the main results of our work is that we derive such an EFT, for the  four-fold contour from the microscopic description of the SYK lattice, including the determination of all transport coefficients. The groundwork for this result was laid in our previous paper~\cite{Bucca:2026xmh}, where the nonlinear SK EFT for energy diffusion to high orders in derivatives was derived and analysed for the  SYK lattice.
The OTOC results will be explicitly computed for the SYK lattice and match the results already in the literature for bi-local operators. We will also obtain novel results for local operators, such as the energy density. 

As checks of our methods, we apply our technology to several OTOC computations in a variety of models and obtain the expected results: 
\begin{itemize}
\item By decoupling the lattice sites in the SYK lattice, we obtain decoupled SYK dots. Hence by sending the diffusion constant $D\to 0$ in the EFT, on the one hand, the OTOC of generic operators on the same site is expected to agree with the SYK dot result. On the other hand, the energy density of the SYK lattice becomes the Hamiltonian of an individual dot, and being a conserved quantity, should have time-independent correlators. We reproduce both of these results.

\item The OTOC of generic operators was computed for the SYK lattice in~\cite{Gu:2016oyy,Guo:2019csw,Choi:2020tdj}, and our real-time SK EFT calculation matches those results.

\item In~\cite{Roberts:2014ifa} the OTOC of scalar primary operators was computed in (1+1)-dimensional conformal field theories (CFTs) in the identity Virasoro block approximation, amounting to capturing the dynamics of energy-momentum tensor exchange.\footnote{This approximation is appropriate for CFTs with a large gap in their primary spectrum. Such theories are prime candidates to have a simple holographic dual.} In all (1+1)d CFTs, the correlation functions of energy-momentum tensors is universal, and a simple calculation shows that the OTOC of the energy-momentum tensor does not show Lyapunov growth. We reproduce all these results by using the EFT of energy-momentum tensor dynamics constructed in~\cite{Alekseev:1988ce,Cotler:2018zff,Haehl:2018izb}; we refer to this EFT as ASCJ theory after the authors of these papers. In the case of the energy-momentum tensor OTOC, we go beyond just isolating (the absence of) the growing piece,  and carry out the full computation of four-point function using real time methods and Feynman diagrams, reproducing the analytic continuation of the well-known Euclidean result (usually derived from Ward identities).

\item It is known that in maximally chaotic theories, while a generic OTOC in \cref{OTOCexpectation} exhibits Lyapunov growth, the special ``commutator-squared'' OTOC does not grow~\cite{Blake:2021wqj}, given as
\begin{calign}\label{eq:comm-square}
    \bigg\langle
    \Big[\hat V(t+t_1),\hat W(t_3)\Big]
    \Big[\hat V(t+t_2),\hat W(t_4)\Big]
    \bigg\rangle.
\end{calign}
Up to additional TOC contributions, this correlator is equivalent to the symmetric combination of the OTOCs $\langle VWVW\rangle + \langle WVWV\rangle$ in \cref{OTOCexpectation}. We reproduce this result using our four-fold SK EFT. To extract the growing part, we find it convenient to define the following ``anti-symmetric'' OTOC which is most directly accessible in our SK EFT:
\begin{calign}\label{eq:nested-comm}
    \bigg\langle\Big[\hat W(t_4),
    \Big[\hat V(t+t_1),
    \Big[\hat V(t+t_2),\hat W(t_3)\Big]\Big] \Big]
    \bigg\rangle.
\end{calign}
Up to TOC contributions, it is equivalent to the antisymmetric combination of OTOCs in \cref{OTOCexpectation}, $\langle VWVW\rangle - \langle WVWV\rangle$. We will compute this explicitly for the energy current operator in our work and show that it grows. 

\end{itemize}

\paragraph*{\bf Local shift symmetry.} An important ingredient in our construction is a spatially-local  $\SL(2,\mathbb{R})$ symmetry acting as\footnote{Technically, this symmetry constitutes the projective $\PSL(2,\bbR)$ group. By a slight abuse of notation prevalent in the literature, we will keep referring to it as $\SL(2,\bbR)$.}
\es{SL2Sch}{
\tanh(\pi F)\, \to\,  {A \,\tanh(\pi F)+B\ov C \,\tanh(\pi F)+D}\,,\qquad  AD-BC=1\,.
}
The symmetry acts non-locally in time, but locally in space: $A,B,C,D$ are functions of $\bx$. We have dropped the contour label above to avoid clutter.
For our purposes it will also be very important that this transformation has to be the same on all part of the OTOC contour in order to constitute a symmetry.
We also note that it is best thought of as a gauge symmetry that physical observables have to be invariant under \cite{Maldacena:2016upp}.
The same local symmetry appears in the Alekseev-Shatashvili-Cotler-Jensen (ASCJ) theory~\cite{Alekseev:1988ce,Witten:1988,Cotler:2018zff} of the Virasoro identity block encoding the energy-momentum tensor dynamics of a (1+1)d CFT. 

It is the $\SL(2,\mathbb{R})$ symmetry that leads to exponential growth~\cite{Maldacena:2016upp,Kitaev:2017awl,Lin:2019qwu}. To see how, we select a certain infinitesimal transformation from~\cref{SL2Sch} and act on the saddle point $F(t) = t/\beta$: 
\es{ExpGrowth}{
 F(t)=\frac{t}{\beta}\, \to\, \frac{t}{\beta}
- \frac{b_+}{2\pi}\, \E{{2\pi/\beta} \,t}
+ O(b_+^2)\,,
}
where $b_+$ is an infinitesimal parameter corresponding to an  $\SL(2,\bbR)$ generator.
 (There is also an exponentially decreasing shift with an independent parameter $b_-$, which we set to zero to avoid clutter.) 
This exponential growth shows up in the propagator $\langle\delta F(t) \delta F(0)\rangle \supset \# \E{{2\pi/\beta} \,t}$. However $F$ is not gauge invariant and the exponential growth is a pure gauge mode. Nevertheless, it can leave its imprints on certain observables. The role of this symmetry is twofold: it guarantees the fast scrambling of OTOCs and protects
time-ordered correlators (TOCs) from growing.

In the SYK example, the Lyapunov exponent is maximal, i.e. $\lambda = 2\pi/\beta$.  It was argued in~\cite{Blake:2021wqj} that the $\SL(2,\mathbb{R})$ symmetry in general gives rise to maximal chaos.

\begin{figure}[t]
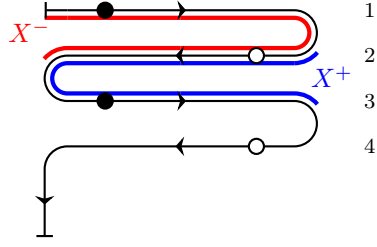

\centering
\tikzsetnextfilename{shockwaves}
\raisebox{-16mm}{
\tikz[thick]{
    \draw[red,ultra thick] (-0.3,0-0.1) -- (0,0-0.1);
    \draw[red,ultra thick] (0,0-0.1) -- (1.5,0-0.1);
    \draw[red,ultra thick] (1.5,0-0.1) -- (3,0-0.1) arc (90:-90:.2) -- (1.5,-.6+0.1);
    \draw[red,ultra thick] (1.5,-.6+0.1) -- (0,-.6+0.1) arc (90:140:.4);
    \draw[blue,ultra thick] (3,-0.6-0.1) arc (-90:-40:.4);
    \draw[blue,ultra thick] (3,-0.6-0.1) -- (1.5,-.6-0.1);
    \draw[blue,ultra thick] (1.5,-.6-0.1) -- (0,-.6-0.1) arc (90:270:.2) -- (1.5,-1.2+0.1);
    \draw[blue,ultra thick] (1.5,-1.2+0.1) -- (3,-1.2+0.1) arc (90:40:.4);
    \draw[|-] (-0.3,0) -- (0,0);
    \draw[|-,right] (0,0) -- (1.5,0);
    \draw[right] (1.5,0) -- (3,0) arc (90:-90:.3) -- (1.5,-.6);
    \draw[right] (1.5,-.6) -- (0,-.6) arc (90:270:.3) -- (1.5,-1.2);
    \draw[right] (1.5,-1.2) -- (3,-1.2) arc (90:-90:.3) -- (1.5,-1.8);
    \draw[right] (1.5,-1.8) -- (0,-1.8) arc (90:180:.3) -- (-0.3,-2.5);
    \draw[-|] (-0.3,-2.5) -- (-0.3,-3);
    \node[red] at (-.5,-.28) {\small $X^-$};
    \node[blue] at (3.5,-.88) {\small $X^+$};
    \draw [fill=white] (2.5,-0.6) circle (1mm);
    \draw [fill=white] (2.5,-1.8) circle (1mm);
    \draw [fill=black] (.5,0) circle (1mm);
    \draw [fill=black] (.5,-1.2) circle (1mm);
    \node at (4,-1.8) {\scriptsize 4};
    \node at (4,-1.2) {\scriptsize 3};
    \node at (4,-.6) {\scriptsize 2};
    \node at (4,0) {\scriptsize 1};
}
}
\caption{Growing (red) shockwave on folds ``1,2'' and decaying (blue) shockwave on folds ``2,3''. The two shockwaves are related to each other by a time-reversal transformation. The growing shockwave vanishes at $t\to-\infty$ while the decaying one vanishes at $t\to +\infty$, constituting a valid continuous field configuration on the four-fold contour. Neither of these are symmetry transformations of the theory as they do not act simultaneously on the entire contour. Such non-trivial configurations are not possible on the two-fold contour (with thermal boundary conditions).}
\label{fig:shocks}
\end{figure}

\paragraph*{Scramblon}
A final check of our results is that it matches the shockwave~\cite{tHooft:1987vrq,Shenker:2013pqa,Roberts:2014isa,Shenker:2014cwa} (often also referred to as scramblon) effective description valid in the double scaling limit $N,t\to\infty, \, {\E{\lam t}/ N}=\text{fixed}$ (and $t_i$ held fixed)~\cite{Kitaev:2017awl,Gu:2018jsv,Gu:2021xaj,Stanford:2021bhl,Choi:2023mab,Stanford:2023npy}. 
Briefly, the key idea behind shockwaves is that
there are two near-zero modes---one exponentially growing, the other exponentially decaying in time---that mediate scrambling~\cite{Kitaev:2017awl,Gu:2018jsv,Gu:2021xaj,Stanford:2021bhl}. 
The growing shockwave can be understood as a configuration where the saddle on folds ``1,2''  of the four-fold contour is shifted according to~\cref{ExpGrowth} with parameter $b_+=X^-$, while  folds ``3,4'' are left unchanged. Similarly, the decaying shockwave acts on  folds ``2,3'' with parameter $b_-=X^+$, while folds ``1,4'' are left unchanged.\footnote{Note that decaying shockwave is denoted $X^+$, while the growing one is denoted $X^-$. This notation is inspired from holography, where they correspond to null shifts on the future and past black hole horizons respectively~\cite{Maldacena:2016upp}. The shockwave formula \eqref{scramblonPrescription} has a negative sign in the exponent if the growing shock is defined on folds ``3,4'' instead.} Note that neither of these constitute an $\SL(2,\bbR)$ symmetry transformation because the shifts are not acting on the entire four-fold contour, see \cref{fig:shocks}.
Pairs of operators at early time source the growing mode and the decaying one acts on them, and vice versa for a pair of operators at late time. If both modes are present, they cost a non-zero, but small action due to their different profiles in time. In a formula:
 \es{scramblonPrescription}{
 &{\Big\langle \hat V(t+t_1)  \hat W(t_3)  \hat V(t+t_2)  \hat W(t_4)\Big\rangle} \\
 &\hspace{4em}\, \to\, \int {\df X^\pm\ov { Z}} \,
 {\cal G}_V\lb t_1,t_2\vert X^-\rb\,
 {\cal G}_W\lb t_3,t_4\vert X^+\rb \,
 \exp\le({iN\ov \alpha\, \E{\lam t}}\, X^+ X^-\ri)\,,
 }
where $X^\pm$ are the shockwave amplitudes, $\alpha$ is the coupling strength between the two shocks, and ${ Z}=\int {\df X^\pm} \,
 \exp\le({iN\ov \alpha\, \E{\lam t}}\, X^+ X^-\ri)$ is the measure factor setting  $ {\langle 1\rangle}=1$. In this formula, $ {\cal G}_V(t_1,t_2 \vert X^-)$ is the two-point function of $V$ in the presence of a growing shockwave, while $ {\cal G}_W(t_3,t_4 \vert X^+)$ is the two-point function of $W$ with a decaying shockwave.
 The shockwave prescription \eqref{scramblonPrescription} is natural: the two-point function in a background field is equal to the resummation of amputated $(n+2)$-point functions with the background field inserted on the $n$-external legs. We note that both $X^\pm$ are allowed to be spatially dependent and $\alpha$ can include spatial derivatives. In the main text, we will determine $\alpha$ explicitly for the SYK lattice.

At $t\ll t_\text{scr}$, the path integral in \cref{scramblonPrescription} can be evaluated using  saddle point, leading to the expectation ~\cref{OTOCexpectation}. In its full form in~\cref{scramblonPrescription}, it also extends to any $t\sim t_\text{scr}$, although for spatially local systems and away from maximal chaos loop corrections are anomalously large~\cite{Stanford:2023npy}.

\paragraph*{A paradox and its resolution.} 
There is a curious observation regarding the OTOC of the energy density in the shockwave picture. In SYK-like models, the energy density operator is local: the energy density $\cE^t_i$ on the $i$th SK fold is simply the Schwarzian of the reparametrisation field $F(t)$ on that fold. In particular, it is invariant under independent $\SL(2,\bbR)$ transformations on each fold. One might therefore expect the two-point function of the energy density in such models to be insensitive to the presence of the shockwave and that the growing part of its OTOC should vanish. This is, however, misleading because the effective action of the theory $S[\{F_i\}]$ is generically only invariant under $\SL(2,\bbR)$ transformations acting the same way on each fold. This means that, even when the energy density operator only has support on individual folds, its correlators $\langle \cE^t_i \cE^t_j \rangle = \int \{\cD F_i\}\,\cE^t_i \cE^t_j\, \E{iS[\{F_i\}]}$ can indeed be sensitive to shockwaves. In fact, we will show  that this prescription correctly reproduces the energy-density OTOC obtained from our four-fold SK EFT. The exception to this conclusion are theories where the effective action is also invariant under $\SL(2,\bbR)$ transformations on each fold; such is the case for the ordinary SYK dot model without spatial interactions or the ASCJ (1+1)d CFT model. Indeed, in both these cases we find that the energy-density OTOC does not admit a growing part.

\paragraph*{Outline.} 

The organisation of the paper is as follows. We begin with a general discussion of the two-fold SK EFT for energy diffusion in \cref{sec:two-fold}. Herein, we discuss the general form of the SK effective action up to cubic order in fluctuations around the thermal state, but all orders in the derivative expansion. We explicitly identify the free parameters in the effective action in terms of the classical constitutive relations for the energy current. Then in \cref{sec:four-fold-theory}, we extend this construction to the four-fold SK EFT. We find that, to cubic order in fluctuations, the four-fold effective action is completely fixed by its two-fold counterpart. We also introduce a convenient basis on the four-fold closed-time contour, which streamlines the subsequent computation of OTOCs.  In \cref{sec:energy-otoc}, we use our four-fold SK EFT to reproduce the OTOC of generic local operators in SYK-like models as well as to obtain new results for the OTOC of energy current. We verify these results with the shockwave description. We finish with some discussion in \cref{sec:discussion}. The paper has several appendices where we provide details of EFT calculations.

The paper is written from a general bottom-up EFT point of view. To make the discussion more concrete, we specialise to the two microscopic models where we have total analytic control, the SYK lattice and the ASCJ theory in shaded boxes throughout the text.

\paragraph*{Notation and conventions.} We will use the boldface characters ($\bx,\by,\ldots$) to denote spatial coordinates (even in one spatial dimension) and reserve the regular characters ($x,y,\ldots$) to denote spacetime coordinates. We will mostly work in $d$ spatial dimensions, while some explicit results for the SYK chain and the ASCJ theory are presented for $d=1$.

\section{Two-fold SK EFT for energy diffusion}
\label{sec:two-fold}

We start with a brief review of the conventional two-fold SK EFT for energy diffusion. As discussed in the introduction, in order to access the scramblon modes together with the hydrodynamic diffusion mode, we will need to work non-perturbatively in derivatives. 
An important ingredient in this construction is a pair of non-perturbative shift symmetries~\cite{Blake:2017ris}, which allow OTOCs to exhibit the characteristic exponential growth at intermediate times, while preventing TOCs from growth. The discussion presented here is based on the bottom up SK EFT construction of~\cite{Blake:2017ris}, derived explicitly in the context of SYK lattices in our recent work~\cite{Bucca:2026xmh}.

\subsection{General structure of the SK EFT}
\label{sec:general-2fold}

\paragraph*{Dynamical fields.}

SK EFTs on a two-fold contour feature two copies of dynamical fields, one for each Lorentzian fold of the closed-time contour. For the SK EFT of energy diffusion, these are a pair of time-reparametrisation fields $F_1(x)$ and $F_2(x)$. In equilibrium, these fields align with the time coordinate $F_{1,2}(x) = t/\beta $, where $\beta $ denotes the constant inverse-temperature of the thermal state.
It is convenient to work in an average-difference basis of the dynamical fields, $F=(F_1+F_2)/2$ and $Q=F_1-F_2$. Here $F$ is identified as the ``physical'' field, while $Q$ as the associated ``noise'' field.

The effective action $S[F_1,F_2]$ of the SK EFT is required to satisfy the following three unitarity conditions~\cite{Crossley:2015evo}:\footnote{Heuristically, these unitarity conditions arise from fundamental properties of density matrices: $\tr\rho = 1$, $\rho^*=\rho$, and $1\geq {\rm Eigenvalues}(\rho)\geq 0$. See~\cite{Liu:2018kfw} for more details.}
\begin{calign}\label{eq:unitarity}
    S[F,F] = 0, \qquad 
    S[F_1,F_2]^* = -S[F_2,F_1], \qquad 
    \Im S[F_1,F_2]\geq 0.
\end{calign}
Due to the first condition above, the effective action vanishes when the difference field $Q$ is set to zero. Consequently, extremising it with respect to $F$ sets $Q$ to zero on-shell, i.e.\footnote{The $Q=0$ solution corresponds to the classical hydrodynamic configurations. More generally, the SK EFT may admit non-trivial saddle point solutions with $Q\neq 0$ as well.}
\begin{calign}
    \frac{\delta S}{\delta F} = 0
    \quad\implies\quad
    Q\Big|_{\rm on-shell} = 0.
\end{calign}
On the other hand, extremising the effective action with respect to $Q$ yields the classical equation of motion for $F$. For energy diffusion SK EFT, this reproduces the energy conservation equation
\begin{calign}\label{eq:classical-energy-conservation}
    \frac{\delta S}{\delta Q}
    = 0
    \quad\implies\quad
    \dow_\mu\cE^\mu\Big|_{\rm on-shell} = 0.
\end{calign}
Here $\cE^\mu$ is the classical energy current of the theory. This conservation form of the equation of motion is ensured by the symmetries of the theory outlined below.

\paragraph*{Symmetries.}

In addition to the unitarity conditions, the SK EFT for energy diffusion must be invariant under the following symmetries:
\begin{subequations}\label{eq:symmetries-at-glance}%
\begin{alignat}{2}
    \text{Doubled time-translations:} &
    &\quad 
    F_{1,2}(t,\bx) &\mapsto F_{1,2}(t+\xi_{1,2},\bx), 
    \label{eq:time-translations} \\
    \text{Space-translations:} &
    &\quad 
    F_{1,2}(t,\bx) &\mapsto F_{1,2}(t,\bx+{\bf c}), 
    \label{eq:space-translations} \\
    \text{Spatially-local diagonal shifts:} &
    &\quad 
    F_{1,2}(t,\bx) &\mapsto F_{1,2}(t,\bx) + a(\bx), 
    \label{eq:diagonal-shifts} \\
    \text{$\bbZ_2$ KMS shifts \cite{Crossley:2015evo,Bucca:2026xmh}:} &
    &\quad 
    F_{1,2}(t,\bx) &\mapsto -F_{1,2}(-t,\eta_\rmP\bx) \pm i/2,
    \label{eq:KMS}
\end{alignat}
\end{subequations}
where $\eta_\rmP = +1$ if the underlying microscopic theory is invariant under time-reversal, while $\eta_\rmP = -1$ if it is instead invariant under spacetime-reversal. For time-reversal-invariant theories like the SYK model, we can choose either sign $\eta_\rmP = \pm 1$ for parity-preserving theories, though we must choose $\eta_\rmP = +1$ in the presence of parity-violating interactions.

The doubled time-translation symmetries in \cref{eq:time-translations} lead to two conserved energy current operators in the SK EFT, $\cE_1^\mu$ and $\cE_2^\mu$. Out of these, the average combination $\cE_r^\mu = (\cE_1^\mu+\cE_2^\mu)/2$ is identified with the physical energy current, while the difference $\cE_a^\mu = \cE_1^\mu-\cE_2^\mu$ with the associated noise. Using Noether procedure, one can check that $\cE_r^\mu$ is associated with difference time-translations $\xi_1-\xi_2$, while $\cE_a^\mu$ is associated with average time-translations $(\xi_1+\xi_2)/2$. In fact, the classical energy current in \cref{eq:classical-energy-conservation} is obtained as $\cE^\mu = \cE_r^\mu|_{Q\to 0}$ after setting the noise field $Q$ to zero.

Unlike the doubled time-translations, the theory is only invariant under average space translations in \cref{eq:space-translations}. This is a general feature of homogeneous physical systems without momentum conservation. Correspondingly, the theory does not feature a physical momentum current, but does admit a noise momentum current. We will not explore this in greater detail in this work.

Looking at the dynamical fields more closely, we find that the fluctuations $\delta F_{1,2} = F_{1,2} - t/\beta$  transform as Goldstones for time translations: $\delta F_{1,2} \mapsto \delta F_{1,2} + \xi_{1,2}\dow_t\delta F_{1,2} + \xi_{1,2}/\beta  + O(\xi_{1,2}^2)$.
In average-difference basis, these imply
\begin{calign}
    \delta F
    &\mapsto \delta F 
    + \xi\,\dow_t\delta F
    + \frac{\zeta}{4} \dow_t\delta Q
    + \frac{\xi}{\beta } + O(\xi_{1,2}^2), \nn\\
    \delta Q
    &\mapsto \delta Q 
    + \xi\,\dow_t\delta Q
    + \zeta\,\dow_t\delta F
    + \frac{\zeta}{\beta } + O(\xi_{1,2}^2),
\end{calign}
where we have defined $\xi_{1,2} = \xi \pm \zeta/2$. Therefore, $\delta F$ and $\delta Q$ can effectively be thought of as Goldstone fields for average and difference time-translations respectively~\cite{Akyuz:2023lsm}. However, we would like to describe physical systems that do not break the physical time translations. To do this, we make spatially-local shifts a redundancy, i.e.~we gauge the shift symmetry as in \cref{eq:diagonal-shifts}. This,  in effect, restores the average combination of the symmetry to be unbroken.
 \Cref{eq:diagonal-shifts} enforces that all dependence on $\delta F$ in the theory must arise via $\dow_t\delta F$, whose transformation no longer contains a non-linear shift. On the other hand, note that the difference time-translation symmetry remains spontaneously broken, with the noise field $\delta Q$ acting as a Goldstone. This phenomenon is known as strong-to-weak symmetry breaking~\cite{Buca:2012zz,Lessa:2024gcw,Sala:2024ply,Gu:2024wgc,Liu:2024mme,Huang:2024rml,Feng:2025ngy}, where a doubled symmetry on the two-fold SK contour spontaneously breaks to its diagonal subgroup.

Lastly, the $\bbZ_2$ KMS shift symmetry in \cref{eq:KMS} ensures that the correlation functions obtained from the SK EFT are consistent with fluctuation-dissipation relations of thermal field theory~\cite{Sieberer:2015hba,Crossley:2015evo,Glorioso:2017fpd}. Together with other symmetries and the unitarity constraints, it also ensures the existence of a local second law of thermodynamics~\cite{Glorioso:2016gsa}. See Appendix E of~\cite{Bucca:2026xmh} for more details.

\paragraph*{Spatially-local Lyapunov shift symmetries.}

The unitarity conditions \eqref{eq:unitarity} and symmetries \eqref{eq:symmetries-at-glance} outlined above apply universally to any homogeneous system in local thermal equilibrium with energy conservation. In this work, we will further specialise to a class of maximally-chaotic systems: known examples include theories with a holographic gravity dual and the SYK lattice studied in~\cite{Bucca:2026xmh}. These feature a pair of spatially-local non-perturbative shift symmetries, whose infinitesimal form is
\begin{calign}\label{eq:Lyapunov-shifts}
    F_{1,2}(t,\bx) \to F_{1,2}(t,\bx) 
    - \frac{1}{2\pi} b_+(\bx)\E{2\pi F_{1,2}(t,\bx)}
    + \frac{1}{2\pi} b_-(\bx)\E{-2\pi F_{1,2}(t,\bx)}
    + \cO(b_\pm^2).
\end{calign}
These symmetries are non-perturbative in the sense that they map  homogeneous thermal states with $1/\dow_t F = \beta $ to highly inhomogeneous states with $1/\dow_t F = \beta + \beta  b_\pm(\bx) \E{\pm\omega_\beta t}$, where $\omega_\beta = 2\pi/\beta $. This is basically the saddle shift we saw in the introduction in \cref{ExpGrowth}. As a consequence, they cannot be realised within a derivative expansion.
These symmetries result in new exponentially-growing and exponentially-decaying modes in the spectrum, which we will refer to as ``scramblon modes''. It is precisely these modes that are responsible for the characteristic exponential growth of OTOCs.

The Lyapunov shift symmetries \eqref{eq:Lyapunov-shifts} together with the diagonal shift symmetry in \cref{eq:diagonal-shifts} form a spatially local $\SL(2,\bbR)$ symmetry in \cref{SL2Sch}, with
\begin{calign}
    A,D &= \cosh(\pi a) + \half \E{\pi a} b_+ b_- 
    \mp \half \lb b_++b_-\rb, \nn\\
    B,C &= \sinh(\pi a) + \half \E{\pi a} b_+ b_- 
    \mp \half\lb b_+-b_-\rb.
\end{calign}

Such shift symmetries were suggested by~\cite{Blake:2017ris} to appear in a generic theory of energy diffusion for maximally chaotic quantum systems.\footnote{Interestingly, unlike the implementation of~\cite{Blake:2017ris}, we note that the two Lyapunov-shift symmetries are not actually independent. In fact, they are related to each other by KMS-conjugation:
\begin{calign}
    \cR^- = \rmK\,\cR^+\rmK, \qquad 
    \cR^+ = \rmK\,\cR^-\rmK,
\end{calign}
where $\rmK$ is the generator of KMS transformations and $\cR^\pm$ are generators of the Lyapunov shift symmetries.}
Together, these are responsible for early-time Lyapunov growth of out-of-time ordered correlation functions (OTOCs), while preventing any growth in ``causal'' correlation functions defined on the 2-fold closed-time contour.

\paragraph*{Time-reversal-odd dissipation.}

For concreteness, we will impose a further condition on the SK EFT inspired from SYK lattices. We require that the imaginary part of the effective action is invariant under time-reversal (or spacetime-reversal), i.e.
\begin{calign}\label{eq:cond-odd-dissipation}
    \Im S[-\tilde F,-\tilde Q] = \Im S[F,Q],
\end{calign}
where tilde denotes the field evaluated at the $\Theta$-conjugate point $(-t,\eta_\rmP\bx)$.
Due to time-translation and KMS symmetries, this is equivalent to the requirement that the time-reversal-even real part of the effective action is entirely factorised
\begin{calign}
    \half \Big( \Re S[F,Q]
    + \Re S[-\tilde F,-\tilde Q] \Big)
    = S_0[F_1]
    - S_0[F_2],
\end{calign}
while the non-factorised dissipative part must be time-reversal odd. We have provided a proof of this equivalence in \cref{app:factorisation}. We note that this is not a generic feature of SK EFTs; we only focus on theories of this type for simplicity.\footnote{All standard diffusion-type theories with equations of motion $\dow_t \epsilon = \cF(\epsilon,\dow_i\epsilon,\dow_i\dow_j\epsilon,\ldots)$ fall under this class. Counter-examples include theories where the right-hand side involves terms with an odd number of time-derivatives.}

\begin{myshaded}
\paragraph*{Microscopic examples.}

We give two examples so that the abstract general formulas in the following discussion become more tangible. The large-$N$ and large-$p$ limit of the SYK chain at low temperatures is dominated by a soft mode action that we can regard as a non-linear all-order hydrodynamic EFT~\cite{Bucca:2024zrx, Bucca:2026xmh}. Its SK EFT is nonlocal in time:
\es{eq:SYK-2fold-start}{
S[F_1,F_2]
&= S_\text{loc}[F_1] - S_\text{loc}[F_2] 
- S_\text{biloc}[F_1,F_2]
- S_\text{biloc}[F_2,F_1] \\
&\qquad 
+ S_\text{biloc}[F_1,F_1]
+ S_\text{biloc}[F_2,F_2] .
}
where we have
\es{eq:final_action}{
    S_{\rm loc}[F]
    &=
    -N 
     \frac{\alpha_S}{\cJ} \int \df t\,\df^d\bx\, 
    \Sch\!\big[\tanh(\pi F(t,\bx)),t\big], \\
    S_{\rm biloc}[F_j,F_k]&=-
    ig\pi^2 N \int \df t_1\df t_2\,\df^d\bx\,
    \sqrt{  \frac{\dow_{t_1} F_j \left(t_1,\bx\right) 
    \dow_{t_2} F_k \left(t_2,\bx\right) }{
    \sinh^2\Big(\pi F_j \left(t_1,\bx\right)
    -\pi F_k \left(t_2,\bx\right)-(j-k)i\epsilon\Big) } } \\
    &\hspace{3em}\times
    \sqrt{  \frac{\dow_{t_1} F_j \left(t_1,\bx+a\right) 
    \dow_{t_2} F_k \left(t_2,\bx+a\right) }{
    \sinh^2\Big(\pi F_j \left(t_1,\bx+a\right)
    -\pi F_k \left(t_2,\bx+a\right)-(j-k)i\epsilon\Big) } }.
}
Here the Schwarzian derivative is defined as $\Sch[h,t] = \dow_t\lb \dow_t^2h/\dow_t h \rb - \half (\dow^2_th/\dow_th)^2$.
In the SYK chain the terms in the second line in \cref{eq:SYK-2fold-start} are, in fact, are integrals of total derivatives, which do not affect the Feynman diagram computations.
It can be shown that this action obeys all the symmetries in~\cref{eq:symmetries-at-glance}, has a spatially local $\SL(2,\bbR)$ gauge symmetry, and has time-reversal-odd dissipation. However, this action is cumbersome to work with. To ameliorate this situation,  in~\cite{Bucca:2026xmh} we worked out its derivative expansion but did not expand in the amplitudes of deviations away from equilibrium. There we also 
performed the expansion in amplitudes to quadratic order (while not expanding in derivatives), which we extend to cubic order in this work. For later use, we note that the classical energy density operator in this theory is just the Schwarzian 
\begin{calign}
    \cE^t_i 
    &= - \frac{\beta c_V}{4\pi^2}
    \Sch\!\big[\tanh(\pi F_i),t\big], \qquad 
    c_V = \frac{4\pi^2\alpha_S N}{\beta\cJ}.
\end{calign}
while the energy flux is given as
\begin{calign}
    \cE^\bx_i 
    &= - \frac{4D}{a^2\dow_\bx} \sinh^2\lb \frac{a\dow_\bx}{2} \rb \cE^t_i 
    + \text{noise terms involving other folds}, \qquad 
    D = \frac{\pi g\cJ}{6\alpha_S}.
\end{calign}
Unlike the energy density, we find that the energy flux is not localised on individual Lorentzian folds and couples fields across different folds as well~\cite{Bucca:2026xmh}.

The ASCJ theory describes the dynamics of the stress tensor in (1+1)d CFT~\cite{Cotler:2018zff,Haehl:2018izb} and hence can be regarded as the EFT of energy-momentum dynamics. Focusing on the holomorphic sector, it takes the form
\begin{calign}\label{ASCJaction-Lorentzian}
    S[F_1,F_2] &= S_{\rm loc}[F_1] - S_{\rm loc}[F_2]
\end{calign}
where
\begin{calign}\label{eq:ASCJ-local-action}
    S_{\rm loc}[F]
    = {C\ov 24\pi}
    \int \df t\,\df \bx\,
    \lB
    2\pi^2 \dow_tF\, \lb \dow_t + \dow_\bx \rb F
    + {\dow_t^2F\, \lb \dow_t + \dow_\bx \rb
    \dow_{t}F\ov 2(\dow_t F)^2}
    \rB\,.
\end{calign}
Note that, unlike SYK, the ASCJ theory does not have any bilocal piece in the effective action. This theory also features all the symmetries outlined in \cref{eq:symmetries-at-glance}, as well as the spatially-local $\SL(2,\bbR)$ symmetry. Here $c=C+13$ is the central charge of the CFT~\cite{Cotler:2018zff}. However, since this theory violates parity and time-reversal individually, while only preserving full spacetime parity, so we must take $\eta_\rmP = -1$.
The holomorphic component of the energy-momentum tensor is
\begin{calign}
    \label{ASCJstress}
    \cE_i =-\frac{\beta c_V}{4\pi^2 }
    \Sch\!\big[\tanh(\pi F_i),t\big]
    \quad\implies\quad 
    \lb \dow_t + \dow_\bx\rb\cE_i
    = 0, \qquad 
    c_V = \frac{\pi C}{6\beta}.
\end{calign}
Note that, in this case, the conservation equation entirely decouples on each Lorentzian fold. A quick note that the holomorphic energy-momentum tensor is often defined with an additional factor of $2\pi$, but we skip it here for homogeneity with the SYK model.

\end{myshaded}

\subsection{Quadratic terms in the effective action}

We wish to construct the general SK effective action for energy diffusion consistent with the symmetry structure outlined above. For our purposes of computing correlation functions using Feynman diagrams (with loops suppressed by entropy),\footnote{In large-$N$ models, such as the SYK lattice, the entropy scales with $N$ and hence loops are $1/N$ suppressed. In a generic finite $N$ model, small wavevector $\bk$ suppresses loops, which we can convert to entropy through the relation $S\sim s/\bk^{d}$ with $s$ the entropy density.} it is sufficient to work perturbatively in fluctuations around the thermal equilibrium solution $F_\eqb = t/\beta $ and $Q_\eqb = 0$.
At quadratic order in fluctuations, the SK effective action can schematically be expressed as
\begin{calign}\label{eq:action-free}
    S_{\rm free}
    &= \int \df^{d+1}x
    \lB \delta Q\,K(\dow)\,\delta F
    + \frac{i}{4}\delta Q\,\cot\bfrac{\beta \dow_t}{2}
    \Big( K(-\dow) - K(\dow) \Big)
    \,\delta Q
    \rB,
\end{calign}
where $K(\dow)$ is some linear differential operator dependent on the underlying physical system under consideration. Note that there is no $\delta F^2$ term here because the effective action is required to vanish when $Q$ is set to zero. The form of the $\delta Q^2$ term is entirely fixed in terms of the $\delta Q\delta F$ term by the KMS symmetry.
The form of the operator $K(\dow)$ is not entirely fixed, but is constrained by the symmetries of the SK EFT. We find
\begin{calign}
    K(\dow)
    &= - \dow_t \lb 1 - \frac{\dow_t^2}{\omega_\beta^2} \rb
    \Delta(\dow),
\end{calign}
for some linear operator $\Delta(\dow)$ that satisfies $\tilde\Delta(\dow)=\Delta(-\dow)$ and $\Delta(\dow=0)=0$. Here tilde denotes operators evaluated at $(\dow,\dow_\bx)\to(-\dow_t,\eta_\rmP\dow_{\bf x})$.
In particular, $K(\dow)$ admits the scramblon modes $\dow_t \sim \pm \omega_\beta$ enforced by the Lyapunov shift symmetries.
The details of the derivation have been relegated to \cref{app:Fourier-constraints}.

\begin{myshaded}
\paragraph*{Microscopic examples.}

For the SYK chain we  find:
\es{DeltaSYKLattice}{
\Delta(\dow)
    = \beta  c_V \lb \dow_t - \frac{4 D}{a^2} \sinh^2\bfrac{a\dow_\bx}{2}\rb.
}
This result was already derived in~\cite{Bucca:2026xmh}, and can be reproduced by linearising~\cref{eq:final_action} around the saddle point and performing the relative time integral in $t_1-t_2$.

The ASCJ theory describing the energy-momentum tensor dynamics in (1+1)d CFTs matches the general form discussed in the main text, with
\begin{calign}\label{eq:ASCJ-maps}
    \Delta(\dow) = \beta  c_V 
    \lb \dow_t + \dow_\bx \rb.
\end{calign}
Note that, in this case, $\tilde\Delta(\dow)=\Delta(-\dow)$ only if we take $\eta_\rmP=-1$.
\end{myshaded}

\paragraph*{Free propagators.}

We can invert the quadratic part of the SK effective action to obtain the free propagators. Ordinarily, the propagators are best expressed in Fourier-domain, parametrised by the frequency $\omega$ and wavevector $\bk$, which is often technically simpler for Feynman diagram computations. However, the differential operator $K(\dow)$ in our theory admits an unstable exponentially-growing Lyapunov mode, which renders the Fourier-transform of its Green's function in time direction to be ill-defined. See \cref{app:full-props} for more details. Nonetheless, Fourier-transform in spatial directions is perfectly well defined, which we will employ to our advantage in later discussions.

Using the free SK effective action in \cref{eq:action-free}, we find the propagators in position-domain
\begin{calign}\label{eq:twofold-props}
    \renewcommand{\arraystretch}{1.5}
    \Big\langle
    \begin{pmatrix}
        \delta F(x) \\ \delta Q(x)
    \end{pmatrix}
    \begin{pmatrix}
        \delta F(0) \\ \delta Q(0)
    \end{pmatrix}\Bigg.^{\!\!\rmT}
    \Big\rangle
    =
    \begin{pmatrix}
        K^{-1}_\rmS(x)
        \equiv \frac{1}{2} \cot\bfrac{\beta \dow_t}{2}
        \Big( K^{-1}(x) - \tilde K^{-1}(x) \Big)
        & iK^{-1}(x) \\
        i\tilde K^{-1}(x) & 0
    \end{pmatrix},
\end{calign}
where $K^{-1}(x)\propto \Theta(t)$ is the retarded Green's function associated with the differential operator $K(\dow)$, supported only for $t>0$. Whereas, $K^{-1}_\rmS(x)$ is the symmetric Green's function, supported for all $t$. In fact, these are precisely the retarded and symmetric Green's functions of reparametrisation fields
\begin{subequations}\label{eq:deltaFcorrelators}
\begin{alignat}{3}
    G^{\rm R}_{\delta F\delta F}(x)
    &\equiv i\,\Theta(t)\Big\langle \big[\delta \hat F(x),\delta \hat F(0)\big] \Big\rangle
    &&= i\Big\langle \delta F(x)\delta Q(0) \Big\rangle
    &&= -K^{-1}(x), \\
    G^{\rm S}_{\delta F\delta F}(x)
    &\equiv \half\Big\langle \big\{\delta \hat F(x),\delta \hat F(0)\big\} \Big\rangle
    &&= \Big\langle \delta F(x)\delta F(0) \Big\rangle
    &&= K^{-1}_\rmS(x),
\end{alignat}
Here hat denote operators in the Hilbert space. 
\end{subequations}

The Green's functions $K^{-1}(x)$ and $K^{-1}_\rmS(x)$ are not themselves physical because the reparametrisation fields transform under the spatially-local $\SL(2,\bbR)$ symmetry. However, we can work with $\SL(2,\bbR)$-invariant fields: the dynamical temperature $T = (1-\dow_t^2/\omega_\beta^2)\dow_t F +\ldots$ and the associated noise field $\chi_a = -Q/\dow_t F + \ldots$, where ellipses denote non-linear corrections. We will discuss these variables in more detail in \cref{sec:TOCdontgrow}.
In terms of these, the propagators are given as
\begin{calign}\label{eq:SL2R-invariant-props}
    \renewcommand{\arraystretch}{1.5}
    &\Big\langle
    \begin{pmatrix}
        \delta T(x) \\ \delta \chi_a(x)
    \end{pmatrix}
    \begin{pmatrix}
        \delta T(0) \\ \delta \chi_a(0)
    \end{pmatrix}\bigg.^{\!\!\rmT}
    \Big\rangle \nn\\
    &\qquad 
    =
    \begin{pmatrix}
        \Delta^{-1}_\rmS(x)
        \equiv \frac{1}{2} \cot\bfrac{\beta \dow_t}{2}
        \lb 1 - \dow_t^2/\omega_\beta^2 \rb \dow_t
        \Big( \Delta^{-1}(x) + \tilde\Delta^{-1}(x) \Big)
        & i\beta \Delta^{-1}(x) \\
        i\beta \tilde\Delta^{-1}(x) & 0
    \end{pmatrix},
\end{calign}
where $\Delta^{-1}(x) = - \dow_t (1 - \dow_t^2/\omega_\beta^2) K^{-1}(x) \propto \Theta(t)$ is the retarded Green's function of the differential operator $\Delta(\dow)$.

We require that the propagators of $\SL(2,\bbR)$-invariant fields on the two-fold contour do not admit any unstable exponentially growing modes. In particular, this means that the differential operator $\Delta(\dow)$ only admits stable quasinormal modes, i.e. $\Delta^{-1}(x)$ falls off sufficiently fast as $t\to\infty$. We will see later in \cref{sec:TOCdontgrow} that this property of $\Delta(\dow)$ also ensures that TOCs of all $\SL(2,\bbR)$-invariant local operator in the theory do not exhibit exponential growth.
By contrast, we find that
\begin{cgather}\label{eq:growing-kernel}
    K^{-1}(x) = \Theta(t)\int_{0}^t \df t' 
    \Big(\cosh\!\big(\omega_\beta(t-t') \big)-1\Big)\Delta^{-1}(t',\bx).
\end{cgather}
The factor of $\cosh(\omega_\beta(t-t'))$ in the integrand arises from the scramblon modes and implies that the retarded Green's function $K^{-1}(x)$ of the reparametrisation field grows exponentially at late times. 

\paragraph*{Diffusion theories.}

For simplicity, we will look at theories where $\Delta(\dow)$ only admits the hydrodynamic diffusion mode. 
Our discussion can be easily generalised to additional non-hydrodynamic modes in the spectrum, which, at most, will change the short-time behaviour of the following results.\footnote{More generally, in the presence of non-hydrodynamic modes, the wavevector-dependent specific heat $c_\bk$ gets promoted to a differential operator $c_\bk(\dow_t)$.} To wit, invoking Fourier-domain in the spatial directions, we take 
\begin{calign}\label{eq:general-Delta}
    \Delta(\dow_t,i\bk) = \beta c_\bk(\dow_t + \cD_\bk),  
\end{calign}
where $c_\bk$ is the wavevector-dependent specific heat, while $\dow_t = -\cD_\bk$ is the energy diffusion mode. In general, we have $\cD_\bk=D \bk^2 + O(\bk^4)$ and $c_\bk = c_V + O(\bk^2)$, where $D$ is the diffusion constant and $c_V$ is the homogeneous specific heat. Specifically for the SYK chain model, we have $\cD_\bk = 4D/a^2\,\sin^2(a\bk/2)$ and $c_\bk = c_V$; see \cref{DeltaSYKLattice}. This form also applies to the right-moving  and left-moving sectors of the ASCJ (1+1)d CFT model, with $\cD_\bk = \pm i\bk$; see~\cref{eq:ASCJ-maps}.

With \cref{eq:general-Delta}, we can explicitly compute 
\begin{calign}\label{eq:DeltaandKfunctions}
    \Delta^{-1}(t,i\bk) 
    &= \frac{\Theta(t)}{\beta  c_\bk}\, \E{-\cD_\bk t}, \nn\\
    K^{-1}(t,i\bk) 
    &= \frac{\Theta(t)}{\beta  c_\bk}
    \lb
    \frac{\E{-\cD_\bk t}}{\cD_\bk (1-\cD_\bk^2/\omega_\beta^2)}
    - \frac{1}{\cD_\bk}
    + \frac{\E{-\omega_\beta t}}{2(\cD_\bk-\omega_\beta)} 
    + \frac{\E{\omega_\beta t}}
    {2(\cD_\bk+\omega_\beta)} 
    \rb,
\end{calign}
where we have ignored certain contact terms. We can see that $\Delta^{-1}(t,i\bk)$ is only sensitive to the hydrodynamic diffusive mode. However, $K^{-1}(t,i\bk)$ is also sensitive to the growing and decaying scramblon modes with maximal rate $\pm\omega_\beta$, corresponding to the usual Euclidean reparametrisation propagator $1/(\omega_E^2(\omega_E^2-\omega_\beta^2))$ of a single SYK dot (at $\bk=0$). The wavevector-dependent  coefficient of the growing mode $\sim 1/(c_\bk(\omega_\beta+\cD_\bk))$ precisely characterises the scramblon action appearing in the shockwave formula in \cref{scramblonPrescription}. We will explore this in more detail in \cref{sec:shockwaves}.

Note that, while $\Delta^{-1}(t,i\bk)$ decays to zero at large $t$, the full retarded Green's function $K^{-1}(t,i\bk)$ grows exponentially as ${\sim}\,\E{\omega_\beta t}$. It is precisely this growing mode that is responsible for the growth of OTOCs. Using analytic continuation to complex time, we can find the symmetric functions $K^{-1}_\rmS(t,i\bk)$ and $\Delta^{-1}_\rmS(t,i\bk)$. The full analytic form is given in \cref{app:full-props}. For our purposes, it is sufficient to note the asymptotic forms
\begin{calign}
    \dow_t \lb 1 - \frac{\dow_t^2}{\omega_\beta^2} \rb
    K^{-1}(t,i\bk) \Bigg|_{t\gg \beta }
    = -\Delta^{-1}(t,i\bk) \Big|_{t\gg \beta }
    \to 0, \nn\\
    \lB\dow_t \lb 1 - \frac{\dow_t^2}{\omega_\beta^2} \rb\rB^2
    K^{-1}_\rmS(t,i\bk)
    \Bigg|_{|t|\gg \beta }
    = -\Delta^{-1}_\rmS(t,i\bk) \Big|_{|t|\gg \beta }
    \to 0.
\end{calign}
See \cref{app:full-props} for more details. In other words, both $\Delta^{-1}(t,i\bk)$ and $\Delta^{-1}_\rmS(t,i\bk)$ do not admit any exponential growth at late times.

\subsection{Three-point interactions}

We can similarly fix the form of interactions in the SK EFT based on symmetries and unitarity constraints. The general form of the three-point interactions can be parametrised as
\begin{calign}\label{eq:action-expansion}
    S_{\text{int-3}}
    &= \int \df^{d+1}x
    \Bigg(
    \cV_{F^2Q}(\dow_1,\dow_2;\dow_3)\, 
    \delta F\,\delta F\,\delta Q \nn\\[-1em]
    &\hspace{6em}
    + i\cV_{FQ^2}(\dow_1;\dow_2,\dow_3)\, 
    \delta F\,\delta Q\, \delta Q 
    + \cV_{Q^3}(\dow_1,\dow_2,\dow_3)\, 
    \delta Q\,\delta Q\,\delta Q
    \Bigg) + \ldots,
\end{calign}
where $\cV_{F^2Q}$, $\cV_{FQ^2}$, $\cV_{Q^3}$ are tri-linear differential operators. The subscripts on the derivatives denote which field they act on from left to right. As defined, $\cV_{F^2Q}$ is symmetric in its first-two arguments, $\cV_{FQ^2}$ is symmetric in its last-two arguments, while $\cV_{Q^3}$ is completely symmetric. 

The Lyapunov shift symmetries require that the $\cV_{F^2Q}$ vertex is parametrised as
\begin{calign}\label{eq:three-point-ansatz}
    \cV_{F^2Q}(\dow_1,\dow_2;\dow_3)
    &= \frac{\beta }{2}
    \dow_{t1}\dow_{t2} \lb 1 - \frac{\dow_{t1}^2}{\omega_\beta^2} \rb
    \Delta(\dow_1)
    + \frac{\beta }{2} \dow_{t1}\dow_{t2} \lb 1 - \frac{\dow_{t2}^2}{\omega_\beta^2} \rb \Delta(\dow_2) \nn\\
    &\qquad 
    - \frac{\beta }{2}\dow_{t1}\dow_{t2}
    \lb 1+\frac{\dow_{t1}^2 + \dow_{t2}^2 + 3\dow_{t1}\dow_{t2}}{\omega_\beta^2} \rb 
    \Delta(-\dow_3)
    \nn\\
    &\qquad 
    - \dow_{t1} \lb 1 - \frac{\dow_{t1}^2}{\omega_\beta^2} \rb
    \dow_{t2} \lb 1 - \frac{\dow_{t2}^2}{\omega_\beta^2} \rb
    \cR(\dow_1,\dow_2),
\end{calign}
for some unfixed bilinear differential operator $\cR(\dow_1,\dow_2)$, satisfying
\begin{calign}
    \cR(\dow_1,\dow_2)=\cR(\dow_2,\dow_1), \qquad 
    \cR(\dow,-\dow) = 0, \qquad 
    \tilde\cR(0,\dow) = \cR(0,-\dow).
\end{calign}
We have discussed the implementation of symmetries constraints in more detail in \cref{app:Fourier-constraints}. In the SYK lattice and the ASCJ theory we find that $\cR(\dow_1,\dow_2)=0$. We provide an explanation for this result at the end of~\cref{sec:energydiff}.

The KMS symmetry entirely fixes the $\cV_{FQ^2}$ and $\cV_{Q^3}$ vertices in terms of the $\cV_{F^2Q}$ vertex. Explicitly, we find
\begin{calign}\label{eq:KMS-relations-3pt}
    \cV_{FQ^2}(\dow_1;\dow_2,\dow_3) 
    &= -\frac{1}{2}
    \Big( \cV_{F^2Q}(\dow_1,\dow_2;\dow_3) 
    + \tilde\cV_{F^2Q}(\dow_2,\dow_3;\dow_1) \Big)
     \cot\bfrac{\beta \dow_{t2}}{2}
     + (\dow_2\leftrightarrow\dow_3)
     , \nn\\
    \cV_{Q^3}(\dow_1,\dow_2,\dow_3) 
    &=
    - \frac{1}{36}
    \cV_{F^2Q}(\dow_1,\dow_2;\dow_3) 
    - \frac{1}{18}
     \tilde\cV_{F^2Q}(\dow_1,\dow_2;\dow_3)
    + (\dow_2\leftrightarrow\dow_3)
    + (\dow_1\leftrightarrow\dow_3).
\end{calign}
In deriving the second relation here, we have assumed that the theory only has time-reversal-odd dissipation, i.e. the SK effective action satisfies the condition in \cref{eq:cond-odd-dissipation}. This leads to the constraint that $\cV_{F^2Q}(\dow_1,\dow_2;\dow_3) - \tilde\cV_{F^2Q}(\dow_1,\dow_2;\dow_3)$ is completely symmetric in its three arguments.\footnote{For theories where $\cR(\dow_1,\dow_2)=0$, this further requires a consistency condition that $\Delta(\dow)-\tilde \Delta(\dow)$ is at most linear in derivatives. As we have seen, this holds for both the SYK lattice and ASCJ models. In particular, for theories with $\Theta$-odd dissipation and $\cR(\dow_1,\dow_2)=0$, we must also choose the specific heat $c_\bk=c_V$ defined in \cref{eq:general-Delta} to be a constant. We will present most of the explicit results for this special class of theories.\label{foot:cV-footnote}} The more general expression of the vertex factors without this assumption is given in \cref{app:Fourier-constraints}.

\subsection{\texorpdfstring{$\SL(2,\bbR)$}{SL(2,R)}-invariant fields}
\label{sec:TOCdontgrow}

Similar to our discussion of free propagators around \cref{eq:SL2R-invariant-props}, we can also cast the interactions in terms of the $\SL(2,\bbR)$-invariant fields. To this end, we need the non-linear definitions of the invariant fields $T$ and $\chi_a$. Let us first define the coordinate fields $X_{1,2}$ as the point-wise inverses of the reparametrisation fields $F_{1,2}$, i.e.
\begin{calign}
    X_{1,2}\big(F_{1,2}(t,\bx),\bx\big) = t.
\end{calign}
We further define the field $\sigma$ as the pointwise inverse of the average combination of $X_{1,2}$, i.e.
\begin{calign}
    \frac{X_1\big(\sigma(t,\bx),\bx\big) 
    + X_2\big(\sigma(t,\bx),\bx\big)}{2}
    &= t.
\end{calign}
Following through these manipulations, we find that $\sigma = F + O(Q^2)$. 

Tracing how the $\SL(2,\bbR)$ transformations act through these definitions, we can check that $\sigma(x)$ transforms in the same way as $F_{1,2}(x)$ in \cref{eq:diagonal-shifts,eq:Lyapunov-shifts}. Therefore, the simplest $\SL(2,\bbR)$-invariant one can construct using this is the Schwarzian derivative $\Sch[\tanh(\pi\sigma),t]$, familiar from the discussion of the SYK chain (\ref{eq:SYK-2fold-start}). In fact, every local in time $\SL(2,\bbR)$-invariant combination of $\sigma$ can be expressed as a function of the Schwarzian and its spacetime derivatives. Hence we can use the Schwarzian to define a local temperature field
\begin{calign}
    T(t,\bx) 
    &= \lb \frac{-1}{2\pi^2} 
    \Sch\!\big[\tanh(\pi\sigma(t,\bx)),t \big]
    \rb^{1/2} \nn\\
    &= \lb \frac{-1}{2\pi^2} 
    \Sch\!\big[\tanh(\pi F(t,\bx)),t \big]
    \rb^{1/2}
    + O(Q^2),
    \label{eq:Tdef}
\end{calign}
which evaluates to $1/\beta$ in equilibrium.\footnote{In~\cite{Bucca:2026xmh}, we identified the local temperature as simply $\dow_tF$. This agrees with the definition in \cref{eq:Tdef} at leading order in derivatives, however the former is not $\SL(2,\bbR)$-invariant. 

Furthermore, \cref{eq:Tdef} is not unique as a definition of local temperature. Another natural definition could be obtained by taking the average of $\Sch[\tanh(\pi F_{1,2}),t]$. However, unlike these, our definition of $T$ in \cref{eq:Tdef} has the advantage that it is invariant under difference time-translations. Even in this sense, though, the definition in \cref{eq:Tdef} is not strictly unique: we can always modify it with higher-derivative terms $T \to T + O(\dow_\mu T,\ldots)$ (known as hydrodynamic frame redefinitions) or with terms involving derivatives of the noise field $T \to T + O(\dow_\mu\chi_a,\ldots)$, or a combination of both. Since this amounts to a simple field redefinition in the SK EFT, such ambiguities are merely conventional and do not affect physical correlation functions.

Finally, a possible confusion arises from the fact that in both of our microscopic examples the energy density is given by the Schwarzian featuring in~\cref{eq:Tdef}. However, we are not using this fact to write this equation, which provides a valid definition of temperature in all models described by the SK EFT.} We can further define the $\SL(2,\bbR)$-invariant noise field through the difference combination of $X_{1,2}$, i.e.
\begin{calign}\label{eq:chi_a-def}
    \chi_a(t,\bx) 
    &= X_1\big(\sigma(t,\bx),\bx\big) - X_2\big(\sigma(t,\bx),\bx\big) \nn\\
    &= 
    - \frac{Q}{\dow_t F}
    + O(Q^3)
    .
\end{calign}
After doing a linearised expansion around the equilibrium configuration, we can check that these are the same fields that we introduced around \cref{eq:SL2R-invariant-props}.

For completeness, we note that the doubled time-translation and space-translation symmetries act on these fields as
\begin{calign}\label{eq:goldstone-trans}
    T(t,\bx) &\mapsto T(t+\xi,\bx+{\bf c}), \nn\\
    \chi_{a}(t,\bx) &\mapsto \chi_{a}(t+\xi,\bx+{\bf c})
    - \zeta, 
\end{calign}
where $\xi_{1,2} = \xi \pm \zeta/2$.
The diagonal time and space translations act on these fields as usual. On the other hand, interestingly, we find that $T$ is invariant under difference time translations, while $\chi_a$ transforms like a Goldstone.

Since the SK EFT is $\SL(2,\bbR)$-invariant, it can be organised entirely in terms of $T$ and $\chi_a$. See \cref{app:Goldstone-basis} for more details. 
The propagators of these fields are insensitive to the scramblon modes, therefore it follows that the scramblon modes are pure gauge on the two-fold SK contour and cannot contribute to TOCs of local $\SL(2,\bbR)$-invariant operators. This result agrees with the general argument presented in~\cite{Blake:2017ris}.

\subsection{Mapping to energy diffusion}\label{sec:energydiff}

Up to three point interactions, the two-fold SK effective action is entirely fixed up to the differential operators $\Delta(\dow)$ and $\cR(\dow_1,\dow_2)$. These, in turn, are theory-specific and can be matched to the particular form of the classical energy diffusion equation. The energy current in the SK EFT arises as the Noether current of difference time translation symmetry in \cref{eq:time-translations}. In \cref{eq:goldstone-trans,eq:chi_a-def}, we saw that $-Q/\dow_t F$ acts as a Goldstone field for difference time translations, at least to linear order in $Q$. Therefore, we can derive the respective classical conservation equation as
\begin{calign}\label{eq:EOM-fourier}
    {\rm EoM} 
    &= -\dow_t F\,\frac{\delta S}{\delta Q} \bigg|_{Q\to 0} \nn\\
    &= -\frac{1}{\beta } K(\dow)\,\delta F
    - \lb \dow_{t1} K(\dow_{t2}) 
    + \frac{1}{\beta }\cV_{F^2Q}(\dow_1,\dow_2;-\dow_1-\dow_2)\rb 
    \delta F\,\delta F + \ldots.
\end{calign}

The above equation must be matched with energy conservation equation $\dow_\mu \cE^\mu$, where the energy current $\cE^\mu$ is constructed out of the local temperature $T$ and its derivatives. Using \cref{eq:Tdef}, we find
\begin{calign}\label{eq:T-def-Sch}
    T
    &=
    \frac{1}{\beta }
    + \lb 1 - \frac{\dow_t^2}{\omega_\beta^2} \rb \dow_t\delta F
    + \frac{2\beta }{\omega_\beta^2}\, \dow_t\delta F\, \dow_t^3\delta F 
    + \frac{3\beta }{2\omega_\beta^2}
    \left(\dow_t^2\delta F\right)^2
    - \frac{\beta }{2\omega_\beta^4} (\dow_t^3\delta F)^2
    + \ldots
    + O(Q).
\end{calign}
Defining $\delta T = T - 1/\beta $, the energy current operator can generically be expressed as
\begin{subequations}\label{eq:linear-expansion-E}
\begin{calign}
    \cE^\mu_i &= \cE_0\delta^\mu_t
    + \cD^\mu_{(1)}(\dow)\,\delta T 
    + \cD^\mu_{(2)}(\dow_1,\dow_2)\, \delta T\,\delta T + \ldots
    + \text{noise terms}
    \nn\\
    &= 
    \cE_0\delta^\mu_t
    + \cO^\mu_{(1)}(\dow)\,\delta F
    + \cO^\mu_{(2)}(\dow_1,\dow_2) \delta F\,\delta F
    + \ldots
    + \text{noise terms},
\end{calign}
where $\cD_{(n)}^\mu$ are some $n$-linear differential operators and
\begin{calign}
    \cO^\mu_{(1)}(\dow)
    &= \dow_t\lb 1 - \frac{\dow_t^2}{\omega_\beta^2} \rb
    \cD^\mu_{(1)}(\dow), \nn\\
    \cO^\mu_{(2)}(\dow_1,\dow_2)
    &= \frac{\beta }{2}
    \dow_{t1}\dow_{t2} \lb 1 + \frac{\dow_{t1}^2 + \dow_{t2}^2
    + 3\dow_{t1}\dow_{t2}}{\omega_\beta^2} \rb
    \cD^\mu_{(1)}(\dow_1+\dow_2) \nn\\
    &\qquad
    + \dow_{t1}\lb 1-\frac{\dow_{t1}^2}{\omega_\beta^2} \rb
    \dow_{t2}\lb 1-\frac{\dow_{t2}^2}{\omega_\beta^2} \rb 
    \lb \cD^\mu_{(2)}(\dow_1,\dow_2)
    - \frac{\beta}{2} \cD^\mu_{(1)}(\dow_1+\dow_2)\rb.
\end{calign}
\end{subequations}
Plugging in the expression for $\delta T$ in terms of $\delta F$ from \cref{eq:T-def-Sch} and comparing with \cref{eq:EOM-fourier}, we can read off
\begin{calign}\label{eq:Delta-R-from-EOM}
    \Delta(\dow) 
    &= \beta  \dow_\mu\cD^\mu_{(1)}(\dow), \nn\\
    \cR(\dow_1,\dow_2)
    &= \beta  \lb \dow_{\mu 1}+\dow_{\mu 2}\rb
    \lb \cD^\mu_{(2)}(\dow_1,\dow_2) 
    - \frac{\beta }{2} \cD^\mu_{(1)}(\dow_1+\dow_2) \rb.
\end{calign}
This enables us to read off the free propagators and three-point interaction vertices in the SK EFT corresponding to a particular model of energy diffusion.

\begin{myshaded}
\paragraph*{Microscopic examples.}

Specifically for the one-dimensional SYK chain~\cite{Bucca:2026xmh}, the energy density and current are given as
\begin{calign}\label{eq:SYK-energy}
    \cE^t
    &= \frac{\beta c_V}{2} T^2 
    = \frac{c_V}{2\beta }
    + c_V \delta T
    + \frac{\beta  c_V}{2} \delta T^2, \nn\\ 
    \cE^\bx 
    &= - \frac{4D }{a^2\dow_\bx}
    \sinh^2\left(\frac{a\dow_\bx}{2}\right) {\cal E}^t 
    = - \frac{4D c_V}{a^2\dow_\bx}
    \sinh^2\left(\frac{a\dow_\bx}{2}\right) \delta T 
    - \frac{2D \beta  c_V}{a^2\dow_\bx}
    \sinh^2\left(\frac{a\dow_\bx}{2}\right) \lb\delta T^2\rb.
\end{calign}
This enables us to read off
\begin{calign}\label{eq:SYK-energy-ops}
    \cD^t_{(1)}(\dow) &= c_V, \nn\\
    \cD^t_{(2)}(\dow_1,\dow_2) &= \frac{\beta c_V}{2}, \nn\\
    \cD^\bx_{(1)}(\dow) 
    &= -\frac{4D c_V}{a^2\dow_\bx}
    \sinh^2\left(\frac{a\dow_\bx}{2}\right), \nn\\
    \cD^\bx_{(2)}(\dow_1,\dow_2) 
    &= -\frac{2D \beta  c_V}{a^2(\dow_{\bx1}+\dow_{\bx2})}
    \sinh^2\left(\frac{a(\dow_{\bx1}+\dow_{\bx2})}{2}\right).
\end{calign}
Plugging these into \cref{eq:Delta-R-from-EOM}, we find the $\Delta(\dow)$ operator for SYK chain presented in~\cref{DeltaSYKLattice}. We also find $\cR(\dow_1,\dow_2) = 0$.
In fact, $\cR(\dow_1,\dow_2)$ identically vanishes for all models whose energy density and flux scale as $\propto T^2$, i.e. $\cE^\mu$ is given as a differential operator acting on $T^2$, including the ASCJ model. We have also explicitly verified the propagators and three-point interactions against the explicit derivation of the two-fold SK effective action for the SYK lattice in \cite{Bucca:2026xmh}.

\end{myshaded}

\section{Four-fold SK EFT for energy diffusion}
\label{sec:four-fold-theory}

Having constructed the two-fold SK EFT for energy diffusion, we will now discuss how to extend it to a four-fold closed-time contour to compute OTOCs. The discussion here is inspired from the recent construction of hydrodynamic four-fold SK EFT for charge diffusion in \cite{Mishra:2025vlf}, while extending it to energy current and non-perturbatively in derivatives for our purposes.  

\subsection{General structure of the four-fold SK EFT}

Similar to the two-fold unitarity constraints in \cref{eq:unitarity}, the four-fold unitarity constraints are given as the following~\cite{Mishra:2025vlf}\footnote{It is not clear to us at the moment of writing whether there is a similar non-negativity constraint on the four-fold contour.}
\begin{cgather}\label{eq:unitarity-4fold}
    S[F_1,F_2,F,F] = S[F_1,F_2], \quad 
    S[F_1,F,F,F_4] = S[F_1,F_4], \quad 
    S[F,F,F_3,F_4] = S[F_3,F_4], \nn\\[0.5em]
    S[F_1,F_2,F_3,F_4]^* = -S[F_4,F_3,F_2,F_1],
\end{cgather}
There are also symmetry constraints analogous to \cref{eq:symmetries-at-glance}. We have
\begin{subequations}\label{eq:symmetries-at-glance-four}%
\begin{alignat}{2}
    \text{Quadrupled time-translations:} &
    &\quad 
    F_{1,2,3,4}(t,\bx) &\mapsto F_{1,2,3,4}(t+\xi_{1,2,3,4},\bx), 
    \label{eq:time-translations-four} \\
    \text{Space-translations:} &
    &\quad 
    F_{1,2,3,4}(t,\bx) &\mapsto F_{1,2,3,4}(t,\bx+{\bf c}), \\
    \text{Spatially-local diagonal shifts:} &
    &\quad 
    F_{1,2,3,4}(t,\bx) &\mapsto F_{1,2,3,4}(t,\bx) + a(\bx), 
    \label{eq:diagonal-shifts-four}.
\end{alignat}
\end{subequations}
The KMS shift symmetry is more intricate on the four-fold closed-time contour. It is comprised of two mutually commuting $\bbZ_2$ symmetries
\begin{subequations}\label{eq:KMS-four}%
\begin{alignat}{2}
    \bbZ_2~\text{KMS$_{(1)}$ shifts}:&\qquad &
    F_{1}(t,\bx) &\mapsto - F_3(-t,\eta_\rmP\bx) + i/2, \nn\\
    && F_{2}(t,\bx) &\mapsto - F_2(-t,\eta_\rmP\bx) + i/2, \nn\\
    && F_{3}(t,\bx) &\mapsto - F_1(-t,\eta_\rmP\bx) + i/2, \nn\\
    && F_{4}(t,\bx) &\mapsto - F_4(-t,\eta_\rmP\bx) - i/2, \label{eq:KMS-4-1}\\[0.5em]
    \bbZ_2~\text{KMS$_{(2)}$ shifts}:&\qquad &
    F_{1}(t,\bx) &\mapsto F_3(t,\bx) + i/2, \nn\\
    && F_{2}(t,\bx) &\mapsto F_4(t,\bx) + i/2, \nn\\
    && F_{3}(t,\bx) &\mapsto F_1(t,\bx) - i/2, \nn\\
    && F_{4}(t,\bx) &\mapsto F_2(t,\bx) - i/2. \label{eq:KMS-4-2}
\end{alignat}
\end{subequations}
The KMS$_{(1)}$ symmetry arises from taking one Lorentzian fold of the closed-time contour across the Euclidean thermal circle, combined with a $\Theta$ transformation. On the other hand, the KMS$_{(2)}$ symmetry arises from simultaneously taking two Lorentzian folds across the Euclidean thermal circle. The latter does not require a $\Theta$ transformation to constitute a symmetry. See~\cite{Mishra:2025vlf} for an analogous discussion for the four-fold SK EFT for charge diffusion. Interestingly, on general $2n$-fold closed-time contour, the KMS symmetry constitutes a dihedral group $\rmD_{2n} = \bbZ_n\rtimes \bbZ_2$; see \cref{app:KMS} for more details.

For the growing part of the OTOCs computed below, it is sufficient to obtain the effective action up to cubic order in fluctuations.\footnote{The full OTOC also receives contact contributions from quartic vertices; see \cref{app:loops}.} As it turns out, the unitarity constraints \eqref{eq:unitarity-4fold}, the KMS symmetry \eqref{eq:KMS-four}, and the $\Theta$-odd dissipation condition in \cref{eq:cond-odd-dissipation} are sufficient to fix the effective action to this order, as given in \cref{eq:4fold-action}.\footnote{In general, we can add a term at cubic order ${\propto} (\delta F_1-\delta F_2)(\delta F_2-\delta F_3)(\delta F_3-\delta F_4)$. However, such a term is not invariant under the KMS symmetry. For theories with $\Theta$-odd dissipation, the factorised form of the four-fold effective action in \cref{eq:4fold-action} is already KMS-invariant and thus such a term does not arise; see more details in \cref{app:check-kms}.}
Starting from four-point interactions, the four-fold effective action can contain genuinely new independent terms that do not appear in the two-fold case; see~\cite{Mishra:2025vlf} for a detailed discussion  and the classification of such terms at leading derivative order.

\begin{myshaded}
\paragraph*{Microscopic examples.}

In the case of the SYK lattice, to leading order in large-$N$, the structure of the four-fold effective action in~\cref{eq:4fold-action} extends to the fully nonlinear theory. To see this, start with the Schwarzian lattice effective action that dominates the low temperature dynamics. It takes the bilocal form~\cite{Altland:2019lne,Bucca:2024zrx}:
\es{eq:bilocalform}{
S[F]_\cC
&=\int_{{\cal C}} \df s\,{\cal L}_\text{loc}\big[F(s)\big]
+\int_{{\cal C}} \df s_1\df s_2\,{\cal L}_\text{biloc}\big[F(s_1),F(s_2)\big]\,,
}
where $s_i$ are  parameters along the multi-fold SK contour ${\cal C}$, and the local and bilocal parts are given in \cref{eq:final_action}. Assuming the number of Lorentzian folds in ${\cal C}$ to be $2n$, we can express the right-hand side in terms of fold labels as
\es{eq:bilocalform2}{
S^{(2n)}\big[\{F_i\}\big]
&=\sum_{i=1}^{2n}(-1)^{i+1}\int_{-\infty}^{\infty} \df t\,{\cal L}_\text{loc}\big[F_i(t)\big] \\
&\qquad 
+ \sum_{i,j=1}^{2n}(-1)^{i+j}\int_{-\infty}^{\infty} \df t_1 \df t_2\,{\cal L}_\text{biloc}\big[F_i(t_1),F_j(t_2)\big] \\
&\equiv \sum_{i=1}^{2n}(-1)^{i+1} S_\text{loc}[F_i]+\sum_{i,j=1}^{2n}(-1)^{i+j}S_\text{biloc}[F_i,F_j].
}
For $n=1$, this gives the two-fold SK effective action in \cref{eq:SYK-2fold-start}.
In terms of this, one can check that \cref{eq:bilocalform2} can generally be expressed as
\begin{calign}\label{eq:bilocalform3}
    S^{(2n)}\big[\{F_i\}\big]
    &= \sum_{i<j=1}^{2n}(-1)^{i+j+1}S[F_i,F_j].
\end{calign}
For $n=2$, we recover the form of the four-fold SK effective action in \cref{eq:4fold-action}, without any additional interactions. 

The ASCJ theory only has a local term in its action in \cref{eq:ASCJ-local-action}, hence its four-fold SK effective action is $S_{\rm loc}[F_1]-S_{\rm loc}[F_2]+S_{\rm loc}[F_3]-S_{\rm loc}[F_4]$.

\end{myshaded}

\subsection{Quadratic effective action}

We can compute the free four-fold effective action by substituting \cref{eq:action-free} into \cref{eq:4fold-action}. Similar to the average-difference basis on the two-fold contour, a helpful basis of fields to simplify the propagators on the four-fold contour is\footnote{This is slightly different from the basis introduced in~\cite{Mishra:2025vlf}
\es{FiBasis}{
    F_R = 4F - 2\bar F, \qquad 
    F_A = Q, \qquad 
    F_+ = Q + 2\bar Q, \qquad 
    F_- = 2\bar F.
}
Our basis has the advantage that the propagators factorise between $F,Q$ and $\bar F,\bar Q$.
}
\begin{subequations}\label{eq:basis-of-fields}
\begin{calign}
    F_1 &= F + \frac{1}{2} Q + \frac{1}{2}\bar Q, \nn\\
    F_2 &= F + \frac{1}{2}\bar Q - \bar F, \nn\\
    F_3 &= F - \frac{1}{2}\bar Q - \bar F, \nn\\
    F_4 &= F - \frac{1}{2} Q - \frac{1}{2}\bar Q,
\end{calign}
or equivalently,
\begin{calign}
    F &= \frac{F_1+F_4}{2}, \nn\\
    Q &= F_1 - F_2 + F_3 - F_4, \nn\\
    \bar Q &= F_2 - F_3, \nn\\
    \bar F &= \frac{F_1 - F_2 - F_3 + F_4}{2}.
\end{calign}
\end{subequations}

Substituting this basis into \cref{eq:4fold-action}, the free part of the four-fold SK effective action takes a particularly simple form
\begin{calign}\label{eq:action4-free}
    S_{\rm free}
    &= \int\df^{d+1}x
    \bigg(
    \delta Q\,K(\dow)\,\delta F
    + \frac{i}{4}\delta Q\,\cot\bfrac{\beta \dow_t}{2}
    \Big( K(-\dow) - K(\dow) \Big)
    \,\delta Q
    + \delta \bar F\,K(\dow)\,\delta \bar Q
    \bigg).
\end{calign}
We see that the free effective action nicely factorises between the $F$, $Q$ and $\bar F$, $\bar Q$ sectors. This structure is also visible in the propagators:
\begin{calign}\label{eq:props-4}
    \renewcommand{\arraystretch}{1.5}
    \Big\langle
    \begin{pmatrix}
        \delta F(x) \\ \delta Q(x) \\ \delta\bar Q(x) \\ \delta\bar F(x)
    \end{pmatrix}
    \begin{pmatrix}
        \delta F(0) \\ \delta Q(0) \\ \delta\bar Q(0) \\ \delta\bar F(0)
    \end{pmatrix}^{\!\!\!\rmT}
    \Big\rangle
    &=
    \begin{pmatrix}
        K_\rmS^{-1}(x)
        & iK^{-1}(x)
        & 0 & 0 \\
        i\tilde K^{-1}(x) & 0 & 0 & 0 \\
        0 & 0 & 0 & iK^{-1}(x) \\
        0 & 0 & i\tilde K^{-1}(x) & 0 \\
    \end{pmatrix}.
\end{calign}
We see that $\langle \bar Q\bar F\rangle$ propagator is a copy of the retarded $\langle FQ\rangle$ propagator. However, all other propagators involving the new four-fold noise fields are zero.
In particular, the fluctuations of $F$ and $Q$ only talk to those of $\bar F$ and $\bar Q$ through non-linear interactions.

\subsection{Three-point interactions}

Similarly, we can obtain the three-point interactions on the four-fold contour. They are schematically represented as
\begin{calign}\label{eq:general-four-fold-interactions}
    S_{\text{int-3}}
    &= \int\df^{d+1}x
    \bigg[
    \cV_{F^2Q}\, \delta F\,\delta F\,\delta Q
    + i \cV_{FQ^2}\, \delta F\,\delta Q\,\delta Q
    + \cV_{Q^3}\, \delta Q\,\delta Q\,\delta Q \nn\\
    &\hspace{6em}
    + \cV_{Q^2\bar Q}\, \delta Q\,\delta Q\,\delta\bar Q
    + \cV_{F\bar Q\bar F}\, \delta F\, \delta\bar Q\, \delta\bar F
    + i\cV_{Q\bar Q\bar F}\, \delta Q\, \delta\bar Q\, \delta\bar F \nn\\
    &\hspace{6em}
    + \cV_{Q\bar Q^2}\, \delta Q\, \delta\bar Q\, \delta\bar Q
    + \cV_{\bar Q \bar F^2}\, \delta \bar Q\, \delta\bar F\, \delta\bar F
    \bigg],
\end{calign}
where we have dropped the explicit derivative arguments of the vertex factors for clarity. All other potential interactions turn out to identically be zero. The interaction vertices for $\cV_{F^2Q}$, $\cV_{FQ^2}$, and $\cV_{Q^3}$ in the four-fold theory take the same form as in the two-fold theory. 
We will focus on theories with exclusively time-reversal-odd dissipation, where we can further use the condition in \cref{eq:cond-odd-dissipation} to simplify some of the interactions, and the nonzero ones are given by:
\begin{calign}\label{eq:theta-odd-four-fold-interactions}
    \cV_{Q^2\bar Q}(\dow_1,\dow_2;\dow_3) 
    &= 
    - \frac14 \tilde\cV_{F^2Q}(\dow_1,\dow_2;\dow_3), \nn\\
    \cV_{F\bar Q\bar F}(\dow_1;\dow_2;\dow_3) 
    &= 2 \cV_{F^2Q}(\dow_1,\dow_2;\dow_3), \nn\\
    \cV_{Q\bar Q\bar F}(\dow_1;\dow_2,\dow_3)
    &=
    - \frac{1}{2} \bigg(
    \cV_{F^2Q}(\dow_1,\dow_2;\dow_3) 
    + \tilde\cV_{F^2Q}(\dow_1,\dow_2;\dow_3)
      \bigg) \cot\lb \frac{\beta \dow_{t1}}{2} \rb
    + (\dow_2\leftrightarrow\dow_3)
      , \nn\\
    \cV_{Q\bar Q^2}(\dow_1;\dow_2,\dow_3) 
    &= \frac14 \cV_{F^2Q}(\dow_2,\dow_3;\dow_1), \nn\\
    \cV_{\bar Q\bar F^2}(\dow_1;\dow_2,\dow_3) 
    &= \tilde\cV_{F^2Q}(\dow_2,\dow_3;\dow_1) .
\end{calign}
These will be useful for us later. 

\subsection{\texorpdfstring{$\SL(2,\bbR)$}{SL(2,R)}-invariant fields}

Similar to the two-fold case, we can express the four-fold effective action in terms of $\SL(2,\bbR)$-invariant fields. Unlike the two-fold case, however, we will find that the scramblon mode is not pure gauge on the four-fold contour, which ultimately mediates the exponential growth of OTOCs. To this end, following the same logic as in \cref{sec:TOCdontgrow}, we define the fields
\begin{calign}
    X_{1,2,3,4}\big(F_{1,2,3,4}(t,\bx),\bx\big) = t,
\end{calign}
and further
\begin{calign}
    \frac{X_1\big(\sigma(t,\bx),\bx\big) 
    + X_4\big(\sigma(t,\bx),\bx\big)}{2}
    &= t.
\end{calign}
This enables us to define the invariant fields
\begin{calign}
    T(t,\bx) 
    &= \lb \frac{-1}{2\pi^2} 
    \Sch\!\big[\tanh(\pi\sigma(t,\bx)),t \big]
    \rb^{1/2}, \nn\\
    \chi_a(t,\bx) 
    &= X_1\big(\sigma(t,\bx),\bx\big) 
    - X_2\big(\sigma(t,\bx),\bx\big)
    + X_3\big(\sigma(t,\bx),\bx\big) 
    - X_4\big(\sigma(t,\bx),\bx\big), \nn\\
    \chi_{\bar a}(t,\bx) 
    &=  X_2\big(\sigma(t,\bx),\bx\big)
    - X_3\big(\sigma(t,\bx),\bx\big), \nn\\
    \chi_{\bar r}(t,\bx) 
    &= \half 
    \bigg( 
    X_1\big(\sigma(t,\bx),\bx\big) 
    - X_2\big(\sigma(t,\bx),\bx\big)
    - X_3\big(\sigma(t,\bx),\bx\big) 
    + X_4\big(\sigma(t,\bx),\bx\big)
    \bigg).
\end{calign}
Note that the linear combinations for the noise fields here are the same as in \cref{eq:basis-of-fields}.

 In terms of these $\SL(2,\bbR)$ invariant fields, we find the propagators
\begin{calign}
    \renewcommand{\arraystretch}{1.5}
    \Big\langle
    \begin{pmatrix}
        \delta T(x) \\ \delta\chi_a(x) \\ \delta\chi_{\bar a}(x)
        \\ \delta\chi_{\bar r}(x)
    \end{pmatrix}
    \begin{pmatrix}
        \delta T(0) \\ \delta\chi_a(0) \\ \delta\chi_{\bar a}(0) 
        \\ \delta\chi_{\bar r}(0)
    \end{pmatrix}^{\!\!\!\rmT}
    \Big\rangle
    =
    \begin{pmatrix}
        \Delta_\rmS^{-1}(x)
        & i\beta \Delta^{-1}(x)
        & 0 & 0 \\
        i\beta \Delta^{-1}(-x) & 0 & 0 & 0 \\
        0 & 0 & 0 & i\beta ^2K^{-1}(x) \\
        0 & 0 & i\beta ^2K^{-1}(-x) & 0 \\
    \end{pmatrix}.
\end{calign}
We see that, indeed, the scramblon modes contained within $K^{-1}(x)$ survive in the $\SL(2,\bbR)$-invariant representation of the four-fold SK EFT in the ($\chi_{\bar a}$,$\chi_{\bar r}$) sector. By contrast, the ($\delta T$,$\chi_a$) sector only involves $\Delta^{-1}(x)$ and $\Delta^{-1}_\rmS(x)$, which are not sensitive to the scramblon modes. Physically, the reason is that the $\SL(2,\bbR)$ symmetry acts simultaneously on the four folds of the closed-time contour, which gauges the scramblon mode in propagators involving the field $F$.
However, since relative $\SL(2,\bbR)$ transformations between folds do not constitute a symmetry, the scramblon mode in the $(\bar Q,\bar F)$ sector survives. In the next section, we will see how it is related to the scramblon propagator. Recall that the scramblon shockwave is precisely defined using relative  $\SL(2,\bbR)$ transformations between folds; see \cref{fig:shocks}.

\section{OTOCs of generic local operators and of energy currents}
\label{sec:energy-otoc}

In this section, we use our four-fold SK EFT to compute the OTOC of local operators. We discuss two qualitatively different cases. First, we compute OTOCs of generic local operators which themselves are not $\SL(2,\bbR)$ invariant, but pairs of their insertions are represented by $\SL(2,\bbR)$-invariant bilocal operators in the SK EFT. For concreteness, we focus on SYK-like theories, where such bilocals are parametrised only by the scaling dimension of local operators. The OTOC for this case has already been computed in the literature and we find complete agreement with these results. Later, we derive new results for OTOCs of collective local operators in the EFT that are themselves $\SL(2,\bbR)$-invariant. The prominent application is to the OTOCs of conserved currents like the energy density and flux operators. Lastly, we verify our computation using the scramblon shockwave prescription.

\subsection{OTOC of generic local operators}
\label{sec:generic-otoc}

\subsubsection{Bilocal representation of correlation functions}

A peculiarity of the all-order EFT framework is that pairs of generic operators are represented as bilocal expressions in the EFT. This can be most directly derived in the SYK family of models, but has been argued for in general in~\cite{Blake:2017ris}. In SYK, this can be derived using the disorder averaging; for a pair of operators we have a chain of representations 
\begin{calign}\label{eq:chainofreps}
    V_i(x_1)V_j(x_2) 
    &\underset{\substack{\text{collective field}\\\text{description}}}{\sim}\, \mu_V \E{\Delta_V g_{ij}(x_1,x_2)} \nn\\
    &\hspace{1.6em}\underset{\substack{\text{EFT}}}{\sim} 
    \mu_V\le(
    \E{-i\pi\phi_{ij}(t_1-t_2)}
    \frac{\dow_{t_1}F_i(x_1)\, \dow_{t_2} F_j(x_2)}
    {\sinh^2\!\big(\pi F_i(x_1)-\pi F_j(x_2)\big)}\ri)^{\Delta_V}
    \delta^{(d)}(\bx_1-\bx_2)\,,
\end{calign}
where the phase comes from the appropriate analytic continuation of the Euclidean bilocal, given as
\begin{calign}
    \phi_{ij}(t) &= 
    \begin{cases}
        \sgn(i-j)\sgn(t) &\quad (i\neq j), \nn\\
        (-1)^{i+1} &\quad (i=j).
    \end{cases}
\end{calign}
Note that 
\begin{calign}
    G^V_{ij}(x_1-x_2)
    &= \Big\langle V_i(x_1)V_j(x_2) \Big\rangle 
    = \frac{\mu_V}{\beta^{2\Delta_V}}
    \frac{\E{-i\pi\phi_{ij}(t_1-t_2)\Delta_V}}
    {\sinh^{2\Delta_V}\!\big(\pi|t_1-t_2|/\beta\big)}
    \delta^{(d)}(\bx_1-\bx_2).
\end{calign}
We can use these to read off the retarded, symmetric, and unordered correlation functions.

One special feature of the bilocal is that it is (gauge) invariant under the spatially local $\SL(2,\bbR)$ symmetry. 
If no operators are inserted between $V(x_1)$ and $V(x_2)$, we can expand the bilocal around the midpoint to 
\begin{calign}\label{OPE}
    \E{\Delta_V g_{ij}(x_1,x_2)}
    &= {\E{-i\phi_{ij}(t_{12})\Delta_V} \ov |\pi t_{12}|^{2\Delta_V}}\le[1+{\Delta_V\ov 6}\,\Sch\!\big[\tanh(\pi F),t\big]\,
    t_{12}^2
    + O(t_{12}^4)\ri]
    + \text{noise terms}
    \,,
\end{calign}
where $t=(t_1+t_2)/2$ and $t_{12} = t_1-t_2$. Here we can recognise the OPE in conformal quantum mechanics with the energy-momentum tensor---the energy density given by the $\SL(2,\bbR)$ invariant Schwarzian---playing the role as the first correction to the identity. 
However, for the OTOC kinematics we do have an operator inserted in between the legs of the bilocal and~\cref{OPE} does not apply. 
Note that the $\SL(2,\bbR)$ symmetry of $F$ is the global conformal symmetry of the two-point function of $V$. This is appropriate to describe simple operators in SYK. In more general theories without an approximately conformal sector, we may need to represent bilocals with more complicated  $\SL(2,\bbR)$ invariants. 

We are interested in the four-point correlator, which takes the form
\begin{calign}\label{eq:general-propagators1}
    C^{VW}_{ij|kl}(x_1,x_2;x_3,x_4)
    &= \Big\langle V_i(x_1) V_j(x_2) 
    W_k(x_3) W_l(x_4) \Big\rangle \nn\\
    &= \mu_V\mu_W \Big\langle \E{\Delta_V g_{ij}(x_1,x_2)} 
    \E{\Delta_W g_{kl}(x_3,x_4)} \Big\rangle\,,
\end{calign}
where the labels $i,j,\ldots$ denote the contour-fold on which the operators are inserted.
There are, in fact, only three qualitatively independent such correlators
\begin{equation}\label{eq:general-propagators}
\begin{alignedat}{2}
    C^{VW}_{43|21}(x_1,x_2;x_3,x_4)
    &= \Big\langle \hat V(x_1)\, \hat V(x_2)\, 
    \hat W(x_3)\, \hat W(x_4) \Big\rangle, 
    &&\qquad \boxed{\text{TOC}}\\
    C^{VW}_{41|32}(x_1,x_2;x_3,x_4)
    &= \Big\langle \hat V(x_1)\, \hat W(x_3)\,
    \hat W(x_4)\, \hat V(x_2) \Big\rangle,  
    &&\qquad \boxed{\text{TOC (using KMS)}} \\
    C^{VW}_{42|31}(x_1,x_2;x_3,x_4)
    &= \Big\langle \hat V(x_1)\, \hat W(x_3)\, \hat V(x_2)\, \hat W(x_4) \Big\rangle.
    &&\qquad \boxed{\text{OTOC}}
\end{alignedat}
\end{equation}
All the remaining correlators can be obtained from these either by shuffling the coordinate labels or by exchanging the $V$ and $W$ operators. On the right-hand side in \cref{eq:general-propagators}, we have expressed the correlation functions in terms of quantum operators, denoted with a hat. To get eq. \cref{eq:general-propagators} from eq. (\ref{eq:general-propagators1}), note that the operators labelled ``1'' are inserted first from the right, then ``2'', then ``3'', and then ``4''.  In particular, if we take $t_1>t_2>t_3>t_4$, the first two correlators in \cref{eq:general-propagators} are TOCs, while the third one is the OTOC.\footnote{The identity of the correlators in \cref{eq:general-propagators} as TOC or OTOC changes for different orderings of time coordinates.} Note that the OTOC takes the same form as we prescribed in \cref{OTOCexpectation}. In fact, for thermal systems, the second correlator in \cref{eq:general-propagators} is related to the first one by taking one of the operator insertions across the Euclidean thermal circle, leading to only one genuinely independent TOC and OTOC.

We will compute these correlators at the leading order in perturbations around the equilibrium thermal state $F_i(x) = t/\beta $. To this end, let us expand
\begin{calign}\label{eq:delta-g}
    \delta\lb\E{g_{ij}(x_1,x_2)}\rb
    = \Big\langle \E{g_{ij}(x_1,x_2)}\Big\rangle\,
    \delta g_{ij}(x_1,x_2)
    + \ldots,
\end{calign}
in terms of which, we can obtain the normalised four-point connected correlator as
\begin{calign}\label{eq:general-correlators-normalised}
    \hat C^{VW}_{ij|kl}(x_1,x_2;x_3,x_4)
    &= \frac{C^{VW}_{ij|kl}(x_1,x_2;x_3,x_4)}
    {G^V_{ij}(x_1-x_2)G^W_{kl}(x_3-x_4)}
    - 1
    \nn\\
    &= \Delta_V\Delta_W\Big\langle 
    \delta g_{ij}(x_1,x_2)
    \,\delta g_{kl}(x_3,x_4) \Big\rangle
    + \ldots.
\end{calign}

\subsubsection{Four-point TOCs and OTOC using SK EFT}

Let us note the expression for the bilocals $\delta g_{ij}$ in terms of the reparametrisation fields $\delta F_i$:
\begin{calign}\label{eq:bilocal-pert}
    \frac{1}{\beta }\delta g_{ij}(x_1,x_2)
    &=\Big( \partial_{t_1} 
    - \omega_\beta \coth\!\big(\omega_\beta t_{12}/2\big)
    \Big)
    \delta F_i(x_1) 
    + \Big(\partial_{t_2}
    + \omega_\beta \coth\!\big(\omega_\beta t_{12}/2\big) \Big)
    \delta F_j(x_2).
\end{calign}
Note that the phase $\phi_{ij}$ does not contribute to $\delta g_{ij}$. Here we have used that the spatial points of the bilocal coincide, $\bx_1=\bx_2$.
Note that $\delta g_{ij}(t_1,t_2) = \delta g_{ji}(t_2,t_1)$.

We can evaluate the four-point correlators by substituting the expression for $\delta g$ from \cref{eq:bilocal-pert} into \cref{eq:general-correlators-normalised} and using the four-fold propagators from \cref{eq:props-4}. For the first correlator (TOC) in \cref{eq:general-propagators}, we find
\begin{calign}
    &\hat C^{VW}_{43|21}(x_1+x,x_2+x;x_3,x_4)
    = - \Delta_V\Delta_W\beta ^2
    \lb \E{t_{12}\dow_t} - 1 \rb
    \lb \E{t_{34}\dow_t} - 1 \rb
    \nn\\
    &\hspace{4em} 
    \Big( \coth\lb t_{12}\dow_t/2 \rb \partial_{t}
    - \omega_\beta \coth\!\big(\omega_\beta t_{12}/2\big) 
     \Big)
    \Big( \coth\lb t_{34}\dow_t/2 \rb \partial_{t}
    - \omega_\beta \coth\!\big(\omega_\beta t_{34}/2\big) 
     \Big)
    \nn\\
    &\hspace{14em} 
    \lb
    K_S^{-1}(x_{23}+x)
    + \frac{i}{2}K^{-1}(x_{23}+x)
    - \frac{i}{2}\tilde K^{-1}(x_{23}+x)
    \rb,
\end{calign}
The important point to note is that both the differential operators in parenthesis in the second line have zeros at the scramblon modes $\dow_t=\pm\omega_\beta$, which identically cancel the scramblon modes in the propagators. Therefore, this correlator does not exhibit Lyapunov growth at late times. The same conclusion holds for the second correlator in \cref{eq:general-propagators}, which, as we mentioned, is related to first one through KMS symmetry.

The interesting Lyapunov growth behaviour arises in the third correlator (OTOC) in \cref{eq:general-propagators}. It yields
\begin{calign}\label{eq:bilocal-otoc-general}
    &\hat C^{VW}_{42|31}(x_1+x,x_2+x;x_3,x_4)
    - \hat C^{VW}_{43|21}(x_1+x,x_2+x;x_3,x_4) \nn\\
    &\hspace{4em}
    = \Delta_V\Delta_W\beta ^2
    \Big( \partial_{t} + \omega_\beta \coth\!\big(\omega_\beta t_{12}/2\big) \Big)
    \Big( \partial_{t} + \omega_\beta \coth\!\big(\omega_\beta t_{34}/2\big) \Big) \nn\\
    &\hspace{16em}
    \Big( iK^{-1}(x_{23}+x) - i\tilde K^{-1}(x_{23}+x) \Big),
\end{calign}
where we have subtracted a TOC contribution to isolate the interesting growth behaviour of the OTOC. Indeed, we see that the differential operators in this case do not have a zero at the scramblon modes. Consequently, this correlator does grow exponentially at late times. This object is referred to as the ``scramblon propagator'' in \cite{Stanford:2023npy}. Noting the growing part of $K^{-1}(x)$ from \cref{eq:DeltaandKfunctions}, we can extract the large $t$ behaviour of the OTOC as
\begin{calign}\label{eq:bilocal-otoc2}
    &\hat C^{VW}_{42|31}(x_1+x,x_2+x;x_3,x_4) \nn\\
    &\qquad \sim 
    \int \frac{\df^d\bk}{(2\pi)^d}
    \frac{\Delta_V\,
    \E{\omega_\beta(t_{1}+t_2)/2}\,
    \E{i\bk\cdot\bx_2}
    }{\sinh\!\big(\omega_\beta t_{12}/2\big)} 
    \frac{\Delta_W\,
    \E{-\omega_\beta(t_3+t_4)/2}\,
    \E{-i\bk\cdot\bx_3}
    }{\sinh\!\big(\omega_\beta t_{34}/2\big)} 
    \lb 
    \frac{i\pi\omega_\beta\,
    }{c_\bk(\omega_\beta+\cD_\bk)}
    \E{\omega_\beta t+i\bk\cdot\bx}
    \rb.
\end{calign}
We have presented the full expression in \cref{app:generic-OTOC}. 
We can explicitly see that the factor of $\E{\omega_\beta t}$ grows exponentially at late times.  For $D\to 0$, this result matches the known SYK dot result as reviewed in~\cite{Sarosi:2017ykf}, with the prefactor $c_V = 2\pi C$ (in units of $\beta=2\pi$).

The result in position-domain depends on the explicit form of the dispersion relations $\cD_\bk$ and the specific heat $c_\bk$. Let us assume that the system is isotropic, so $\cD_\bk$ and $c_\bk$ only depend on $|\bk|$. If we assume that these functions are sufficiently well-behaved in the complex plane, we can evaluate the integral over $k\equiv|\bk|$ using a contour deformation (after extending $k\to-k$), picking up a pole at $\cD_{i k_*}=-\omega_\beta$, as in~\cite{Gu:2016oyy,Mezei:2019dfv,Choi:2020tdj}. The remaining angular integral in momentum space yields a Bessel function, which we approximate with its asymptotics for large $|\bx|$. We obtain the final result
\begin{calign}\label{eq:BilocalOTOC}
    \hat C^{VW}_{42|31}(x_1+x,x_2+x;x_3,x_4)
    \sim 
    \frac{-\pi\omega_\beta\,k_*^{(d-1)/2}}{c_{ik_*}\cD'_{ik_*}}\,
    \frac{\Delta_V\,\Delta_W\,
    \E{\omega_\beta(t_{1}+t_2-t_3-t_4)/2}\,
    }{\sinh\!\big(\omega_\beta t_{12}/2\big)
    \sinh\!\big(\omega_\beta t_{34}/2\big)} \,
\frac{\E{\omega_\beta(t-|\bx|/v_b)}}{|2\pi\bx|^{(d-1)/2}},
\end{calign}
where $\cD'_{ik_*} = \dow_k \cD_k|_{k=ik_*}$ and we identified the butterfly velocity $v_b\equiv {\omega_\beta/k_*}$.
Recall that the two legs of a bilocal is restricted to be at the same point in space, $\bx_1=\bx_2$ and $\bx_3=\bx_4$. Without loss of generality, we have further set $\bx_{23}=0$.  We see that the OTOC exhibits maximal growth with rate $\lambda=\omega_\beta$. The $1/N$ correction is large within the butterfly cone $|\bx| \leq v_b |t|$ and small outside it. 

\begin{myshaded}
\paragraph*{Microscopic examples.}
In the 1d SYK chain with $\cD_\bk = 4D/a^2\sin^2(a\,\bk/2)$ and $c_\bk=c_V$, we find that the butterfly velocity is given as
\es{eq:replacements}{
v_b = {\omega_\beta a \ov\cosh^{-1}\le(1+ \omega_\beta a^2/(2D)\ri)}=\sqrt{\omega_\beta D} + O(a^2)\,.
}
This latter relation is familiar in a host of contexts~\cite{Gu:2016oyy,Blake:2016wvh,Blake:2016sud,Blake:2016jnn,Davison:2016ngz,Gu:2017ohj,Blake:2017qgd,Hartman:2017hhp,Blake:2017ris}. 
The prefactor of \cref{eq:BilocalOTOC} becomes
\es{eq:prefactor}{
    \frac{-\pi\omega_\beta}{c_{ik_*}\cD'_{ik_*}} 
    &= \frac{i\pi^2}{\beta c_V}
    \frac{a/D}{\sinh(a\omega_\beta/ v_b)}
    = \frac{i\pi^2}{\beta c_V v_b}
    + O(a^2)
    \,.
}
To obtain this result, we also took into account that the $\bk$ integral only goes over the Brillouin zone. These results agree with~\cite{Gu:2016oyy,Choi:2020tdj}.

For the ASCJ theory, we have simply $\cD_\bk=i\bk$ (where $\bk$ is a number). In this case, we can perform the position space answer in \cref{eq:bilocal-otoc-general} directly, yielding
\begin{align}
    &\hat C^{VW}_{42|31}(x_1,x_2;x_3,x_4)
    - \hat C^{VW}_{43|21}(x_1,x_2;x_3,x_4) \nn\\
    &= - \frac{24i\pi}{C}\Delta_V\Delta_W
    \Big( \Theta(\bx_{23}) - \Theta(\bx_{23}-t_{23}) \Big) \nn\\
    &\qquad 
    \lb
    \frac{1
    }{\tanh\!\big(\omega_\beta t_{12}/2\big) 
    \tanh\!\big(\omega_\beta t_{34}/2\big)}
    - \frac{
    \cosh\lb\omega_\beta\frac{t_{1}+t_2-t_3-t_4}{2}-\omega_\beta\bx_{23}\rb 
    }{\sinh\!\big(\omega_\beta t_{12}/2\big) 
    \sinh\!\big(\omega_\beta t_{34}/2\big)
    }
    \rb.
\end{align}
We see that butterfly velocity $v_b=1$ in this case.

\end{myshaded}

\subsection{Commutator-squared and antisymmetric OTOC}

The basis of fields $F,Q,\bar Q,\bar F$ on the four-fold closed-time contour is quite convenient for calculations as the propagators factorise between $F,Q$ and $\bar F,\bar Q$. However, directly accessing the OTOCs of the kind ``42|31'' in \cref{eq:general-propagators} is quite non-trivial in this basis. We need to define a different OTOC, perhaps by combining the ``42|31'' OTOC with other TOCs, that is more easily accessible in the factorised basis. This situation is similar to how the average-difference basis on the two-fold contour gives direct access to symmetric and retarded correlation functions, but not necessarily the unordered (Wightman) or time-ordered correlation functions.

\subsubsection{OTOCs in the four-fold basis}

To this end, similar to the basis of fields in \cref{eq:basis-of-fields}, let us define a basis of local operators on the four-fold contour
\begin{calign}\label{eq:operator-basis}
    V_1 &= V_r + \frac{1}{2} V_a
    + \frac{1}{2}V_{\bar a}, \nn\\
    V_2 &= V_r + \frac{1}{2}V_{\bar a} - V_{\bar r}, \nn\\
    V_3 &= V_r - \frac{1}{2}V_{\bar a} - V_{\bar r}, \nn\\
    V_4 &= V_r - \frac{1}{2} V_a - \frac{1}{2}V_{\bar a},
\end{calign}
for some local operator $V$. Using these, we define the following four-point correlation functions of four arbitrary local operators $V$ and $W$:
\begin{calign}\label{eq:new-basis-otocs}
    C^{VW}_{\bar a \bar a |\bar r \bar r}(x_1,x_2;x_3,x_4)
    &\equiv \Big\langle
    V_{\bar a}(x_1)\,
    V_{\bar a}(x_2)\,
    W_{\bar r}(x_3)\,
    W_{\bar r}(x_4)
    \Big\rangle, \nn\\
    C^{VW}_{\bar a \bar a |\bar r a}(x_1,x_2;x_3,x_4)
    &\equiv \Big\langle
    V_{\bar a}(x_1)\,
    V_{\bar a}(x_2)\,
    W_{\bar r}(x_3)\,
    W_{a}(x_4)
    \Big\rangle.
\end{calign}
One can check that both of these are OTOCs and cannot be represented on a two-fold contour. The generalisation of these definitions to distinct operators on $x_{1,2}$ and $x_{3,4}$ is straightforward. Note also that while the bilocals defined in \cref{sec:generic-otoc} were restricted to have $\bx_1=\bx_2$ and $\bx_3=\bx_4$ and vanished otherwise, such a restriction does not apply for operators that are locally $\SL(2,\bbR)$-invariant.
Taking $x_{1,2}=x$ and $x_{3,4}=0$ these are given in operator language as
\begin{calign}\label{eq:OTOC-combos}
    C^{VW}_{\bar a \bar a |\bar r \bar r}(x,x;0,0)
    &=
    2\Theta(t)
    \bigg\langle\Big[\hat V(x),\hat W(0)\Big]^2\bigg\rangle , \nn\\
    C^{VW}_{\bar a \bar a |\bar r a}(x,x;0,0) 
    &=
    \Theta(t)
    \bigg\langle\Big[\hat W(0),\Big[\hat V(x),\Big[\hat V(x),\hat W(0)\Big]\Big] \Big] \bigg\rangle.
\end{calign}
The analogous expression for distinct operator insertion points is given in \cref{app:loops}.
In particular, $C_{\bar a \bar a |\bar r \bar r}(x,x,0,0)$ is the conventional ``commutator-squared'' OTOC. We find that $C_{\bar a \bar a |\bar r a}$ generically exhibits growth, while $C_{\bar a \bar a |\bar r \bar r}$ does not. We will look at these results in more detail for the energy current operator below and for arbitrary local operators in~\cref{app:loops}.

Note that we cannot normalise these correlators in the same way as \cref{eq:general-correlators-normalised}, because the noise two-point correlators $G_{\bar a\bar a}$, $G_{\bar r\bar r}$, and $G_{\bar r a}$ are all zero. A convenient normalisation can be obtained using the expectation value of the commutator, also known as the spectral function:
\begin{align}
    \rho^V(x_1-x_2) &= \bigg\langle \Big[\hat V(x_1),\hat V(x_2)\Big]\bigg\rangle
    = G^V_{21}(x_1-x_2) - G^V_{12}(x_1-x_2).
\end{align}
We define
\begin{align}\label{eq:spectral-renormalisation}
    \hat C^{VW}_{\bar a \bar a |\bar r a}(x_1,x_2;x_3,x_4)
    &= \frac{C^{VW}_{\bar a \bar a |\bar r a}(x_1,x_2;x_3,x_4)}
    {\rho^V(x_1,x_2)\,\rho^W(x_3,x_4)},
\end{align}
which tames part of the singularities arising from taking $x_1\approx x_2$ and $x_3\approx x_4$.

\subsubsection{Commutator-squared and antisymmetric OTOC for generic operators}

To get a better understanding of these OTOC definitions, let us compute them for generic operators discussed in \cref{sec:generic-otoc}. First, using \cref{eq:chainofreps,eq:delta-g,eq:bilocal-pert}, we find that to leading order in fluctuations bilocals in this basis are given as
\begin{align}
    \frac{V_{\bar a}(x_1)V_{\bar a}(x_2)}{\rho^V(x_1-x_2)}
    &= -\beta\Delta_V
    \Big( \partial_{t}
    + \omega_\beta 
    \coth\!\big(\omega_\beta|t_{12}|/2\big) \Big) 
    \Big( \Theta(t_{12})\,\delta \bar Q(x_2)
    - \Theta(t_{21})\,\delta \bar Q(x_1) \Big)
    + \ldots, \nn\\
    \frac{W_{\bar r}(x_3)W_{\bar r}(x_4) }{\rho^W(x_3-x_4)}
    &= -\frac{\beta\Delta_W}{4}
    \Big( {-}\partial_{t} + \omega_\beta 
    \coth(\omega_\beta|t_{34}|/2)\Big) 
    \Big( \Theta(t_{34})\,\delta Q(x_3)
    - \Theta(t_{43})\,\delta Q(x_4)
    \Big)
    + \ldots, \nn\\
    \frac{W_{\bar r}(x_3)W_{a}(x_4) }{\rho^W(x_3-x_4)}
    &= -\beta\Delta_W
    \Big( {-}\partial_{t} + \omega_\beta 
    \coth(\omega_\beta|t_{34}|/2)\Big) 
    \Big( \Theta(t_{34})\,\delta\bar F(x_3)
    - \Theta(t_{43})\,\delta\bar F(x_4)
    \Big)
    + \ldots.
\end{align}
In deriving this, the phase $\phi_{ij}$ in \cref{eq:chainofreps} plays an important role. We have also utilised normalisation using spectral functions as in \cref{eq:spectral-renormalisation}, given as
\begin{calign}
    \rho^V(x_1-x_2)
    &= \frac{-2i\mu_V}{\beta^{2\Delta_V}}
    \frac{\sgn(t_{12})\sin(\pi\Delta_V)}
    {\sinh^{2\Delta_V}\!\big(\pi|t_{12}|/\beta\big)}.
\end{calign}
We have skipped writing out the spatial delta-function for clarity.

Using this, we see that the $C_{\bar a\bar a|\bar ra}$ correlator is mediated exclusively by the $\langle \bar Q \bar F\rangle$ propagator. Assuming the ordering of times $t_1>t_2>t_3>t_4$, we find
\begin{calign}\label{eq:bilocal-diagram-EFT}
\tikzsetnextfilename{bilocal-diagram-EFT}
\raisebox{-8mm}{
\tikz[thick,scale=0.7]{
    \draw [dashed] (20mm,0mm) arc (0:160:10mm);
    \draw [dashed] (20mm,0mm) arc (0:-160:10mm);
    \draw [sol2] (20mm,0) -- (30mm,0);
    \draw [aux2] (30mm,0) -- (40mm,0);
    \draw [dashed] (40mm,0mm) arc (180:20:10mm);
    \draw [dashed] (40mm,0mm) arc (-180:-20:10mm);
    \draw [fill=black] (0.6mm,3.4mm) circle (1.25mm);
    \draw [fill=black] (0.6mm,-3.4mm) circle (1.25mm);
    \draw [fill=white] (59.4mm,3.4mm) circle (1.25mm);
    \draw [fill=white] (59.4mm,-3.4mm) circle (1.25mm);
    \node at (-11mm,5mm) {$W_{a}(x_4)$};
    \node at (-11mm,-5mm) {$W_{\bar r}(x_3)$};
    \node at (70mm,-5mm) {$V_{\bar a}(x_2)$};
    \node at (70mm,5mm) {$V_{\bar a}(x_1)$};
}
}
\end{calign}
\begin{calign}
    \hat C^{VW}_{\bar a\bar a|\bar r a}&(x_1,x_2;x_3,x_4) \nn\\
    &= \beta^2\Delta_V\Delta_W
    \Big( \partial_{t}
    + \omega_\beta 
    \coth\!\big(\omega_\beta t_{12}/2\big) \Big)
    \Big(\partial_{t} + \omega_\beta 
    \coth(\omega_\beta t_{34}/2)\Big)
    iK^{-1}(x_{23}).
\end{calign}
Here we have denoted $\delta \bar Q$ with a double wavy line and $\delta \bar F$ with double solid line. 
This is the SK EFT analogue of the schematic bilocal diagram \eqref{eq:bilocal-diagram} from the Introduction. 
This coincides with the OTOC defined in \cref{eq:bilocal-otoc-general} for $t_2>t_3$.

We also see that the $C_{\bar a\bar a|\bar r\bar r}$ correlator would be mediated by the $\langle \bar Q Q\rangle$ propagator. However, since this propagator is identically zero in the four-fold SK EFT, we see that, at least at linear order in fluctuations, the commutator-squared OTOC is zero for generic local operators. While this stronger conclusion is not necessarily true in the presence of interactions; we illustrate in \cref{app:symmeric-otoc} that the commutator-squared OTOC of the energy density does not admit a growing part.

\subsection{Energy-current OTOCs from four-fold SK EFT}

In \cref{sec:four-fold-theory}, we constructed a four-fold SK EFT for computing OTOCs of local operators. We used this theory to recover the OTOCs of generic few-body local operators whose EFT representation is given by $\SL(2,\bbR)$-invariant bilocals. At tree level, the OTOC of such operators is mediated by a single field exchange, as in \cref{eq:bilocal-diagram}. As explained in the introduction, computing the OTOC in this limit does not require the full machinery of the four-fold SK EFT and can instead be carried out directly within a two-fold SK EFT~\cite{Blake:2017ris}. The four-fold theory becomes necessary, however, once one considers loop corrections involving multiple field exchanges.

While these observations apply to generic few-body operators, collective hydrodynamic operators, such as the energy-momentum tensor or charge current, are special. To see why, recall that the generic local operators considered in \cref{sec:generic-otoc} need not themselves be $\SL(2,\bbR)$-invariant; it is only their bilocal representation in the EFT that is required to be $\SL(2,\bbR)$-invariant. Consequently, the bilocal couples directly to the scramblon mode at linear order in fluctuations, as in \cref{eq:bilocal-pert}, and the tree-level OTOC can be obtained without invoking interactions in the four-fold SK EFT. By contrast, collective hydrodynamic operators are themselves locally $\SL(2,\bbR)$-invariant. Their associated bilocals, therefore, couple to the scramblon modes only at non-linear order in fluctuations. Their OTOCs consequently probe interactions between fluctuations on the different SK folds, making the full four-fold SK EFT essential even at tree level.

In the following, we will illustrate this principle using the OTOC of the energy current operator in the four-fold SK EFT of energy diffusion. We find that the commutator-squared OTOC is not sensitive to Lyapunov growth, as generically expected for maximally chaotic theories~\cite{Blake:2021wqj}, though the antisymmetric energy current OTOC does feature a growing part. The growth in time is controlled by the maximal Lyapunov exponent $\lambda = \omega_\beta$. However, unlike generic operators discussed in \cref{eq:BilocalOTOC}, the characteristic front velocity for spatial scrambling is found to either be $v_b$ or $v_b/\sqrt{2}$ depending on the short-distance behaviour of the operator insertions at $x_1\approx x_2$ and $x_3\approx x_4$. We also match our results with a shockwave computation.

Let us compute the antisymmetric OTOC $C_{\bar a \bar a |\bar r a}$ defined in \cref{eq:OTOC-combos} at tree level using our four-fold SK EFT. In particular, we will be interested in extracting the part of $C_{\bar a \bar a |\bar r a}$ that is sensitive to the scramblon modes and grows exponentially at late times.
The computation of the symmetric OTOC $C_{\bar a \bar a |\bar r \bar r}$ has been relegated to \cref{app:loops}; we find that it is insensitive to the growing modes.
The possible tree-level diagrams contributing to four-point correlation functions of local operators can be classified into s-channel, t-channel, u-channel, and contact diagrams. Schematically, for $C_{\bar a\bar a\bar ra}$, they are given as
\begin{cgather}
\tikzsetnextfilename{2loop-diagram-s-1-main}
\tikz[thick,scale=0.7]{
    \draw [aux] (10mm,10mm) arc (90:160:10mm);
    \draw [] (10mm,10mm) arc (90:0:10mm);
    \draw [aux2] (20mm,0) arc (0:-90:10mm);
    \draw [sol2] (10mm,-10mm) arc (-90:-160:10mm);
    \draw [sol2] (20mm,0) -- (30mm,0);
    \draw [aux2] (30mm,0) -- (40mm,0);
    \draw [sol2] (40mm,0) arc (180:90:10mm);
    \draw [aux2] (50mm,10mm) arc (90:20:10mm);
    \draw [aux2] (50mm,-10mm) arc (-90:-20:10mm);
    \draw [sol2] (50mm,-10mm) arc (-90:-180:10mm);
    \draw [fill=black] (0.6mm,3.4mm) circle (1.25mm);
    \draw [fill=black] (0.6mm,-3.4mm) circle (1.25mm);
    \draw [fill=white] (59.4mm,3.4mm) circle (1.25mm);
    \draw [fill=white] (59.4mm,-3.4mm) circle (1.25mm);
    \node at (-11mm,5mm) {$W_{a}(x_4)$};
    \node at (-11mm,-5mm) {$\cO^{W'}_{\bar r}(x_3)$};
    \node at (70mm,-5mm) {$\cO^{V'}_{\bar a}(x_2)$};
    \node at (70mm,5mm) {$V_{\bar a}(x_1)$};
    \node at (30mm,-16mm) {s-channel};
} \nn\\
\tikzsetnextfilename{2loop-diagram-t-1-main}
\tikz[thick,scale=0.7]{
    \draw [aux] (4.4mm,10.6mm) arc (135:165:15mm);
    \draw [] (15mm,15mm) arc (90:135:15mm);
    \draw [sol2] (15mm,15mm) arc (90:45:15mm);
    \draw [aux2] (25.6mm,10.6mm) arc (45:15:15mm);
    \draw [aux2] (15mm,15mm) -- (15mm,0);
    \draw [sol2] (15mm,0) -- (15mm,-15mm);
    \draw [sol2] (4.4mm,-10.6mm) arc (-135:-165:15mm);
    \draw [aux2] (15mm,-15mm) arc (-90:-135:15mm);
    \draw [sol2] (15mm,-15mm) arc (-90:-45:15mm);
    \draw [aux2] (25.6mm,-10.6mm) arc (-45:-15:15mm);
    \draw [fill=black] (0.6mm,3.4mm) circle (1.25mm);
    \draw [fill=black] (0.6mm,-3.4mm) circle (1.25mm);
    \draw [fill=white] (29.4mm,3.4mm) circle (1.25mm);
    \draw [fill=white] (29.4mm,-3.4mm) circle (1.25mm);
    \node at (-11mm,5mm) {$W_{a}(x_4)$};
    \node at (-11mm,-5mm) {$\cO^{W'}_{\bar r}(x_3)$};
    \node at (40mm,-5mm) {$\cO^{V'}_{\bar a}(x_2)$};
    \node at (40mm,5mm) {$V_{\bar a}(x_1)$};
    \node at (15mm,-22mm) {t-channel};
}  \qquad
\tikzsetnextfilename{2loop-diagram-u-1-main}
\tikz[thick,scale=0.7]{
    \draw [aux] (4.4mm,10.6mm) arc (135:165:15mm);
    \draw [] (15mm,15mm) arc (90:135:15mm);
    \draw [sol2] (15mm,15mm) -- (22.5mm,5.5mm);
    \draw [aux2] (22.5mm,5.5mm) -- (30mm,-4mm);
    \draw [aux2] (15mm,15mm) -- (15mm,0);
    \draw [sol2] (15mm,0) -- (15mm,-15mm);
    \draw [sol2] (4.4mm,-10.6mm) arc (-135:-165:15mm);
    \draw [aux2] (15mm,-15mm) arc (-90:-135:15mm);
    \draw [fill=white, draw=none, ,rotate around={50:(27mm,0)}] (26.9mm-1.5mm,-1.5mm) rectangle +(3mm,3mm);
    \draw [sol2] (15mm,-15mm) -- (22.5mm,-5.5mm);
    \draw [aux2] (22.5mm,-5.5mm) -- (30mm,4mm);
    \draw [fill=black] (0.6mm,3.4mm) circle (1.25mm);
    \draw [fill=black] (0.6mm,-3.4mm) circle (1.25mm);
    \draw [fill=white] (29.4mm,3.4mm) circle (1.25mm);
    \draw [fill=white] (29.4mm,-3.4mm) circle (1.25mm);
    \node at (-11mm,5mm) {$W_{a}(x_4)$};
    \node at (-11mm,-5mm) {$\cO^{W'}_{\bar r}(x_3)$};
    \node at (40mm,-5mm) {$\cO^{V'}_{\bar a}(x_2)$};
    \node at (40mm,5mm) {$V_{\bar a}(x_1)$};
    \node at (15mm,-22mm) {u-channel};
}
\nn\\[1em]
\tikzsetnextfilename{2loop-diagram-contact-1-main}
\raisebox{-6mm}{
\tikz[thick,scale=0.8]{
    \draw [sol2] (20mm,0) arc (180:90:10mm);
    \draw [aux2] (30mm,10mm) arc (90:20:10mm);
    \draw [aux2] (30mm,-10mm) arc (-90:-20:10mm);
    \draw [sol2] (30mm,-10mm) arc (-90:-180:10mm);
    \draw [aux] (10mm,10mm) arc (90:160:10mm);
    \draw [] (10mm,10mm) arc (90:0:10mm);
    \draw [aux2] (20mm,0) arc (0:-90:10mm);
    \draw [sol2] (10mm,-10mm) arc (-90:-160:10mm);
    \draw [fill=black] (0.6mm,3.4mm) circle (1.25mm);
    \draw [fill=black] (0.6mm,-3.4mm) circle (1.25mm);
    \draw [fill=white] (39.4mm,3.4mm) circle (1.25mm);
    \draw [fill=white] (39.4mm,-3.4mm) circle (1.25mm);
    \node at (-10mm,5mm) {$W_{a}(x_4)$};
    \node at (-10mm,-5mm) {$\cO^{W'}_{\bar r}(x_3)$};
    \node at (49mm,-5mm) {$\cO^{V'}_{\bar a}(x_2)$};
    \node at (49mm,5mm) {$V_{\bar a}(x_1)$};
    \node at (20mm,-17mm) {Contact};
}}
\end{cgather}
We have denoted $\delta F$ with solid lines, $\delta Q$ with wavy lines, $\delta \bar Q$ with double wavy lines, and $\delta \bar F$ with double solid lines. The two operator insertions at $x_1,x_2$ are denoted with white circles, while the two insertions at $x_3,x_4$ with black circles.
We will work in position space, therefore we need to integrate over the spacetime positions of all the interaction vertices in a diagram. Since the operators of interest are generically non-linear in the fundamental fields, as for the energy current in \cref{eq:linear-expansion-E}, there are also several related diagrams in every channel involving multiple propagators connected to the operator insertions. We go over these in detail in \cref{app:loops}.

Having gone through the computation, we find that only the s-channel diagrams contribute to growth. Performing a partial Fourier transform in the spatial directions, we find that the growing part is given as
\begin{calign}\label{eq:OTOC-result}
    C^{VW}_{\bar a \bar a |\bar r a}&(x_1,x_2;x_3,x_4) \nn\\
    &\sim \frac{\omega_\beta^5}{4\pi i} \E{\omega_\beta (t_2-t_3)}
    \int \frac{\df^d\bk}{(2\pi)^d}
    \frac{\E{i\bk\cdot(\bx_2-\bx_3)}c_\bk}{\cD_\bk+\omega_\beta} \,
    \cI^V_d(x_1-x_2;\bk) \,
    \cI^W_d(x_3-x_4;\bk).
\end{calign}
where we have assumed the time-ordering of operator insertions $t_1>t_2>t_3>t_4$ and used $\Delta(\dow_t,i\bk) = \beta c_\bk(\dow_t+\cD_\bk)$. The coefficients $\cI_d$ appearing above depend on the precise form the energy diffusion SK EFT. It also changes depending on the operators involved in the OTOC.

\subsubsection{OTOC of energy density}

In particular, for SYK-like models with $\cR(\dow_1,\dow_2)=0$ (which also requires us to set $c_\bk=c_V$; see \cref{foot:cV-footnote}) and choosing the operators of interest in the OTOC to be energy density, we find
\begin{calign}\label{eq:general-integral-energydensity}
    \cI_d(y;\bk)
    &= \int\frac{\df^d\bq}{(2\pi)^d}
    \E{i\bq\cdot\by}
    \E{-\cD_{\bq}|y^0|} 
    \frac{\cD_{\bq}}{\omega_\beta}
    \lb 1-\frac{\cD_{\bq}^2}{\omega_\beta^2}\rb 
    \lb 2 + \frac{\cD_{\bq}}{\omega_\beta} \rb
    \frac{\cD_{\bq} + \cD_{\bk-\bq} - \cD_{\bk}}
    {\cD_{\bq} + \cD_{\bk-\bq} + \omega_\beta}.
\end{calign}
The more general result is given in \cref{eq:general-integral}.

A systematic analysis of the Fourier transform to position space is beyond the scope of this paper. Instead we will analyse the example of continuum theory in the limit, where the regime $x_1\approx x_2$ and $x_3\approx x_4$, which amounts to taking $y\to 0$ in the integral above. This limit is expected to correspond to $\bq\to\infty$, in which case
\es{eq:smalldist}{
\cI_d(y;\bk) 
&\sim - \int\frac{\df^d\bq}{(2\pi)^d}
    \E{i\bq\cdot\by}
    \E{-\cD_{\bq}|y^0|} 
    \frac{\cD_{\bq}^4}{\omega_\beta^4}\,,
}
independently of $\bk$. Then the $\bk$ integral is the same as we encountered in the discussion of the OTOC of generic operators around~\cref{eq:BilocalOTOC}.

There is one potential subtlety. Depending on precisely how we take the $y\to 0$ limit, it could be that the above leading expression for $\cI_d(y;\bk)$ vanishes, in which case we need to analyse corrections to the formula. We present an example below. 

\begin{myshaded}
\paragraph*{Microscopic examples.}

A consistency check is provided by (1+1)d CFT, where $\cD_\bk = i\bk$ and $c_\bk=c_V$ is a constant, thus the coefficient $\cI_d$ identically vanishes. Indeed, the analytic structure of the energy-momentum correlators in (1+1)d CFT  is entirely fixed by the infinite dimensional Virasoro symmetry, and from the explicit form it is not hard to conclude that the associated OTOC does not grow. In \cref{app:2dcft}, we apply our SK EFT to recover the full analytic structure of the energy-momentum OTOC in (1+1)d CFT.

Next, we analyse pure diffusion with dispersion relation $\cD_\bk = D\bk^2$, as well as choose $c_\bk=c_V$. Since in the SYK lattice $D$ is independent of the temperature and for scrambling only wavenumbers obeying $|\cD_\bk|\lesssim\omega_\beta$ are important, at very low temperatures the physics of the SYK lattice is well-approximated by pure diffusion.
In this case, \cref{eq:general-integral-energydensity} yields
\begin{calign}\label{eq:diff-OTOC-result}
    \cI_d(y;\bk)
    &= 
    \int \frac{\df^d\bq}{(2\pi)^d}\,
    \E{i\bq\cdot\by-D|y^0|\bq^2} 
    \frac{\bq^2}{k_*^2}
    \lb 1- \frac{\bq^4}{k_*^4}\rb 
    \lb 2 + \frac{\bq^2}{k_*^2} \rb
    \frac{2\bq^2 -2 \bk\cdot\bq}
    {2\bq^2 -2 \bk\cdot\bq + \bk^2 + k_*^2},
\end{calign}
where $k_*=\sqrt{\omega_\beta/D} = \omega_\beta/v_b$ is the characteristic wavevector scale of the theory.
We will focus on the
$y\to 0$ limit in the integral above. The qualitative behaviour of the integral depends on the order in which we take the spatial or temporal component of $y$ to zero. 

First we take $y_0\to 0$ first,  followed by taking $\by$ to be small. In this case,~\cref{eq:smalldist} takes the form
\es{eq:smalldist2}{
\cI_d(y;\bk) 
&\sim - \int\frac{\df^d\bq}{(2\pi)^d}
    \E{i\bq\cdot\by}
    \frac{{\bq}^8}{k_*^8}\,,
}
which is a pure contact term, and vanishes for any infinitesimal $\by\neq0$. We analyse the subleading behaviour of $\cI_d(y;\bk)$ in  \cref{eq:energy-otoc-computation} for this case, and find that it leads to a narrower spatial profile of growth of the OTOC with front velocity $v_b/\sqrt{2}$.

Next, we choose to take $\by\to 0$ first, followed by taking $y^0$ to be small. Note that in the bilocal case, there was only the possibility of time separation, so this is natural.
In this case, we find that $\cI_d(y;\bk)$ is dominated by
\begin{calign}
\cI_d(y;\bk) 
&\sim
- \left(\frac{\omega_\beta}{4\pi D}\right)^{d/2}
\frac{d(d+2)(d+4)(d+6)}
{16(\omega_\beta |y^0|)^{4+d/2}}.
\end{calign}
It diverges as $y^0\to 0$ and dominates the magnitude of the OTOC. We can tame part of the singularity by normalising with the energy-density spectral function 
\begin{calign}
    \rho^{\cE^t}(y) 
    &= \frac{c_V\omega_\beta}{i\beta} 
    \sgn(y^0) \int \frac{\df^d\bq}{(2\pi)^d}\E{i\bq\cdot\by}
    \E{-D\bq^2|y^0|} 
    \frac{\bq^2}{k_*^2}
    \lb 1 - \frac{\bq^4}{k_*^4} \rb  \nn\\
    &\sim 
    \frac{ic_V\omega_\beta}{\beta} 
    \sgn(y^0)
    \left(\frac{\omega_\beta}{4\pi D}\right)^{d/2}
\frac{d(d+2)(d+4)}
{8(\omega_\beta |y^0|)^{3+d/2}} .
\end{calign}
where we have assumed $c_\bk=c_V$. With this prescription, we find
\begin{calign}
    \hat C^{\cE^t\cE^t}_{\bar a \bar a |\bar r a}(x_1,x_2,x_3,x_4)
    &\sim \frac{i\beta(d+6)^2}{8c_V|t_{12}t_{34}|}
    \int \frac{\df^d\bk}{(2\pi)^d}
    \frac{\E{\omega_\beta t+i\bk\cdot\bx}}{\omega_\beta +D\bk^2} \,.
\end{calign}
where $t=t_2-t_3$ and $\bx=\bx_2-\bx_3$. In analogy with the bilocal discussion around \cref{eq:bilocal-otoc2}, this integral is controlled by the pole at $|\bk|= ik_* = i\sqrt{\omega_\beta/D}$, leading to the Lyapunov growth $\E{\omega_\beta\lb t-|\bx|/v_b\rb}$, with the butterfly velocity $v_b = \sqrt{\omega_\beta D}$. We explicitly find the large distance behaviour
\begin{calign}
    \hat C^{\cE^t\cE^t}_{\bar a \bar a |\bar r a}(x_1,x_2,x_3,x_4)
    &\sim 
    \frac{i\beta^2 (d+6)^2 k_*^{d/2+1/2}}{32\pi c_V|t_{12}t_{34}|}
\frac{
\E{\omega_\beta(t-|\bx|/v_b)}
}{|2\pi \bx|^{(d-1)/2}}.
\end{calign}
While in 1 spatial dimension the spatial profile of growth is purely exponential, in higher-dimensions there is an additional polynomial suppression at large distances. Notice that this has the same qualitative behaviour as for generic operators in \cref{eq:BilocalOTOC}.
\end{myshaded}

\subsubsection{OTOCs of energy flux}

The above computation of the antisymmetric OTOC can be extended to arbitrary local operators. A detailed derivation appears in \cref{app:loops}.
In particular, for the full energy density and flux, we find the generalisation of \cref{eq:OTOC-result} as follows
\begin{calign}\label{eq:energy-current-OTOC-result}
    &\Big\langle 
    \cE^\mu_{\bar a}(x_1)
    \cE^\nu_{\bar a}(x_2)
    \cE^\rho_{\bar r}(x_3)
    \cE^\sigma_{a}(x_4)
    \Big\rangle \nn\\
    &\qquad 
    \sim \frac{\omega_\beta^5}{4\pi i} \E{\omega_\beta (t_2-t_3)}
    \int \frac{\df^d\bk}{(2\pi)^d}
    \frac{\E{i\bk\cdot(\bx_2-\bx_3)}c_\bk}{\cD_\bk+\omega_\beta} \,
    \cI^{\mu\nu}_d(x_1-x_2;\bk) \,
    \cI^{\sigma\rho}_d(x_3-x_4;\bk),
\end{calign}
where we have assumed $t_1>t_2>t_3>t_4$. For SYK-like models with $\cR(\dow_1,\dow_2)=0$ and $c_\bk=c_V$, the coefficients appearing above are given as
\begin{calign}\label{eq:general-integral-energycurrent}
    \cI^{\mu\nu}_d(y;\bk)
    &= \int\frac{\df^d\bq}{(2\pi)^d}
    \E{i\bq\cdot\by}
    \E{-\cD_{\bq}|y^0|} \,
    \,
    \frac{\cD_{\bq}}{\omega_\beta}
    \lb 1-\frac{\cD_{\bq}^2}{\omega_\beta^2}\rb 
    \frac{\cD^\mu_{(1)}(\omega_\beta,i\bq)}{c_V} \nn\\
    &
    \Bigg[
    \lb 2 + \frac{\cD_{\bq}}{\omega_\beta} \rb
    \frac{\cD_{\bq} + \cD_{\bk-\bq} - \cD_{\bk}}
    {\cD_{\bq} + \cD_{\bk-\bq} + \omega_\beta}
    \frac{\cD^\nu_{(1)}(\omega_\beta,i\bk-i\bq)}{c_V}
    + \frac{\cY^{\nu}_{-\bq,{\bk}}}{\pi c_V} \Bigg],
\end{calign}
where the coefficient $\cY^\mu=(0,\cY^\bx)$ is given in \cref{eq:Ymu-simple}.
Recall that $\cD_{(1)}^\mu$ denotes the linear part of the energy current operator defined in \cref{eq:linear-expansion-E}, which takes the specific form for SYK chain in \cref{eq:SYK-energy-ops}. In particular, under our assumption of the form of $\Delta(\dow)$ in \cref{eq:general-Delta}, note that $\cD^t_{(1)}(\omega_\beta,i\bk) = c_V$ and $i\bk\cdot\cD^\bx_{(1)}(\omega_\beta,i\bk) = c_V\cD_\bk$. Therefore, \cref{eq:general-integral-energycurrent} agrees with the special case of energy density in \cref{eq:general-integral-energydensity}. Using these relations, it is also easy to check that the energy current OTOC in \cref{eq:energy-current-OTOC-result} satisfies the Ward identities.

\subsection{The shockwave description}
\label{sec:shockwaves}

\subsubsection{OTOC of bilocal operators}

We will now recover our results for the energy density OTOC using the scramblon shockwave prescription. To this end, it is helpful to first revisit the OTOC of generic operators computed from  bilocal insertions in the EFT. 
The result \cref{eq:bilocal-otoc2} is already of the correct shockwave form given in the Introduction. For that we need to notice that for times $t$ before $t_\text{scr}\sim \beta\log(N)$ the factor in the exponent in~\cref{scramblonPrescription} is large, so we can perform saddle point integration. The leading order answer is just a product of two two-point functions. The first $1/N$ correction comes from expanding the perturbed Green's function to first order in $X^\pm$ leading to
\es{eq:shockwaveMain}{
    &\hat C^{VW}_{42|31}(x_1+x,x_2+x;x_3,x_4) \\
    &\sim \int \frac{\df^d \bk}{(2 \pi)^d}\,
    {\dow_{X^-(\bk) }{\cal G}_V(x_1,x_2\vert X^-) \ov {\cal G}_V(x_1,x_2\vert X^-)}\Bigg\vert_{X^-=0}\,
    {\dow_{X^+(-\bk)}{\cal G}_W(x_3,x_4\vert X^+)\ov {\cal G}_W(x_3,x_4\vert X^+) }\Bigg\vert_{X^+=0} \, 
    \frac{i\alpha(\bk)}{N}\,
    \E{\lambda(\bk) t+ i \bk\cdot\bx}
    \,,
}
where $i\alpha(\bk)/N\,\E{\lambda(\bk) t}$ is the propagator $\langle X^+(-\bk)X^-(\bk)\rangle$.
Structurally this matches~\cref{eq:bilocal-otoc2}. Below we show that the ingredients indeed match precisely. 

\begin{figure}[t]
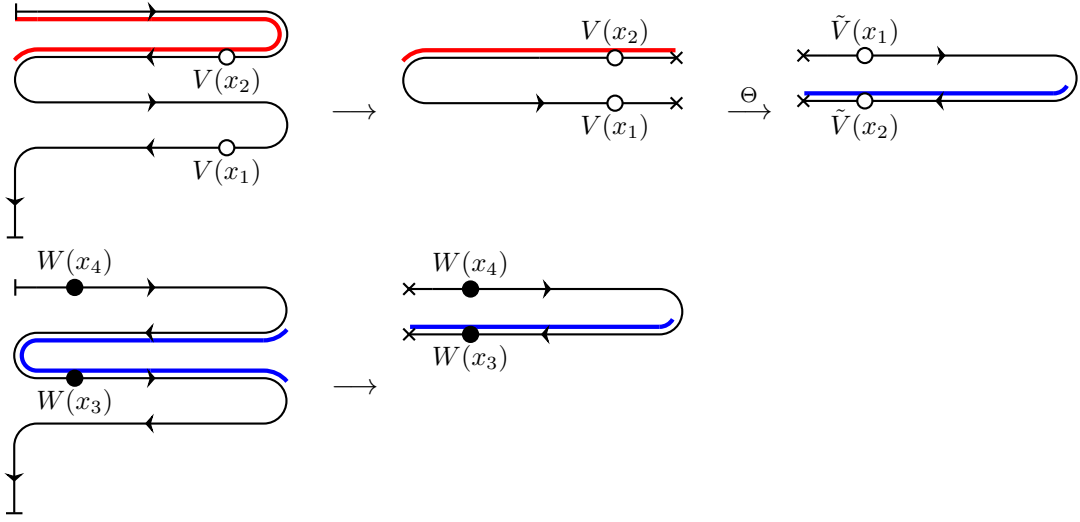

\centering
\begin{tabular}{ccc}
\tikzsetnextfilename{shockwaves-growing-1}
\raisebox{-16mm}{
\tikz[thick]{
    \draw[red,ultra thick] (-0.3,0-0.1) -- (0,0-0.1);
    \draw[red,ultra thick] (0,0-0.1) -- (1.5,0-0.1);
    \draw[red,ultra thick] (1.5,0-0.1) -- (3,0-0.1) arc (90:-90:.2) -- (1.5,-.6+0.1);
    \draw[red,ultra thick] (1.5,-.6+0.1) -- (0,-.6+0.1) arc (90:140:.4);
    \draw[|-] (-0.3,0) -- (0,0);
    \draw[|-,right] (0,0) -- (1.5,0);
    \draw[right] (1.5,0) -- (3,0) arc (90:-90:.3) -- (1.5,-.6);
    \draw[right] (1.5,-.6) -- (0,-.6) arc (90:270:.3) -- (1.5,-1.2);
    \draw[right] (1.5,-1.2) -- (3,-1.2) arc (90:-90:.3) -- (1.5,-1.8);
    \draw[right] (1.5,-1.8) -- (0,-1.8) arc (90:180:.3) -- (-0.3,-2.5);
    \draw[-|] (-0.3,-2.5) -- (-0.3,-3);
    \draw [fill=white] (2.5,-0.6) circle (1mm);
    \draw [fill=white] (2.5,-1.8) circle (1mm);
    \node at (2.5,-.6-.32) {\small $V(x_2)$};
    \node at (2.5,-1.8-.32) {\small $V(x_1)$};
}
}
&$\longrightarrow$\hspace{-.2em}
\tikzsetnextfilename{shockwaves-growing-2}
\raisebox{-4.2mm}{
\tikz[thick]{
    \draw[red,ultra thick] (3.3,-.6+0.1) -- (0,-.6+0.1) arc (90:140:.4);
    \draw[{Rays}-] (3.4,-.6) -- (1.5,-.6);
    \draw[right] (1.5,-.6) -- (0,-.6) arc (90:270:.3) -- (1.5,-1.2);
    \draw[-{Rays}] (1.5,-1.2) -- (3.4,-1.2);
    \draw [fill=white] (2.5,-0.6) circle (1mm);
    \draw [fill=white] (2.5,-1.2) circle (1mm);
    \node at (2.5,-.6+.33) {\small $V(x_2)$};
    \node at (2.5,-1.2-.32) {\small $V(x_1)$};
}
}
&$\overset{\Theta}{\longrightarrow}$\hspace{-.2em}
\tikzsetnextfilename{shockwaves-growing-3}
\raisebox{-4.2mm}{
\tikz[thick]{
    \draw[blue,ultra thick] (3,-.6+0.1) arc (-90:-30:.2);
    \draw[blue,ultra thick] (3,-.6+0.1) -- (-0.3,-.6+0.1);
    \draw[{Rays}-] (-0.4,0) -- (0,0);
    \draw[|-,right] (0,0) -- (1.5,0);
    \draw[right] (1.5,0) -- (3,0) arc (90:-90:.3) -- (1.5,-.6);
    \draw[-{Rays}] (1.5,-.6) -- (-0.4,-.6);
    \draw [fill=white] (.5,-0) circle (1mm);
    \draw [fill=white] (.5,-.6) circle (1mm);
    \node at (.5,0+.33) {\small $\tilde V(x_1)$};
    \node at (.5,-.6-.32) {\small $\tilde V(x_2)$};
}
} \\[1em]
\tikzsetnextfilename{shockwaves-4}
\raisebox{-16mm}{
\tikz[thick]{
    \draw[blue,ultra thick] (3,-0.6-0.1) arc (-90:-40:.4);
    \draw[blue,ultra thick] (3,-0.6-0.1) -- (1.5,-.6-0.1);
    \draw[blue,ultra thick] (1.5,-.6-0.1) -- (0,-.6-0.1) arc (90:270:.2) -- (1.5,-1.2+0.1);
    \draw[blue,ultra thick] (1.5,-1.2+0.1) -- (3,-1.2+0.1) arc (90:40:.4);
    \draw[|-] (-0.3,0) -- (0,0);
    \draw[|-,right] (0,0) -- (1.5,0);
    \draw[right] (1.5,0) -- (3,0) arc (90:-90:.3) -- (1.5,-.6);
    \draw[right] (1.5,-.6) -- (0,-.6) arc (90:270:.3) -- (1.5,-1.2);
    \draw[right] (1.5,-1.2) -- (3,-1.2) arc (90:-90:.3) -- (1.5,-1.8);
    \draw[right] (1.5,-1.8) -- (0,-1.8) arc (90:180:.3) -- (-0.3,-2.5);
    \draw[-|] (-0.3,-2.5) -- (-0.3,-3);
    \draw [fill=black] (.5,0) circle (1mm);
    \draw [fill=black] (.5,-1.2) circle (1mm);
    \node at (.5,0+.33) {\small $W(x_4)$};
    \node at (.5,-1.2-.32) {\small $W(x_3)$};
}
}
&$\longrightarrow$\hspace{-.2em}
\tikzsetnextfilename{shockwaves-growing-5}
\raisebox{1.7mm}{
\tikz[thick]{
    \draw[blue,ultra thick] (3,-.6+0.1) arc (-90:-30:.2);
    \draw[blue,ultra thick] (3,-.6+0.1) -- (-.3,-.6+0.1);
    \draw[{Rays}-] (-0.4,0) -- (0,0);
    \draw[right] (0,0) -- (1.5,0);
    \draw[right] (1.5,0) -- (3,0) arc (90:-90:.3) -- (1.5,-.6);
    \draw[-{Rays}] (1.5,-.6) -- (-0.4,-.6);
    \draw [fill=black] (.5,-0) circle (1mm);
    \draw [fill=black] (.5,-.6) circle (1mm);
    \node at (.5,0+.33) {\small $W(x_4)$};
    \node at (.5,-.6-.32) {\small $W(x_3)$};
}
}
\end{tabular}
\caption{We can map the two-fold correlators on the growing and decaying shockwave configurations to the two-fold contour by moving around operators along the closed time contour (without crossing boundaries of the shockwave). 
In the growing shockwave case, we need to also use $\Theta$ time-reversal symmetry to map to the conventional two-fold contour, which maps the growing shockwave to a decaying shockwave instead. Note that the decaying shockwave on a two-fold contour preserves the trace boundary conditions at $+\infty$, but non-trivially modify the initial state. Therefore the two-fold representation corresponds to a non-trivial non-thermal initial state.}
\label{fig:shocks-maps}
\end{figure}

In the Introduction we also explained that the shockwave $X^-$ is nothing but an $\SL(2,\bbR)$ transformation of the thermal saddle $F_\eqb(t)={t/ \beta}$ on the folds ``12'' and no transformation on ``34'', while $X^+$ is active on ``23'' and not on ``14''; see \cref{fig:shocks}. At leading order in $1/N$, we get ${\cal G}_W$ by evaluating the bilocal on a saddle point shifted by a finite decaying $\SL(2,\bbR)$ transformation on fold ``2'' as $\E{2\pi F_\eqb(t)} = \E{\omega_\beta t} + X^+(\bx)$, but not on ``$1$''. We can also obtain ${\cal G}_V$ through the same decaying shockwave configuration of fold ``2'', but with $\Theta$-transformed parameter $\tilde X^-(\bx)=X^-(\eta_\rmP\bx)$; see \cref{fig:shocks-maps}. We have~\cite{Maldacena:2016upp}
\begin{calign}
    {\cal G}_V(x_1,x_2\vert X^-)
    &= \Big\langle \tilde V_1(x_1) \tilde V_2(x_2)  \Big\rangle_{\tilde X^-}, \nn\\
    &= \mu_V\le(
    \frac{\E{-i\pi\sgn(t_{12})}/\beta^2}{\left(\sinh(\omega_\beta t_{12}/2)
    + \half X^-(\bx_2) \E{\omega_\beta(t_1+t_2)/2}\right)^2}
    \ri)^{\Delta_V}
    \delta^{(d)}(\bx_1-\bx_2)\,, \nn\\
    {\cal G}_W(x_3,x_4\vert X^+)
    &= \Big\langle W_2(x_3) W_1(x_4)  \Big\rangle_{X^+} \nn\\
    &= \mu_W\le(
    \frac{\E{-i\pi\sgn(t_{34})}/\beta^2}{\left(\sinh(\omega_\beta t_{34}/2)
    + \half X^+(\bx_3) \E{-\omega_\beta(t_3+t_4)/2}\right)^2}
    \ri)^{\Delta_W}
    \delta^{(d)}(\bx_3-\bx_4).
\end{calign}
Recall that $\tilde V(x) = V(-t,\eta_\rmP\bx)$. This yields
\begin{calign}
    {\dow_{X^-(\bk) }{\cal G}_V(x_1,x_2\vert X^-) \ov {\cal G}_V(x_1,x_2\vert X^-)}\Bigg\vert_{X^-=0}\,
    &= - \frac{
    \Delta_V \E{\omega_\beta(t_1+t_2)/2}\,
    \E{i\bk\cdot\bx_2}}{\sinh(\omega_\beta t_{12}/2)} \nn\\
    {\dow_{X^+(-\bk) }{\cal G}_W(x_3,x_4\vert X^+) \ov {\cal G}_W(x_3,x_4\vert X^+)}\Bigg\vert_{X^+=0}\,
    &= - \frac{
    \Delta_W \E{-\omega_\beta(t_3+t_4)/2}\,
    \E{-i\bk\cdot\bx_3}}{\sinh(\omega_\beta t_{34}/2)}.
\end{calign}
Hence the vertex factors indeed reproduce the dependence of~\cref{eq:bilocal-otoc2}  on $t_i$. This enables us to read off the coefficient $\alpha(\bk)$ as
\es{eq:match}{
    \frac{\alpha(\bk)}{N}&= 
    \frac{\pi}{c_\bk}\,
    \frac{\omega_\beta}{\omega_\beta+\cD_{\bk}}\,,
}
while the Lyapunov exponent takes the $\bk$-independent maximal value $\lambda(\bk)=\omega_\beta$. Therefore, we find a precise agreement between the Feynman diagram computation~\eqref{eq:bilocal-otoc2} and the shockwave description~\eqref{eq:shockwaveMain}. 

The last ingredient, $\alpha(\bk)$ is determined by the action of the shockwaves. It can be obtained by plugging the shockwave configuration into the action and working to quadratic order in the amplitudes. However, this computation requires one to keep careful track of boundary terms in time, which we have been dropping freely throughout this paper as they do not affect the computation of OTOCs using Feynman diagrams. While it would be interesting to recover the shockwave action purely from the SK EFT, we will only obtain it from more microscopic starting points.

\begin{myshaded}
\paragraph*{Microscopic examples.}

For the Schwarzian lattice reparametrization model~\eqref{eq:final_action} the details are worked out in~\cref{app:scramblon_action}, producing exactly~\cref{eq:match}. Curiously, this computation requires the contribution from the thermal (Euclidean) part of the time-contour. We expect that this contribution can be moved around by adding total derivatives (in an appropriate sense applicable to non-local theories), so that the full contribution comes from the Lorentzian part of the time-contour.

The shockwave action defined in~\cref{scramblonPrescription} was computed for the large-$p$ SYK chain for all temperatures using its microscopic collective fields~\cite{Choi:2023mab}. Taking  the low temperature limit of (5.10) of this reference, we find  
\es{510}{
{\alpha(\bk)\ov N}={p^2\ov N}\,{2\sqrt{\pi}i\E{-i\pi\kappa/2}\ov \cos\le(\pi\kappa/2\ri)\cos\le(\pi\kappa/v\ri)\Gamma\le(1+{\kappa\ov v}\ri)\Gamma\le(\frac12-{\kappa\ov v}\ri)}\, \to\, \frac{\pi}{c_V}\,
    \frac{\omega_\beta}{\omega_\beta+\cD_\bk} \,,
}
where $v$ is a proxy for $\beta\cJ$ and $\kappa=\lambda(\bk)/\omega_\beta$. In taking the limit, we expand the all-temperature expressions of~\cite{Choi:2023mab} to obtain at low temperatures and weak inter-site coupling:
\es{lowTexp}{
v&=1-{2\ov \beta \cJ}+\dots\,, \qquad \kappa=v\le(1-{2\ov \beta \cJ}\, {\cD_\bk\ov \omega_\beta}+\dots\ri)\,, \qquad c_V={\pi^2 N\ov p^2\beta{\cal J}}\,, 
}
We find that \cref{510} provides an exact agreement with our result~\cref{eq:match} for $c_\bk=c_V$.
\end{myshaded}

\subsubsection{OTOC of local operators}

The above results were all known in prior literature. A stringent test of the shockwave framework is whether it also applies to local $\SL(2,\bbR)$ current operators. We have already seen that the energy current OTOC grows as $\E{\omega_\beta t}$ and that its spatial structure is determined by $v_b$. Using the $\alpha(\bk)$ identified above, we immediately see that the OTOC result in~\cref{eq:OTOC-result} organises in the form
\begin{calign}\label{eq:OTOC-resultSketch}
    &C^{VW}_{\bar a \bar a |\bar r a}(x_1+x,x_2+x;x_3,x_4) \nn\\
    &\hspace{4em}\sim 
    \int \frac{\df^d\bk}{(2\pi)^d}
    \lb \frac{\omega_\beta^2\,c_\bk\,\E{\omega_\beta t_2}\E{i\bk\cdot\bx_2}}{2\pi i} \cI^V_d(x_1-x_2,\bk)\rb \nn\\
    &\hspace{10em}
    \lb \frac{\omega_\beta^2\,c_\bk\,\E{-\omega_\beta t_3}\E{-i\bk\cdot\bx_3}}{2\pi i} \cI^W_d(x_3-x_4;\bk)\rb
    \frac{i\alpha(\bk)}{N}
    \E{\omega_\beta t+ i\bk\cdot\bx}.
\end{calign}
This should be compared to the shockwave result obtained from \cref{eq:shockwaveMain}, i.e.
\begin{calign}\label{eq:energy-shockwave}
    &C^{\cE^t\cE^t}_{\bar a \bar a |\bar r a}(x_1+x,x_2+x;x_3,x_4) \nn\\
&\sim \int {\df^d \bk\ov (2\pi)^d}\, 
\frac{\dow{\cal G}^V_{\bar a \bar a}\lb x_1,x_2\vert X^-\rb}{\dow X^-(\bk) }\Bigg\vert_{X^-=0}\,
\frac{\dow{\cal G}^W_{\bar r a}\lb x_3,x_4\vert X^+\rb}{\dow X^+(-\bk) }\Bigg\vert_{X^+=0} \, 
{i\alpha(\bk)\ov N}
    \E{\omega_\beta t+ i\bk\cdot\bx},
\end{calign}
where ${\cal G}^V_{\bar a \bar a}$ is the $\langle V_{\bar a}V_{\bar a} \rangle$ correlator in the presence of a growing shockwave on folds ``12'', while ${\cal G}^W_{\bar ra}$ is the $\langle W_{\bar r}W_{a} \rangle$ correlator in the presence of a decaying shockwave on folds ``23''. Interestingly, note that both these correlators are identically zero on the trivial thermal background in the absence of the shock. This is why, we have not normalised these OTOCs. The remaining challenge is to establish that the vertex factors in \cref{eq:OTOC-resultSketch} are indeed the coupling of these pairs of operators to the shockwaves. 

Since the vertex factors are expected to vanish on the thermal background, the natural two-fold candidate for these is the noise-noise two-point correlator $\langle V_{a}V_{a} \rangle$. Indeed, using manipulations similar to \cref{fig:shocks-maps}, we show in~\cref{app:shocks} that these vertex factors can be evaluated using the noise-noise correlator on the two-fold contour as in a configuration where the saddle on fold ``2'' is shifted by a decaying $\SL(2,\bbR)$ transformation. To wit
\begin{calign}
    {\cal G}^V_{\bar a \bar a}\lb x_1,x_2\vert X^-\rb
    &= \Big\langle V_a(-x_1)V_a(-x_2)\Big\rangle_{\tilde X^-}, \nn\\
    {\cal G}^W_{\bar r a}\lb x_3,x_4\vert X^+\rb
    &= \half \Big\langle W_a(x_3)W_a(x_4)\Big\rangle_{X^+}
    - \half \Big\langle W_a(x_3)W_a(x_4)\Big\rangle_{-X^+}
    .
\end{calign}
We also show in \cref{app:shocks} that the latter precisely reproduces the vertex factors identified in \cref{eq:OTOC-resultSketch} at leading order in $X^\pm$. Therefore, our energy density OTOC exactly agrees with the scramblon prescription, constituting a highly non-trivial consistency check for our framework.

There are some comments in order. At first, it may look like the shockwaves should not couple to the energy current at all: the shockwaves are $\SL(2,\bbR)$ gauge transformations on each fold (but not a valid gauge transformation globally, as those act on each fold simultaneously). Whereas, the SYK lattice energy density is the Schwarzian, an $\SL(2,\bbR)$-invariant operator localised on one fold, which hence should be insensitive to a gauge transformation.\footnote{This argument should then extend to the energy flux as well, which is related to the energy density through a Ward identity, even though its expression involves an inter-fold term~\cite{Bucca:2026xmh}. } However, this argument treats the phrase gauge transformation too loosely: the shockwave is a \emph{gauge transformation on individual folds}, not of the full theory. In other words, it is not a symmetry of the four-fold effective action $S[\{F_{i}\}]$. Therefore, even when the energy density operator is formally insensitive to shockwaves, its correlation functions $\langle \cE^t_i \cE^t_j \rangle = \int \{\cD F_i\}\,\cE^t_i \cE^t_j\, \E{iS[\{F_i\}]}$ can become sensitive to shocks due to the weighting by the action.

\section{Discussion and outlook}
\label{sec:discussion}

In this work, we have extended the SK EFT of energy diffusion to a four-fold time contour needed to describe OTOCs that probe scrambling. We have shown that the same degrees of freedom that govern conventional hydrodynamic correlation functions can also capture the exponential growth of OTOCs. Since the rate of this exponential growth is proportional to the EFT cutoff scale, following the lead of refs.~\cite{Blake:2017ris,Blake:2021wqj}, we had to work non-perturbatively in derivatives, but using tree level Feynman diagrams allowed us to expand in the amplitude of fluctuations. A crucial ingredient in the construction was the spatially-local $\SL(2,\mathbb{R})$ symmetry, which protects time-ordered correlators from exponential growth while allowing scrambling to manifest in out-of-time-order observables. 

Our results extend the quantum hydrodynamic description of chaos proposed in~\cite{Blake:2017ris,Blake:2021wqj} in several directions: While the scrambling of generic local operators can be understood through the exchange of hydrodynamic modes on a two-fold contour, computing OTOCs of energy currents requires the full four-fold SK EFT. The resulting correlation functions provide a particularly direct probe of scrambling through observables that are universally defined in systems with diffusive energy transport.

We have gathered very strong evidence for this EFT: We derived it from the low temperature, maximally chaotic limit of the SYK lattice following the philosophy of our earlier derivation of the two-fold EFT in~\cite{Bucca:2026xmh}. We have also shown that the framework (with minor modifications) can incorporate the stress tensor dynamics of (1+1)d CFTs. We have shown that the EFT results match the shockwave/scramblon effective description of scrambling explored in~\cite{Maldacena:2016upp,Stanford:2021bhl,Gu:2021xaj,Choi:2023mab,Stanford:2023npy}. While shockwaves have been constructed in special microscopic models, obtaining them within the SK EFT framework takes us closer to providing a universal description for them.

Our work explored the behaviour of OTOCs before the scrambling time. Going beyond this time requires the resummation of diagrams, which has mainly been explored in the shockwave description in prior work; see however the SYK analysis in~\cite{Bagrets:2017pwq}. Hence, we leave open the question what is the ultimate long-time behaviour of OTOCs. This was the question investigated in the framework of charge diffusion in~\cite{Mishra:2025vlf}. Note, however, that the SK EFT of charge diffusion is not expected to give rise to exponential scrambling even in large-$N$ systems, hence it is unclear how the conclusions carry over to the energy diffusion case analysed in this work. Perhaps, after the OTOC has reached close to its saturation value, likely involving some condensate of shockwaves, the degrees of freedom reorganise into an EFT with no manifest $\SL(2,\bbR)$ symmetry and hence no exponentially growing propagator.

There are several interesting open questions that our work motivates. 
For simplicity, we focussed on systems where dissipation is exclusively odd under generalised time-reversal transformations that are used to define the KMS symmetry (e.g. time-reversal or spacetime-reversal). This indeed applies to the two prime examples discussed in this work: the SYK lattice and ASCJ (1+1)d CFT. As a future direction, it will be interesting to identify models where this assumption fails and what that implies for the growth of OTOCs. We leave this exploration for future work.

It would be interesting to build a four-fold SK EFT for the transport of conserved energy and momentum. Momentum conservation leads to propagating hydrodynamic modes and a potentially rich interplay between transport and scrambling. Recent holographic computations~\cite{Knysh:2024asf, Blake:2026jue, Blake:2026axf} have proposed the generalisation of spatially-local $\SL(2,\bbR)$ symmetry to maximally-chaotic momentum conserving systems. It will be interesting to explore what this richer symmetry structure implies for the structure of OTOCs and their exponential growth behaviour.

The extension of hydrodynamics to a four-fold contour is part of a broader programme of describing probes of many-body chaos through EFTs. R\'enyi entropies and the spectral form factor require multiple copies of the original system with a very similar time contour, but different boundary conditions. We expect that similar constructions to this paper's would elucidate the role of hydrodynamic degrees of freedom in these quantities.  
Additionally, our results should be applicable to operator entanglement \cite{Zanardi:2001zza,Prosen:2007gfp} and closely related ``entanglement in time'' \cite{Fullwood:2022rjd,Doi:2023zaf,Milekhin:2025ycm}, since the  second R\'enyi operator entanglement can be obtained as a sum over OTOCs of different operators \cite{Hosur:2015ylk}. It would be interesting to see if our results and \cite{Jonay:2018yei,Friedman:2019gyi,Winer:2020gdp,vonKeyserlingk:2017dyr,Alba:2020npy} admit a universal description.

Finally, it would be important to understand how to extend the hydrodynamic EFT framework away from maximal chaos~\cite{Shenker:2014cwa,Gu:2018jsv,Choi:2019bmd,Mezei:2019dfv,Choi:2020tdj,Gao:2023wun,Choi:2023mab}.   It was argued in~\cite{Blake:2021wqj} that the $\SL(2,\mathbb{R})$ symmetry in general gives rise to maximal chaos, indeed this is the case in the prime microscopic example in this paper, the low temperature SYK lattice. A potential loophole to this argument may be offered by the work~\cite{Lin:2023trc}, where they find a non-geometric action of $\SL(2,\mathbb{R})$ in a certain special corner of parameters in the SYK model, which can however be geometrized through a ``fake thermal circle''. It would be interesting to promote this setup to an EFT on the fake SK contour. In general, non-maximally chaotic models are not expected to possess an $\SL(2,\mathbb{R})$ symmetry and it would be very interesting to understand the corresponding four-fold EFTs, especially the mechanism for how they protect the TOCs from growing.

Our results supply additional evidence that the universality of hydrodynamics extends far beyond its traditional domain of time-ordered correlation functions. Exploring multi-copy observables and probing the generic non-maximally chaotic landscape of theories offer promising directions for future work.

\subsection*{AI disclosure}

We have used ChatGPT 5.6 to initially compute Fourier integrals, which we then verified by Mathematica and by hand. All other computations were done by hand and by Mathematica. We have also used both ChatGPT 6 and Claude Opus 5.5 to proofread the draft, but no output produced by these models was included in this manuscript.

\acknowledgements

We thank Mike Blake, Luca Delacr\'etaz, Felix Haehl, Hong Liu, Natalia Pinzani-Fokeeva, and Riccardo Rattazzi for useful discussions. MB is supported by the Maryam Mirzakhani Graduate Scholarship.
AJ and MM are supported by the ERC Consolidator Grant GeoChaos-101169611. 
AM is supported by NSF award PHY 2610025.

This work is funded by the European Union. Views and opinions expressed are however those of the authors only and do not necessarily reflect those of the European Union or the European Research Council Executive Agency. Neither the European Union nor the granting authority can be held responsible for them.

For the purpose of open access, the authors have applied a CC BY public copyright licence to any Author Accepted Manuscript (AAM) version arising from this submission.

\appendix

\section{Details of position-space propagators}
\label{app:full-props}

In this appendix, we derive the retarded and symmetric propagators in position space by analytical continuation of the Euclidean propagator.

\subsection*{Euclidean to Lorentzian propagators}

To begin, let us consider the Euclidean propagator corresponding to the differential operator 
\begin{calign}
    K_\rmE(\dow_\tau,i\bk) = -i\dow_\tau\lb 1+\frac{\dow_\tau^2}{\omega_\beta^2}\rb 
    \beta  c_\bk (i\dow_\tau+\cD_\bk).
\end{calign}
To wit,
\begin{calign}
    G_\rmE(\tau,i\bk) 
    &= \frac{1}{\beta ^2 c_\bk}
    \sum_{n>1}
    \frac{\E{-in\omega_\beta \tau}}
    {n\omega_\beta\lb 1 - n^2 \rb\lb n\omega_\beta + \cD_{\bk} \rb}
    + \frac{1}{\beta ^2 c_{-\bk}}
    \sum_{n<1}
    \frac{\E{-in\omega_\beta \tau}}
    {n\omega_\beta\lb 1 - n^2 \rb\lb n\omega_\beta -\cD_{-\bk} \rb}.
\end{calign}
The Euclidean propagator satisfies $G_\rmE(-\tau,-i\bk)=G_\rmE(\tau,i\bk)$ and $G_\rmE(\tau+\beta ,i\bk)=G_\rmE(\tau,i\bk)$. We can perform the Matsubara sums in terms of the Lerch transcendent $\Phi$, leading to
\begin{calign}
    G_\rmE(\tau,i\bk) 
    &= \frac{-1}{4\pi^2 c_\bk}
    \Bigg[
    \frac{\E{-iu}}{a}
    +\frac{1+\frac12\E{-iu}}{2(1-a)}
    + \lb \frac{1}{a}
    - \frac{\E{-iu}}{2(a+1)}
    - \frac{\E{iu}}{2(a-1)}\rb
    \ln\lb 1-\E{-iu}\rb \nn\\
    &\hspace{10em}
    + \frac{\E{-2iu}}{a(1-a^2)}\,\Phi\lb \E{-iu},1,a+2\rb
    \Bigg] \nn\\
    &\qquad 
    + (t\leftrightarrow -t,\bk\leftrightarrow-\bk),
\end{calign}
where $u=\omega_\beta\tau$ and $a = \cD_\bk/\omega_\beta$. We can read off the unordered (Wightman) function by analytic continuation
\begin{calign}
    G_>(t,i\bk)
    &= G_\rmE(it+0^+,i\bk) \nn\\
    &= G_\rmS(t,i\bk) 
    - \frac{i}{2} G_\rmR(t,i\bk)
    + \frac{i}{2} G_\rmR(-t,-i\bk).
\end{calign}
where the retarded and symmetric functions are
\begin{calign}
    G_\rmR(t,i\bk) 
    &= \frac{-\Theta(t)}{2\pi c_\bk}\lb 
    \frac{ \E{-a s}}{a(1-a^2)}
    - \frac{1}{a}
    + \frac{\E{s}}{2(1+a)}
    - \frac{\E{-s}}{2(1-a)}
    \rb, \nn\\
    G_\rmS(t,i\bk)
    &=
    \frac{-1}{4\pi^2c_\bk}\Bigg[
    \frac{\E{-|s|}}{a}
    - \frac{1+\frac12 \E{-|s|}}{2(\sgn(s)+a)}
    + \lb \frac{1}{a}
    - \frac{\E{s}}{2(1+a)}
    + \frac{\E{-s}}{2(1-a)}\rb
    \ln\lb 1-\E{-|s|}\rb \nn\\
    &\hspace{6em}
    + \frac{\E{-2|s|}}{a(1-a^2)}
    \Phi\lb \E{-|s|},1,2-\sgn(s)a\rb \nn\\
    &\hspace{6em} 
    + \Theta(t)
    \lb \frac{\pi\E{-as}}{a(1-a^2)} \cot (\pi  a)
    + \frac{5-3 a}{4 (1-a)^2} \E{-s}
    + \frac{s\E{-s}}{2(1-a)}
    \rb 
    \Bigg] \nn\\
    &\qquad 
    - \frac{\Theta(t)}{4\pi^2c_\bk}\lb 
    \frac{5+3 a}{4 (1+a)^2} \E{s}
    - \frac{s}{2(1+a)}\E{s}
    + \frac{s}{a}
    - \frac{1}{a^2}
    \rb 
    \nn\\
    &\qquad 
    + (t\leftrightarrow -t,\bk\leftrightarrow-\bk),
\end{calign}
where $s=\omega_\beta t$. Identifying $K^{-1}(t,i\bk) = -G_\rmR(t,i\bk)$, we recover \cref{eq:DeltaandKfunctions}. Whereas, the symmetric Green's function in \cref{eq:twofold-props} is given as $K^{-1}_\rmS(t,i\bk) = G_\rmS(t,i\bk)$.\footnote{To match with \cref{eq:twofold-props}, it is useful to note the explicit form of trigonometric differential operators. The basic trigonometric differential operators are defined as follows:
\begin{alignat}{2}
    \sin(\lambda\dow_t) h(t) 
    &= \frac{1}{2i}\E{i\lambda\dow_t} h(t)
    - \frac{1}{2i}\E{-i\lambda\dow_t} h(t)
    &&= \Im h(t+i\lambda), \nn\\
    \cos(\lambda\dow_t) h(t) 
    &= \frac{1}{2}\E{i\lambda\dow_t} h(t)
    + \frac{1}{2}\E{-i\lambda\dow_t} h(t)
    &&= \Re h(t+i\lambda).
\end{alignat}
The other trigonometric differential operators are actually integro-differential operators
\begin{calign}
    \csc(\lambda\dow_t) h(t)
    &= \frac{1}{2\lambda} \int_{-\infty}^\infty
    \df t' \tanh\lb\frac{\pi(t-t')}{2\lambda} \rb h(t'), \nn\\
    \sec(\lambda\dow_t) h(t)
    &= \frac{1}{2\lambda}
    \int_{-\infty}^{\infty}\df t'\,
    \sech\left(
    \frac{\pi(t-t')}{2\lambda}
    \right)h(t'), \nn\\
    \tan(\lambda\partial_t)h(t)
    &=
    -\frac{1}{2\lambda}
    \operatorname{PV}
    \int_{-\infty}^{\infty}\df t'\,
    \csch\left(
    \frac{\pi(t-t')}{2\lambda}
    \right)h(t'), \nn\\
    \cot(\lambda\partial_t)h(t)
    &=
    \frac{1}{2\lambda}
    \operatorname{PV}
    \int_{-\infty}^{\infty}\df t'\,
    \coth\left(
    \frac{\pi(t-t')}{2\lambda}
    \right)h(t').
\end{calign}
Note that, even though the integral Kernels of these operators seem to have a growing part in $t$, these are all perfectly convergent at large-$t$ for any sufficiently decaying function $h(t')$. In fact, using $h(t)=\E{-\gamma |t|}$, the first and fourth integrals saturate to $1/(\lambda\gamma)$, while the second and third decay to $0$. This is qualitatively different from the Kernel we encountered in the definition of $K^{-1}(x)$ in \cref{eq:growing-kernel}, which grows exponentially at late times.

}
One can check that the terms in the square brackets in $G_\rmS(t,i\bk)$ fall off to zero as $t\to\pm\infty$, and hence are not important for us. 
Explicitly, we can write the remaining important terms as
\begin{calign}
    G_\rmS(t,i\bk) \Big|_{|t|\gg \beta }
    &\sim 
    - \frac{\Theta(t)}{\beta ^2 c_\bk}\lb 
    \frac{5+3\cD_\bk/\omega_\beta}{4 (\omega_\beta+\cD_\bk)^2} \E{\omega_\beta t}
    - \frac{t}{2(\omega_\beta+\cD_\bk)}
    \E{\omega_\beta t}
    + \frac{t}{\cD_\bk}
    - \frac{1}{\cD_\bk^2}
    \rb \nn\\
    &\qquad
    + (t\leftrightarrow -t,\bk\leftrightarrow-\bk),
\end{calign}
which is valid for $|t|\gg \beta $. 
Note that 
\begin{calign}
    \dow_t
    \lb 1 - \frac{\dow_t^2}{\omega_\beta^2} \rb^2
    G_\rmS(t,i\bk) \Big|_{|t|\gg\beta }
    &\sim 
    - \frac{\Theta(t)}{\beta ^2 c_V}
    \frac{1}{\cD_\bk}
    + \frac{\Theta(-t)}{\beta ^2 c_V}
    \frac{1}{\cD_{-\bk}},
\end{calign}
so the growing mode goes away when $G_\rmS(t,i\bk)$ is acted upon by the operator $\dow_t(1 - \dow_t^2/\omega_\beta^2)^2$ and reproduces the asymptotic behaviour in \cref{eq:DeltaandKfunctions}.

Using the expressions for the retarded and symmetric propagators above, we can compute
\begin{calign}
    \Delta^{-1}(t,i\bk) 
    &= \frac{\Theta(t)}{\beta  c_\bk}\, \E{-\cD_\bk t}, \nn\\
    \Delta^{-1}_\rmS(t,i\bk)
    &= 
    \frac{1}{\beta^2c_\bk}\Bigg[ 
    \frac{2\sgn(s)a \left(1-a^2\right)}{\left(1-\E{|s|}\right)^3}
    \left(\frac{1}{a} -\frac{\E{s}}{2 (a+1)} + \frac{\E{-s}}{2 (1-a)}\right) 
    - \frac{3\lb \sgn(s) +a\rb}{\left(1-\E{|s|}\right)^2}
    \nn\\
    &\hspace{6em}
    + a \left(1-a^2\right) \E{-2|s|} \Phi \left(\E{-|s|},1,2-\sgn(s)a\right)
    + \Theta(t)a \left(1-a^2\right) \pi  e^{-a s} \cot (\pi  a)
    \Bigg] 
    \nn\\
    &\qquad 
    + (t\leftrightarrow -t,\bk\leftrightarrow-\bk).
\end{calign}
Note that both of these vanish asymptotically for large $t$. In deriving these, we have dropped certain contact terms proportional to $\delta(t)$, $\delta'(t)$, etc.

\paragraph*{Why not frequency-domain?} 

Let us briefly explain why we have opted to only perform a partial Fourier transform in the spatial directions, instead of working in the full frequency-wavevector-domain which is often technically simpler. In full Fourier-domain representation, $\Delta(\dow)$ is represented as $\Delta(-i\omega,i\bk) = \beta c_\bk(-i\omega + \cD_\bk)$. In fact, one can check that the inverse-Fourier-transform of $1/\Delta(-i\omega,i\bk)$ is precisely the one given in \cref{eq:DeltaandKfunctions}. On the other hand, the operator $K(\dow)$ is represented in full Fourier-domain as $K(-i\omega,i\bk) = \beta c_\bk i\omega(1+\omega^2/\omega_\beta^2)(-i\omega + \cD_\bk)$. Firstly, it has a pole on the real axis, so its Fourier-transform is ill-defined. Though, we can nudge it slightly in the lower-half complex-$\omega$ plane to find
\begin{calign}
    \int \frac{\df\omega}{2\pi}
    \frac{\E{-i\omega t}}{K(-i\omega,i\bk)}
    &= \frac{\Theta(t)}{\beta c_\bk}
    \lb 
    \frac{\E{-\cD_\bk t}}{\cD_\bk (1-\cD_\bk^2/\omega_\beta^2)}
    - \frac{1}{\cD_\bk}
    + \frac{\E{-\omega_\beta t}}{
    2(\cD_\bk-\omega_\beta)}
    \rb \nn\\
    &\qquad
    - \frac{\Theta(-t)}{\beta c_\bk}
    \frac{\E{\omega_\beta t}}{
    2(\cD_\bk+\omega_\beta)}.
\end{calign}
This is different from the one in \cref{eq:DeltaandKfunctions}: it falls off exponentially fast at large $t$ and is not retarded because it also has support for $t<0$. This is to be expected because exponentially growing functions do not admit well-defined Fourier-transforms.

A clever trick pointed out in~\cite{Blake:2017ris} involves shifting the contour of frequency integrals above the pole at $\omega=i\omega_\beta$, say along the line $C=(-\infty + i\omega_\beta+i\varepsilon,\infty + i\omega_\beta+i\varepsilon)$. By performing a change of variables $\tilde\omega = \omega - i\omega_\beta -i\varepsilon$, this can be achieved by defining the shifted Fourier transform of $F$ as $\delta F(t) = \frac{1}{2\pi}\int \df\tilde\omega\,\E{-i\tilde\omega t+\omega_\beta t+\varepsilon t} \delta \hat F(\tilde\omega+i\omega_\beta+i\varepsilon)$. However, the set of functions $\E{-i\tilde\omega t+\omega_\beta t+\varepsilon t}$ do not comprise an orthonormal basis. To wit, plugging this into the free action in \cref{eq:action-free}, we find
\begin{calign}
    S_{\rm free}
    &= 
    \int_C \frac{\df\omega\,\df^d\bk}{(2\pi)^{d+1}}\,
     \delta \hat Q(-\omega,-\bk')
    \,K(-i\tilde\omega+\omega_\beta+\varepsilon,i\bk)\,\,\delta \hat F(\omega,\bk) \nn\\
    &\quad 
    - 
    \int_C \frac{\df\omega\,\df^d\bk}{(2\pi)^{d+1}}
    \int_C \frac{\df\omega'}{2\pi}
    \int\df t\,
     \E{i(\omega+\omega') t} \,\nn\\
    &\hspace{6em}
    \delta \hat Q(-\omega',-\bk')
     \frac14 \coth\bfrac{\beta \omega}{2}
    \Big( K(i\omega,i\bk) 
    - K(-i\omega,-i\bk) \Big)
    \delta \hat Q(-\omega,\bk'),
\end{calign}
where we have defined $\delta Q(t) = \frac{1}{2\pi}\int \df\tilde\omega\,\E{-i\tilde\omega t-\omega_\beta t-\varepsilon t} \delta \hat Q(\tilde\omega-i\omega_\beta-i\varepsilon)$. With these choices, the classical $QF$ part of the action does yield the frequency integrals with shifted frequency contour, but the noise $QQ$ part does not. The integral $\int\df t\,\E{i(\omega+\omega') t}$ does not yield a delta-function for complex values of frequencies, which is why we are unable to collapse $t$ and $\omega'$ integrals in the second term. As a result, we are unable to pass onto a full Fourier representation even with a shifted frequency contour.

\subsection*{Energy current correlation functions}

For later use, let us illustrate how we can use this formalism to compute various two-point correlation functions of energy current. They can be defined in the same way as the $\delta F$ correlation functions in \cref{eq:deltaFcorrelators}. To this end, we need the average and difference combinations of the energy current operator, $\cE^\mu_{r}= (\cE^\mu_1 + \cE^\mu_2)/2$ and $\cE^\mu_{a}= \cE^\mu_1 - \cE^\mu_2$. To leading order in fluctuations, we can use the KMS symmetry to fix
\begin{calign}
    \cE^\mu_r 
    &= \cO^\mu_{(1)}(\dow)\delta F 
    - \frac{i}{2} \cot\left(\frac{\beta \partial_t}{2}\right) \Big( \cO^\mu_{(1)}(\dow) 
    + \eta^{\cE^\mu}_\Theta\tilde\cO^\mu_{(1)}(\dow) 
    \Big)\delta Q, \nn\\
    \cE^\mu_{a}
    &= -\eta^{\cE^\mu}_\Theta\tilde\cO^\mu_{(1)}(\dow) \delta Q,
\end{calign}
where $\eta_\Theta=+1$ for energy density and $\eta_\Theta=-\eta_\rmP$ for the energy flux. Using this, we find
\begin{subequations}
\begin{calign}
    G^{\rm R}_{\delta\cE^\mu\delta\cE^\nu}(x)
    &= i\Big\langle \delta\cE^\mu_r(x)\delta\cE^\nu_a(0) \Big\rangle \nn\\
    &= -\eta^{\cE^\nu}_\Theta\, \cD^\mu_{(1)}(\dow) 
    \tilde\cD^\nu_{(1)}(-\dow)
    \lb 1 - \frac{\dow_t^2}{\omega_\beta^2} \rb\dow_t \Delta^{-1}(x)
    , \nn\\
    G^{\rm S}_{\delta\cE^\mu\delta\cE^\nu}(x)
    &= \Big\langle \delta\cE^\mu_r(x)\delta\cE^\nu_r(0) \Big\rangle \nn\\
    &= 
    \frac{1}{2} \cot\bfrac{\beta \dow_t}{2} 
    \lb 1 - \frac{\dow_t^2}{\omega_\beta^2} \rb\dow_t \nn\\
    &\qquad 
    \lb \eta^{\cE^\nu}_\Theta\cD^\mu_{(1)}(\dow)
    \tilde\cD^\nu_{(1)}(-\dow) 
    \Delta^{-1}(x) 
    + \eta^{\cE^\mu}_\Theta\tilde\cD^\mu_{(1)}(\dow) 
    \cD^\nu_{(1)}(-\dow)
    \tilde\Delta^{-1}(x) \rb,
\end{calign}
together with the unordered function
\begin{calign}
    \Big\langle 
    \delta \hat \cE(x)\delta \hat \cE(0) \Big\rangle
    &= \frac{1}{2} 
    \lb \cot\bfrac{\beta \dow_t}{2} + i \rb
    \lb 1 - \frac{\dow_t^2}{\omega_\beta^2} \rb\dow_t \nn\\
    &\qquad 
    \lb \eta^{\cE^\nu}_\Theta\cD^\mu_{(1)}(\dow)
    \tilde\cD^\nu_{(1)}(-\dow) 
    \Delta^{-1}(x) 
    + \eta^{\cE^\mu}_\Theta\tilde\cD^\mu_{(1)}(\dow) 
    \cD^\nu_{(1)}(-\dow)
    \tilde\Delta^{-1}(x) \rb.
\end{calign}
\end{subequations}
We emphasise that these formulas are only valid up to contact terms. To fix the contact terms, we need to know the expressions for the energy currents in the presence of background fields; see~\cite{Bucca:2026xmh}. The correlation functions satisfy the fluctuation-dissipation relations and the Onsager's relations
\begin{calign}
    G^{\rm S}_{\delta\cE^\mu\delta\cE^\nu}(x)
    &=  -
    \frac{1}{2} \cot\bfrac{\beta \dow_t}{2} 
    \Big( G^{\rm R}_{\delta\cE^\mu\delta\cE^\nu}(x)
    - \eta_\Theta^\mu \eta_\Theta^\nu  \tilde G^{\rm R}_{\delta\cE^\mu\delta\cE^\nu}(x) \Big), \nn\\
    \tilde G^{\rm R}_{\delta\cE^\mu\delta\cE^\nu}(x)
    &= \eta_\Theta^\mu \eta_\Theta^\nu  G^{\rm R}_{\delta\cE^\nu\delta\cE^\mu}(-x)
\end{calign}
Recall that tilde denotes the function evaluated at $(-t,\eta_\rmP\bx)$. 

In fact, since the physical energy correlators do not exhibit unstable exponential growth, we can represent them in their conventional full Fourier form. To wit
\begin{calign}
    G^{\rm R}_{\delta\cE^\mu\delta\cE^\nu}(-i\omega,i\bk)
    &= \eta^{\cE^\nu}_\Theta\, \cD^\mu_{(1)}(-i\omega,i\bk) 
    \cD^\nu_{(1)}(-i\omega,-i\eta_\rmP \bk)
    \frac{i\omega\lb 1 + {\omega^2/\omega_\beta^2} \rb }{\Delta(-i\omega,i\bk)}, \nn\\
    G^{\rm S}_{\delta\cE^\mu\delta\cE^\nu}(-i\omega,i\bk)
    &= 
    \frac{1}{2} 
    \omega\coth\bfrac{\beta \omega}{2} 
    \lb 1 + \frac{\omega^2}{\omega_\beta^2} \rb \nn\\
    &\hspace{-1em}
    \lb 
    \eta^{\cE^\nu}_\Theta\frac{\cD^\mu_{(1)}(-i\omega,i\bk)
    \cD^\nu_{(1)}(-i\omega,-i\eta_\rmP\bk) }{\Delta(-i\omega,i\bk)}
    + \eta^{\cE^\mu}_\Theta\frac{\cD^\mu_{(1)}(i\omega,i\eta_\rmP\bk)
    \cD^\nu_{(1)}(i\omega,-i\bk) }{\Delta(i\omega,i\eta_\rmP\bk)}\rb
    ,
\end{calign}
up to contact terms polynomial in $\omega$, $\bk$. We can simplify these using our simpler form of $\Delta(\dow_t,i\bk)$ in \cref{eq:general-Delta}, which amounts to fixing $\cD^t(\dow_t,i\bk) = c_\bk$ and $i\bk\cdot\cD^\bx(\dow_t,i\bk) = c_\bk\cD_\bk$. Further using $\tilde\Delta(\dow_t,i\bk) = \Delta(-\dow_t,-i\bk)$, we get that $c_{\eta_\rmP\bk} = c_{-\bk}$ and $\cD^\bx(\dow_t,i\eta_\rmP\bk) = -\cD^\bx(\dow_t,-i\bk)$. This leads to the energy density correlators
\begin{calign}\label{eq:energy-correlator}
    G^{\rm R}_{\delta\cE^t\delta\cE^t}(-i\omega,i\bk)
    &= \frac{c_\bk\cD_\bk}{\beta}
    \frac{1 + {\omega^2/\omega_\beta^2}}{-i\omega + \cD_\bk}
    , \nn\\
    G^{\rm S}_{\delta\cE^t\delta\cE^t}(-i\omega,i\bk)
    &= 
    \frac{1}{2i\beta} 
    \coth\bfrac{\beta \omega}{2} 
    \lb 1 + \frac{\omega^2}{\omega_\beta^2} \rb
    \lb 
    \frac{c_\bk \cD_\bk}{-i\omega+\cD_\bk}
    - 
    \frac{c_{-\bk}\cD_{-\bk}}{i\omega + \cD_{-\bk}}\rb.
\end{calign}
Here we have added some contact term improvements to bring the correlators into conventional form (i.e. the energy density correlators should vanish when $\bk\to 0$). We can similarly obtain the correlators involving energy flux.

The retarded energy correlator in \cref{eq:energy-correlator} also enables us to see the phenomenon of pole skipping~\cite{Grozdanov:2017ajz,Blake:2017ris,Blake:2018leo,Grozdanov:2018kkt,Haehl:2018izb,Blake:2019otz,Ahn:2019rnq,Haehl:2019eae,Ramirez:2020qer,Choi:2020tdj,Knysh:2024asf}. We see that at the special (generically complex) value of wavevector $\bk_b$ determined by $\cD_{\bk_b} = \pm \omega_\beta$, the pole of the retarded correlator cancels with its numerator. Notice that one of these coincide with the wavevector pole of the OTOC in \cref{eq:OTOC-result}, which leads to the butterfly velocity. Indeed, the butterfly velocity is related to the pole skipping point as simply $v_b = \omega_\beta/|\bk_b|$. In particular, for the SYK model, we find $\bk_b = \sqrt{\pm\omega_\beta/D}$, while for the ASCJ model, we have $\bk_b = \mp i\omega_\beta$.

\section{Details of the SK effective action}
\label{app:Fourier-constraints}

In this appendix we discuss in more detail the derivation of the free propagators and three-point interactions in the SK EFT for energy diffusion.

\subsection{Symmetry constraints on the two-fold SK effective action}

We begin with the general form of the effective action up to cubic order in perturbations, permitted by the unitarity constraints:
\begin{calign}
    S
    &= \int \df^{d+1}x
    \bigg[ \delta Q\,K(\dow)\,\delta F
    + i\delta Q\,K_{QQ}(\dow)\,\delta Q
     \nn\\ 
    &\hspace{8em}
    + 
    \cV_{F^2Q}(\dow_1,\dow_2;\dow_3)\, 
    \delta F\,\delta F\,\delta Q \nn\\
    &\hspace{8em}
    + i\cV_{FQ^2}(\dow_1;\dow_2,\dow_3)\, 
    \delta F\,\delta Q\, \delta Q 
    + \cV_{Q^3}(\dow_1,\dow_2,\dow_3)\, 
    \delta Q\,\delta Q\,\delta Q
    + \ldots
    \bigg].
\end{calign}

\subsubsection*{Time and space translation symmetries}

We noted in \cref{sec:general-2fold} that the SK EFT for energy diffusion is invariant under average and difference time-translation symmetries, while only average space-translation symmetry. The average spacetime-translation symmetries are already manifest when the effective action is represented in terms of a local Lagrangian density. On the other hand, the action of difference time-translations from \cref{eq:time-translations} on linearised fields is given as
\begin{calign}\label{eq:diff-trans-fourier}
    \rmH^{\rm diff}_\zeta:&\quad 
    \delta F
    \to \delta F + \frac{\zeta}{4} \dow_t\delta Q + O(\zeta^2), \nn\\
    \rmH^{\rm diff}_\zeta:&\quad 
    \delta Q
    \to \delta Q + \zeta \dow_t \delta F 
    + \frac{\zeta}{\beta }
    + O(\zeta^2),
\end{calign}
where $\zeta = \xi_1-\xi_2$.
Note that the transformation of the noise field $\delta Q$ involves a constant shift $\zeta/\beta $ when expanded around $F = t/\beta  + \delta F$. Requiring the invariance of the quadratic part of the action in \cref{eq:action-free}, denoted $S_{\rm free}$, under this constant part results in the constraints on the coefficients: 
\begin{calign}
    K(\dow=0) = 0, \qquad K_{QQ}(\dow=0)=0.
\end{calign}

The remaining linear piece of the transformation constrains the interactions: e.g. the transformation of $S_{\rm free}$ under the linear part of \cref{eq:diff-trans-fourier} must compensate for the transformation of the three-point interactions $S_{\text{int-3}}$ under the constant part of \cref{eq:diff-trans-fourier}. We find
\begin{calign}
    \cV_{F^2Q}(\dow,-\dow;\dow_3=0) 
    &= \frac{\beta \dow_t}{2} 
    \Big( K(\dow) - K(-\dow) \Big), \nn\\ 
    \cV_{FQ^2}(\dow;-\dow,\dow_3=0)
    &= -\beta  \dow_t K_{QQ}(\dow), \nn\\
    \cV_{Q^3}(\dow,-\dow,\dow_3=0)
    &= - \frac{\beta \dow_t}{24} 
    \Big( K(\dow) - K(-\dow) \Big).
\end{calign}

\subsubsection*{Dynamical KMS symmetry}

The dynamical KMS symmetry (whereby the KMS shift symmetry is conjugated with a difference time-translation) acts as
\begin{calign}\label{eq:KMS-fourier}
    \rmK:&\quad 
    \delta F \to - \sfC \delta \tilde F 
    + \frac{i}{2} \sfS\delta \tilde Q, \nn\\
    \rmK:&\quad 
    \delta Q \to - \sfC \delta \tilde Q + 2i\sfS\delta \tilde F.
\end{calign}
where $\delta \tilde F(t,x) = \delta F(-t,\eta_\rmP\bx)$ and similarly for $\delta Q$, while we have defined the operators
\begin{calign}
    \sfC = \cos\bfrac{\beta \dow_t}{2}, \qquad 
    \sfS = \sin\bfrac{\beta \dow_t}{2}.
\end{calign}

Requiring the quadratic action to be invariant under this transformation results in the constraints
\begin{calign}\label{eq:FDT-fourier}
    K_{QQ}(\dow) = \frac{-1}{4} \frac{\sfC}{\sfS}
    \Big( K(\dow) - K(-\dow) \Big), \qquad 
    \tilde K(-\dow) = K(\dow),
\end{calign}
resulting in the free effective action in \cref{eq:action-free}. Note that the KMS symmetry completely fixes the quadratic noise coefficient $K_{QQ}$ in terms of the linear coefficient $K$.
The second constraint above is trivial for $\eta_\rmP = -1$, i.e. when the dynamical KMS symmetry is based on the microscopic spacetime-reversal invariance. However, for $\eta_\rmP = 1$, i.e. when the dynamical KMS symmetry is based on the microscopic time-reversal invariance, it enforces $K(\dow)$ to be an even function of $\dow_{\bx}$. \Cref{eq:FDT-fourier} are nothing but the well-known fluctuation-dissipation relations and Onsager constraints.

The analysis for cubic interactions is more involved. We find the constraints
\begin{calign}
    &
    \bigg[\frac{1}{4} \cV_{F^2Q}(\dow_1,\dow_2;\dow_3) 
    \sfC_1\sfC_2\sfS_3 
    + \frac{1}{2} \cV_{FQ^2}(\dow_1;\dow_2,\dow_3) 
    \sfC_1\sfS_2\sfS_3
    - \cV_{Q^3}(\dow_1,\dow_2,\dow_3)
    \sfS_1\sfS_2\sfS_3\bigg]_{(123)}
    = 0, \nn\\
    &
    \bigg[\cV_{F^2Q}(\dow_1,\dow_2;\dow_3) 
    \sfC_1\sfC_2\sfC_3
    - 2\cV_{F^2Q}(\dow_1,\dow_3;\dow_2) 
    \sfC_1\sfS_2\sfS_3 \nn\\
    &\qquad 
    - 2\cV_{FQ^2}(\dow_3;\dow_2,\dow_1) 
    \sfS_1\sfS_2\sfS_3
    + 4\cV_{FQ^2}(\dow_1;\dow_2,\dow_3) 
    \sfC_1\sfS_2\sfC_3 
    - 12 \cV_{Q^3}(\dow_1,\dow_2,\dow_3)
    \sfS_1\sfS_2\sfC_3\bigg]_{(12)}
    \nn\\
    &\hspace{25em}
    = -\tilde\cV_{F^2Q}(\dow_1,\dow_2;\dow_3), \nn\\
    &
    \bigg[ 
    - \cV_{F^2Q}(\dow_1,\dow_2;\dow_3) 
    \sfC_1\sfS_2\sfC_3
    + \frac{1}{2}\cV_{F^2Q}(\dow_3,\dow_2;\dow_1) 
    \sfS_1\sfS_2\sfS_3  \nn\\
    &\qquad 
    - 2\cV_{FQ^2}(\dow_2;\dow_1,\dow_3) 
    \sfS_1\sfS_2\sfC_3 
    + \cV_{FQ^2}(\dow_1;\dow_2,\dow_3) 
    \sfC_1\sfC_2\sfC_3 
    - 6 \cV_{Q^3}(\dow_1,\dow_2,\dow_3)
    \sfS_1\sfC_2\sfC_3
    \bigg]_{(23)}
    \nn\\
    &\hspace{25em}
    = -\tilde\cV_{FQ^2}(\dow_1;\dow_2,\dow_3), \nn\\
    &
    \bigg[
    - \frac{1}{4}\cV_{F^2Q}(\dow_1,\dow_2;\dow_3) 
    \sfS_1\sfS_2\sfC_3 
    + \frac{1}{2}\cV_{FQ^2}(\dow_1;\dow_2,\dow_3) 
    \sfS_1\sfC_2\sfC_3 
    + \cV_{Q^3}(\dow_1,\dow_2,\dow_3) \sfC_1\sfC_2\sfC_3 \bigg]_{(123)} \nn\\
    &\hspace{25em}
    = - \tilde\cV_{Q^3}(\dow_1,\dow_2,\dow_3).
\end{calign}
The notation $[\ldots]_{(ij\ldots)}$ means that the inside the square brackets are averaged over permutations of the indicated labels.
These can be solved to give
\begin{calign}
    \cV_{FQ^2}(\dow_1;\dow_2,\dow_3) 
    &= \bigg[
    - \cV_{F^2Q}(\dow_1,\dow_2;\dow_3) 
    \frac{\sfC_2}{\sfS_2}
    -
    \tilde \cV_{F^2Q}(\dow_2,\dow_3;\dow_1) 
     \frac{\sfC_2}{\sfS_2}
    \bigg]_{(23)}, \nn\\
    \cV_{Q^3}(\dow_1,\dow_2,\dow_3) 
    &= 
    \frac{1}{4}  \bigg[
    - \cV_{F^2Q}(\dow_1,\dow_2;\dow_3) 
    \frac{\sfC_1\sfC_2}{\sfS_1\sfS_2}  
    + 
     \tilde\cV_{F^2Q}(\dow_1,\dow_2;\dow_3) 
     \frac{\sfC_3}{\sfS_1\sfS_2}
    \bigg]_{(123)}.
\end{calign}
This results in the relations mentioned in \cref{eq:KMS-relations-3pt}. Once again, the quadratic and cubic noise vertices are entirely fixed by the linear one. In deriving the equations above, it is useful to note that $\dow_1+\dow_2+\dow_3=0$ at each vertex up to total derivative terms. This implies, for instance, that $\sfC_1 = \sfC_2\sfC_3 - \sfS_2\sfS_3$ and $\sfS_1 = -\sfS_2\sfC_3 - \sfC_2\sfS_3$, etc.

\subsubsection*{Diagonal shift symmetry}

The diagonal spatially-local shift symmetry acts by shifting $F$ with an arbitrary time-independent function. In terms of fluctuations, this results in
\begin{calign}\label{eq:diagonal-shift-fourier}
    \cH_a:&\quad 
    \delta F \to \delta F + a(\bx) \nn\\
    \cH_a:&\quad 
    \delta Q \to \delta Q.
\end{calign}
It simply implies
\begin{calign}
    K(\dow)\propto \dow_t, \qquad 
    \cV_{F^2Q}(\dow_1,\dow_2;\dow_3) \propto \dow_{t1}\dow_{t2}, \qquad 
    \cV_{FQ^2}(\dow_1;\dow_2,\dow_3) \propto \dow_{t1}.
\end{calign}

\subsubsection*{Lyapunov shift symmetries}

Lastly, we need to impose the pair of Lyapunov-shift symmetries. Unlike all the other symmetry transformations discussed so far, the Lyapunov-shift symmetries act non-linearly in perturbations, making them slightly trickier to implement. However, for our purposes, it suffices to expand up to linear order in fields. We find
\begin{calign}\label{eq:pert-lyapunov}
    \cR^\pm_{b_\pm}:&\quad \delta F \to \delta F
    \mp \frac{1}{2\pi}b_\pm(\bx)\, \E{\pm\omega_\beta t}
    \Big[
    1 \pm 2\pi \delta F
    + \ldots
    \Big]
    + \cO(b_\pm^2), \nn\\
    \cR^\pm_{b_\pm}:&\quad \delta Q \to \delta Q
    - b_\pm(\bx)\, \E{\pm \omega_\beta t}
    \Big[
    \delta Q
    + \ldots
    \Big]
    + \cO(b_\pm^2).
\end{calign}
where $\omega_\beta = 2\pi/\beta $.
Here ellipses denote non-linear terms in field perturbations.
Note that, similar to difference time-translations and diagonal shift transformations, this symmetry transformation involves a non-perturbative shift of $\delta F(p)$, together with a linear as well as higher non-linear pieces.

Applying the shift part of the transformation in \cref{eq:pert-lyapunov} on the quadratic action $S_{\rm free}$ generates a linear term, setting which to zero gives the simple constraint
\begin{calign}
    K(\dow) \Big|_{\dow_t = \pm\omega_\beta} = 0.
\end{calign}
This, together with the diagonal shift symmetry constraint, can be implemented simply by ensuring that $K(\dow)\propto \dow_t (1-\dow_t^2/\omega_\beta^2)$. Next, the transformation of $S_{\rm free}$ arising from the linear part of \cref{eq:pert-lyapunov}  must compensate for the transformation of $S_{\text{int-3}}$ arising from the non-perturbative shift part. This results in more non-trivial constraints
\begin{calign}\label{eq:SL2R-constraint-2pt}
    \cV_{F^2Q}(\dow_1,\dow;-\dow_1-\dow) \Big|_{\dow_{t1} 
    = \pm \omega_\beta}
    &= \mp \pi\Big(
    K(\dow+\dow_1) + K(\dow) \Big) \Big|_{\dow_{t1} = \pm \omega_\beta}, \nn\\
    \cV_{FQ^2}(\dow_1;\dow,-\dow_1-\dow) \Big|_{\dow_{t1} = \pm \omega_\beta}
    &= 
    \mp 2\pi \Big(K_{QQ}(\dow+\dow_1) + K_{QQ}(\dow) \Big) \Big|_{\dow_{t1} = \pm \omega_\beta}.
\end{calign}
The most general solution of these constraints is given in \cref{eq:three-point-ansatz}.
The constraints become more complicated for higher-point interactions. For instance, the transformation of $S_{\rm free}$ under the quadratic part of \cref{eq:pert-lyapunov}, combined with the transformation of $S_{\text{int-3}}$ under the linear part and that of $S_{\text{int-4}}$ under the non-perturbative shift part must mutually cancel. This will result in a complicated constraint involving four-point and three-point vertices as well as the free propagators.

\subsection{\texorpdfstring{$\SL(2,\bbR)$}{SL(2,R)}-invariant fields}
\label{app:Goldstone-basis}

Let us express the two-fold SK EFT in terms of $\SL(2,\bbR)$-invariant fields. \Cref{eq:Tdef} leads to
\begin{calign}
    \delta \sigma
    &= \delta F 
    - \lB \frac{\beta}{8}\lb \dow_{t1}+\dow_{t2}\rb \rB 
    \delta Q\delta Q
    + \ldots, \nn\\
    \delta T 
    &= \lb 1 -\frac{\dow_t^2}{\omega_\beta ^2}
    \rb \dow_t\delta \sigma 
    + \lB \beta  \dow_{t1} \dow_{t2}
    \lb 
    \frac{\dow_{t1}^2+\dow_{t2}^2}{\omega_\beta ^2} 
    + \frac{3\dow_{t1}\dow_{t2}}{2 \omega_\beta ^2}
    -\frac{\dow_{t1}^2\dow_{t2}^2}{2 \omega_\beta ^4}
    \rb\rB
    \delta\sigma
    \delta\sigma
    + \ldots \nn\\
    &= \lb 1 -\frac{\dow_t^2}{\omega_\beta ^2}
    \rb \dow_t\delta F
    + \lB \beta  \dow_{t1} \dow_{t2}
    \lb 
    \frac{\dow_{t1}^2+\dow_{t2}^2}{\omega_\beta ^2} 
    + \frac{3\dow_{t1}\dow_{t2}}{2 \omega_\beta ^2}
    -\frac{\dow_{t1}^2\dow_{t2}^2}{2 \omega_\beta ^4}
    \rb\rB
    \delta F \delta F \nn\\
    &\qquad 
    - \lB \frac{\beta }{8}
    \lb 1 -\frac{\lb \dow_{t1}+\dow_{t2}\rb^2}{\omega_\beta ^2} \rb \lb \dow_{t1}+\dow_{t2}\rb^2 \rB \delta Q\delta Q
    + \ldots, \nn\\
    \delta\chi_a 
    &= -\beta \delta Q 
    + \Big[\beta ^2 \dow_{t1}\Big]\delta F \delta Q 
    + \ldots 
\end{calign}
The free part of the action in this basis takes the form
\begin{calign}\label{eq:SL2R-invariant-free-action}
    \hat S_{\rm free}
    &= \int \df^{d+1}x
    \bigg( \delta\chi_a\,
    \frac{\Delta(\dow)}{\beta }\delta T
    + \frac{i}{4\beta ^2}\delta\chi_a\,\cot\bfrac{\beta \dow_t}{2}
    \Big( K(-\dow) - K(\dow) \Big)
    \,\delta\chi_a
    \bigg),
\end{calign}
which leads to the propagators given in \cref{eq:SL2R-invariant-props}.
The three-point interactions take the form
\begin{calign}\label{eq:action-expansion-goldstone}
    \hat S_{\text{int-3}}
    &= \int \df^{d+1}x
    \Bigg(
    \cV_{T^2\chi_a}(\dow_1,\dow_2;\dow_3)\, 
    \delta T\,\delta T\,\delta\chi_a \nn\\[-1em]
    &\hspace{6em}
    + i\cV_{T\chi_a^2}(\dow_1;\dow_2,\dow_3)\, 
    \delta T\,\delta\chi_a\, \delta\chi_a
    + \cV_{\chi_a^3}(\dow_1,\dow_2,\dow_3)\, 
    \delta\chi_a\,\delta\chi_a\,\delta\chi_a
    \Bigg) + \ldots.
\end{calign}
where we have identified the vertex factors
\begin{calign}
    \cV_{T^2\chi_a}(\dow_1,\dow_2;\dow_3)
    &= \frac{1}{2}\Delta(\dow_1+\dow_2)
    + \frac{1}{\beta }\cR(\dow_1,\dow_2), \nn\\
    \cV_{T\chi_a^2}(\dow_1;\dow_2,\dow_3)
    &= 
    \lB \frac{K(-\dow_2) - K(\dow_2)}{2\beta \lb 1 - \frac{\dow_{t1}^2}{\omega_\beta^2} \rb}
    \cot\bfrac{\beta \dow_{t2}}{2} \rB_{(23)}
    + \frac{\cV_{FQ^2}(\dow_1;\dow_2,\dow_3)}{\beta ^2\dow_{t1}\lb 1 - \frac{\dow_{t1}^2}{\omega_\beta^2} \rb} \nn\\
    &= - \Bigg[ \Bigg(
    \frac{1}{2\beta }
    \dow_{t2} \Delta(\dow_1) 
    - \frac{1}{2\beta } 
    \lb\dow_{t2}-\dow_{t3}\rb 
    \Delta(-\dow_3) 
    \nn\\
    &\hspace{6em}
    - 
    \frac{1}{\beta ^2} \dow_{t2} \lb 1 - \frac{\dow_{t2}^2}{\omega_\beta^2} \rb
    \cR(\dow_1,\dow_2) \Bigg)
     \cot\bfrac{\beta \dow_{t2}}{2} \Bigg]_{(23)} \nn\\
     &\quad
     + \frac{1}{2\beta ^2} \Bigg[ 
     - \frac{\beta }{2} \dow_{t1}\dow_{t3}
    \lb 1 - \frac{\dow_{t3}^2}{\omega_\beta^2}\rb \Delta(-\dow_3) 
    -\frac{\beta }{2} \dow_{t1}\dow_{t2}
    \lb 1 - \frac{\dow_{t2}^2}{\omega_\beta^2}\rb 
    \Delta(-\dow_2) \nn\\
    &\hspace{6em}
    + \tilde\cV_{F^2Q}(\dow_2,\dow_3;\dow_1)
     \Bigg]
     \frac{\sin\bfrac{\beta \dow_{t1}}{2}}{
     \sin\bfrac{\beta \dow_{t2}}{2}
     \sin\bfrac{\beta \dow_{t3}}{2}
     \dow_{t1}\lb 1 - \frac{\dow_{t1}^2}{\omega_\beta^2} \rb
     }, \nn\\
     \cV_{\chi_a^3}(\dow_1,\dow_2,\dow_3)
     &= 
    \lB \frac{1}{8\beta ^2}
    \lb 1 -\frac{\dow_{t1}^2}{\omega_\beta^2} \rb \dow_{t1}^2 
    \Delta(-\dow_{1})
    \rB_{(123)}
    - \frac{1}{\beta ^3}\cV_{Q^3}(\dow_1,\dow_2,\dow_3).
\end{calign}
Note that while we have divided the definition of $\cV_{T\chi_a^2}$ vertex with $\dow_{t1}(1-\dow_{t1}^2/\omega_\beta^2)$, it is actually completely regular near the growing modes and near $\dow_{t1}=0,\pm\omega_\beta$. This illustrates that the action can be expressed entirely in terms of $\delta T$ and $\delta\chi_a$ (at least until cubic interactions).

\subsection{Expansion of local operators}
\label{app:operator-expansion}

We are interested in correlation functions of local $\SL(2,\bbR)$-invariant operators $\cO$ inserted along the SK closed-time contour. Classically, such an operator expansion can be expanded in fields as
\begin{calign}
    \cO_i 
    &= \cO_0 
    + \cD_{(1)}(\dow)\delta T 
    + \cD_{(2)}(\dow_1,\dow_2)\delta T\delta T
    + \ldots 
    + {\rm noise~terms} \nn\\
    &= \cO_0 
    + \cO_{(1)}(\dow)\delta F 
    + \cO_{(2)}(\dow_1,\dow_2)\delta F\delta F
    + \ldots 
    + {\rm noise~terms},
\end{calign}
where ``$i$'' denotes the contour label, while $\cD_{(n)}$ and $\cO_{(n)}$ are $n$-linear differential operators. Note that the contour labels are classically irrelevant when the noise fields are set to zero. We have
\begin{calign}\label{eq:operator-operators}
    \cO_{(1)}(\dow) 
    &= 
    \dow_t\lb 1 - \frac{\dow_t^2}{\omega_\beta^2} \rb
    \cD_{(1)}(\dow), \nn\\
    \cO_{(2)}(\dow_1,\dow_2)
    &= \dow_{t1}\dow_{t2}
    \Bigg[
    \frac{\beta }{2}
    \lb 1 + \frac{\dow_{t1}^2 + \dow_{t2}^2+ 3 \dow_{t1}\dow_{t2}}{\omega_\beta^2} \rb
    \cD_{(1)}(\dow_1+\dow_2) \nn\\
    &\hspace{6em}
    + \lb 1-\frac{\dow_{t1}^2}{\omega_\beta^2} \rb
    \lb 1-\frac{\dow_{t2}^2}{\omega_\beta^2} \rb 
    \Big(\cD_{(2)}(\dow_1,\dow_2) 
    - \frac{\beta }{2} \cD_{(1)}(\dow_1+\dow_2)
    \Big)
    \Bigg].
\end{calign}

The noise terms in the expansion above can be fixed using the unitarity and symmetry constraints. To this end, we can employ the following algorithm. We first derive the noise terms on the two-fold contour by defining the source action $S_\phi = \int \lb \cO_1\phi_1-\cO_2\phi_2 \rb$ and fix the noise contributions by requiring it to be invariant under the two-fold KMS symmetry and other symmetries of the energy diffusion SK EFT. We then define the four-fold source action analogous to the combination given in \cref{eq:4fold-action}, which enables us to read off the operators on the four-fold contour. To wit, generic operator-source terms in the two-fold effective action are given as
\begin{calign}
    S_{\rm source}[F_1,F_2;\phi_1,\phi_2]
    &= \delta\phi_a \cO_r
    + \delta\phi_r \cO_a ,
\end{calign}
where we have identified
\begin{calign}
    \cO_r 
    &= \cO_0 + \cO_{(1)}(\dow) \delta F
    + iM_{a}(\dow) \delta Q \nn\\
    &\qquad
    + \cO_{(2)}(\dow_1,\dow_2) \delta F \delta F
    + iM_{ra}(\dow_1,\dow_2) \delta F \delta Q
    + M_{aa}(\dow_1,\dow_2) \delta Q \delta Q
    + \ldots, \nn\\
    \cO_a 
    &= N_{a}(\dow) \delta Q
    + N_{ra}(\dow_1,\dow_2) \delta F \delta Q
    + iN_{aa}(\dow_1,\dow_2) \delta Q \delta Q
    + \ldots,
\end{calign}
for some differential operators $M$'s and $N$'s.
Requiring that the source action is KMS-invariant enables us to fix
\begin{calign}
    M_{a}(\dow)
    &= \frac{-1}{2} \frac{\sfC}{\sfS} \Big( \cO_{(1)}(\dow) 
    + \eta_\Theta\tilde\cO_{(1)}(\dow) 
    \Big), \nn\\ 
    N_{a}(\dow) 
    &= -\eta_\Theta\tilde\cO_{(1)}(\dow), \nn\\
    M_{ra}(\dow_1,\dow_2)
    &= \frac{\sfC_3}{\sfS_3}
    \Big( \cO_{(2)}(\dow_1,\dow_2)-\eta_\Theta\tilde\cO_{(2)}(\dow_1,\dow_2)\Big) 
    + 
    \lb 
    \frac{\sfC_3}{\sfS_3}
    Y(\dow_1,\dow_2)
    + \frac{\sfS_1}{\sfS_2\sfS_3} \eta_\Theta
    \tilde Y(\dow_2,\dow_1) \rb, \nn\\
    M_{aa}(\dow_1,\dow_2)
    &= \frac{1}{4} \cO_{(2)}(\dow_1,\dow_2)
    + \frac{\sfC_1}{4\sfS_1}
    \lb \frac{\sfC_3}{\sfS_3} Y(\dow_1,\dow_2) 
    + \frac{\sfS_1}{\sfS_2\sfS_3} 
    \eta_\Theta\tilde Y(\dow_2,\dow_1)
    \rb \nn\\
    &\hspace{6em}
    + \frac{\sfC_2}{4\sfS_2}
    \lb \frac{\sfC_3}{\sfS_3} Y(\dow_2,\dow_1) 
    + \frac{\sfS_2}{\sfS_1\sfS_3} 
    \eta_\Theta \tilde Y(\dow_1,\dow_2)
    \rb, \nn\\
    N_{ra}(\dow_1,\dow_2)
    &=
    2 \eta_\Theta \tilde\cO_{(2)}(\dow_1,\dow_2)
    - 2 Y(\dow_1,\dow_2), \nn\\
    N_{aa}(\dow_1,\dow_2)
    &= \frac{\sfC_1}{2\sfS_1}  Y(\dow_1,\dow_2)
    + \frac{\sfC_2}{2\sfS_2}  Y(\dow_2,\dow_1),
\end{calign}
where $\eta_\Theta$ is the time-reversal eigenvalue of $\cO$ and we have identified an unfixed differential operator $ Y(\dow_1,\dow_2)$. To get some intuition behind these expressions, let us note the representation of the operators in contour-basis
\begin{calign}\label{eq:generic-two-fold-op}
    &\cO_{1,2}
    = \cO_0 
    + \cO_{(1)}(\dow) \delta F_{1,2}
    + \cO_{(2)}(\dow_1,\dow_2) \delta F_{1,2}\delta F_{1,2}
    + \ldots
    \nn\\
    &\qquad
    - \half \lb
    \frac{i\sfC}{\sfS} 
    \pm 1
    \rb \Big( \cO_{(1)}(\dow) 
    + \eta_\Theta\tilde\cO_{(1)}(\dow) 
    \Big) \delta Q
    \nn\\
    &\qquad 
    + \lb \frac{i\sfC_3}{\sfS_3} \mp 1 \rb
    \Big( \cO_{(2)}(\dow_1,\dow_2) 
    - \eta_\Theta \tilde\cO_{(2)}(\dow_1,\dow_2) \Big)
    \delta F \delta Q
    \nn\\
    &\qquad 
    + \lB \frac{i\sfC_3}{\sfS_3}
    \lb 
    \frac{1}{\sfS_1}
     Y(\dow_1,\dow_2)
    + \frac{\eta_\Theta}{\sfS_2\sfC_3}
    \tilde Y(\dow_2,\dow_1) \rb
    \mp \frac{1}{\sfS_1} Y(\dow_1,\dow_2) 
    \rB 
    \Big( \sfS_1 \delta F \delta Q
    - i \sfC_1\delta Q \delta Q
    \Big)
    + \ldots.
\end{calign}
We see that the operator $\cO_i$ is localised on the $i$th fold of the closed-time contour if and only if $\cO_{(n)}$ has time-reversal eigenvalue $(-1)^n\eta_\Theta$ and $ Y$ is zero. In other words, $ Y$ and the part of $\cO_{(n)}$ with eigenvalue $(-1)^{n+1}\eta_\Theta$ parametrise the coupling of the operator across different folds of the closed-time contour. The $\SL(2,\bbR)$ symmetry further requires that
\begin{calign}
    Y(\dow_1,\dow_2) = \dow_{t1} \lb 1 - \frac{\dow_{t1}^2}{\omega_\beta^2} \rb
    \cY(\dow_1,\dow_2).
\end{calign}

Furthermore, for theories with exclusively time-reversal-odd dissipation, i.e. imposing \cref{eq:cond-odd-dissipation}, we further find that 
\begin{calign}
    \tilde Y(\dow_1,\dow_2) &= -\eta_\Theta  Y(\dow_1,\dow_2), \nn\\
    \implies 
    \tilde\cY(\dow_1,\dow_2) &= \eta_\Theta\cY(\dow_1,\dow_2)
\end{calign}

We can find the representation of the operators on four-fold contour by defining the source action
\begin{calign}
    S_{\rm source}&[F_1,F_2,F_3,F_4;\phi_1,\phi_2,\phi_3,\phi_4] \nn\\
    &= S_{\rm source}[F_1,F_2;\phi_1,\phi_2]
    - S_{\rm source}[F_1,F_3;\phi_1,\phi_3]
    + S_{\rm source}[F_1,F_4;\phi_1,\phi_4] \nn\\
    &\qquad 
    + S_{\rm source}[F_2,F_3;\phi_2,\phi_3]
    - S_{\rm source}[F_2,F_4;\phi_2,\phi_4]
    + S_{\rm source}[F_3,F_4;\phi_3,\phi_4]
    + \ldots
    \nn\\
    &= 
    \delta\phi_a \cO_r
    + \delta\phi_r \cO_a 
    + \delta\phi_{\bar r} \cO_{\bar a}
    + \delta\phi_{\bar a} \cO_{\bar r}.
\end{calign}
Going through the calculations, we find
\begin{calign}\label{eq:general-operator-expansion}
    \cO_r 
    &= \cO_{(1)}(\dow)\delta F 
    - \frac{i}{2} \frac{\sfC}{\sfS} \Big( \cO_{(1)}(\dow) 
    + \eta_\Theta\tilde\cO_{(1)}(\dow) 
    \Big)\delta Q \nn\\
    &\qquad 
    + \cO_{(2)}(\dow_1,\dow_2)
    \left(
    \delta F \delta F
    + \frac{1}{4} \delta Q \delta Q
    + \frac{1}{4} \delta \bar Q\delta \bar Q
    + \frac{1}{2} \delta \bar Q\delta Q
    \right) \nn\\
    &\qquad 
    + \frac{i\sfC_3}{\sfS_3} 
    \lb \cO_{(2)}(\dow_1,\dow_2)-\eta_\Theta \tilde\cO_{(2)}(\dow_1,\dow_2)\rb 
    \Big( \delta F \delta Q + \delta \bar F \delta \bar Q \Big)\nn\\
    &\qquad 
    + \left(
    \frac{\sfC_3}{\sfS_3} 
    Y(\dow_1,\dow_2)
    + \frac{\sfS_1}{\sfS_2\sfS_3}
    \eta_\Theta\tilde Y(\dow_2,\dow_1)
    \right)
    \left(\frac{1}{2} \frac{\sfC_1}{\sfS_1}\delta Q \delta Q
    + i\delta F \delta Q \right)  \nn\\
    &\qquad 
    +  Y(\dow_1,\dow_2) 
    \left(\frac{1}{2} \delta Q \delta \bar Q
    + \frac{i \sfC_3}{\sfS_3} \Big(\delta \bar Q \delta \bar F
    + \delta \bar F \delta \bar Q\Big)\right) \nn\\
    &\qquad 
    + \lbr \frac{ Y(\dow_1,\dow_2)+\eta_\Theta \tilde Y(\dow_1,\dow_2)}{\sfS_1 \sfS_3}
    \left(\frac{\sfC_2}{4} \Big(\delta Q\delta \bar Q
    + \delta \bar Q \delta Q \Big)
    +\frac{i\sfS_2}{2} \Big(\delta \bar Q \delta \bar F
    + \delta \bar F \delta \bar Q \Big)\right) \rbr
    + \ldots, \nn\\
    \cO_{a}
    &= -\eta_\Theta\tilde\cO_{(1)}(\dow) \delta Q \nn\\
    &\qquad 
    + 2 \Big( \eta_\Theta\tilde\cO_{(2)}(\dow_1,\dow_2)
    -  Y(\dow_1,\dow_2)\Big)
    \Big( \delta \bar Q \delta \bar F
    + \delta F \delta Q\Big)
    + \frac{i\sfC_1}{\sfS_1}  Y(\dow_1,\dow_2) \delta Q \delta Q
    + \ldots, \nn\\
    \cO_{\bar a}
    &= \cO_{(1)}(\dow) \delta\bar Q \nn\\
    &\qquad 
    + 2 \cO_{(2)}(\dow_1,\dow_2) \Big(\delta \bar Q\delta F
    - \delta \bar Q \delta \bar F \Big)
    -  Y(\dow_1,\dow_2) 
    \lb 2 \delta \bar F \delta \bar Q
    - \frac{i\sfC_1}{\sfS_1} \delta Q\delta \bar Q\rb \nn\\
    &\qquad 
    - \lbr \frac{ Y(\dow_1,\dow_2)+\eta_\Theta \tilde Y(\dow_1,\dow_2)}{\sfS_1 \sfS_3}
    \left(\sfC_2 \Big(\delta \bar Q \delta \bar F
    + \delta \bar F \delta \bar Q\Big)
    +\frac{i\sfS_2}{2}\Big(\delta \bar Q \delta Q -\delta Q\delta \bar Q\Big)\right) \rbr
    + \ldots, \nn\\
    \cO_{\bar r}
    &= -\eta_\Theta\tilde\cO_{(1)}(\dow)\delta\bar F \nn\\
    &\qquad
    + \eta_\Theta \tilde\cO_{(2)}(\dow_1,\dow_2) \left(
    2 \delta F \delta \bar F
    - \delta \bar F \delta \bar F
    + \frac{1}{2} \delta \bar Q \delta Q
    + \frac{1}{4} \delta Q \delta Q\right) \nn\\
    &\qquad 
    - Y(\dow_1,\dow_2) \left(
    2 \delta F \delta \bar F
    + \frac{1}{2} \delta \bar Q \delta Q
    - \frac{i\sfC_1}{\sfS_1} \delta Q \delta \bar F
    \right) \nn\\
    &\qquad 
    + \lbr \frac{ Y(\dow_1,\dow_2)+\eta_\Theta \tilde Y(\dow_1,\dow_2)}{\sfS_1 \sfS_3}
    \left(
    \sfC_2 \lb \frac{1}{4} \delta Q \delta Q
    - \delta \bar F \delta \bar F \rb
    -\frac{i\sfS_2}{2} \Big(\delta \bar F \delta Q
    - \delta Q \delta \bar F \Big)
    \right) \rbr
    + \ldots.
\end{calign}
The terms in curly-braces above drop out for theories with $\Theta$-odd dissipation.

In general, the quadratic noise terms in \cref{eq:general-operator-expansion} involve an arbitrary differential operator $Y(\dow_1,\dow_2)= \dow_{t1}( 1 - \dow_{t1}^2/\omega_\beta^2)\cY(\dow_1,\dow_2)$ left unfixed by symmetries. The energy current operator can be obtained exactly, including noise terms, by using the equations of motion of the SK EFT, up to total derivative terms. This fixes the divergence of the associated $\cY^\mu(\dow_1,\dow_2)$ operator for the energy current:
\begin{calign}
    &\Big(\dow_{1\mu}+\dow_{2\mu}\Big) \cY^\mu(\dow_1,\dow_2) \nn\\
    &= 
    \lb \dow_{t1}+2\dow_{t2}\rb
    \left(-\frac{1}{2} \Delta(\dow_1)
    - \frac{1}{2} \Delta(-\dow_1-\dow_2)
    + \frac{1}{2} \Delta(-\dow_2) 
    \right) \nn\\
    &\hspace{2em}
    - \frac{1}{\beta }\dow_{t2} \left(1-\frac{\dow_{t2}^2}{\omega_\beta ^2}\right) 
    \tilde\cR(\dow_1,\dow_2)
    + \frac1\beta\lb \dow_{t1}+\dow_{t2} \rb 
    \left(1+\frac{(\dow_{t1}+\dow_{t2})^2}{\omega_\beta ^2}\right) 
    \cR(\dow_1,-\dow_1-\dow_2).
\end{calign}
One can check that the right-hand side is indeed a total derivative in $\dow_{1\mu}+\dow_{2\mu}$, i.e. it vanishes when we take $\dow_{1\mu}=-\dow_{2\mu}$. We can solve for $\cY^\mu$ using the expressions for $\Delta$ and $\cR$ in \cref{eq:Delta-R-from-EOM}. In addition, since $\Delta(\dow_1)-\Delta(-\dow_2)$ vanishes when $\dow_1=-\dow_2$, we can generically factorise it as $\Delta(\dow_1)+\Delta(-\dow_1-\dow_2)-\Delta(-\dow_2) = (\dow_{1\mu}+\dow_{2\mu})Z^\mu(\dow_1,\dow_2)$. Formally, we can write,
\begin{align}
    Z^\mu(\dow_1,\dow_2)
    &= \frac{\Delta(\dow_1)+\Delta(-\dow_1-\dow_2)-\Delta(-\dow_2)}{\dow_{1\mu}+\dow_{2\mu}}.
\end{align}
Using these definitions, we can solve
\begin{calign}\label{eq:Ymu-def}
    \cY^\mu(\dow_1,\dow_2) 
    &= -\half \lb \dow_{t1}+2\dow_{t2}\rb Z^\mu(\dow_1,\dow_2) \nn\\
    &\hspace{4em}
    + \dow_{t2} \left(1-\frac{\dow_{t2}^2}{\omega_\beta ^2}\right) 
    \lb \eta_\Theta \tilde\cD^\mu_{(2)}(\dow_1,\dow_2) 
    - \frac{\beta }{2}\cD^\mu_{(1)}(-\dow_1-\dow_2) \rb \nn\\
    &\hspace{4em}
    - \delta^\mu_t
    \left(1+\frac{(\dow_{t1}+\dow_{t2})^2}{\omega_\beta ^2}\right) 
    \dow_{2\nu}
    \lb \cD^\nu_{(2)}(\dow_1,-\dow_1-\dow_2) 
    - \frac{\beta }{2} \cD^\nu_{(1)}(-\dow_2) \rb.
\end{calign}
We can use this to read off the full physical and noise energy current operators up to quadratic order in fluctuations. In the special case of SYK like models where $\Delta(\dow_t,i\bk) = \beta c_\bk\lb \dow_t + \cD_\bk\rb$ and $\cR(\dow_1,\dow_2)=0$, we find
\begin{calign}\label{eq:Ymu-def-basic}
    \cY^t(\dow_1,\dow_2)
    &= 0, \nn\\
    \cY^\bx(\dow_1,\dow_2)
    &= -\frac{1}{2} 
    \lb \dow_{t1}+2\dow_{t2}\rb
    Z^\bx(\dow_1,\dow_2) \nn\\
    &= 
    \beta Dc_V 
    \lb \dow_{t1}+2\dow_{t2}\rb
    \dow_{\bx1}
    + \text{higher-order in derivatives}.
\end{calign}
Note that since $\Delta$ is linear in $\dow_t$, we find that $Z^t=0$ in this case. In other words, only the energy flux couples across different Lorentzian folds in such theories, while the energy density is completely localised to individual folds.
For the ASCJ theory, $\Delta$ is entirely linear in $\dow_\mu$, so $Z^\mu$ and hence $Y^\mu$ vanishes identically. This is to be expected because the energy-momentum tensor in ASCJ theory is just the Schwarzian on individual folds.

\section{Details of OTOC computation}
\label{app:loops}

In this appendix we give the details of OTOC calculation for local operators in the four-fold SK EFT. We are interested in computing OTOCs of local operators using SK EFT. To this end, we will employ the basis defined in \cref{eq:new-basis-otocs}, slightly generalised to distinct operators at all four insertion points
\begin{calign}\label{eq:new-basis-otocs-general}
    C_{\bar a \bar a |\bar r \bar r}(x_1,x_2;x_3,x_4)
    &\equiv \Big\langle
    V_{\bar a}(x_1)\,
    V'_{\bar a}(x_2)\,
    W'_{\bar r}(x_3)\,
    W_{\bar r}(x_4)
    \Big\rangle, \nn\\
    C_{\bar a \bar a |\bar r a}(x_1,x_2;x_3,x_4)
    &\equiv \Big\langle
    V_{\bar a}(x_1)\,
    V'_{\bar a}(x_2)\,
    W'_{\bar r}(x_3)\,
    W_{a}(x_4)
    \Big\rangle.
\end{calign}
These are represented in operator-language as
\begin{calign}
    C_{\bar a \bar a |\bar r \bar r}
    &(x_1,x_2;x_3,x_4) \nn\\
    &=
    \half\Theta(t_1,t_2,t_3,t_4) 
    \bigg\langle
    \Big\{ 
    \Big[\hat V(x_1),\hat W'(x_3)\Big],
    \Big[\hat V'(x_2),\hat W(x_4)\Big]
    \Big\} \nn\\
    &\hspace{14em}
    + \Big\{ 
    \Big[\hat V(x_1),\hat W(x_4)\Big],
    \Big[\hat V'(x_2),\hat W'(x_3)\Big]
    \Big\}
    \bigg\rangle \nn\\
    &\qquad 
    + \half \Theta(t_1,t_4,t_2,t_3) 
    \bigg\langle
    \Big\{ 
    \Big[\hat V(x_1),\hat W(x_4)\Big],
    \Big[\hat V'(x_2),\hat W'(x_3)\Big]
    \Big\}
    \bigg\rangle
    \nn\\
    &\qquad 
    + (x_1\leftrightarrow x_2)
    + (x_3\leftrightarrow x_4)
    + (x_1\leftrightarrow x_2,
    x_3\leftrightarrow x_4), \nn\\
    C_{\bar a \bar a |\bar r a}
    &(x_1,x_2;x_3,x_4) \nn\\
    &=
    \Theta(t_1,t_2,t_3,t_4)
    \bigg\langle\Big[\hat W(x_4),
    \Big[\hat V(x_1),
    \Big[\hat V'(x_2),\hat W'(x_3)\Big]\Big] \Big] 
    \bigg\rangle
    \nn\\
    &\qquad 
    + \Theta(t_1,t_4,t_2,t_3)
    \bigg\langle\Big[\Big[\hat V'(x_2),\hat W'(x_3)\Big],
    \Big[\hat V(x_1),\hat W(x_4)\Big]\Big]
    \bigg\rangle
    \nn\\
    &\qquad 
    + \Theta(t_1,t_2,t_4,t_3)
    \bigg\langle
    \Big[\hat W(x_4),
    \Big[\hat V(x_1),
    \Big[\hat V'(x_2),\hat W'(x_3)\Big]\Big] \Big]   \nn\\
    &\hspace{14em}
    + \Big[\hat V(x_1),
    \Big[\hat V'(x_2),
    \Big[\hat W'(x_3),\hat W(x_4)\Big]\Big] \Big]  
    \bigg\rangle
    \nn\\
    &\qquad 
    + (x_1\leftrightarrow x_2),
\end{calign}
where $\Theta(t_i,t_j,t_k,t_l)= \Theta(t_i-t_j)\Theta(t_j-t_k)\Theta(t_k-t_l)$. One can see that $C_{\bar a \bar a |\bar r\bar r}$ is a linear combination of products of commutators in \cref{eq:comm-square} for any ordering of insertions times. On the other hand, $C_{\bar a \bar a |\bar ra}$ cannot generically be expressed as simply a product of commutators and requires the nested commutator structure mentioned in \cref{eq:nested-comm}.
Note that $C_{\bar a \bar a |\bar r a}$ is zero if $t_3 > t_1$ or $t_3>t_2$. On the other hand, $C_{\bar a \bar a |\bar r \bar r}$ is zero if either $t_1$ or $t_2$ is the smallest, or either $t_3$ or $t_4$ is the largest.

There are four classes of diagrams for both cases: contact diagrams, $s$-channel, $t$-channel, and $u$-channel. We will compute these in detail in the following.

\subsection{Antisymmetric OTOC in SK EFT}

\subsubsection{Contact diagrams}
\label{sec:contact-diagrams}

The typical contact diagram for linear operator insertions involves a four-point interaction, which for our case is given as
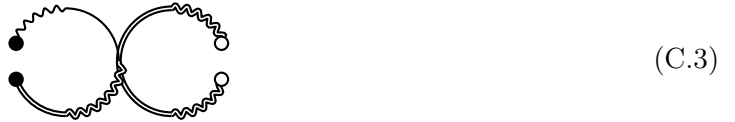
\begin{calign}
\tikzsetnextfilename{2loop-diagram-contact-1}
\raisebox{-6mm}{
\tikz[thick,scale=0.7,xscale=-1]{
    \draw [aux2] (10mm,10mm) arc (90:160:10mm);
    \draw [sol2] (10mm,10mm) arc (90:0:10mm);
    \draw [sol2] (20mm,0) arc (0:-90:10mm);
    \draw [aux2] (10mm,-10mm) arc (-90:-160:10mm);
    \draw [] (20mm,0) arc (180:90:10mm);
    \draw [aux] (30mm,10mm) arc (90:20:10mm);
    \draw [sol2] (30mm,-10mm) arc (-90:-20:10mm);
    \draw [aux2] (30mm,-10mm) arc (-90:-180:10mm);
    \draw [fill=white] (0.6mm,3.4mm) circle (1.25mm);
    \draw [fill=white] (0.6mm,-3.4mm) circle (1.25mm);
    \draw [fill=black] (39.4mm,3.4mm) circle (1.25mm);
    \draw [fill=black] (39.4mm,-3.4mm) circle (1.25mm);
}}
\end{calign}
However, since our operators are generically non-linear, we have 4 related diagrams involving the cubic parts of operator insertions as well, obtained by collapsing one of the external propagators in the diagram above. To wit
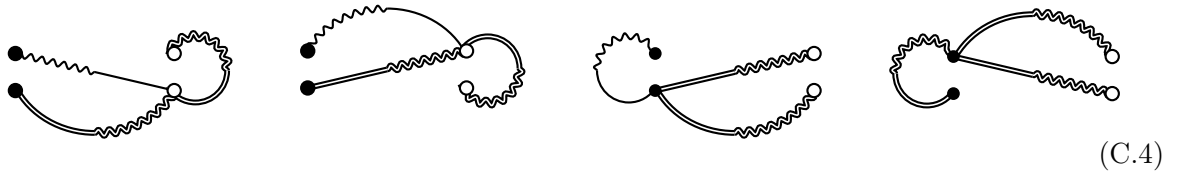
\begin{cgather}
\tikzsetnextfilename{2loop-diagram-contact-2-1}
\raisebox{-7mm}{
\tikz[thick,scale=0.7,xscale=-1]{
\draw [aux2] (-10mm,-3.25mm) arc (180:30:6mm);
\draw [sol2] (-10mm,-3.25mm) arc (-180:-30:6mm);
\draw [] (0,-7mm) -- (15mm,-3.5mm);
\draw [aux] (15mm,-3.5mm) -- (30mm,0);
\draw [aux2] (15mm,-15mm) arc (-90:-155:16.5mm and 15mm);
\draw [sol2] (15mm,-15mm) arc (-90:-25:16.5mm and 15mm);
\draw [fill=white] (0,0) circle (1.25mm);
\draw [fill=white] (0,-7mm) circle (1.25mm);
\draw [fill=black] (30mm,0) circle (1.25mm);
\draw [fill=black] (30mm,-7mm) circle (1.25mm);
}} \qquad 
\tikzsetnextfilename{2loop-diagram-contact-2-3}
\raisebox{-3mm}{
\tikz[thick,scale=0.7,xscale=-1]{
\draw [sol2] (-10mm,-3.25mm) arc (180:30:6mm);
\draw [aux2] (-10mm,-3.25mm) arc (-180:-30:6mm);
\draw [aux2] (0,0) -- (15mm,-3.5mm);
\draw [sol2] (15mm,-3.5mm) -- (30mm,-7mm);
\draw [] (15mm,8mm) arc (90:155:16.5mm and 15mm);
\draw [aux] (15mm,8mm) arc (90:25:16.5mm and 15mm);
\draw [fill=white] (0,0) circle (1.25mm);
\draw [fill=white] (0,-7mm) circle (1.25mm);
\draw [fill=black] (30mm,0) circle (1.25mm);
\draw [fill=black] (30mm,-7mm) circle (1.25mm);
}} \qquad 
\tikzsetnextfilename{2loop-diagram-contact-2-2}
\raisebox{-7mm}{
\tikz[thick,scale=0.7,xscale=-1]{
\draw [aux2] (0,0) -- (15mm,-3.5mm);
\draw [sol2] (15mm,-3.5mm) -- (30mm,-7mm);
\draw [aux2] (15mm,-15mm) arc (-90:-155:16.5mm and 15mm);
\draw [sol2] (15mm,-15mm) arc (-90:-25:16.5mm and 15mm);
\draw [aux] (41mm,-3.25mm) arc (0:150:6mm);
\draw [] (41mm,-3.25mm) arc (0:-150:6mm);
\draw [fill=white] (0,0) circle (1.25mm);
\draw [fill=white] (0,-7mm) circle (1.25mm);
\draw [fill=black] (30mm,0) circle (1mm);
\draw [fill=black] (30mm,-7mm) circle (1mm);
}} \qquad
\tikzsetnextfilename{2loop-diagram-contact-2-4}
\raisebox{-3mm}{
\tikz[thick,scale=0.7,xscale=-1]{
\draw [aux2] (0,-7mm) -- (15mm,-3.5mm);
\draw [sol2] (15mm,-3.5mm) -- (30mm,0);
\draw [aux2] (15mm,8mm) arc (90:155:16.5mm and 15mm);
\draw [sol2] (15mm,8mm) arc (90:25:16.5mm and 15mm);
\draw [aux2] (41mm,-3.25mm) arc (0:150:6mm);
\draw [sol2] (41mm,-3.25mm) arc (0:-150:6mm);
\draw [fill=white] (0,0) circle (1.25mm);
\draw [fill=white] (0,-7mm) circle (1.25mm);
\draw [fill=black] (30mm,0) circle (1mm);
\draw [fill=black] (30mm,-7mm) circle (1mm);
}}
\end{cgather}
None of these diagrams contribute to the OTOC growth. This is because every propagator in these diagrams is connected to a linear operator insertion, which means that every propagator is accompanied by the structure factor $\cO_{(1)}(\dow)$ or $\tilde\cO_{(1)}(\dow)$ at the operator insertion. Due to the $\SL(2,\bbR)$ symmetry, the structure factor at the linear operator insertion behaves as $\cO_{(1)}(\dow) \sim \dow_t(1-\dow_t^2/\omega_\beta^2)$ in \cref{eq:operator-operators}, which exactly cancels the growing part of the propagator. To wit
\begin{calign}
    \cO_{(1)}(\dow) iK^{-1}(x-y)
    = -\cD^\cO_{(1)}(\dow) i\Delta^{-1}(x-y).
\end{calign}
Since $\Delta^{-1}(x-y)$ decays to zero at late times, these diagrams cannot contribute to the characteristic growth of OTOCs. We will compute these explicitly later for the case of (1+1)d CFT in \cref{app:2dcft}.

\subsubsection{t-channel and u-channel diagrams}

The next set of diagrams to consider are the t-channel and u-channel diagrams, given as
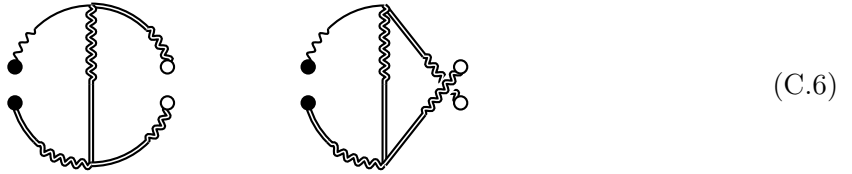
\begin{cgather}
\tikzsetnextfilename{2loop-diagram-t-1}
\raisebox{-10mm}{
\tikz[thick,scale=0.7,xscale=-1]{
    \draw [aux2] (4.4mm,10.6mm) arc (135:165:15mm);
    \draw [sol2] (15mm,15mm) arc (90:135:15mm);
    \draw [] (15mm,15mm) arc (90:45:15mm);
    \draw [aux] (25.6mm,10.6mm) arc (45:15:15mm);
    \draw [aux2] (15mm,15mm) -- (15mm,0);
    \draw [sol2] (15mm,0) -- (15mm,-15mm);
    \draw [aux2] (4.4mm,-10.6mm) arc (-135:-165:15mm);
    \draw [sol2] (15mm,-15mm) arc (-90:-135:15mm);
    \draw [aux2] (15mm,-15mm) arc (-90:-45:15mm);
    \draw [sol2] (25.6mm,-10.6mm) arc (-45:-15:15mm);
    \draw [fill=white] (0.6mm,3.4mm) circle (1.25mm);
    \draw [fill=white] (0.6mm,-3.4mm) circle (1.25mm);
    \draw [fill=black] (29.4mm,3.4mm) circle (1.25mm);
    \draw [fill=black] (29.4mm,-3.4mm) circle (1.25mm);
}}  \qquad\qquad 
\tikzsetnextfilename{2loop-diagram-u-1}
\raisebox{-10mm}{
\tikz[thick,scale=0.7]{
    \draw [aux] (4.4mm,10.6mm) arc (135:165:15mm);
    \draw [] (15mm,15mm) arc (90:135:15mm);
    \draw [sol2] (15mm,15mm) -- (22.5mm,5.5mm);
    \draw [aux2] (22.5mm,5.5mm) -- (30mm,-4mm);
    \draw [aux2] (15mm,15mm) -- (15mm,0);
    \draw [sol2] (15mm,0) -- (15mm,-15mm);
    \draw [sol2] (4.4mm,-10.6mm) arc (-135:-165:15mm);
    \draw [aux2] (15mm,-15mm) arc (-90:-135:15mm);
    \draw [fill=white, draw=none, ,rotate around={50:(27mm,0)}] (26.9mm-1.5mm,-1.5mm) rectangle +(3mm,3mm);
    \draw [sol2] (15mm,-15mm) -- (22.5mm,-5.5mm);
    \draw [aux2] (22.5mm,-5.5mm) -- (30mm,4mm);
    \draw [fill=black] (0.6mm,3.4mm) circle (1.25mm);
    \draw [fill=black] (0.6mm,-3.4mm) circle (1.25mm);
    \draw [fill=white] (29.4mm,3.4mm) circle (1.25mm);
    \draw [fill=white] (29.4mm,-3.4mm) circle (1.25mm);
}}
\end{cgather}
Similar to before, there are 8 related diagrams in the t-channel involving the non-linear parts of operator insertions, obtained by contracting one of the external propagators
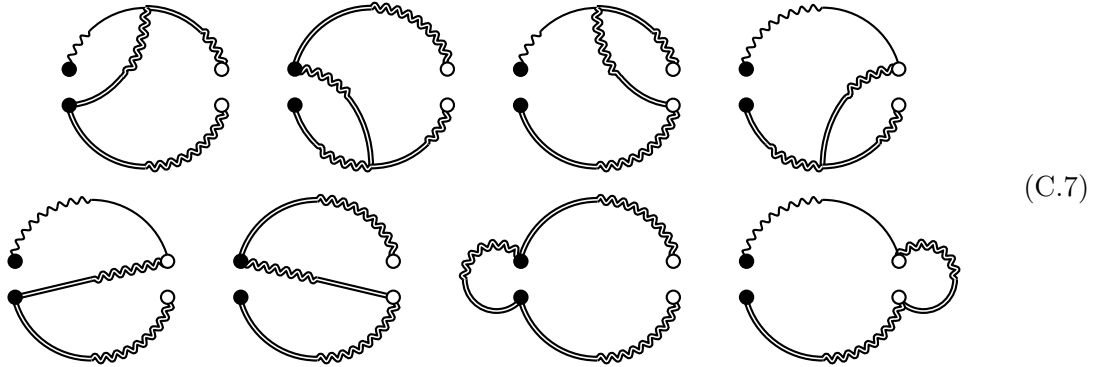
\begin{cgather}
\tikzsetnextfilename{2loop-diagram-t-2-1}
\tikz[thick,scale=0.7,xscale=-1]{
    \draw [aux2] (4.4mm,10.6mm) arc (135:165:15mm);
    \draw [sol2] (15mm,15mm) arc (90:135:15mm);
    \draw [] (15mm,15mm) arc (90:45:15mm);
    \draw [aux] (25.6mm,10.6mm) arc (45:15:15mm);
    \draw [aux2] (15mm,15mm) arc (180:225:15mm and 18.5mm);
    \draw [sol2] (30mm,-3.4mm) arc (-90:-135:15mm and 18.5mm);
    \draw [aux2] (15mm,-15mm) arc (-90:-165:15mm);
    \draw [sol2] (15mm,-15mm) arc (-90:-15:15mm);
    \draw [fill=white] (0.6mm,3.4mm) circle (1.25mm);
    \draw [fill=white] (0.6mm,-3.4mm) circle (1.25mm);
    \draw [fill=black] (29.4mm,3.4mm) circle (1.25mm);
    \draw [fill=black] (29.4mm,-3.4mm) circle (1.25mm);
}\qquad 
\tikzsetnextfilename{2loop-diagram-t-2-2}
\tikz[thick,scale=0.7,xscale=-1]{
    \draw [aux2] (15mm,15mm) arc (90:165:15mm);
    \draw [sol2] (15mm,15mm) arc (90:15:15mm);
    \draw [sol2] (15mm,-15mm) arc (180:135:15mm and 18.5mm);
    \draw [aux2] (30mm,3.4mm) arc (90:135:15mm and 18.5mm);
    \draw [aux2] (4.4mm,-10.6mm) arc (-135:-165:15mm);
    \draw [sol2] (15mm,-15mm) arc (-90:-135:15mm);
    \draw [aux2] (15mm,-15mm) arc (-90:-45:15mm);
    \draw [sol2] (25.6mm,-10.6mm) arc (-45:-15:15mm);
    \draw [fill=white] (0.6mm,3.4mm) circle (1.25mm);
    \draw [fill=white] (0.6mm,-3.4mm) circle (1.25mm);
    \draw [fill=black] (29.4mm,3.4mm) circle (1.25mm);
    \draw [fill=black] (29.4mm,-3.4mm) circle (1.25mm);
} \qquad 
\tikzsetnextfilename{2loop-diagram-t-2-3}
\tikz[thick,scale=0.7,xscale=-1]{
    \draw [aux2] (4.4mm,10.6mm) arc (135:165:15mm);
    \draw [sol2] (15mm,15mm) arc (90:135:15mm);
    \draw [] (15mm,15mm) arc (90:45:15mm);
    \draw [aux] (25.6mm,10.6mm) arc (45:15:15mm);
    \draw [aux2] (15mm,15mm) arc (0:-45:15mm and 18.5mm);
    \draw [sol2] (0mm,-3.4mm) arc (-90:-45:15mm and 18.5mm);
    \draw [aux2] (15mm,-15mm) arc (-90:-165:15mm);
    \draw [sol2] (15mm,-15mm) arc (-90:-15:15mm);
    \draw [fill=white] (0.6mm,3.4mm) circle (1.25mm);
    \draw [fill=white] (0.6mm,-3.4mm) circle (1.25mm);
    \draw [fill=black] (29.4mm,3.4mm) circle (1.25mm);
    \draw [fill=black] (29.4mm,-3.4mm) circle (1.25mm);
}\qquad 
\tikzsetnextfilename{2loop-diagram-t-2-4}
\tikz[thick,scale=0.7,xscale=-1]{
    \draw [] (15mm,15mm) arc (90:165:15mm);
    \draw [aux] (15mm,15mm) arc (90:15:15mm);
    \draw [sol2] (15mm,-15mm) arc (0:45:15mm and 18.5mm);
    \draw [aux2] (0mm,3.4mm) arc (90:45:15mm and 18.5mm);
    \draw [aux2] (4.4mm,-10.6mm) arc (-135:-165:15mm);
    \draw [sol2] (15mm,-15mm) arc (-90:-135:15mm);
    \draw [aux2] (15mm,-15mm) arc (-90:-45:15mm);
    \draw [sol2] (25.6mm,-10.6mm) arc (-45:-15:15mm);
    \draw [fill=white] (0.6mm,3.4mm) circle (1.25mm);
    \draw [fill=white] (0.6mm,-3.4mm) circle (1.25mm);
    \draw [fill=black] (29.4mm,3.4mm) circle (1.25mm);
    \draw [fill=black] (29.4mm,-3.4mm) circle (1.25mm);
} \nn\\
\tikzsetnextfilename{2loop-diagram-t-3-1}
\tikz[thick,scale=0.7,xscale=-1]{
    \draw [] (15mm,15mm) arc (90:165:15mm);
    \draw [aux] (15mm,15mm) arc (90:15:15mm);
    \draw [sol2] (15mm,-15mm) arc (-90:-15:15mm);
    \draw [aux2] (15mm,-15mm) arc (-90:-165:15mm);
    \draw [aux2] (0.6mm,3.4mm) -- (15mm,0);
    \draw [sol2] (15mm,0) -- (29.4mm,-3.4mm);
    \draw [fill=white] (0.6mm,3.4mm) circle (1.25mm);
    \draw [fill=white] (0.6mm,-3.4mm) circle (1.25mm);
    \draw [fill=black] (29.4mm,3.4mm) circle (1.25mm);
    \draw [fill=black] (29.4mm,-3.4mm) circle (1.25mm);
} \qquad 
\tikzsetnextfilename{2loop-diagram-t-3-2}
\tikz[thick,scale=0.7,xscale=-1]{
    \draw [aux2] (15mm,15mm) arc (90:165:15mm);
    \draw [sol2] (15mm,15mm) arc (90:15:15mm);
    \draw [sol2] (15mm,-15mm) arc (-90:-15:15mm);
    \draw [aux2] (15mm,-15mm) arc (-90:-165:15mm);
    \draw [sol2] (0.6mm,-3.4mm) -- (15mm,0);
    \draw [aux2] (15mm,0) -- (29.4mm,3.4mm);
    \draw [fill=white] (0.6mm,3.4mm) circle (1.25mm);
    \draw [fill=white] (0.6mm,-3.4mm) circle (1.25mm);
    \draw [fill=black] (29.4mm,3.4mm) circle (1.25mm);
    \draw [fill=black] (29.4mm,-3.4mm) circle (1.25mm);
} \qquad
\tikzsetnextfilename{2loop-diagram-t-4-1}
\tikz[thick,scale=0.7,xscale=-1]{
    \draw [aux2] (15mm,15mm) arc (90:165:15mm);
    \draw [sol2] (15mm,15mm) arc (90:15:15mm);
    \draw [sol2] (15mm,-15mm) arc (-90:-15:15mm);
    \draw [aux2] (15mm,-15mm) arc (-90:-165:15mm);
    \draw [aux2] (40mm,0) arc (0:150:6mm);
    \draw [sol2] (40mm,0) arc (0:-150:6mm);
    \draw [fill=white] (0.6mm,3.4mm) circle (1.25mm);
    \draw [fill=white] (0.6mm,-3.4mm) circle (1.25mm);
    \draw [fill=black] (29.4mm,3.4mm) circle (1.25mm);
    \draw [fill=black] (29.4mm,-3.4mm) circle (1.25mm);
} \qquad 
\tikzsetnextfilename{2loop-diagram-t-4-2}
\tikz[thick,scale=0.7,xscale=-1]{
    \draw [aux2] (-10mm,0) arc (180:30:6mm);
    \draw [sol2] (-10mm,0) arc (-180:-30:6mm);
    \draw [] (15mm,15mm) arc (90:165:15mm);
    \draw [aux] (15mm,15mm) arc (90:15:15mm);
    \draw [sol2] (15mm,-15mm) arc (-90:-15:15mm);
    \draw [aux2] (15mm,-15mm) arc (-90:-165:15mm);
    \draw [fill=white] (0.6mm,3.4mm) circle (1.25mm);
    \draw [fill=white] (0.6mm,-3.4mm) circle (1.25mm);
    \draw [fill=black] (29.4mm,3.4mm) circle (1.25mm);
    \draw [fill=black] (29.4mm,-3.4mm) circle (1.25mm);
} 
\end{cgather}
Similarly, we have 8 non-linear diagrams in the u-channel, which we will not draw here explicitly. These can be obtained by simply exchanging the two operator insertions on the left. Collectively, these diagrams also do not contribute the OTOC growth, however the argument is slightly more involved. In position space, the 9 t-channel diagrams together yield
\begin{calign}\label{eq:t-channel-full}
    C^{\rm t}_{\bar a\bar a \bar r a}
    &= \int \df^{d+1} y_1\df^{d+1} y_2\,
    \cM^{VW}_{\rm t}(x_1-y_1,y_1-x_4)\,
    \overline\cM^{V'W'}_{\rm t}(x_2-y_2,y_2-x_3)\,
    i K^{-1}(y_1-y_2).
\end{calign}
where we have defined
\begin{calign}\label{eq:M-def-minus-otoc-tchannel}
    &\cM^{VW}_{\rm t}(x_1-y_1,y_1-x_4) \nn\\
    &= i\cV_{F\bar Q\bar F}(\dow_4,\dow_1-\dow_4,-\dow_1)\,
    \Big(i\cO^V_{(1)}(\dow_1) K^{-1}_1(x_1-y_1)\Big)\,
    \Big({-}i\eta_\Theta\tilde\cO^W_{(1)}(-\dow_4) K^{-1}_4(y_1-x_4)\Big)
     \nn\\
    &\qquad
    + 2\Big(\eta_\Theta^W\tilde\cO^W_{(2)}(-\dow_1,\dow_1-\dow_4)
    -  Y^W(\dow_1-\dow_4,-\dow_1)
    \Big)\,
    \Big(i\cO^V_{(1)}(\dow_1)K^{-1}_1(x_1-x_4)\Big)\,
    \delta^{(d+1)}_4(y_1-x_4)  \nn\\
    &\qquad
    + 2\cO^V_{(2)}(\dow_4,\dow_1-\dow_4)\,
    \Big({-}i\eta_\Theta\tilde\cO^ W_{(1)}(-\dow_4)K^{-1}_4(x_1-x_4)\Big)
    \delta^{(d+1)}_1(x_1-y_1), \nn\\
    &\overline\cM^{V'W'}_{\rm t}(x_2-y_2,y_2-x_3) \nn\\
    &= 2i\cV_{\bar Q\bar F^2}(\dow_3;-\dow_2,\dow_2-\dow_3)\,
    \Big(i\cO^{V'}_{(1)}(\dow_2)K^{-1}_2(x_2-y_2)\Big)\,
    \Big({-}i\eta_\Theta\tilde \cO^{W'}_{(1)}(-\dow_3)K^{-1}_3(y_2-x_3)\Big) \nn\\
    &\qquad
    - 2\eta^{W'}_\Theta\tilde \cO^{W'}_{(2)}(-\dow_2,\dow_2-\dow_3)\,
    \Big(i\cO^{V'}_{(1)}(\dow_2)K^{-1}_2(x_2-x_3)\Big)\,
    \delta^{(d+1)}_3(y_2-x_3) \nn\\
    &\qquad
    - 2\Big(\cO^{V'}_{(2)}(\dow_3,\dow_2-\dow_3)
    +  Y^{V'}(\dow_2-\dow_3,\dow_3)
    \Big)\,
    \Big({-}i\eta_\Theta\tilde \cO^{W'}_{(1)}(-\dow_3)K^{-1}_3(x_2-x_3)\Big)
    \delta^{(d+1)}_2(x_2-y_2).
\end{calign}
Here we have labelled the external propagators at $x_1,x_2$ on the right as ``1'' and ``2'', while those at $x_3,x_4$ on the left as ``3'' and ``4'', and the internal propagator as ``5''. Correspondingly, we have labelled the differential operators from ``1--5'' depending on which propagator they act on. We have assumed $\Theta$-odd dissipation for the non-linear operator structure factors.
Further using the Feynman rules for the vertex factors and the operator insertions, we can obtain
\begin{calign}\label{eq:M-def-minus-otoc-tchannel-result}
    &\cM^{VW}_{\rm t}(x_1-y_1,y_1-x_4) \nn\\
    &= 2i\eta^W_\Theta
    \lb \dow_{t1}-\dow_{t4} \rb \lb 1 - \frac{\lb \dow_{t1}-\dow_{t4} \rb^2}{\omega_\beta^2} \rb \nn\\
    &\quad 
    \Bigg[
    \frac{\beta }{2} \dow_{t4} \Delta(\dow_1-\dow_4) 
    \Big(\cD^V_{(1)}(\dow_1) \Delta^{-1}_1(x_1-y_1)\Big)\,
    \Big(\tilde\cD^W_{(1)}(-\dow_4) \Delta^{-1}_4(y_1-x_4)\Big)\nn\\
    &\qquad 
    - \dow_{t4} \lb 1 - \frac{\dow_{t4}^2}{\omega_\beta^2} \rb
    \cR(\dow_4,\dow_1-\dow_4)
    \Big(\cD^V_{(1)}(\dow_1) \Delta^{-1}_1(x_1-y_1)\Big)\,
    \Big(\tilde\cD^W_{(1)}(-\dow_4) \Delta^{-1}_4(y_1-x_4)\Big)
     \nn\\
    &\qquad
    + \dow_{t1} \lb 1-\frac{\dow_{t1}^2}{\omega_\beta^2} \rb
    \Big(\tilde\cD^W_{(2)}(-\dow_1,\dow_1-\dow_4) 
    - \frac{\beta }{2} \tilde\cD^W_{(1)}(-\dow_4)
    \Big)
    \Big(\cD^V_{(1)}(\dow_1)\Delta^{-1}_1(x_1-x_4)\Big)\,
    \delta_4^{(d+1)}(y_1-x_4)  \nn\\
    &\qquad
    + \dow_{t4}
    \lb 1-\frac{\dow_{t4}^2}{\omega_\beta^2} \rb
    \Big(\cD^V_{(2)}(\dow_4,\dow_1-\dow_4) 
    - \frac{\beta }{2} \cD^V_{(1)}(\dow_1)
    \Big)
    \Big(\tilde\cD^W_{(1)}(-\dow_4)\Delta^{-1}_4(x_1-x_4)\Big)
    \delta_1^{(d+1)}(x_1-y_1) \nn\\
    &\qquad
    + \frac{\beta }{2} 
    \lb \dow_{t1} + \dow_{t4} \rb
    \tilde\cD^W_{(1)}(-\dow_4)
    \Big(\cD^V_{(1)}(\dow_1) \Delta^{-1}_1(x_1-x_4)\Big)\,
    \delta_4^{(d+1)}(y_1-x_4) \nn\\
    &\qquad 
    + \eta^W_\Theta\cY^W(\dow_1-\dow_4,-\dow_1)
    \Big(\cD^V_{(1)}(\dow_1)\Delta^{-1}_1(x_1-x_4)\Big)\,
    \delta_4^{(d+1)}(y_1-x_4) 
    \Bigg] , \nn\\
    &\overline\cM^{V'W'}_{\rm t}(x_2-y_2,y_2-x_3)
    = \Theta_\rmP\cM^{W'V'}_{\rm t}(y_2-x_3,x_2-y_2).
\end{calign}
Here $\Theta_\rmP = \rmP\rmT\Theta$, which is identity for $\Theta = \rmT\rmP$ and $\rmP$ for $\Theta=\rmT$. The above results imply that
\begin{calign}
    \cM^{VW}_{\rm t}(x_1-y_1,y_1-x_4)\,
    \overline\cM^{V'W'}_{\rm t}(x_2-y_2,y_2-x_3)
    = \dow_{s_1} \lb 1 - \frac{\dow_{s_1}^2}{\omega_\beta^2} \rb
    \dow_{s_2} \lb 1 - \frac{\dow_{s_2}^2}{\omega_\beta^2} \rb
    \#,
\end{calign}
for some expression $\#$, where we have used $y_i = (s_i,\by_i)$. It follows that
\begin{calign}
    C^{\rm t}_{\bar a\bar a \bar r a}
    &= \int \df y_1\df y_2\,
    \#\,
    \dow_{s_1} \lb 1 - \frac{\dow_{s_1}^2}{\omega_\beta^2} \rb
    \dow_{s_2} \lb 1 - \frac{\dow_{s_2}^2}{\omega_\beta^2} \rb
    K^{-1}(y_1-y_2) \nn\\
    &= \int \df y_1\df y_2\,
    \#\,
    \dow_{s_1} \lb 1 - \frac{\dow_{s_1}^2}{\omega_\beta^2} \rb
    \Delta^{-1}(y_1-y_2).
\end{calign}
Consequently, the t-channel diagrams do not contribute to growth.

The u-channel contribution is exactly the same as the t-channel, except for $x_1\leftrightarrow x_2$ and $V\leftrightarrow V'$, leading to
\begin{calign}\label{eq:u-channel-full}
    C^{\rm u}_{\bar a\bar a\bar r a}
    &= \int \df^{d+1} y_1\df^{d+1} y_2\,
    \cM^{V'W}_{\rm t}(x_2-y_1,y_1-x_4)\,
    \overline\cM^{VW'}_{\rm t}(x_1-y_2,y_2-x_3)\,
    i K^{-1}(y_1-y_2).
\end{calign}
Therefore, the u-channel diagrams also do not contribute to growth.

Note that these results only assume $\Theta$-odd dissipation for simplicity, but otherwise are entirely general for the kind of operator or the explicit form of the energy diffusion SK EFT.

\subsubsection{s-channel}

The interesting growth behaviour of the OTOC comes from the s-channel diagrams. The basic s-channel diagram is given as
\begin{calign}
\tikzsetnextfilename{2loop-diagram-s-1}
\tikz[thick,scale=0.7,xscale=-1]{
    \draw [aux2] (10mm,10mm) arc (90:160:10mm);
    \draw [sol2] (10mm,10mm) arc (90:0:10mm);
    \draw [sol2] (20mm,0) arc (0:-90:10mm);
    \draw [aux2] (10mm,-10mm) arc (-90:-160:10mm);
    \draw [aux2] (20mm,0) -- (30mm,0);
    \draw [sol2] (30mm,0) -- (40mm,0);
    \draw [] (40mm,0) arc (180:90:10mm);
    \draw [aux] (50mm,10mm) arc (90:20:10mm);
    \draw [sol2] (50mm,-10mm) arc (-90:-20:10mm);
    \draw [aux2] (50mm,-10mm) arc (-90:-180:10mm);
    \draw [fill=white] (0.6mm,3.4mm) circle (1.25mm);
    \draw [fill=white] (0.6mm,-3.4mm) circle (1.25mm);
    \draw [fill=black] (59.4mm,3.4mm) circle (1.25mm);
    \draw [fill=black] (59.4mm,-3.4mm) circle (1.25mm);
}
\end{calign}
This is qualitatively similar to the bilocal computation in \cref{eq:bilocal-diagram}, where the bilocal operator insertions at $x_1,x_2$ and $x_3,x_4$ excite a local fluctuation respectively, connected by a local on-shell propagator. The above diagrams only accounts for the linear part of operator insertions. We have 8 more related diagrams involving the non-linear parts of operator insertions:
\begin{cgather}%
\tikzsetnextfilename{2loop-diagram-s-2-1}
\tikz[thick,scale=0.7,xscale=-1]{
    \draw [aux2] (10mm,10mm) arc (90:160:10mm);
    \draw [sol2] (10mm,10mm) arc (90:0:10mm);
    \draw [sol2] (20mm,0) arc (0:-90:10mm);
    \draw [aux2] (10mm,-10mm) arc (-90:-160:10mm);
    \draw [aux2] (20mm,0) -- (30mm,-1.75mm);
    \draw [sol2] (30mm,-1.75mm) -- (40mm,-3.5mm);
    \draw [aux] (50mm,0) arc (0:150:6mm);
    \draw [] (50mm,0) arc (0:-150:6mm);
    \draw [fill=white] (0.6mm,3.4mm) circle (1.25mm);
    \draw [fill=white] (0.6mm,-3.4mm) circle (1.25mm);
    \draw [fill=black] (39.4mm,3.4mm) circle (1.25mm);
    \draw [fill=black] (39.4mm,-3.4mm) circle (1.25mm);
} \quad 
\tikzsetnextfilename{2loop-diagram-s-2-2}
\tikz[thick,scale=0.7,xscale=-1]{
    \draw [aux2] (10mm,10mm) arc (90:160:10mm);
    \draw [sol2] (10mm,10mm) arc (90:0:10mm);
    \draw [sol2] (20mm,0) arc (0:-90:10mm);
    \draw [aux2] (10mm,-10mm) arc (-90:-160:10mm);
    \draw [aux2] (20mm,0) -- (30mm,1.75mm);
    \draw [sol2] (30mm,1.75mm) -- (40mm,3.5mm);
    \draw [aux2] (50mm,0) arc (0:150:6mm);
    \draw [sol2] (50mm,0) arc (0:-150:6mm);
    \draw [fill=white] (0.6mm,3.4mm) circle (1.25mm);
    \draw [fill=white] (0.6mm,-3.4mm) circle (1.25mm);
    \draw [fill=black] (39.4mm,3.4mm) circle (1.25mm);
    \draw [fill=black] (39.4mm,-3.4mm) circle (1.25mm);
} \quad
\tikzsetnextfilename{2loop-diagram-s-2-3}
\tikz[thick,scale=0.7,xscale=-1]{
    \draw [aux2] (10mm,0) arc (180:30:6mm);
    \draw [sol2] (10mm,0) arc (-180:-30:6mm);
    \draw [aux2] (20mm,-3.5mm) -- (30mm,-1.75mm);
    \draw [sol2] (30mm,-1.75mm) -- (40mm,0);
    \draw [] (40mm,0) arc (180:90:10mm);
    \draw [aux] (50mm,10mm) arc (90:20:10mm);
    \draw [sol2] (50mm,-10mm) arc (-90:-20:10mm);
    \draw [aux2] (50mm,-10mm) arc (-90:-180:10mm);
    \draw [fill=white] (20.6mm,3.4mm) circle (1.25mm);
    \draw [fill=white] (20.6mm,-3.4mm) circle (1.25mm);
    \draw [fill=black] (59.4mm,3.4mm) circle (1.25mm);
    \draw [fill=black] (59.4mm,-3.4mm) circle (1.25mm);
} \quad 
\tikzsetnextfilename{2loop-diagram-s-2-4}
\tikz[thick,scale=0.7,xscale=-1]{
    \draw [sol2] (10mm,0) arc (180:30:6mm);
    \draw [aux2] (10mm,0) arc (-180:-30:6mm);
    \draw [aux2] (20mm,3.5mm) -- (30mm,1.75mm);
    \draw [sol2] (30mm,1.75mm) -- (40mm,0);
    \draw [] (40mm,0) arc (180:90:10mm);
    \draw [aux] (50mm,10mm) arc (90:20:10mm);
    \draw [sol2] (50mm,-10mm) arc (-90:-20:10mm);
    \draw [aux2] (50mm,-10mm) arc (-90:-180:10mm);
    \draw [fill=white] (20.6mm,3.4mm) circle (1.25mm);
    \draw [fill=white] (20.6mm,-3.4mm) circle (1.25mm);
    \draw [fill=black] (59.4mm,3.4mm) circle (1.25mm);
    \draw [fill=black] (59.4mm,-3.4mm) circle (1.25mm);
} \nn\\
\tikzsetnextfilename{2loop-diagram-s-3-1}
\tikz[thick,scale=0.7,xscale=-1]{
    \node at (15mm,8mm) {~};
    \draw [aux2] (-10mm,0) arc (180:30:6mm);
    \draw [sol2] (-10mm,0) arc (-180:-30:6mm);
    \draw [aux2] (0mm,-3.4mm) -- (15mm,-3.4mm);
    \draw [sol2] (15mm,-3.4mm) -- (30mm,-3.4mm);
    \draw [aux] (40mm,0) arc (0:150:6mm);
    \draw [] (40mm,0) arc (0:-150:6mm);
    \draw [fill=white] (0.6mm,3.4mm) circle (1.25mm);
    \draw [fill=white] (0.6mm,-3.4mm) circle (1.25mm);
    \draw [fill=black] (29.4mm,3.4mm) circle (1.25mm);
    \draw [fill=black] (29.4mm,-3.4mm) circle (1.25mm);
} \quad 
\tikzsetnextfilename{2loop-diagram-s-3-2}
\tikz[thick,scale=0.7,xscale=-1]{
    \node at (15mm,8mm) {~};
    \draw [sol2] (-10mm,0) arc (180:30:6mm);
    \draw [aux2] (-10mm,0) arc (-180:-30:6mm);
    \draw [aux2] (0mm,3.4mm) -- (15mm,0);
    \draw [sol2] (15mm,0) -- (30mm,-3.4mm);
    \draw [aux] (40mm,0) arc (0:150:6mm);
    \draw [] (40mm,0) arc (0:-150:6mm);
    \draw [fill=white] (0.6mm,3.4mm) circle (1.25mm);
    \draw [fill=white] (0.6mm,-3.4mm) circle (1.25mm);
    \draw [fill=black] (29.4mm,3.4mm) circle (1.25mm);
    \draw [fill=black] (29.4mm,-3.4mm) circle (1.25mm);
} \quad
\tikzsetnextfilename{2loop-diagram-s-3-3}
\tikz[thick,scale=0.7,xscale=-1]{
    \node at (15mm,8mm) {~};
    \draw [aux2] (-10mm,0) arc (180:30:6mm);
    \draw [sol2] (-10mm,0) arc (-180:-30:6mm);
    \draw [aux2] (0mm,-3.4mm) -- (15mm,0);
    \draw [sol2] (15mm,0) -- (30mm,3.4mm);
    \draw [aux2] (40mm,0) arc (0:150:6mm);
    \draw [sol2] (40mm,0) arc (0:-150:6mm);
    \draw [fill=white] (0.6mm,3.4mm) circle (1.25mm);
    \draw [fill=white] (0.6mm,-3.4mm) circle (1.25mm);
    \draw [fill=black] (29.4mm,3.4mm) circle (1.25mm);
    \draw [fill=black] (29.4mm,-3.4mm) circle (1.25mm);
}  \quad 
\tikzsetnextfilename{2loop-diagram-s-3-4}
\tikz[thick,scale=0.7,xscale=-1]{
    \node at (15mm,8mm) {~};
    \draw [sol2] (-10mm,0) arc (180:30:6mm);
    \draw [aux2] (-10mm,0) arc (-180:-30:6mm);
    \draw [aux2] (0mm,3.4mm) -- (15mm,3.4mm);
    \draw [sol2] (15mm,3.4mm) -- (30mm,3.4mm);
    \draw [aux2] (40mm,0) arc (0:150:6mm);
    \draw [sol2] (40mm,0) arc (0:-150:6mm);
    \draw [fill=white] (0.6mm,3.4mm) circle (1.25mm);
    \draw [fill=white] (0.6mm,-3.4mm) circle (1.25mm);
    \draw [fill=black] (29.4mm,3.4mm) circle (1.25mm);
    \draw [fill=black] (29.4mm,-3.4mm) circle (1.25mm);
} 
\end{cgather}
These can be thought of as arising from sequentially contracting one of the external propagators in exchange for non-linearities at the operator insertions. 

The total contribution of the s-channel diagrams is given as
\begin{calign}\label{eq:s-channel-full}
    C^{\rm s}_{\bar a\bar a \bar r a}
    &= \int \df^{d+1} y_1\df^{d+1} y_2\,
    \cM^{VV'}_{\rm s}(x_1-y_1,x_2-y_1)\,
    \overline\cM^{W'W}_{\rm s}(y_2-x_3,y_2-x_4)\,
    i K^{-1}(y_1-y_2),
\end{calign}
where we have defined
\begin{calign}
    &\cM^{VV'}_{\rm s}(x_1-y_1,x_2-y_1) \nn\\
    &=  
    2i\cV_{\bar Q\bar F^2}(\dow_1+\dow_2;-\dow_1,-\dow_2)\,
    \Big(i\cO^V_{(1)}(\dow_1)K^{-1}_1(x_1-y_1)\Big)\,
    \Big(i\cO^{V'}_{(1)}(\dow_2)K^{-1}_2(x_2-y_1)\Big) \nn\\
    &\qquad
    - 2\Big( \cO^{V'}_{(2)}(-\dow_1,\dow_1+\dow_2)
    + Y^{V'}(-\dow_1,\dow_5)
    \Big)\,
    \Big(i\cO^V_{(1)}(\dow_1) K^{-1}_1(x_1-x_2)\Big)\,
    \delta^{(d+1)}_2(x_2-y_1)  \nn\\
    &\qquad
    - 2\Big( \cO^V_{(2)}(-\dow_2,\dow_5)
    + Y^V(-\dow_2,\dow_1+\dow_2)
    \Big)\,
    \Big(i\cO^{V'}_{(1)}(\dow_2)K^{-1}_2(x_2-x_1)\Big)\,
    \delta_1^{(d+1)}(x_1-y_1), \nn\\
    &\overline\cM^{W'W}_{\rm s}(y_2-x_3,y_2-x_4) \nn\\
    &=
    i\cV_{F\bar Q\bar F}(\dow_4,\dow_3,-\dow_3-\dow_4)\,
    \Big({-}i\eta_\Theta\tilde \cO^{W'}_{(1)}(-\dow_3)K^{-1}_3(y_2-x_3)\Big)
    \Big({-}i\eta_\Theta\tilde \cO^W_{(1)}(-\dow_4)K^{-1}_4(y_2-x_4)\Big) \nn\\
    &\qquad
    + 2\Big(\eta_\Theta\tilde \cO^W_{(2)}(\dow_3,-\dow_3-\dow_4)
    - Y^W(\dow_3,-\dow_3-\dow_4)
    \Big)\nn\\
    &\hspace{16em}
    \Big({-}i\eta_\Theta\tilde \cO^{W'}_{(1)}(-\dow_3)K^{-1}_3(x_4-x_3)\Big)
    \delta_4^{(d+1)}(y_2-x_4) \nn\\
    &\qquad
    + 2\Big(\eta_\Theta\tilde \cO^{W'}_{(2)}(\dow_4,-\dow_3-\dow_4)
    - Y^{W'}(\dow_4,-\dow_3-\dow_4)
    \Big)\nn\\
    &\hspace{16em}
    \Big({-}i\eta_\Theta\tilde \cO^W_{(1)}(-\dow_4)K^{-1}_4(x_3-x_4)\Big)
    \delta_3^{(d+1)}(y_2-x_3)
    \Bigg].
\end{calign}
Using the explicit form of the interaction vertices and operator factors, we find
\begin{calign}\label{eq:M-def-minus-otoc-schannel-result}
    &\cM^{VV'}_{\rm s}(x_1-y_1,x_2-y_1) \nn\\
    &=  
    2i\Bigg[
    \frac{\beta }{2}
    \lb 1+\frac{\dow_{t1}^2 + \dow_{t2}^2 + 3\dow_{t1}\dow_{t2}}{\omega_\beta^2} \rb 
    \Delta(\dow_1+\dow_2)
    + \lb 1 - \frac{\dow_{t1}^2}{\omega_\beta^2} \rb
    \lb 1 - \frac{\dow_{t2}^2}{\omega_\beta^2} \rb
    \cR(\dow_1,\dow_2)
    \Bigg] \nn\\
    &\hspace{6em}
    \Big(\cD^V_{(1)}(\dow_1)\dow_{t1}\Delta^{-1}_1(x_1-y_1)\Big)\,
    \Big(\cD^{V'}_{(1)}(\dow_2)\dow_{t2}\Delta^{-1}_2(x_2-y_1)\Big) \nn\\
    &\qquad 
    - 2i
    \Bigg[  
    \frac{\beta }{2} 
    \lb \dow_{t1} + 2\dow_{t2} \rb
    \cD^{V'}_{(1)}(\dow_2) \nn\\
    &\qquad\qquad 
    + \lb \dow_{t1}+\dow_{t2}\rb
    \lb 1-\frac{\lb \dow_{t1}+\dow_{t2}\rb^2}{\omega_\beta^2} \rb 
    \Big(\cD^{V'}_{(2)}(-\dow_1,\dow_1+\dow_2) 
    - \frac{\beta }{2} \cD_{(1)}(\dow_2)
    \Big) \nn\\
    &\qquad\qquad
    + \cY^{V'}(-\dow_1,\dow_1+\dow_2)
    \Bigg]\,
    \lb \cD^V_{(1)}(\dow_1) 
    \lb 1 - \frac{\dow_{t1}^2}{\omega_\beta^2}  \rb 
    \dow_{t1}
    \Delta^{-1}_1(x_1-x_2)\rb\,
    \delta_2^{(d+1)}(x_2-y_1)  \nn\\
    &\qquad 
    - 2i
    \Bigg[  
    \frac{\beta }{2} 
    \lb \dow_{t2} + 2\dow_{t1} \rb
    \cD^V_{(1)}(\dow_1) \nn\\
    &\qquad\qquad
    + \lb \dow_{t1}+\dow_{t2}\rb
    \lb 1-\frac{\lb \dow_{t1}+\dow_{t2}\rb^2}{\omega_\beta^2} \rb 
    \Big(\cD^V_{(2)}(-\dow_2,\dow_1+\dow_2) 
    - \frac{\beta }{2} \cD^V_{(1)}(\dow_1)
    \Big) \nn\\
    &\qquad\qquad
    + \cY^V(-\dow_2,\dow_1+\dow_2)
    \Bigg]\,
    \lb \cD^{V'}_{(1)}(\dow_2) 
    \lb 1 - \frac{\dow_{t2}^2}{\omega_\beta^2}  \rb 
    \dow_{t2}
    \Delta^{-1}_2(x_2-x_1)\rb\,
    \delta_1^{(d+1)}(x_1-y_1), \nn\\
    &\overline\cM^{W'W}_{\rm s}(y_2-x_3,y_2-x_4)
    = \Theta_\rmP\cM^{W'W}_{\rm s}(y_2-x_3,y_2-x_4).
\end{calign}
This result is indeed sensitive to the exponentially growing part of $C^{\rm s}_{\bar a\bar a \bar r a}$.

\subsubsection{Extracting the growing part}

This is as far as we can tractably go in complete position-domain. To proceed further, we pass on to the Fourier-domain in spatial coordinates. We will assume that none of the external operators are inserted at the same precise point in space or time, so as to avoid the contact terms. We will also assume the energy-diffusion-related coefficients in the SK EFT to take the form
\begin{calign}
    \cD^t_{(1)}(\dow_t,i\bk) 
    &= c_\bk, \nn\\
    i\bk\cdot\cD^\bx_{(1)}(\dow_t,i\bk) 
    &= c_\bk\cD_{\bk}, \nn\\
    \cD^t_{(2)}(\dow_{t1},i\bk_1,\dow_{t2},i\bk_2) 
    - \frac{\beta }{2} \cD^t_{(1)}(\dow_{t1}+\dow_{t2},i\bk_1+i\bk_2) 
    &= r_{\bk_1,\bk_2}, \nn\\
    i(\bk_1+\bk_2)\cdot
    \lb \cD^\bx_{(2)}(\dow_{t1},i\bk_1,\dow_{t2},i\bk_2)
    - \frac{\beta }{2} \cD^\bx_{(1)}(\dow_{t1}+\dow_{t2},i\bk_1+i\bk_2)\rb
    &= \frac{1}{\beta }\cR_{\bk_1,\bk_2}
    - \omega_\beta r_{\bk_1,\bk_2}.
\end{calign}
In particular, we have taken all of these operators to be independent of $\dow_t$.
This results in
\begin{calign}
    \Delta(\dow_t,i\bk) 
    &= \beta  c_\bk \lb \dow_t
    + \cD_\bk \rb, \nn\\
    \cR(\dow_{t1},i\bk_1,\dow_{t2},i\bk_2)
    &= \beta  \lb \dow_{t1}+\dow_{t2}-\omega_\beta\rb
    r_{\bk_1,\bk_2}
    + \cR_{\bk_1,\bk_2}.
\end{calign}
Recall that $\cR=0$ for SYK-like models, which correspondingly leads to vanishing $r_{\bk_1,\bk_2}$ and $\cR_{\bk_1,\bk_2}$. For theories with $\Theta$-odd dissipation, it also forces $c_\bk=c_V$ to be a constant.

Using \cref{eq:s-channel-full} and \cref{eq:DeltaandKfunctions}, we find
\begin{calign}
    C^{\rm s}_{\bar a\bar a \bar r a}
    &= i\int \frac{\df^d\bk}{(2\pi)^d}
    \int \df^{d+1} y_1\df^{d+1} y_2\,
    \cM^{VV'}_{\rm s}(x_1-y_1,x_2-y_1)\,
    \overline\cM^{W'W}_{\rm s}(y_2-x_3,y_2-x_4)\nn\\
    &\hspace{4em} 
    \lB \frac{\Theta(s_1-s_2)}{\beta  c_\bk}
    \lb
    \frac{\E{-\cD_\bk(s_1-s_2)}}{\cD_\bk (1-\cD_\bk^2/\omega_\beta^2)}
    - \frac{1}{\cD_\bk}
    + \frac{\E{-\omega_\beta(s_1-s_2)}}{2(\cD_\bk-\omega_\beta)} 
    + \frac{\E{\omega_\beta(s_1-s_2)}}
    {2(\cD_\bk+\omega_\beta)} 
    \rb\rB.
\end{calign}
Here we have used $y_i = (s_i,\by_i)$. Since we are interested in the growing part of the answer, we only care about the final term in the parenthesis in the second line. We find
\begin{calign}\label{eq:s-channel-reduced}
    C_{\bar a \bar a |\bar r a}
    &\sim 
    i\int \frac{\df^d\bk}{(2\pi)^d}
    \int \df^{d+1} y_1\df^{d+1} y_2\,
    \cM^{VV'}_{\rm s}(x_1-y_1,x_2-y_1)\,
    \overline\cM^{W'W}_{\rm s}(y_2-x_3,y_2-x_4)\nn\\
    &\hspace{20em} 
    \E{i\bk\cdot(\by_1-\by_2)}
    \frac{\Theta(s_1-s_2)}{\beta  c_\bk}
    \frac{\E{\omega_\beta(s_1-s_2)}}
    {2(\cD_\bk+\omega_\beta)} \nn\\
    &= i\int \frac{\df^d\bk}{(2\pi)^d}
    \frac{1}{2\beta  c_\bk(\cD_\bk+\omega_\beta)}
    \int_{-\infty}^\infty \df s_1
    \int_{-\infty}^{s_1} \df s_2\,
    \E{\omega_\beta(s_1-s_2)} \nn\\
    &\hspace{16em} 
    \cM^{VV'}_{\rm s}(x_1,x_2,s_1;i\bk)\,
    \overline\cM^{W'W}_{\rm s}(-x_3,-x_4,-s_2;i\bk),
\end{calign}
where
\begin{calign}
    \cM^{VV'}_{\rm s}(x_1,x_2,s_1;i\bk)
    &= \int \df^d \by_1\,
    \E{i\bk\cdot\by_1}
    \cM^{VV'}_{\rm s}(x_1-y_1,x_2-y_1), \nn\\
    \overline\cM^{W'W}_{\rm s}(-x_3,-x_4,-s_2;i\bk)
    &= \int \df^d \by_2\,
    \E{-i\bk\cdot\by_2}
    \overline\cM^{W'W}_{\rm s}(y_2-x_3,y_2-x_4).
\end{calign}
Recall that the contact, t-channel, or u-channel diagrams do not contribute to the Lyapunov growth of $C_{\bar a \bar a |\bar r a}$, so the above expression is complete as far as the growing part is concerned. Note that $\cM_{\rm s}$ and $\overline\cM_{\rm s}$ are the same functions.

Similar to the t-channel, the operator $1-(\dow_{t1}+\dow_{t2})^2/\omega_\beta^2$ in $\cM_{\rm s}$ is equal to $1-\dow_{s_1}^2/\omega_\beta^2$, so the corresponding terms do not contribute to the growing part in \cref{eq:s-channel-reduced}. This enables us to extract
\begin{calign}
    \cM^{VV'}_{\rm s}(x_1,x_2,s_1;i\bk)
    &\sim
    \int\frac{\df^d\bk_1}{(2\pi)^d}
    \E{i\bk_1\cdot\bx_1}
    \E{i(\bk-\bk_1)\cdot\bx_2}
    \nn\\
    &\hspace{4em}
    \Bigg[
    \Theta(t_1-s_1)
    \Theta(t_2-s_1)
    \E{-\cD_{\bk_1}(t_1-s_1)}
    \E{-\cD_{\bk-\bk_1}(t_2-s_1)} 
    G^{VV'}_\Theta(\bk_1,\bk-\bk_1)
    \nn\\
    &\hspace{6em}
    + \Theta(t_1-t_2)
    \delta(t_2-s_1)
    \E{-\cD_{\bk_1}(t_1-t_2)}
    G^{VV'}_\delta(\bk_1,\bk-\bk_1)\,  \nn\\
    &\hspace{6em} 
    + \Theta(t_2-t_1)
    \delta(t_1-s_1)
    \E{-\cD_{\bk-\bk_1}(t_2-t_1)}
    G^{V'V}_\delta(\bk-\bk_1,\bk_1)\Bigg] \nn\\
    &\qquad 
    + \E{-\omega_\beta s_1} \dow_{s_1}(\ldots),
\end{calign}
where
\begin{calign}
    G^{VV'}_{\Theta}(\bk_1,\bk_2)
    &= 
    \frac{2ic^V_{\bk_1}c^{V'}_{\bk_2}}{\beta  c_{\bk_1}c_{\bk_2}} 
    \cD_{\bk_1}\cD_{\bk_2} 
    \nn\\
    &\hspace{-2em}
    \Bigg\{
    \frac{\beta  c_{\bk_1+\bk_2}}{2}
    \lb \cD_{\bk_1+\bk_2} - \cD_{\bk_1} - \cD_{\bk_2} \rb 
    \lb 1+\frac{\cD_{\bk_1}^2+\cD_{\bk_2}^2+3\cD_{\bk_1}\cD_{\bk_2}}{\omega_\beta^2} \rb \nn\\
    &\hspace{0em}
    + \frac{1}{\beta }
    \lb 1 - \frac{\cD_{\bk_1}^2}{\omega_\beta^2} \rb
    \lb 1 - \frac{\cD_{\bk_2}^2}{\omega_\beta^2} \rb
    \cR_{\bk_1,\bk_2}
    \Bigg\}, \nn\\
    G^{VV'}_\delta(\bk_1,\bk_2)
    &= 
    \frac{2ic^V_{\bk_1}}{\beta  c_{\bk_1}}
    \cD_{\bk_1} \nn\\
    &\hspace{-2em}
    \Bigg\{
    \frac{\beta }{2}
    c^{V'}_{\bk_2}
    \lb 1 - \frac{c_{\bk_1+\bk_2}}{c_{\bk_2}} \rb
    \lb 2\omega_\beta + \cD_{\bk_1} \rb
    \lb 1 - \frac{\cD_{\bk_1}^2}{\omega_\beta^2}  \rb \nn\\
     &\hspace{-0em} 
     + \frac{\beta  }{2} c^{V'}_{\bk_2}\frac{c_{\bk_1+\bk_2}}{c_{\bk_2}} 
    \lb 
    2 
    - \frac{\cD_{\bk_1}^2-\cD_{\bk_2}^2-2\cD_{\bk_1} \cD_{\bk_2}
    +\omega_\beta\lb \cD_{\bk_1} + \cD_{\bk_2}\rb}{\omega_\beta^2}
    \rb  \nn\\
    &\hspace{20em}
    \times \lb \cD_{\bk_1}+\cD_{\bk_2}-\cD_{{\bk_1+\bk_2}}\rb 
    \nn\\
    &\hspace{-0em}
    + \frac{c^{V'}_{\bk_2}}{\beta c_{\bk_2}} 
    \cR_{\bk_1,\bk_2} 
    \left(1-\frac{\cD_{\bk_1}^2}{\omega_\beta^2}\right) 
    \left(\frac{\cD_{\bk_1}^2+\cD_{\bk_2}^2-\cD_{\bk_1}\cD_{\bk_2}
    + \omega_\beta\lb 2\cD_{\bk_1}-\cD_{\bk_2}\rb
    }{\omega_\beta^2}
    \right)
      \nn\\
    &\hspace{-0em}
    + \cY^{V'}_{-\bk_1,{\bk_1+\bk_2}}
    \lb 1 - \frac{\cD_{\bk_1}^2}{\omega_\beta^2}  \rb
    \Bigg\}.
\end{calign}
In deriving this, we have further assumed that 
\begin{calign}\label{eq:Yk}
    \cD_{(1)}^\cO(\dow_t,i\bk) 
    &= c^\cO_\bk \qquad \text{(independent of $\dow_t$)}, \nn\\
    \cY^\cO(\cD_{\bk_1},i\bk_1,\omega_\beta,i\bk_2)
    &= \cY^\cO_{\bk_1,\bk_2}.
\end{calign}

Using this into \cref{eq:s-channel-reduced}, we find
\begin{calign}
    C_{\bar a \bar a |\bar r a}
    &\sim 
    i\int \frac{\df^d\bk}{(2\pi)^d}
    \frac{\E{i\bk\cdot(\bx_2-\bx_3)}}{2\beta  c_\bk(\cD_\bk+\omega_\beta)}
    \int_{-\infty}^\infty \df s_1
    \int_{-\infty}^{s_1} \df s_2\,
    \E{\omega_\beta(s_1-s_2)} \nn\\
    &\hspace{2em}
    \int\frac{\df^d\bk_1}{(2\pi)^d}
    \E{i\bk_1\cdot(\bx_1-\bx_2)}
    \int\frac{\df^d\bk_4}{(2\pi)^d}
    \E{-i\bk_4\cdot(\bx_3-\bx_4)}
    \nn\\
    &\hspace{4em}
    \Bigg[
    \Theta(t_1-s_1)
    \Theta(t_2-s_1)
    \E{-\cD_{\bk_1}(t_1-s_1)}
    \E{-\cD_{\bk-\bk_1}(t_2-s_1)} 
    G^{VV'}_\Theta(\bk_1,\bk-\bk_1)
    \nn\\
    &\hspace{6em}
    + \Theta(t_1-t_2)
    \delta(t_2-s_1)
    \E{-\cD_{\bk_1}(t_1-t_2)}
    G^{VV'}_\delta(\bk_1,\bk-\bk_1)\,  \nn\\
    &\hspace{6em} 
    + \Theta(t_2-t_1)
    \delta(t_1-s_1)
    \E{-\cD_{\bk-\bk_1}(t_2-t_1)}
    G^{V'V}_\delta(\bk-\bk_1,\bk_1)\Bigg]  \nn\\
    &\hspace{4em}
    \Bigg[
    \Theta(s_2-t_4)
    \Theta(s_2-t_3)
    \E{-\cD_{-\bk_4}(s_2-t_4)}
    \E{-\cD_{\bk+\bk_4}(s_2-t_3)} 
    G^{WW'}_\Theta(-\bk_4,\bk+\bk_4)
    \nn\\
    &\hspace{6em}
    + \Theta(t_3-t_4)
    \delta(s_2-t_3)
    \E{-\cD_{-\bk_4}(t_3-t_4)}
    G^{WW'}_\delta(-\bk_4,\bk+\bk_4)\,  \nn\\
    &\hspace{6em} 
    + \Theta(t_4-t_3)
    \delta(s_2-t_4)
    \E{-\cD_{\bk+\bk_4}(t_4-t_3)}
    G^{W'W}_\delta(\bk+\bk_4,-\bk_4)\Bigg].
\end{calign}
Further performing the $s_{1,2}$ integrals, we find the result
\begin{calign}
    C_{\bar a \bar a |\bar r a}&(x_1,x_2;x_3,x_4) \nn\\
    &\sim \Theta(t_1,t_2,t_3,t_4)\,
    \E{\omega_\beta (t_2-t_3)}
    \frac{\omega_\beta^5}{4\pi i}
    \int \frac{\df^d\bk}{(2\pi)^d}
    \frac{\E{i\bk\cdot(\bx_2-\bx_3)}c_\bk}{\cD_\bk+\omega_\beta} 
    \cI^{VV'}_d(x_1-x_2,\bk) 
    \cI^{WW'}_d(x_3-x_4;\bk) \nn\\
    &\qquad 
    + (x_1\leftrightarrow x_2)
    + (x_3\leftrightarrow x_4)
    + (x_1\leftrightarrow x_2,x_3\leftrightarrow x_4),
\end{calign}
where $\Theta(t_i,t_j,t_k,t_l)=\Theta(t_i-t_j)\Theta(t_j-t_k)\Theta(t_k-t_l)$ and we have only kept the exponentially growing contribution. We have also identified the wavevector integrals
\begin{calign}
    \cI^{VV'}_d(x_1-x_2;\bk)
    &= \frac{-i}{\omega_\beta^2c_\bk}\int\frac{\df^d\bk_1}{(2\pi)^d}
    \E{i\bk_1\cdot(\bx_1-\bx_2)}
    \E{-\cD_{\bk_1}(t_1-t_2)} \nn\\
    &\hspace{6em}
    \lb 
    \frac{G^{VV'}_\Theta(\bk_1,\bk-\bk_1)}{\omega_\beta + \cD_{\bk_1} + \cD_{\bk-\bk_1}}
    +  G^{VV'}_\delta(\bk_1,\bk-\bk_1)\rb, \nn\\
    \cI^{WW'}_d(x_3-x_4;\bk)
    &= \frac{-i}{\omega_\beta^2c_\bk}\int\frac{\df^d\bk_4}{(2\pi)^d}
    \E{-i\bk_4\cdot(\bx_3-\bx_4)}
    \E{-\cD_{-\bk_4}(t_3-t_4)} \nn\\
    &\hspace{6em}
    \lb 
    \frac{G^{WW'}_\Theta(-\bk_4,\bk+\bk_4)}{\omega_\beta + \cD_{-\bk_4} + \cD_{\bk+\bk_4}}
    + G^{WW'}_\delta(-\bk_4,\bk+\bk_4)
    \rb.
\end{calign}
Note that both the definitions for $\cI_d$ are identical. Explicitly, we find
\begin{calign}\label{eq:general-integral}
    \cI^{VV'}_d(y;\bk)
    &= \int\frac{\df^d\bk_1}{(2\pi)^d}
    \E{i\bk_1\cdot\by}
    \E{-\cD_{\bk_1}y^0} 
    \frac{c^V_{\bk_1}c^{V'}_{\bk-\bk_1}}{c_{\bk_1}c_{\bk-\bk_1}}
    \frac{\cD_{\bk_1}}{\omega_\beta}
    \lb 1-\frac{\cD_{\bk_1}^2}{\omega_\beta^2}\rb 
    \lb 2 + \frac{\cD_{\bk_1}}{\omega_\beta} \rb
    \nn\\
    &\hspace{8em}
    \Bigg[
    \frac{\cD_{\bk_1} + \cD_{\bk-\bk_1} - \cD_{\bk}}{\omega_\beta + \cD_{\bk_1} + \cD_{\bk-\bk_1}}
    + \frac{2}{\beta ^2c_\bk}
    \frac{\cD_{\bk_1}}{\omega_\beta}
    \lb 1 + \frac{\cD_{\bk_1}}{\omega_\beta}\rb
    \frac{\cR_{\bk_1,\bk-\bk_1} }{\cD_{\bk_1} + \cD_{\bk-\bk_1}+\omega_\beta}
    \Bigg]
    \nn\\
    &\qquad 
    + \int\frac{\df^d\bk_1}{(2\pi)^d}
    \E{i\bk_1\cdot\by}
    \E{-\cD_{\bk_1}y^0} 
    \frac{c^V_{\bk_1}}{c_{\bk_1}}
    \frac{\cD_{\bk_1}}{\omega_\beta}
    \lb 1 - \frac{\cD_{\bk_1}^2}{\omega_\beta^2}  \rb
    \nn\\
    &\hspace{8em}
    \Bigg[
    \frac{c^{V'}_{\bk-\bk_1}}{c_\bk}
    \lb 1 - \frac{c_{\bk}}{c_{\bk-\bk_1}} \rb
    \lb 2 + \frac{\cD_{\bk_1}}{\omega_\beta} \rb 
    + \frac{2}{\beta\omega_\beta c_\bk}\cY^{V'}_{-\bk_1,{\bk}}
    \Bigg].
\end{calign}

\subsubsection{Final result for energy current in SYK-like models}
\label{eq:energy-otoc-computation}

\Cref{eq:general-integral} is a complicated integral in general and depends on the precise structure of the SK EFT and the operator under consideration. For the SYK-like model, where $\cR_{\bk_1,\bk_2}=0$ and $c_\bk = c_V$, and using the operator to be energy density, so that $\cY^t_{-\bk_1,\bk}=0$, this simplifies to the expression given in \cref{eq:general-integral-energydensity}. On the other hand, for energy flux, we have to take
\begin{calign}\label{eq:Ymu-simple}
    \cY^\bx_{-\bk_1,\bk}
    &= -\frac{\beta \omega_\beta c_V}{2} 
    \lb 2 + \frac{\cD_{-\bk_1}}{\omega_\beta} \rb
    \frac{\cD_{-\bk_1}+\cD_{\bk_1-\bk}-\cD_{-\bk}}{i\bk-i\bk_1}.
\end{calign}

As generally the case with such wavevector integrals, it splits into a non-analytic part that is independent of the small cutoff $y\approx 0$ and an analytic part that depends on how precisely $y$ is taken to 0. To wit, 
\begin{calign}
    \cI_d(y;\bk)
    &= \lb 1 + \frac{\bk^2}{m^2}\rb \alpha_d(2+\bk^2/m^2)
    + \cI_0(y)
    + \lb 1 + \frac{\bk^2}{m^2}\rb
    \times\text{(polynomial in $\bk$)}.
\end{calign}
The polynomial terms in $\bk$ isolated above only contribute to spatial contact terms in the OTOC \cref{eq:OTOC-result}. They drop out provided that we take the points $\bx_2$ and $\bx_3$ to at least be infinitesimally separated. We find the non-analytic part
\begin{calign}
    \alpha_d(z)
    &= \frac{m^d z^{d/2-1}}{2^{2d+5}\pi^{d/2}\Gamma(d/2+2)}
    \Big( 9 d (d+2) 
    - 20 (d+2) z
    + (10 d+32) z^2
    -12 z^3 
    +3 z^4 \Big)
    \nn\\
    &\qquad\qquad 
    \times 
    \begin{cases}
    \dsp (-1)^{\frac{d-1}{2}} \pi ,
    &\quad \text{$d$ is odd}, \\
    \dsp (-1)^{\frac{d}{2}}
    \log z, 
    &\quad \text{$d$ is even},
    \end{cases}
\end{calign}
On the other hand, the analytic part is given as 
\begin{calign}\label{eq:regular-term}
    \cI_0(y)
    &= \int \frac{\df^d\bq}{(2\pi)^d}\,
    \E{i\bq\cdot\by-Dy^0\bq^2} 
    \frac{\bq^2}{m^2}
    \lb 1- \frac{\bq^4}{m^4}\rb 
    \lb 2 + \frac{\bq^2}{m^2} \rb \nn\\
    &= 
    - \frac{\nabla_\by^2}{m^{2}}
    \lb 1- \frac{\nabla_\by^4}{m^4}\rb 
    \lb 2- \frac{\nabla_\by^2}{m^2} \rb
    \lB 
    \lb \frac{1}{4\pi D y^0} \rb^{d/2}
    \E{-\by^2/(4D y^0)} \rB.
\end{calign}
This is essentially a linear combination of the derivatives of a $d$-dimensional Gaussian distribution centred around $\by=0$ and width $\sqrt{2 D y^0}$. The qualitative behaviour of the analytic part depends on the order in which we take the spatial or temporal component of $y$ to be zero. In the main text, we primarily used the prescription where the spatial cutoffs $\by$ is taken to zero first. Here we will discuss both the order of limits in greater detail.

If we take $y^0\to 0$ first, the Gaussian in \cref{eq:regular-term} collapses into a delta function, turning $\cI_0(0,\by)$ into a combination of spatial contact terms. We can ignore these provided that all the operator insertions are at least infinitesimally separated in space. On the other hand, if we take $\by\to 0$ first, we can evaluate $\cI_0(y^0,0)$ to be
\begin{calign}
\cI_0(y^0,0)
&=
\left(\frac{1}{4\pi Dy^0}\right)^{d/2}
\Bigg(
\frac{d}{\omega_\beta y^0}
+\frac{d(d+2)}{4(\omega_\beta y^0)^2}
-\frac{d(d+2)(d+4)}
{4(\omega_\beta y^0)^3}
-\frac{d(d+2)(d+4)(d+6)}
{16(\omega_\beta y^0)^4}
\Bigg).
\end{calign}
It diverges as $y^0\to 0$ and dominates the magnitude of the OTOC. There are, therefore, three qualitatively independent cases for the OTOC $C_{\bar a \bar a |\bar r a}(x,x,0,0)$ in \cref{eq:OTOC-result}:
\begin{enumerate}
\item We take the temporal separations $t_1-t_2$ and $t_3-t_4$ to zero first, followed by taking the spatial separations $\bx_1-\bx_2$ and $\bx_3-\bx_4$ to infinitesimal (but not strictly zero). In this case, the analytic part $\cI_0$ does not contribute and we get
\begin{calign}
    C^{\cE^t\cE^t}_{\bar a \bar a |\bar r a}(x,x,0,0) 
    &\sim \frac{-i\omega_\beta^3}{2\beta} \E{\omega_\beta t}
    \int \frac{\df^d\bk}{(2\pi)^d}
    \E{i\bk\cdot\bx}c_\bk \,
    \lb 1 + \frac{\bk^2}{m^2}\rb 
    \Big(\alpha_d\lb 2+\bk^2/m^2\rb \Big)^2.
\end{calign}
Therefore, the spatial spread of the OTOC is controlled by the branch point of the integrand at $\bk^2 = -2m^2$. Performing the Fourier transform, we find the Lyapunov growth $\E{\omega_\beta\lb t-\sqrt{2}|\bx|/v_b\rb}$, with the front velocity $v_b/\sqrt{2}$ that is smaller than a generic operator discussed in \cref{sec:generic-otoc}. We explicitly find
\begin{calign}
    C^{\cE^t\cE^t}_{\bar a \bar a |\bar r a}(x,x,0,0)\Big|_{d=1}
    &\sim 
   \#\,
    \E{\omega_\beta\lb t-\sqrt{2}|\bx|/v_b\rb}, \nn\\
    C^{\cE^t\cE^t}_{\bar a \bar a |\bar r a}(x,x,0,0)\Big|_{d\geq 3,\text{odd}}
    &\sim 
    0, \nn\\
    C^{\cE^t\cE^t}_{\bar a \bar a |\bar r a}(x,x,0,0)\Big|_{d\,\text{even}}
    &\sim 
    \#
    \frac{\log(m|\bx|)}
    {(m|\bx|)^{\frac32(d-1)}}
    \E{\omega_\beta\lb t-\sqrt{2}|\bx|/v_b\rb}.
\end{calign}
Note that while in $d=1$ we have pure butterfly spreading, in higher dimensions we also have additional $\bx$-dependent factors in the magnitude.

\item We take the spatial separations $\bx_1-\bx_2$ and $\bx_3-\bx_4$ to zero first, followed by taking the temporal separations $t_1-t_2$ and $t_3-t_4$ to be an infinitesimal number $\varepsilon$ (but not strictly zero). In this case, the analytic part $\cI_0$ dominates the OTOC and we simply get
\begin{calign}
    C^{\cE^t\cE^t}_{\bar a \bar a |\bar r a}(x,x,0,0)
    &\sim \frac{-i\omega_\beta^3}{2\beta} \E{\omega_\beta t}
    \cI_0(\varepsilon,0)^2
    \int \frac{\df^d\bk}{(2\pi)^d}
    \frac{\E{i\bk\cdot\bx}c_\bk}{1+\bk^2/m^2} \,.
\end{calign}
This integral is controlled by the pole at $\bk^2=-m^2$, leading to the Lyapunov growth $\E{\omega_\beta\lb t-|\bx|/v_b\rb}$, with the butterfly velocity $v_b$. We explicitly find
\begin{calign}
    C^{\cE^t\cE^t}_{\bar a \bar a |\bar r a}(x,x,0,0)
    &\sim \#\,
    \frac{\E{\omega_\beta\lb t - \bx/v_b\rb}}{(m|\bx|)^{(d-1)/2}}.
\end{calign}

\item An interesting intermediate case is when we take $t_1-t_2$ to zero first for the first set of operators, while $\bx_3-\bx_4$ to zero first for the second set of operators. This is qualitatively similar to Case 1 and we get the growing part of the OTOC
\begin{calign}
    C^{\cE^t\cE^t}_{\bar a \bar a |\bar r a}(x,x,0,0)
    &\sim \frac{-i\omega_\beta^3}{2\beta}\cI_0(\varepsilon,0)
    \E{\omega_\beta t}
    \int \frac{\df^d\bk}{(2\pi)^d}
    \E{i\bk\cdot\bx}c_\bk\,
    \alpha_d\lb 2+\bk^2/m^2\rb.
\end{calign}
In this case as well, the integrand is dominated by the branch point at $\bk^2 = -2m^2$, leading to the reduced front velocity $v_b/\sqrt{2}$. In this case, we find
\begin{calign}
    C^{\cE^t\cE^t}_{\bar a \bar a |\bar r a}(x,x,0,0) \Big|_{d=1}
    &\sim
    \#\,
    \frac{\E{\omega_\beta \lb t - \sqrt{2}\bx/v_b\rb}}{(m|\bx|)^{d-1/2}}.
\end{calign}
Of course, the answer remain qualitatively the same if we flip to taking $\bx_1-\bx_2$ to zero first for the first set of operators, while $t_3-t_4$ to zero first for the second set of operators.

\end{enumerate}

\subsection{Commutator-squared OTOC in SK EFT}
\label{app:symmeric-otoc}

\subsubsection{Contact diagrams}

We can repeat the computation above for the symmetric OTOC $C_{\bar a \bar a |\bar r \bar r}$. We will find that it does not admit any growing part. Let us start with the contact diagrams. In this case, the core contact diagram is given as
\begin{calign}
\tikzsetnextfilename{2loop-diagram-plus-contact-1}
\raisebox{-6mm}{
\tikz[thick,scale=0.7,xscale=-1]{
    \draw [aux2] (10mm,10mm) arc (90:160:10mm);
    \draw [sol2] (10mm,10mm) arc (90:0:10mm);
    \draw [sol2] (20mm,0) arc (0:-90:10mm);
    \draw [aux2] (10mm,-10mm) arc (-90:-160:10mm);
    \draw [aux2] (20mm,0) arc (180:90:10mm);
    \draw [sol2] (30mm,10mm) arc (90:20:10mm);
    \draw [sol2] (30mm,-10mm) arc (-90:-20:10mm);
    \draw [aux2] (30mm,-10mm) arc (-90:-180:10mm);
    \draw [fill=white] (0.6mm,3.4mm) circle (1.25mm);
    \draw [fill=white] (0.6mm,-3.4mm) circle (1.25mm);
    \draw [fill=black] (39.4mm,3.4mm) circle (1.25mm);
    \draw [fill=black] (39.4mm,-3.4mm) circle (1.25mm);
} }
\end{calign}
together with those with non-linear parts of operator insertions
\begin{cgather}
\tikzsetnextfilename{2loop-diagram-plus-contact-2-1}
\raisebox{-7mm}{
\tikz[thick,scale=0.7,xscale=-1]{
\draw [aux2] (-10mm,-3.25mm) arc (180:30:6mm);
\draw [sol2] (-10mm,-3.25mm) arc (-180:-30:6mm);
\draw [aux2] (0,-7mm) -- (15mm,-3.5mm);
\draw [sol2] (15mm,-3.5mm) -- (30mm,0);
\draw [aux2] (15mm,-15mm) arc (-90:-155:16.5mm and 15mm);
\draw [sol2] (15mm,-15mm) arc (-90:-25:16.5mm and 15mm);
\draw [fill=white] (0,0) circle (1.25mm);
\draw [fill=white] (0,-7mm) circle (1.25mm);
\draw [fill=black] (30mm,0) circle (1.25mm);
\draw [fill=black] (30mm,-7mm) circle (1.25mm);
}} \qquad 
\tikzsetnextfilename{2loop-diagram-plus-contact-2-3}
\raisebox{-3mm}{
\tikz[thick,scale=0.7,xscale=-1]{
\draw [sol2] (-10mm,-3.25mm) arc (180:30:6mm);
\draw [aux2] (-10mm,-3.25mm) arc (-180:-30:6mm);
\draw [aux2] (0,0) -- (15mm,-3.5mm);
\draw [sol2] (15mm,-3.5mm) -- (30mm,-7mm);
\draw [aux2] (15mm,8mm) arc (90:155:16.5mm and 15mm);
\draw [sol2] (15mm,8mm) arc (90:25:16.5mm and 15mm);
\draw [fill=white] (0,0) circle (1.25mm);
\draw [fill=white] (0,-7mm) circle (1.25mm);
\draw [fill=black] (30mm,0) circle (1.25mm);
\draw [fill=black] (30mm,-7mm) circle (1.25mm);
}} \qquad 
\tikzsetnextfilename{2loop-diagram-plus-contact-2-2}
\raisebox{-7mm}{
\tikz[thick,scale=0.7,xscale=-1]{
\draw [aux2] (0,0) -- (15mm,-3.5mm);
\draw [sol2] (15mm,-3.5mm) -- (30mm,-7mm);
\draw [aux2] (15mm,-15mm) arc (-90:-155:16.5mm and 15mm);
\draw [sol2] (15mm,-15mm) arc (-90:-25:16.5mm and 15mm);
\draw [sol2] (41mm,-3.25mm) arc (0:150:6mm);
\draw [aux2] (41mm,-3.25mm) arc (0:-150:6mm);
\draw [fill=white] (0,0) circle (1.25mm);
\draw [fill=white] (0,-7mm) circle (1.25mm);
\draw [fill=black] (30mm,0) circle (1mm);
\draw [fill=black] (30mm,-7mm) circle (1mm);
}} \qquad
\tikzsetnextfilename{2loop-diagram-plus-contact-2-4}
\raisebox{-3mm}{
\tikz[thick,scale=0.7,xscale=-1]{
\draw [aux2] (0,-7mm) -- (15mm,-3.5mm);
\draw [sol2] (15mm,-3.5mm) -- (30mm,0);
\draw [aux2] (15mm,8mm) arc (90:155:16.5mm and 15mm);
\draw [sol2] (15mm,8mm) arc (90:25:16.5mm and 15mm);
\draw [aux2] (41mm,-3.25mm) arc (0:150:6mm);
\draw [sol2] (41mm,-3.25mm) arc (0:-150:6mm);
\draw [fill=white] (0,0) circle (1.25mm);
\draw [fill=white] (0,-7mm) circle (1.25mm);
\draw [fill=black] (30mm,0) circle (1mm);
\draw [fill=black] (30mm,-7mm) circle (1mm);
}}
\end{cgather}
Our comments regarding the contact diagrams for minus OTOC also apply to the plus OTOC. Since all propagators are connected to a linear operator insertion, the structure factor at the operator insertion identically cancels the growing part of the propagator.

\subsubsection{t-channel and u-channel diagrams}

There are 3 basic t-channel diagrams
\begin{calign}\label{basic-t-plus}
\tikzsetnextfilename{2loop-diagram-plus-t-1}
\raisebox{-10mm}{
\tikz[thick,scale=0.7,xscale=-1]{
    \draw [aux2] (4.4mm,10.6mm) arc (135:165:15mm);
    \draw [sol2] (15mm,15mm) arc (90:135:15mm);
    \draw [aux2] (15mm,15mm) arc (90:45:15mm);
    \draw [sol2] (25.6mm,10.6mm) arc (45:15:15mm);
    \draw [] (15mm,15mm) -- (15mm,0);
    \draw [] (15mm,0) -- (15mm,-15mm);
    \draw [aux2] (4.4mm,-10.6mm) arc (-135:-165:15mm);
    \draw [sol2] (15mm,-15mm) arc (-90:-135:15mm);
    \draw [aux2] (15mm,-15mm) arc (-90:-45:15mm);
    \draw [sol2] (25.6mm,-10.6mm) arc (-45:-15:15mm);
    \draw [fill=white] (0.6mm,3.4mm) circle (1.25mm);
    \draw [fill=white] (0.6mm,-3.4mm) circle (1.25mm);
    \draw [fill=black] (29.4mm,3.4mm) circle (1.25mm);
    \draw [fill=black] (29.4mm,-3.4mm) circle (1.25mm);
}} \qquad 
\tikzsetnextfilename{2loop-diagram-plus-t-2}
\raisebox{-10mm}{
\tikz[thick,scale=0.7,xscale=-1]{
    \draw [aux2] (4.4mm,10.6mm) arc (135:165:15mm);
    \draw [sol2] (15mm,15mm) arc (90:135:15mm);
    \draw [aux2] (15mm,15mm) arc (90:45:15mm);
    \draw [sol2] (25.6mm,10.6mm) arc (45:15:15mm);
    \draw [aux] (15mm,15mm) -- (15mm,0);
    \draw [] (15mm,0) -- (15mm,-15mm);
    \draw [aux2] (4.4mm,-10.6mm) arc (-135:-165:15mm);
    \draw [sol2] (15mm,-15mm) arc (-90:-135:15mm);
    \draw [aux2] (15mm,-15mm) arc (-90:-45:15mm);
    \draw [sol2] (25.6mm,-10.6mm) arc (-45:-15:15mm);
    \draw [fill=white] (0.6mm,3.4mm) circle (1.25mm);
    \draw [fill=white] (0.6mm,-3.4mm) circle (1.25mm);
    \draw [fill=black] (29.4mm,3.4mm) circle (1.25mm);
    \draw [fill=black] (29.4mm,-3.4mm) circle (1.25mm);
}} \qquad 
\tikzsetnextfilename{2loop-diagram-plus-t-3}
\raisebox{-10mm}{
\tikz[thick,scale=0.7,xscale=-1]{
    \draw [aux2] (4.4mm,10.6mm) arc (135:165:15mm);
    \draw [sol2] (15mm,15mm) arc (90:135:15mm);
    \draw [aux2] (15mm,15mm) arc (90:45:15mm);
    \draw [sol2] (25.6mm,10.6mm) arc (45:15:15mm);
    \draw [] (15mm,15mm) -- (15mm,0);
    \draw [aux] (15mm,0) -- (15mm,-15mm);
    \draw [aux2] (4.4mm,-10.6mm) arc (-135:-165:15mm);
    \draw [sol2] (15mm,-15mm) arc (-90:-135:15mm);
    \draw [aux2] (15mm,-15mm) arc (-90:-45:15mm);
    \draw [sol2] (25.6mm,-10.6mm) arc (-45:-15:15mm);
    \draw [fill=white] (0.6mm,3.4mm) circle (1.25mm);
    \draw [fill=white] (0.6mm,-3.4mm) circle (1.25mm);
    \draw [fill=black] (29.4mm,3.4mm) circle (1.25mm);
    \draw [fill=black] (29.4mm,-3.4mm) circle (1.25mm);
}}
\end{calign}
There are 8 non-linear diagrams related to the first one of these 
\begin{cgather}
\tikzsetnextfilename{2loop-diagram-plus-t-2-1}
\tikz[thick,scale=0.7,xscale=-1]{
    \draw [aux2] (4.4mm,10.6mm) arc (135:165:15mm);
    \draw [sol2] (15mm,15mm) arc (90:135:15mm);
    \draw [aux2] (15mm,15mm) arc (90:45:15mm);
    \draw [sol2] (25.6mm,10.6mm) arc (45:15:15mm);
    \draw [] (15mm,15mm) arc (180:225:15mm and 18.5mm);
    \draw [] (30mm,-3.4mm) arc (-90:-135:15mm and 18.5mm);
    \draw [aux2] (15mm,-15mm) arc (-90:-165:15mm);
    \draw [sol2] (15mm,-15mm) arc (-90:-15:15mm);
    \draw [fill=white] (0.6mm,3.4mm) circle (1.25mm);
    \draw [fill=white] (0.6mm,-3.4mm) circle (1.25mm);
    \draw [fill=black] (29.4mm,3.4mm) circle (1.25mm);
    \draw [fill=black] (29.4mm,-3.4mm) circle (1.25mm);
}\qquad 
\tikzsetnextfilename{2loop-diagram-plus-t-2-3}
\tikz[thick,scale=0.7,xscale=-1]{
    \draw [aux2] (4.4mm,10.6mm) arc (135:165:15mm);
    \draw [sol2] (15mm,15mm) arc (90:135:15mm);
    \draw [aux2] (15mm,15mm) arc (90:45:15mm);
    \draw [sol2] (25.6mm,10.6mm) arc (45:15:15mm);
    \draw [] (15mm,15mm) arc (0:-45:15mm and 18.5mm);
    \draw [] (0mm,-3.4mm) arc (-90:-45:15mm and 18.5mm);
    \draw [aux2] (15mm,-15mm) arc (-90:-165:15mm);
    \draw [sol2] (15mm,-15mm) arc (-90:-15:15mm);
    \draw [fill=white] (0.6mm,3.4mm) circle (1.25mm);
    \draw [fill=white] (0.6mm,-3.4mm) circle (1.25mm);
    \draw [fill=black] (29.4mm,3.4mm) circle (1.25mm);
    \draw [fill=black] (29.4mm,-3.4mm) circle (1.25mm);
}\qquad 
\tikzsetnextfilename{2loop-diagram-plus-t-2-2}
\tikz[thick,scale=0.7,xscale=-1]{
    \draw [aux2] (15mm,15mm) arc (90:165:15mm);
    \draw [sol2] (15mm,15mm) arc (90:15:15mm);
    \draw [] (15mm,-15mm) arc (180:135:15mm and 18.5mm);
    \draw [] (30mm,3.4mm) arc (90:135:15mm and 18.5mm);
    \draw [aux2] (4.4mm,-10.6mm) arc (-135:-165:15mm);
    \draw [sol2] (15mm,-15mm) arc (-90:-135:15mm);
    \draw [aux2] (15mm,-15mm) arc (-90:-45:15mm);
    \draw [sol2] (25.6mm,-10.6mm) arc (-45:-15:15mm);
    \draw [fill=white] (0.6mm,3.4mm) circle (1.25mm);
    \draw [fill=white] (0.6mm,-3.4mm) circle (1.25mm);
    \draw [fill=black] (29.4mm,3.4mm) circle (1.25mm);
    \draw [fill=black] (29.4mm,-3.4mm) circle (1.25mm);
} \qquad 
\tikzsetnextfilename{2loop-diagram-plus-t-2-4}
\tikz[thick,scale=0.7,xscale=-1]{
    \draw [aux2] (15mm,15mm) arc (90:165:15mm);
    \draw [sol2] (15mm,15mm) arc (90:15:15mm);
    \draw [] (15mm,-15mm) arc (0:45:15mm and 18.5mm);
    \draw [] (0mm,3.4mm) arc (90:45:15mm and 18.5mm);
    \draw [aux2] (4.4mm,-10.6mm) arc (-135:-165:15mm);
    \draw [sol2] (15mm,-15mm) arc (-90:-135:15mm);
    \draw [aux2] (15mm,-15mm) arc (-90:-45:15mm);
    \draw [sol2] (25.6mm,-10.6mm) arc (-45:-15:15mm);
    \draw [fill=white] (0.6mm,3.4mm) circle (1.25mm);
    \draw [fill=white] (0.6mm,-3.4mm) circle (1.25mm);
    \draw [fill=black] (29.4mm,3.4mm) circle (1.25mm);
    \draw [fill=black] (29.4mm,-3.4mm) circle (1.25mm);
} \nn\\
\tikzsetnextfilename{2loop-diagram-plus-t-3-1}
\tikz[thick,scale=0.7,xscale=-1]{
    \draw [aux2] (15mm,15mm) arc (90:165:15mm);
    \draw [sol2] (15mm,15mm) arc (90:15:15mm);
    \draw [sol2] (15mm,-15mm) arc (-90:-15:15mm);
    \draw [aux2] (15mm,-15mm) arc (-90:-165:15mm);
    \draw [] (0.6mm,3.4mm) -- (15mm,0);
    \draw [] (15mm,0) -- (29.4mm,-3.4mm);
    \draw [fill=white] (0.6mm,3.4mm) circle (1.25mm);
    \draw [fill=white] (0.6mm,-3.4mm) circle (1.25mm);
    \draw [fill=black] (29.4mm,3.4mm) circle (1.25mm);
    \draw [fill=black] (29.4mm,-3.4mm) circle (1.25mm);
} \qquad 
\tikzsetnextfilename{2loop-diagram-plus-t-3-2}
\tikz[thick,scale=0.7,xscale=-1]{
    \draw [aux2] (15mm,15mm) arc (90:165:15mm);
    \draw [sol2] (15mm,15mm) arc (90:15:15mm);
    \draw [sol2] (15mm,-15mm) arc (-90:-15:15mm);
    \draw [aux2] (15mm,-15mm) arc (-90:-165:15mm);
    \draw [] (0.6mm,-3.4mm) -- (15mm,0);
    \draw [] (15mm,0) -- (29.4mm,3.4mm);
    \draw [fill=white] (0.6mm,3.4mm) circle (1.25mm);
    \draw [fill=white] (0.6mm,-3.4mm) circle (1.25mm);
    \draw [fill=black] (29.4mm,3.4mm) circle (1.25mm);
    \draw [fill=black] (29.4mm,-3.4mm) circle (1.25mm);
} \qquad
\tikzsetnextfilename{2loop-diagram-plus-t-4-1}
\tikz[thick,scale=0.7,xscale=-1]{
    \draw [aux2] (15mm,15mm) arc (90:165:15mm);
    \draw [sol2] (15mm,15mm) arc (90:15:15mm);
    \draw [sol2] (15mm,-15mm) arc (-90:-15:15mm);
    \draw [aux2] (15mm,-15mm) arc (-90:-165:15mm);
    \draw [] (40mm,0) arc (0:150:6mm);
    \draw [] (40mm,0) arc (0:-150:6mm);
    \draw [fill=white] (0.6mm,3.4mm) circle (1.25mm);
    \draw [fill=white] (0.6mm,-3.4mm) circle (1.25mm);
    \draw [fill=black] (29.4mm,3.4mm) circle (1.25mm);
    \draw [fill=black] (29.4mm,-3.4mm) circle (1.25mm);
} \qquad 
\tikzsetnextfilename{2loop-diagram-plus-t-4-2}
\tikz[thick,scale=0.7,xscale=-1]{
    \draw [] (-10mm,0) arc (180:30:6mm);
    \draw [] (-10mm,0) arc (-180:-30:6mm);
    \draw [aux2] (15mm,15mm) arc (90:165:15mm);
    \draw [sol2] (15mm,15mm) arc (90:15:15mm);
    \draw [sol2] (15mm,-15mm) arc (-90:-15:15mm);
    \draw [aux2] (15mm,-15mm) arc (-90:-165:15mm);
    \draw [fill=white] (0.6mm,3.4mm) circle (1.25mm);
    \draw [fill=white] (0.6mm,-3.4mm) circle (1.25mm);
    \draw [fill=black] (29.4mm,3.4mm) circle (1.25mm);
    \draw [fill=black] (29.4mm,-3.4mm) circle (1.25mm);
} 
\end{cgather}
Similarly, there are 8 non-linear diagrams each for the other two t-channel diagrams in \cref{basic-t-plus}, obtained by appropriately swapping out the internal propagator.

The diagrams for u-channel can be obtained by exchanging the two operator insertions on the right in the diagrams above; we will not draw these explicitly here.

The collective contribution of the t-channel diagrams is given as
\begin{calign}\label{eq:t-channel-full-symmetric}
    C^{\rm t}_{\bar a\bar a\bar r\bar r}
    &= \int \df^{d+1} y_1\df^{d+1} y_2\nn\\
    &\qquad
    \Bigg[
    \cN_{{\rm t}, r}(x_1-y_1,y_1-x_4)\,
    \cN_{{\rm t}, r}(x_2-y_2,y_2-x_3)
    K^{-1}_{\rmS}(y_1-y_2) \nn\\
    &\hspace{4em}
    + 
    \cN_{{\rm t}, r}(x_1-y_1,y_1-x_4)\,
    \cN_{{\rm t}, a}(x_2-y_2,y_2-x_3)\,
    iK^{-1}(y_1-y_2) \nn\\
    &\hspace{4em}
    + 
    \cN_{{\rm t}, a}(x_1-y_1,y_1-x_4)\,
    \cN_{{\rm t}, r}(x_2-y_2,y_2-x_3)\,
    iK^{-1}(y_2-y_1)
    \Bigg],
\end{calign}
where we have identified
\begin{calign}
    &\cN_{{\rm t}, r}(x_1-y_1,y_1-x_4) \nn\\
    &= i\cV_{F\bar Q\bar F}(\dow_1-\dow_4,\dow_4,-\dow_1)\,
    \Big(i\cO^V_{(1)}(\dow_1) K^{-1}_1(x_1-y_1)\Big)\,
    \Big({-}i\eta_\Theta\tilde\cO^W_{(1)}(-\dow_4) K^{-1}_4(y_1-x_4)\Big)
     \nn\\
    &\qquad
    + 2\Big(\eta_\Theta\tilde\cO^W_{(2)}(-\dow_1,\dow_1-\dow_4)
    - Y^W(\dow_1-\dow_4,-\dow_1)
    \Big)\,
    \Big(i\cO^V_{(1)}(\dow_1)K^{-1}_1(x_1-x_4)\Big)\,
    \delta_4^{(d+1)}(y_1-x_4)  \nn\\
    &\qquad
    + 2\cO^V_{(2)}(\dow_4,\dow_1-\dow_4)\,
    \Big({-}i\eta_\Theta\tilde\cO^W_{(1)}(-\dow_4)K^{-1}_4(x_1-x_4)\Big)
    \delta_1^{(d+1)}(x_1-y_1), \nn\\
    &\cN_{{\rm t}, a}(x_1-y_1,y_1-x_4) \nn\\
    &= -\cV_{Q\bar Q\bar F}(\dow_1-\dow_4,\dow_4,-\dow_1)\,
    \Big(i\cO^V_{(1)}(\dow_1) K^{-1}_1(x_1-y_1)\Big)\,
    \Big({-}i\eta_\Theta\tilde\cO^W_{(1)}(-\dow_4) K^{-1}_4(y_1-x_4)\Big) \nn\\
    &\qquad 
    + i\cot\lb \frac{\beta (\dow_{t1}-\dow_{t4})}{2} \rb Y^W(\dow_1-\dow_4,-\dow_1)
    \Big(i\cO^V_{(1)}(\dow_1)K^{-1}_1(x_1-x_4)\Big)\,
    \delta_4^{(d+1)}(y_1-x_4)  \nn\\
    &\qquad
    + i\cot\lb \frac{\beta (\dow_{t1}-\dow_{t4})}{2} \rb Y^V(\dow_1-\dow_4,\dow_4)
    \Big({-}i\eta_\Theta\tilde\cO^W_{(1)}(-\dow_4)K^{-1}_4(x_1-x_4)\Big)
    \delta_1^{(d+1)}(x_1-y_1).
\end{calign}
Looking a little closely, we note that $\cN_{{\rm t}, r}$ is the same as $\cM_{\rm t}$ defined earlier for minus OTOC in \cref{eq:M-def-minus-otoc-tchannel} because $\cV_{F\bar Q\bar F}$ is symmetric in its first two arguments. On the other hand, we have
\begin{calign}
    \cN_{{\rm t}, a}&(x_1-y_1,y_1-x_4) \nn\\
    &= \lb \dow_{t1}-\dow_{t4}\rb
    \lb 1 - \frac{\lb \dow_{t1}-\dow_{t4}\rb^2}{\omega_\beta^2} \rb
    \cot\lb \frac{\beta (\dow_{t1}-\dow_{t4})}{2} \rb \nn\\
    &\quad 
    \Bigg\{-\eta_\Theta
    \Bigg[\frac{\beta }{4}
    \lb \dow_{t1}-\dow_{t4}\rb
    \Big(\Delta(\dow_1-\dow_4) + \Delta(\dow_4-\dow_1) \Big) \nn\\
    &\hspace{8em}
    - \frac{\beta }{4} \lb \dow_{t4}+\dow_{t1}\rb
    \Big( \Delta(\dow_4) - \Delta(-\dow_1)
    + \Delta(-\dow_4) - \Delta(\dow_1)
    \Big) \nn\\
    &\hspace{8em}
    + \half 
    \dow_{t4} \lb 1 - \frac{\dow_{t4}^2}{\omega_\beta^2} \rb
    \Big(\cR(\dow_1-\dow_4,\dow_4) + \tilde\cR(\dow_1-\dow_4,\dow_4)\Big) \nn\\
    &\hspace{8em}
    - \half 
    \dow_{t1} \lb 1 - \frac{\dow_{t1}^2}{\omega_\beta^2} \rb
    \Big(\cR(\dow_1-\dow_4,-\dow_1) + \tilde\cR(\dow_1-\dow_4,-\dow_1)\Big)
    \Bigg] \nn\\
    &\hspace{6em}
    \Big(\cD^V_{(1)}(\dow_1) \Delta^{-1}_1(x_1-y_1)\Big)\,
    \Big(\tilde\cD^W_{(1)}(-\dow_4) \Delta^{-1}_4(y_1-x_4)\Big) \nn\\
    &\qquad 
    - \cY^W(\dow_1-\dow_4,-\dow_1)
    \Big(\cO^V_{(1)}(\dow_1)K^{-1}_1(x_1-x_4)\Big)\,
    \delta_4^{(d+1)}(y_1-x_4)  \nn\\
    &\qquad
    + \eta_\Theta\cY^V(\dow_1-\dow_4,\dow_4)
    \Big(\tilde\cO^W_{(1)}(-\dow_4)K^{-1}_4(x_1-x_4)\Big)
    \delta_1^{(d+1)}(x_1-y_1) \Bigg\} \nn\\
    &= \dow_{s_1}
    \lb 1 - \frac{\dow_{s_1}^2}{\omega_\beta^2} \rb
    \cot\lb \frac{\beta \dow_{s_1}}{2} \rb\#,
\end{calign}
for some expression $\#$. Despite the appearance of the $\cot$, this expression entirely regular on the modes $\dow_{s_1} = 0,\pm\omega_\beta$.

Therefore, using \cref{eq:M-def-minus-otoc-tchannel-result}, it follows that
\begin{calign}
    \cN_{{\rm t}, r}(x_1-y_1,y_1-x_4)\,
    \cN_{{\rm t}, r}(x_2-y_2,y_2-x_3)
    &= \dow_{s_1} \lb 1 - \frac{\dow_{s_1}^2}{\omega_\beta^2} \rb
    \dow_{s_2} \lb 1 - \frac{\dow_{s_2}^2}{\omega_\beta^2} \rb
    \#_1, \nn\\
    \cN_{{\rm t}, r}(x_1-y_1,y_1-x_4)\,
    \cN_{{\rm t}, a}(x_2-y_2,y_2-x_3)
    &= \dow_{s_1} \lb 1 - \frac{\dow_{s_1}^2}{\omega_\beta^2} \rb
    \#_2, \nn\\
    \cN_{{\rm t}, a}(x_1-y_1,y_1-x_4)\,
    \cN_{{\rm t}, r}(x_2-y_2,y_2-x_3)
    &= \dow_{s_2} \lb 1 - \frac{\dow_{s_2}^2}{\omega_\beta^2} \rb
    \#_3,
\end{calign}
for some expressions $\#_{1,2,3}$. Plugging this back into \cref{eq:t-channel-full}, we find
\begin{calign}
    C^{\rm t}_{\bar a\bar a\bar r\bar r}
    &= \int \df^{d+1} y_1\df^{d+1} y_2\nn\\
    &\qquad
    \Bigg[
    \#_1
    \dow_{s_1} \lb 1 - \frac{\dow_{s_1}^2}{\omega_\beta^2} \rb
    \dow_{s_2} \lb 1 - \frac{\dow_{s_2}^2}{\omega_\beta^2} \rb
    K^{-1}_{\rmS}(y_1-y_2) \nn\\
    &\hspace{4em}
    -
    i\#_2
    \dow_{s_1} \lb 1 - \frac{\dow_{s_1}^2}{\omega_\beta^2} \rb
    K^{-1}(y_1-y_2) 
    - i\#_3
    \dow_{s_2} \lb 1 - \frac{\dow_{s_2}^2}{\omega_\beta^2} \rb K^{-1}(y_2-y_1)
    \Bigg] \nn\\
    &= \int \df^{d+1} y_1\df^{d+1} y_2
    \Bigg[
    {-}\#_1 \Delta^{-1}_{\rmS}(y_1-y_2) 
    + i\#_2 \Delta^{-1}(y_1-y_2) 
    + i\#_3 \Delta^{-1}(y_2-y_1)
    \Bigg],
\end{calign}
which is insensitive to the growing mode. The same conclusion also follows for the u-channel.

\subsubsection{s-channel diagrams}

One may check that there are no possible s-channel diagrams for $C_{\bar a \bar a |\bar r \bar r}$. Together with the results from the previous subsections, it follows that the symmetric OTOC $C_{\bar a \bar a |\bar r \bar r}$ admits no Lyapunov growth, at least at tree level. This conclusion holds for any particular form of the energy diffusion SK EFT and any local operator insertions.

\subsection{Details of OTOC of generic operators}
\label{app:generic-OTOC}

The full form of the OTOC for generic operators in \cref{eq:bilocal-otoc-general} is given as
\begin{calign}\label{eq:bilocal-otoc}
    &\hat C^{VW}_{42|31}(x_1,x_2;x_3,x_4)
    - \hat C^{VW}_{43|21}(x_1,x_2;x_3,x_4) \nn\\
    &= - 2i\pi\Delta_V\Delta_W \Theta(t_{23}) 
    \int \frac{\df^d\bk}{(2\pi)^d}
    \frac{\E{i\bk\cdot(\bx_2-\bx_3)}}{c_\bk} \nn\\
    &\hspace{2em}
    \Bigg[
    \frac{\frac{\omega_\beta}{\cD_\bk}\lb 1-\E{-\cD_\bk t_{23}} \rb
    }{\tanh\!\big(\omega_\beta t_{12}/2\big) 
    \tanh\!\big(\omega_\beta t_{34}/2\big)} \nn\\
    &\hspace{3em}
    - \frac{
    \frac{\omega_\beta}{2(\omega_\beta+\cD_\bk)}
    \E{\omega_\beta(t_{12}+t_{34})/2} 
    \lb \E{\omega_\beta t_{23}} - \E{-\cD_\bk t_{23}} \rb
    }{\sinh\!\big(\omega_\beta t_{12}/2\big) 
    \sinh\!\big(\omega_\beta t_{34}/2\big)
    } 
    + \frac{
    \frac{\omega_\beta}{2(\omega_\beta-\cD_\bk)}
    \E{-\omega_\beta(t_{12}+t_{34})/2} 
    \lb \E{-\omega_\beta t_{23}} - \E{-\cD_\bk t_{23}} \rb
    }{\sinh\!\big(\omega_\beta t_{12}/2\big) 
    \sinh\!\big(\omega_\beta t_{34}/2\big)
    }
    \Bigg] \nn\\
    &\qquad 
    - (x_1\leftrightarrow x_4,x_2\leftrightarrow x_3).
\end{calign}
The interesting growing part of this result is given in \cref{eq:bilocal-otoc2}.

\section{Details of the scramblon prescription}

\subsection{Evaluation of the scramblon action for SYK chains}
\label{app:scramblon_action}
Following \cite{Maldacena:2016upp} we consider the configuration shown in~\cref{fig:shocks} on the four-fold OTOC contour. For convenience we introduce the $y$ variable: $y = \tanh(\pi t/\beta)$,
\begin{calign}
y_1 &= y - \frac{(1+y)^2\,X^-}{2+(1+y)X^-}, \\
y_2 &= y - \frac{(1+y)^2\,X^-}{2+(1+y)X^-} 
+ \frac{(1-y)^2\,X^+}{2+(1-y)X^+},  \\
y_3 &= y + \frac{(1-y)^2\,X^+}{2+(1-y)X^+}, \\
y_4 &= y, \\
y_E &= y,
\end{calign}
where $y_E$ denotes the Euclidean part. Here $X^\pm$ can depend on the spatial coordinate $\bx$. In fact, it is crucial for obtaining a non-zero action: since the non-local term in (\ref{eq:final_action}) is marginal, it will reduce to a total derivative if we plug in an $x$-independent configuration. Following the continuum limit of~\cite{Bucca:2026xmh}, we plug in $X^\pm(\bx) - \partial_\bx X^\pm/2$ inside $F_i(\bx)$ on site $\bx$ and $X^\pm(\bx) + \partial_\bx X^\pm/2$ on site $\bx+a$.

Substituting the above configuration into the SYK Schwarzian term in \cref{eq:bilocalform} produces the standard \cite{Maldacena:2016upp} term
\beq
\frac{N\alpha_S}{\cJ}\, \frac{4 \pi}{\beta} X^+ X^-.
\eeq
This term only comes from $y_2$ because this is the only place where both $X^+$ and $X^-$ are turned on.

The evaluation of the non-local piece is much more subtle. The first non-trivial fact is that the Euclidean part of the contour gives a non-zero contribution. The second non-trivial fact is that resulting integrals are order dependent: integrating over $t_1$ first and then over $t_2$ produces a different result compared to the opposite order of integration. We resolve this ambiguity by noticing that in the original integral $\int \df t_1 \df t_2$, the variables go over the whole contour, so, for example, there is a configuration where $t_1$ runs over the first fold and $t_2$ runs over the second and then $t_1$ can run over the second and $t_2$ over first. This naturally leads to the symmetric prescription. The same prescription is valid~\cite{MM_unpublished} for more complicated non-local actions dominating over the Schwarzian in coupled SYK models \cite{Milekhin:2021cou}.
After adopting this prescription, the remaining integrals can be easily evaluated without the use of special functions. 
Working at quadratic order in spatial derivatives, we get the following contributions proportional to $Ng\, \partial_\bx X^+ \partial_\bx X^-/4$:

From pairwise terms $(y_4-y_2) - (y_1 - y_2) + (y_1 - y_3) - (y_2 - y_3) $ we get:
\es{eq:combinations}{
-\frac{\pi ^2 \left(e^{\frac{\pi  \left(2 t_1-t_2\right)}{\beta }}+e^{\frac{\pi  t_2}{\beta }}+2 e^{\frac{\pi  \left(t_2-2 t_1\right)}{\beta }}\right) \left(\tanh \left(\frac{\pi  t_1}{\beta }\right)+1\right) \text{csch}^3\left(\frac{\pi  \left(t_1-t_2\right)}{\beta }\right) \text{sech}\left(\frac{\pi  t_1}{\beta }\right)}{4 \beta ^2}.
}
Since these contours separated by $i \epsilon$ we can safely avoid the pole at $t_1 = t_2$.
Integrating over $t_1 \in (-\infty,+\infty)$ yields zero. In the opposite order, $t_2$ first and then $t_1$, the result is simply $-2$.

For $(y_2 - y_2)$ the integrand is:
\beq
\frac{\pi ^2 \text{sech}^2\left(\frac{\pi  t_1}{\beta }\right) \text{sech}^2\left(\frac{\pi  t_2}{\beta }\right)}{2 \beta ^2}\,\df t_1\,\df t_2\,.
\eeq
It has to be integrated using the anti-time-ordered prescription. However, the integrand is symmetric in $t_1 \leftrightarrow t_2$, giving an ordinary integral equal to $2$. This cancels with the above pairwise terms.

The final contribution, $(y_E - y_2)$ involves the Euclidean part of the contour:
\beq
 -\frac{\pi ^2 \text{sech}^2\left(\frac{\pi  t_2}{\beta }\right) \text{csch}^4\left(\frac{\pi  \left(t_2+i \tau \right)}{\beta }\right) \left(\cos \left(\frac{2 \pi  \tau }{\beta }\right)+2 \cosh \left(\frac{2 \pi  \left(t_2+i \tau \right)}{\beta }\right)-1\right)}{4 \beta ^2}\,i\df\tau \,\df t_2\,,
\eeq
where $t_1 = -i \tau$.
Integrating over $\tau$ first yields zero,
Integrating over $t_2$ and then $\tau$ yields $4 \pi/3$. This is the only place contributing to the final answer: 
\beq
\frac{Ng \pi}{3}\partial_\bx X^+ \partial_\bx X^-.
\eeq
Added together with the Schwarzian contribution they give
\es{eq:Appshockaction}{
S_L= \int \frac{\df^d\bk}{(2\pi)^d}\,\underbrace{\frac{2N\alpha_S\omega_\beta}{\cJ}\le(1+{D\bk^2\ov\omega_\beta}\ri)}_{=N/\alpha(\bk)} X^+(\bk)  X^-(-\bk)\,,
}
leading to precise agreement with~\cref{eq:match} (up to $O(\bk^2)$ that we have been working in this appendix).

\subsection{Scramblon prescription for OTOCs of local operators}
\label{app:shocks}

We wish to compute the two-point correlator of the energy density around the shockwave solution. The shockwave prescription for our $C_{\bar a \bar a |\bar r a}$ OTOC is given in \cref{eq:energy-shockwave}. To evaluate this, let us introduce a more complete notation $\cG^V_{ij}(x,x'|\gamma)_{(+)_{k,l}}$ for the  two-point function $\langle V_i(x)V_j(x')\rangle$ on a background with the growing scramblon shock with parameter $\gamma$ is present on ``$k,l$'' folds, and similarly $\cG^W_{ij}(x,x'|\gamma)_{(-)_{k,l}}$ for the decaying scramblon shock. Using our definitions of four-fold fields, we can check that
\begin{calign}
    {\cal G}^V_{\bar a \bar a}\lb x_1,x_2\vert X^-\rb_{(+)_{1,2}}
    &= {\cal G}^V_{22}\lb x_1,x_2\vert X^-\rb_{(+)_{1,2}}
    + {\cal G}^V_{33}\lb x_1,x_2\vert X^-\rb_{(+)_{1,2}} \nn\\
    &\qquad 
    - {\cal G}^V_{23}\lb x_1,x_2\vert X^-\rb_{(+)_{1,2}}
    - {\cal G}^V_{32}\lb x_1,x_2\vert X^-\rb_{(+)_{1,2}} \nn\\
    &\hspace{-.9em}
    \overset{\rm KMS_{(1)}}{=}
    {\cal G}^V_{22}\big( {-}x_1,-x_2\vert \tilde X^-\big)_{(-)_{3,2}}
    + {\cal G}^V_{11}\big( {-}x_1,-x_2\vert \tilde X^-\big)_{(-)_{3,2}} \nn\\
    &\qquad 
    - {\cal G}^V_{21}\big( {-}x_1,-x_2\vert \tilde X^-\big)_{(-)_{3,2}}
    - {\cal G}^V_{12}\big( {-}x_1,-x_2\vert \tilde X^-\big)_{(-)_{3,2}} \nn\\
    &\hspace{-.7em}
    \overset{\text{2-fold}}{=}
    {\cal G}^V_{aa}\big( {-}x_1,-x_2\vert \tilde X^-\big)_{(-)_2}
\end{calign}
In the second step, we have used the KMS$_{(1)}$ symmetry on the four-fold contour from \cref{eq:KMS-four} to match the correlators to a configuration with decaying scramblon mode (but with the conjugated parameter $\tilde X^-$). Note that, in doing so, no operator is being taken across the thermal circle. In the third step, we have represented the correlators in terms of the two-fold basis. Similarly, we can find
\begin{calign}
    {\cal G}^W_{\bar r a}\lb x_3,x_4\vert X^+\rb_{(-)_{2,3}}
    &= \half {\cal G}^W_{11}\lb x_3,x_4\vert X^+\rb_{(-)_{2,3}}
    + \half {\cal G}^W_{22}\lb x_3,x_4\vert X^+\rb_{(-)_{2,3}} \nn\\
    &\qquad 
    - \half {\cal G}^W_{12}\lb x_3,x_4\vert X^+\rb_{(-)_{2,3}}
    - \half {\cal G}^W_{21}\lb x_3,x_4\vert X^+\rb_{(-)_{2,3}} \nn\\
    &\qquad 
    - \half {\cal G}^W_{33}\lb x_3,x_4\vert X^+\rb_{(-)_{2,3}}
    - \half {\cal G}^W_{44}\lb x_3,x_4\vert X^+\rb_{(-)_{2,3}} \nn\\
    &\qquad 
    + \half {\cal G}^W_{34}\lb x_3,x_4\vert X^+\rb_{(-)_{2,3}}
    + \half {\cal G}^W_{43}\lb x_3,x_4\vert X^+\rb_{(-)_{2,3}} \nn\\
    &= \half {\cal G}^W_{11}\lb x_3,x_4\vert X^+\rb_{(-)_{2,3}}
    + \half {\cal G}^W_{22}\lb x_3,x_4\vert X^+\rb_{(-)_{2,3}} \nn\\
    &\qquad 
    - \half {\cal G}^W_{12}\lb x_3,x_4\vert X^+\rb_{(-)_{2,3}}
    - \half {\cal G}^W_{21}\lb x_3,x_4\vert X^+\rb_{(-)_{2,3}} \nn\\
    &\qquad 
    - \half {\cal G}^W_{11}\lb x_3,x_4\vert X^+\rb_{(-)_{4,1}}
    - \half {\cal G}^W_{22}\lb x_3,x_4\vert X^+\rb_{(-)_{4,1}} \nn\\
    &\qquad 
    + \half {\cal G}^W_{12}\lb x_3,x_4\vert X^+\rb_{(-)_{4,1}}
    + \half {\cal G}^W_{21}\lb x_3,x_4\vert X^+\rb_{(-)_{4,1}} \nn\\
    &\hspace{-.7em}
    \overset{\text{2-fold}}{=}
    \half {\cal G}^W_{aa}\lb x_3,x_4\vert X^+\rb_{(-)_{2}} 
    - \half {\cal G}^W_{aa}\lb x_3,x_4\vert X^+\rb_{(-)_{1}} \nn\\
    &= \half {\cal G}^W_{aa}\lb x_3,x_4\vert X^+\rb_{(-)_{2}} 
    - \half {\cal G}^W_{aa}\lb x_3,x_4\vert {-}X^+\rb_{(-)_{2}}
\end{calign}
In the second step we have used the KMS$_{(2)}$ symmetry on the four-fold contour from \cref{eq:KMS-four} on some of the correlators. In the third step, we have represented the correlators in terms of the two-fold basis, while in the fourth step we have performed a complete $\SL(2,\bbR)$ transformation with parameter $-X^+$ on the second term, which shifts the shock from fold ``1'' to fold ``2''. 
Using these, we find
\begin{calign}
    \frac{\dow {\cal G}^V_{\bar a\bar a}\lb x_1,x_2\vert X^-\rb_{(+)_{1,2}}}{\dow X^-(\bk) }\Bigg\vert_{X^-=0}
    &= \frac{\dow {\cal G}^V_{aa}\big( {-}x_1,-x_2\vert \tilde X^-\big)_{(-)_2}}{\dow X^-(\bk) }\Bigg\vert_{X^-=0}, \nn\\
    \frac{\dow{\cal G}^W_{\bar r a}\lb x_3,x_4\vert X^+\rb_{(-)_{2,3}}}{\dow X^+(\bk) }\Bigg\vert_{X^+=0}
    &=
    \frac{\dow{\cal G}^W_{aa}\lb x_3,x_4\vert X^+\rb_{(-)_{2}}}{\dow X^+(\bk) }\Bigg\vert_{X^+=0}.
\end{calign}

To complete this, we only need to obtain ${\cal G}^V_{aa}( x,x'\vert \tilde X^+)_{(-)_2}$ and ${\cal G}^W_{aa}\lb x,x'\vert X^+\rb_{(-)_2}$ in the two fold SK EFT, where the decaying shock is acting on  fold ``2''. To this end, we will use two prescriptions. For the first, we will use the same non-linear two-fold SK EFT as developed in \cref{sec:two-fold}, but expand it around a shifted saddle point where only the solution $F_{2,\eqb}(t,\bx)$ on fold ``2'' is modified by a decaying $\SL(2,\bbR)$ transformation. For the second, we will perform a non-linear $\SL(2,\bbR)$ transformation on the entire field $F_2(t,\bx)$, not merely its equilibrium value. The two prescriptions differ by simply a non-linear field redefinition of $\delta F_{2}$ in the SK EFT, and thus do not affect correlation functions of physical operators like the energy density. We will explicitly illustrate this in the following.

\subsubsection{Using shift of the on individual folds}

Let us consider the two-fold SK EFT where $F_1$ is expanded around the thermal solution $t/\beta$ but $F_2$ is expanded around the saddle shifted by the decaying scramblon $\E{2\pi F_{2,\eqb}} = \E{2\pi t/\beta} + X^+(\bx)$. In the following, we will denote $g(x) = \frac{1}{2\pi}X^+(\bx) \E{-\omega_\beta t}$.
This operation amounts to performing the following substitution in our two-fold SK EFT:
\begin{calign}\label{eq:shift-saddle}
    \delta F_{2} \to 
    \delta F_{2}
    + g + O(g^2),
\end{calign}
up to leading order in $g$. This leads to the maps
\begin{calign}\label{eq:shift-two-fold-fields}
    \delta F
    &\to \delta F 
    + \half g + O(g^2), \nn\\ 
    \delta Q
    &\to \delta Q
    - g + O(g^2).
\end{calign}
Since these transformations involve a zeroth order shift, they mix $(n+1)$-point interactions in the SK EFT with $n$-point ones. Using this, we find the new free part of the effective action generates an $O(F^2)$ term
\begin{calign}\label{eq:free-action-shock}
    \cL_{\rm free}^{\rm shock}
    &= \delta Q\,K(\dow)\,\delta F
    + \frac{i}{4}\delta Q\,\cot\bfrac{\beta \dow_t}{2}
    \Big( K(-\dow) - K(\dow) \Big)
    \delta Q
    \nn\\
    &\qquad 
    - \cV_{F^2Q}(\dow_1,\dow_2;\dow_3)
    \,\delta F\,\delta F\,g
    + (\ldots) \delta F\,\delta Q\,g
    + (\ldots) \delta Q\,\delta Q\,g
    + O(g^2).
\end{calign}
Due to this, we find that $\langle QQ\rangle$ correlator is nonzero on the shockwave background and takes the form
\begin{calign}\label{eq:noise-propagator-shock}
    \Big\langle\delta Q(x)\delta Q(x') \Big\rangle_{X^+}
    &= 
    2i \int\df^{d+1}y\,\cV_{F^2Q}(\dow_1,\dow_2;\dow_3)
    K^{-1}(y-x)
    K^{-1}(y-x') g(y).
\end{calign}

We are interested in computing the noise correlator of a local operator $\langle V_aV_a\rangle$ on the shockwave background. To this end, we need to know how the $V_a$ operator modifies in the presence of a shockwave. Using \cref{eq:generic-two-fold-op}, we find that the noise part an $\SL(2,\bbR)$-invariant local operator $V_a$ can now excite the average field $\delta F$ as
\begin{calign}
    V_a
    &= - \eta_\Theta\tilde\cO_{(1)}(\dow)\delta Q
    - 2 \Big(\eta_\Theta\tilde\cO_{(2)}(\dow_1,\dow_2)
    -  Y(\dow_1,\dow_2)\Big) \delta F g
    + (\ldots)\delta Q
    + \ldots \nn\\
    &=  - \eta_\Theta\tilde\cO_{(1)}(\dow)\delta Q \nn\\
    &\qquad 
    + 2 \lB   
    \pi \eta_\Theta \dow_{t1}
    \lb 1 - \frac{\dow_{t1}}{\omega_\beta} \rb
    \lb 2 - \frac{\dow_{t1}}{\omega_\beta} \rb
    \tilde\cD_{(1)}(\dow_1+\dow_2)
    + Y(\dow_1,\dow_2)\rB \delta F g
    + (\ldots)\delta Q
    + \ldots.
\end{calign}
The $\delta Q$ term in the expression above is not important for our calculation at the moment at linear order in $g$. We find
\begin{calign}
    {\cal G}^V_{aa}&\lb x,x'\vert X^+\rb_{(-)_2}
    = \Big\langle V_a(x)V_a(x') \Big\rangle_{X^+} \nn\\
    &= \tilde\cO_{(1)}(\dow_1)
    \tilde\cO_{(1)}(\dow_2)
    \Big\langle \delta Q(x)
    \delta Q(x') \Big\rangle_{X^+} \nn\\
    &\qquad 
    + 2 \Big(\tilde\cO_{(2)}(\dow_1,\dow_3)
    -  \eta_\Theta Y(\dow_1,\dow_3)\Big) 
    \tilde\cO_{(1)}(\dow_2) 
    \Big\langle \delta F(x)
    \delta Q(x') \Big\rangle  g(x) \nn\\
    &\qquad 
    + 2 \Big(\tilde\cO_{(2)}(\dow_1,\dow_3)
    -  \eta_\Theta Y(\dow_1,\dow_3)\Big) 
    \tilde\cO_{(1)}(\dow_2) 
    \Big\langle \delta F(x')
    \delta Q(x) \Big\rangle  g(x')
    +\ldots.
\end{calign}
After a detailed calculation, one can check that this precisely gives
\begin{calign}
    {\cal G}^V_{aa}\lb x,x'\vert X^+\rb_{(-)_2}
    &= \frac{\omega_\beta^2}{2\pi i}
    \Theta(t-t')\,\E{-\omega_\beta t}
    \int \frac{\df^{d+1}\bk}{(2\pi)^d}
    X^+(-\bk) 
    \E{-i\bk\cdot\bx} c_\bk\,\cI_{d}(x-x',\bk)
    + \ldots
    \nn\\
    &\qquad 
    + (x\leftrightarrow x'),
\end{calign}
where the same $\cI_d$ coefficient appears as derived from the full SK EFT computation in \cref{eq:general-integral}. This leads to
\begin{calign}\label{eq:shockwave-energy-result}
    \frac{\dow{\cal G}^V_{aa}\lb x,x'\vert X^+\rb_{(-)_2}}{\dow X^+(-\bk)}
    \Bigg|_{X^+=0}
    &= \frac{\omega_\beta^2}{2\pi i}
    \Theta(t-t')\,\E{-\omega_\beta t}
    \E{-i\bk\cdot\bx} c_\bk\,\cI_{d}(x-x',\bk)
    + \ldots
    \nn\\
    &\qquad 
    + (x\leftrightarrow x').
\end{calign}
Interestingly, we see that the same coefficient $\cI_d$ appears here as we found using the full SK EFT.  This is precisely the vertex factor obtained in \cref{eq:OTOC-resultSketch}.

\subsubsection{Using partial \texorpdfstring{$\SL(2,\bbR)$}{SL(2,R)} transformations on individual folds}

We can take an alternate viewpoint on the shockwave configurations. Instead of shifting the saddle around which the correlation functions are being computed, we can perform a fully non-linear $\SL(2,\bbR)$ transformation on individual folds to define the shockwave. This prescription has the  advantage that local operators that depend only on fields on an individual fold, like the Schwarzian energy density in SYK-like models, are left exactly invariant. However, as the $\SL(2,\bbR)$ transformations are not being performed on the entire closed-time contour simultaneously, these are not a symmetry of the non-linear effective action that dictates dynamics. Therefore, the correlation functions of such operators are still modified.

In detail, instead of \cref{eq:shift-saddle}, we perform a decaying $\SL(2,\bbR)$ transformation of the full field $F_2$ as $\E{2\pi F_2} \to \E{2\pi F_2} + X^+(\bx)$. This leads to the infinitesimal form
\begin{align}
    \delta F_2 \to g + (1-2\pi g)\delta F_2 + \ldots + O(g^2),
\end{align}
where ellipses denote subleading order terms in fluctuations. As we can see, this only differs from \cref{eq:shift-saddle} by a redefinition of the fluctaution $\delta F_2$. Recall that we do not perform any transformation on the $F_1$ field. Instead of \cref{eq:shift-two-fold-fields}, we now get
\begin{calign}
    \delta F
    &\to \half g 
    + (1-\pi g)\delta F 
    + \frac{\pi}{2} g\delta Q
    + \ldots 
    + O(g^2), \nn\\ 
    \delta Q
    &\to - g 
    + 2\pi g\delta F
    + (1-\pi g)\delta Q
    + \ldots 
    + O(g^2).
\end{calign}
Similarly, instead of \cref{eq:free-action-shock}, we get
\begin{calign}
    \cL_{\rm free}^{\rm shock}
    &= 
    \delta Q\,K(\dow)\,\delta F
    + \frac{i}{4}\delta Q\,\cot\bfrac{\beta \dow_t}{2}
    \Big( K(-\dow) - K(\dow) \Big) \delta Q \nn\\
    &\hspace{-2em}
    + \Big( {-}
    \cV_{F^2Q}(\dow_1,\dow_2;\dow_3)
    + \pi K(\dow_1)
    + \pi K(\dow_2)
    \Big)\,
    \delta F\,\delta F\,g
    + (\ldots)\, \delta F\,\delta Q\,g
    + (\ldots)\, \delta Q\,\delta Q\,g
    + O(g^2),
\end{calign}
which, in turn, leads to
\begin{calign}
    \Big\langle\delta Q(x)\delta Q(x') \Big\rangle_{X^+}
    &= 
    2i \int\df^{d+1}y\,
    \Big(\cV_{F^2Q}(\dow_1,\dow_2;\dow_3)
    - \pi K(\dow_1)
    - \pi K(\dow_2)
    \Big) \nn\\
    &\hspace{16em}
    K^{-1}(y-x)
    K^{-1}(y-x') g(y).
\end{calign}
This expression is different from \cref{eq:noise-propagator-shock} in the previous prescription because we have changed the definition of fields.

On the other hand, we find that the noise component of an $\SL(2,\bbR)$-invariant local operator $V_a$ excites $\delta F$ in this interpretation of the shockwave as
\begin{align}
    V_a 
    &= -\eta_\Theta \tilde\cO_{(1)}(\dow)\delta Q
    -2\pi \eta_\Theta\tilde\cO_{(1)}(\dow_1+\dow_2) \delta F\, g
    - \lb 2\eta_\Theta \tilde\cO_{(2)}(\dow_1,\dow_2)
    - 2 Y(\dow_1,\dow_2) \rb \delta F\, g
    + \ldots \nn\\
    &= - \eta_\Theta\tilde\cO_{(1)}(\dow)\delta Q
    + 2 Y(\dow_1,\dow_2) \delta F g
    + (\ldots)\delta Q.
\end{align}
Indeed, we see that the term identified above is zero for theories where the operator is purely factorised on individual folds, i.e. where the operator $Y(\dow_1,\dow_2)=0$; see our discussion around \cref{eq:generic-two-fold-op}. Using this, we find
\begin{calign}
    {\cal G}^V_{aa}&\lb x,x'\vert X^+\rb_{(-)_2}
    = \Big\langle V_a(x)V_a(x') \Big\rangle_{X^+} \nn\\
    &= \tilde\cO_{(1)}(\dow_1)
    \tilde\cO_{(1)}(\dow_2)
    \Big\langle \delta Q(x)
    \delta Q(x') \Big\rangle_{X^+} \nn\\
    &\qquad 
    - 2\eta_\Theta Y(\dow_1,\dow_3)
    \tilde\cO_{(1)}(\dow_2) 
    \Big\langle \delta F(x)
    \delta Q(x') \Big\rangle  g(x) \nn\\
    &\qquad 
    - 2\eta_\Theta Y(\dow_1,\dow_3)
    \tilde\cO_{(1)}(\dow_2) 
    \Big\langle \delta F(x')
    \delta Q(x) \Big\rangle  g(x')
    +\ldots
\end{calign}
In fact, with the new expression for the $\langle QQ\rangle$ correlator, one can see that this evaluates to the same value as in \cref{eq:shockwave-energy-result}. Therefore, the two viewpoints of the shockwave yield the same physical result for the correlation functions of $\SL(2,\bbR)$-invariant operators.

\section{Energy-momentum tensor dynamics in (1+1)d CFT}
\label{app:2dcft}

In (1+1)d CFT, the energy-momentum tensor dynamics is fixed by the infinite dimensional conformal symmetry. The energy-momentum correlation functions are known in closed form and are universal up to a simple dependence on the central charge. As a consistency check, in this appendix we use our EFT methods to recover the four-point OTOC of energy momentum tensor in (1+1)d CFT.

\subsection*{Energy-momentum correlation functions}

In a (1+1)d CFT, the Euclidean connected two-point correlation function of the holomorphic part of the energy-momentum tensor $\cE$ is given as
\begin{calign}
    \Big\langle \cT_\tau\, \cE(w_1)\,\cE(w_2) \Big\rangle_{\rm conn}
    = \frac{c}{2(2\pi)^2}\frac{1}{z_{12}^4},
\end{calign}
where $w_i = \bx_i + i\tau_i$, $\cT_\tau$ is the Euclidean time-ordering operator, and $c$ denotes the central charge. 
We have also defined the thermal combination $z_{ij} = (\beta /\pi)\sinh\lb \pi/\beta \,(w_i-w_j) \rb$.  We have also used $\cE$ to denote the holomorphic energy-momentum tensor instead of the usual $T$ to match with the energy density operator in the rest of the paper, as well as to avoid confusion with the local temperature. We have also rescaled it with a factor of $2\pi$ to match with the energy density operator.
Similarly, the Euclidean connected four-point correlation function is given as
\begin{calign}
    &\Big\langle \cT_\tau\, \cE(w_1)\,\cE(w_2)\,\cE(w_3)\,\cE(w_4) \Big\rangle_{\rm conn}
    = \frac{c}{(2\pi)^4}\, \lb \frac{1}{z^2_{12}z^2_{23}z^2_{34}z^2_{14}}
    + \frac{1}{z^2_{12}z^2_{24}z^2_{34}z^2_{13}}
    + \frac{1}{z^2_{13}z^2_{23}z^2_{24}z^2_{14}}
    \rb,
\end{calign}

We can find the real-time unordered correlation functions by analytic continuation, i.e.
\begin{calign}
    \Big\langle \cE(x_1)\,\cE(x_2) \Big\rangle_{\rm conn}
    &= \frac{c}{2(2\pi)^2}\frac{1}{z_{12}^4}, \nn\\
    \Big\langle \cE(x_1)\,\cE(x_2)\,\cE(x_3)\,\cE(x_4) \Big\rangle_{\rm conn}
    &= \frac{c}{(2\pi)^4}\, \lb \frac{1}{z^2_{12}z^2_{23}z^2_{34}z^2_{14}}
    + \frac{1}{z^2_{12}z^2_{24}z^2_{34}z^2_{13}}
    + \frac{1}{z^2_{13}z^2_{23}z^2_{24}z^2_{14}} \rb,
\end{calign}
where we have redefined $z_{ij} = (\beta /\pi)\sinh\lb \pi/\beta \,(\bx_i-t_i-\bx_j+t_j) + i\epsilon \rb$. 

Assuming $t_1>t_2>t_3>t_4$, we can check that
\begin{calign}
    C&{}^{\cE\cE}_{\bar a\bar a|\bar r a}(x_1,x_2;x_3,x_4) \nn\\
    &=
    \Big\langle \cE_{\bar a}(x_1)\,\cE_{\bar a}(x_2)\,
    \cE_{\bar r}(x_3)\,\cE_{a}(x_4) \Big\rangle \nn\\
    &= \Big\langle 
    {-}\cE(x_1)\, \cE(x_2)\, \cE(x_3)\, \cE(x_4)
    + \cE(x_4)\, \cE(x_3)\, \cE(x_2)\, \cE(x_1) \nn\\
    &\hspace{4em}
    - \cE(x_3)\, \cE(x_2)\, \cE(x_1)\, \cE(x_4)
    + \cE(x_4)\, \cE(x_1)\, \cE(x_2)\, \cE(x_3) \nn\\
    &\hspace{4em}
    + \cE(x_1)\, \cE(x_3)\, \cE(x_2)\, \cE(x_4)
    + \cE(x_2)\, \cE(x_3)\, \cE(x_1)\, \cE(x_4) \nn\\
    &\hspace{4em}
    - \cE(x_4)\, \cE(x_1)\, \cE(x_3)\, \cE(x_2)
    - \cE(x_4)\, \cE(x_2)\, \cE(x_3)\, \cE(x_1)
    \Big\rangle \nn\\
    &= \frac{-c}{(2\pi)^4}\lb \frac{1}{z^2_{r24} z^2_{a13} z^2_{a23} z^2_{a14}}
    + \frac{1}{z^2_{r14} z^2_{a13} z^2_{a23} z^2_{a24}}
    + \frac{1}{z^2_{r14} z^2_{a12} z^2_{a23} z^2_{a34}}
    + \frac{1}{z^2_{r34} z^2_{a12} z^2_{a23} z^2_{a14}} \rb,
\end{calign}
where $z^{-2}_{rij} = \Re z^{-2}_{ij}$ and $z^{-2}_{aij} = 2i\Im z^{-2}_{ij}$. Noting the distribution identities
\begin{calign}
    \frac{1}{\sinh^2(s+i\epsilon)}
    - \frac{1}{\sinh^2(s-i\epsilon)}
    &= 2\pi i\, \delta'(s), \nn\\
    \frac{1}{\sinh^4(s+i\epsilon)}
    - \frac{1}{\sinh^4(s-i\epsilon)}
    &= \frac{\pi i}{3}\lb 
    \delta'''(s)
    - 4\delta'(s) \rb,
\end{calign}
we can express this in partial-Fourier representation as
\begin{calign}\label{eq:ASCJ-OTOC-result}
    C&{}^{\cE\cE}_{\bar a\bar a\bar r a}(t_1,k_1,t_2,k_2,t_3,k_3,t_4,k_4) \nn\\
    &= \frac{c\pi}{6\beta^2}
    \E{-ik_1t_1-ik_2t_2-ik_3t_3-ik_4t_4}
     k_4
    \left(k_2-k_3\right)\left(2 k_1+k_4\right)
    \lb 1 + \frac{k_4^2}{\omega_\beta^2}  \rb,
\end{calign}
where $k_1+k_2+k_3+k_4=0$. This answer is recovered in the EFT through a tree-level computation, up to replacing the central charge $c$ appearing in the prefactor with the shifted value $C$ appearing in the ASCJ model \cref{eq:ASCJ-local-action}. We expect the central charge to match upon including loop corrections in the EFT.

\subsection*{Antisymmetric OTOC: t-channel, u-channel, and s-channel diagrams}

For t-channel, the operator $\cM_{\rm t}$ in \cref{eq:M-def-minus-otoc-tchannel-result} is given as
\begin{calign}
    &\cM_{\rm t}(t_1-s_1,k_1,s_1-t_4,-k_4) \nn\\
    &= - c_V \E{-ik_1 t_1-ik_4 t_4}
    \dow_{s_1} \lb 1 - \frac{\dow_{s_1}^2}{\omega_\beta^2} \rb \nn\\
    &\quad 
    \lbr \Bigg[
    k_4\lb \dow_{s_1} - i(k_1+k_4)\rb 
    \Big(\Theta_4(s_1-t_4)\,\circ\Big) 
    + \lb k_1-k_4 \rb \delta(s_1-t_4) 
    \Bigg] 
    \Big( \Theta(t_1-s_1)
    \E{i(k_1+k_4) s_1}
    \Big)\rbr,
\end{calign}
while $\overline\cM_{\rm t}$ is obtained through the relation in \cref{eq:M-def-minus-otoc-tchannel-result}. 
This leads to the t-channel contribution (assuming $t_1>t_2>t_3>t_4$)
\begin{calign}
    C^{\rm t}_{\bar a\bar a \bar r a}
    &= \frac{-c_V}{\beta }
    \E{-ik_1 t_1-ik_2 t_2-ik_3 t_3-ik_4 t_4}
    k_4 
    (k_2+k_3)
     \lb k_2 - k_3 \rb
     \lb 1 + \frac{(k_2+k_3)^2}{\omega_\beta^2} \rb.
\end{calign}
Repeating the calculation with $t_1\leftrightarrow t_2$, $k_1\leftrightarrow k_2$, we find the u-channel contribution
\begin{calign}
    C^{\rm u}_{\bar a\bar a \bar r a}
    &= \frac{-c_V}{\beta } 
    \E{-ik_1 t_1-ik_2 t_2-ik_3 t_3-ik_4 t_4}
    k_1k_4  
    (k_1+k_3)
    \lb 1 + \frac{(k_1+k_3)^2}{\omega_\beta^2} \rb .
\end{calign}
On the other hand, the operator $\cM_{\rm s}$ in \cref{eq:M-def-minus-otoc-schannel-result} in the s-channel is given as
\begin{calign}
    &\cM_{\rm s}(t_1-s_1,k_1,t_2-s_1,k_2) \nn\\
    &=  
    c_V \E{-ik_1t_1-ik_2t_2}
    \dow_{s_1}\lb 1 -\frac{\dow_{s_1}^2}{\omega_\beta^2} \rb \nn\\
    &\qquad 
    \lbr 
    k_1  \Theta(t_1-s_1)
    \E{i(k_1+k_2)s_1}
    \delta_2(t_2-s_1)
    + k_2  \Theta(t_2-s_1)
    \E{i(k_1+k_2)s_1}
    \delta(t_1-s_1)
    \rbr,
\end{calign}
which leads to
\begin{calign}
    C^{\rm s}_{\bar a\bar a \bar r a}
    &= \frac{c_V}{\beta }
    \E{-ik_1t_1-ik_2t_2-ik_3 t_3- ik_4t_4}
    k_1 
    k_4
    \lb 1+\frac{(k_1+k_2)^2}{\omega_\beta^2}
    \rb (k_1+k_2).
\end{calign}

\subsection*{Antisymmetric OTOC: contact diagrams}

To match with the full result of energy OTOC, we also need to compute the contact diagrams in \cref{sec:contact-diagrams}. This requires going further than the analysis presented so far: we need information about four-point interactions in the SK EFT, as well as need to expand the energy-momentum tensor operator to cubic order in field fluctuations. Both of these tasks are straightforward. The complete four-fold SK effective action is given as $S[F_1]-S[F_2]+S[F_3]-S[F_4]$, which we can expand to arbitrary order in field fluctuations. Similarly, the energy-momentum tensor of the theory is essentially just the Schwarzian in \cref{ASCJstress}. This means that the holomorphic energy-momentum tensor on each Lorentzian fold of the closed-time contour is simply given as
\begin{calign}
    \cE_i =-{C\ov24\pi}\Sch\le[\tanh(\pi F_i),t\ri].
\end{calign}
We can use this to obtain the expressions for $\cE_{r,a,\bar a,\bar r}$ to arbitrary order in fluctuations. Schematically, we find
\begin{calign}
    &C_{\bar a \bar a |\bar r a}^{\rm contact}
    = \int\df y\Bigg[
    2i\cV_{F\bar Q\bar F^2}(\dow_3,\dow_4;-\dow_1,-\dow_2)  \nn\\
    &\hspace{6em}
    \Big(\cO_{(1)}(\dow_1) iK^{-1}_1(x+\varepsilon-y)\Big)
    \Big(\cO_{(1)}(\dow_2) iK^{-1}_2(x-\varepsilon-y)\Big) \nn\\
    &\hspace{6em}
    \Big({-}\eta_\Theta\tilde\cO_{(1)}(-\dow_3) iK^{-1}_3(y-0-\varepsilon')\Big)
    \Big({-}\eta_\Theta\tilde\cO_{(1)}(-\dow_4) iK^{-1}_4(y-0+\varepsilon')\Big) \nn\\
    &\hspace{4em}
    - 6\cO_{(3)}(-\dow_2,\dow_3,\dow_4)
    \Big(\cO_{(1)}(\dow_2) iK^{-1}_2(x-\varepsilon-y)\Big) \nn\\
    &\hspace{6em}
    \Big({-}\eta_\Theta\tilde\cO_{(1)}(-\dow_3) iK^{-1}_3(y-0-\varepsilon')\Big)
    \Big({-}\eta_\Theta\tilde\cO_{(1)}(-\dow_4) iK^{-1}_4(y-0+\varepsilon')\Big)\nn\\
    &\hspace{4em}
    - 6\cO_{(3)}(-\dow_1,\dow_3,\dow_4)
    \Big(\cO_{(1)}(\dow_1) iK^{-1}_1(x+\varepsilon-y)\Big) \nn\\
    &\hspace{6em}
    \Big({-}\eta_\Theta\tilde\cO_{(1)}(-\dow_3) iK^{-1}_3(y-0-\varepsilon')\Big)
    \Big({-}\eta_\Theta\tilde\cO_{(1)}(-\dow_4) iK^{-1}_4(y-0+\varepsilon')\Big) \nn\\
    &\hspace{4em}
    - 6\cO_{(3)}(-\dow_1,-\dow_2,\dow_4)
    \Big(\cO_{(1)}(\dow_1) iK^{-1}_1(x+\varepsilon-y)\Big)
    \Big(\cO_{(1)}(\dow_2) iK^{-1}_2(x-\varepsilon-y)\Big) \nn\\
    &\hspace{6em}
    \Big({-}\eta_\Theta\tilde\cO_{(1)}(-\dow_4) iK^{-1}_4(y-0+\varepsilon')\Big) \nn\\
    &\hspace{4em}
    - 6\cO_{(3)}(-\dow_1,-\dow_2,\dow_3)
    \Big(\cO_{(1)}(\dow_1) iK^{-1}_1(x+\varepsilon-y)\Big)
    \Big(\cO_{(1)}(\dow_2) iK^{-1}_2(x-\varepsilon-y)\Big) \nn\\
    &\hspace{6em}
    \Big({-}\eta_\Theta\tilde\cO_{(1)}(-\dow_3) iK^{-1}_3(y-0-\varepsilon')\Big) 
    \Bigg].
\end{calign}
Here $\cV_{F\bar Q\bar F^2}$ denotes the relevant four-point interaction, while $\cO_{(n)}$ denote the differential operators in the expansion of the classical part of the energy-momentum tensor as defined in \cref{eq:linear-expansion-E}. Performing the computation explicitly, we find
\begin{calign}
    C_{\bar a \bar a |\bar r a}^{\rm contact}
    &= \frac{-2c_V}{\beta \omega_\beta^2}
    \E{-ik_1t_1-ik_2t_2-ik_3t_3-ik_4t_4}k_1 k_4 \nn\\
    &\hspace{6em}
    \Bigg[ 
    k_3
    \lb k_1^2 - \frac{3}{2}k_2^2 + k_3^2 + k_4^2
    + 3k_1 k_3
    + 3k_1 k_4
    + 3k_3 k_4
    \rb \nn\\
    &\hspace{6em}
    - k_2
    \lb 
    k_1^2 + k_2^2
    - \frac{3}{2} k_3^2
    + k_4^2
    + 3k_1k_2
    + 3k_1k_4
    + 3k_2k_4
    \rb
    \Bigg].
\end{calign}

The results for t-channel, u-channel, and s-channel diagrams, combined with the contact diagrams, give the full result for the OTOC
\begin{calign}
    C&{}_{\bar a\bar a\bar r a}(t_1,k_1,t_2,k_2,t_3,k_3,t_4,k_4) \nn\\
    &= \frac{c_V}{\beta }
    \E{-ik_1 t_1-ik_2 t_2-ik_3 t_3-ik_4 t_4}\,
    k_4
    \lb k_2 - k_3 \rb
    \lb 2k_1+k_4\rb \lb 1 + \frac{k_4^2}{\omega_\beta^2}\rb \nn\\
    &= \frac{C \pi}{6\beta^2} 
    \E{-ik_1 t_1-ik_2 t_2-ik_3 t_3-ik_4 t_4}\,
    k_4
    \lb k_2 - k_3 \rb
    \lb 2k_1+k_4\rb \lb 1 + \frac{k_4^2}{\omega_\beta^2}\rb.
\end{calign}
In the second line, we have substituted the value of $c_V$ from \cref{ASCJstress}. This matches the expected result in \cref{eq:ASCJ-OTOC-result}, except for the central charge $c$ replaced with $C$. Getting the prefactor to match exactly would require computing one-loop corrections to this result. We leave this exercise for future work.

\section{Miscellaneous details of the SK EFT framework}

\subsection{Factorisation and time-reversal-odd dissipation}
\label{app:factorisation}

In this appendix, we show that requiring the imaginary part of the effective action to be $\Theta$-even is equivalent to its $\Theta$-even real part being factorised. Here $\Theta$ denotes the discrete time-reversal or spacetime-reversal transformation used to define the KMS symmetry. Let us first split the effective action into real/imaginary and $\Theta$-even/odd parts
\begin{calign}
    S 
    = \Re S_{\rm even} 
    + \Re S_{\rm odd} 
    + i\Im S_{\rm even} 
    + i\Im S_{\rm odd}.
\end{calign}
One can check that the KMS (shift) symmetry only mixes $\Re S_{\rm even}$ with $\Im S_{\rm odd}$, while $\Re S_{\rm odd}$ with $\Im S_{\rm even}$. Whereas, the time translation symmetries act on each of these parts independently.

It follows that for SK EFTs where $\Im S_{\rm odd}=0$, the $\Re S_{\rm even}$ part of the effective action must be independently invariant under KMS and time-translation symmetries.
Recall that the KMS shift symmetry is comprised of a $\Theta$-transformation followed by a constant shift of $Q \to Q+i$. Since $\Re S_{\rm even}$ is already even under $\Theta$, the KMS symmetry requires that it must be invariant under $Q\to Q+i$. Combined with time-translation symmetries, it implies that $\Re S_{\rm even}$ is invariant under
\begin{subequations}\label{eq:trans-for-factor}
\begin{calign}
    F_1(t) \to F_1(t-i\beta /2) + i/2, \qquad 
    F_2(t) \to F_2(t+i\beta/2 ) - i/2.
\end{calign}
Since this transformation leaves the state $F_{1,2}=t/\beta $ invariant, it is linear in fluctuations:
\begin{calign}
    \delta F_1(t) \to \E{-i\beta /2\,\dow_t} \delta F_1(t), \qquad 
    \delta F_2(t) \to \E{i\beta /2\,\dow_t} \delta F_2(t).
\end{calign}
\end{subequations}
Consider a general term in the linearised expansion of the effective action in frequency-space, comprised of $p$ factors of $\delta F_1$ and $q$ factors of $\delta F_2$, given as
\begin{calign}\label{eq:general-term}
    G\lb\omega_1,\ldots,\omega_p;\omega'_1,\ldots,\omega'_q\rb\,
    \delta F_1(\omega_1)\ldots 
    \delta F_1(\omega_p)\,
    \delta F_2(\omega'_1)\ldots
    \delta F_2(\omega'_q),
\end{calign}
where $\omega_1+\ldots+\omega_p+\omega'_1+\ldots+\omega'_q = 0$. Invariance under \cref{eq:trans-for-factor} requires that
\begin{calign}\label{eq:argument-even}
    \Big( \E{-\beta (\omega_1+\ldots+\omega_p)/2
    + \beta (\omega'_1+\ldots+\omega'_q)/2} 
    - 1 \Big)
    G\lb\omega_1,\ldots,\omega_p;\omega'_1,\ldots,\omega'_q\rb
    = 0,
\end{calign}
for arbitrary frequencies. Therefore, such a term is only allowed in $\Re S_{\rm even}$ if either $p=0$ or $q=0$. In other words, $\Re S_{\rm even}$ is factorised into terms that depend either only on $F_1$ or $F_2$. Further noting that $\Re S_{\rm even}$ must vanish when $F_1=F_2$, we recover that $\Re S_{\rm even} = S_0[F_1] - S_0[F_2]$.

To show the converse statement, let us start with the condition that $\Re S_{\rm even} = S_0[F_1] - S_0[F_2]$. Since this is already KMS-invariant, we need to impose that $\Im S_{\rm odd}$ is also KMS-invariant. Since $\Im S_{\rm odd}$ is odd under $\Theta$, the KMS symmetry requires that it must be odd under \cref{eq:trans-for-factor}. Every term like \cref{eq:general-term} in $\Im S_{\rm odd}$ must then satisfy
\begin{calign}\label{eq:argument-odd}
    \Big( \E{-\beta (\omega_1+\ldots+\omega_p)/2
    + \beta (\omega'_1+\ldots+\omega'_q)/2} 
    + 1 \Big)
    G\lb\omega_1,\ldots,\omega_p;\omega'_1,\ldots,\omega'_q\rb
    = 0.
\end{calign}
But the prefactor here is manifestly positive, so there are no such possible terms. It follows that $\Im S_{\rm odd}$ is identically zero. To summarise
\begin{calign}
    \boxed{\Im S_{\rm odd} = 0
    \quad\Longleftrightarrow\quad
    \Re S_{\rm even} = S_0[F_1] - S_0[F_2].}
\end{calign}

As an aside, we can also make similar inferences for the other two parts of the effective action. If we impose that $\Re S_{\rm odd}=0$, the same argument as \cref{eq:argument-even} implies that $\Im S_{\rm even}$ must be factorised. However, this is a red herring because $\Im S_{\rm even}$ is also even under $F_1\leftrightarrow F_2$ and vanishes when $F_1=F_2$, which sets it to be identically zero. Conversely, using the same argument as \cref{eq:argument-odd}, $\Im S_{\rm even}=0$ implies that $\Re S_{\rm odd}=0$. In fact, there is an even stronger statement. We know that $\Im S = \Im S_{\rm even} + \Im S_{\rm odd} \geq 0$. However, $\Im S_{\rm odd}$ changes sign under $\Theta$-odd configurations, so we must require that $\Im S_{\rm even} \geq |\Im S_{\rm odd}|$. It follows that $\Im S_{\rm even}=0$ also implies $\Im S_{\rm odd}= 0$, which further implies that $\Re S_{\rm even}$ is factorised. As a result, the effective action must be completely factorised. To summarise
\begin{calign}
    \boxed{\Im S_{\rm even} = 0
    \quad\Longleftrightarrow\quad
    \Re S_{\rm odd} = 0
    \quad\Longleftrightarrow\quad
    S = S_0[F_1] - S_0[F_2].}
\end{calign}
This is a variant of the fluctuation-dissipation theorem.

\subsection{Multi-fold KMS symmetry}
\label{app:KMS}

In this appendix, we give a brief derivation of the KMS symmetry on multi-fold closed-time contours; see also~\cite{Haehl:2017qfl,Haehl:2017eob,Chaudhuri:2018ihk,Mishra:2025vlf}.. On a $2n$-fold contour, the KMS symmetry is characterised by the Dihedral group $\rmD_{2n}$ (i.e. the group of isometries of a regular $n$-sided polygon).\footnote{Note that this group is often also denoted $D_{n}$.} In $n=1,2$, this is isomorphic to $\bbZ_2$ and $\bbZ_2\times\bbZ_2$ respectively.
Consider a general correlation function that can be represented on a $2n$-fold contour:
\begin{calign}\label{eq:2n-corr}
    &\bigg\langle 
    \tilde\cT\Big( \hat A^{(2n)}(t_{2n}) \ldots \Big)
    \cT\Big( \hat A^{(2n-1)}(t_{2n-1}) \ldots \Big)
    \ldots 
    \tilde\cT\Big( \hat A^{(2)}(t_{2})\ldots \Big)
    \cT\Big( \hat A^{(1)}(t_1)\ldots \Big)
    \bigg\rangle \nn\\
    &= \tr\bigg[\rho_\beta
    \tilde\cT\Big( \hat A^{(2n)}(t_{2n}) \ldots \Big)
    \cT\Big( \hat A^{(2n-1)}(t_{2n-1}) \ldots \Big)
    \ldots 
    \tilde\cT\Big( \hat A^{(2)}(t_{2})\ldots \Big)
    \cT\Big( \hat A^{(1)}(t_1)\ldots \Big)
    \bigg],
\end{calign}
where $\rho_\beta$ denotes the thermal density matrix
This is essentially an alternating product of time-ordered and anti-time-ordered correlation functions (as counted from the right), denoted by time-ordering and anti-time-ordering operators $\cT$ and $\tilde\cT$ respectively. We can take the set of operators on the extreme left and use the cyclic property of the trace to move it to the extreme right. In doing so, the operators pick a phase $i\beta$ in time when passed through the thermal density matrix; to wit, $\rho_\beta \hat A(t) = \hat A(t+i\beta)\rho_\beta$. Therefore, the same correlator is equal to
\begin{calign}\label{eq:KMS-one-circle}
    \tr\bigg[\rho_\beta
    \cT\Big( \hat A^{(2n-1)}(t_{2n-1}) \ldots \Big)
    \ldots 
    \tilde\cT\Big( \hat A^{(2)}(t_{2})\ldots \Big)
    \cT\Big( \hat A^{(1)}(t_1)\ldots \Big)
    \tilde\cT\Big( \hat A^{(2n)}(t_{2n}+i\beta) \ldots \Big)
    \bigg].
\end{calign}
This is not a symmetry of the theory just yet, because the new correlator starts with an anti-time-ordered set of operators first from the right, while the original one in \cref{eq:2n-corr} starts with time-ordered ones. So this operation simply maps one set of observables to another. This does become a genuine symmetry if the theory features an anti-unitary time-reversal symmetry $\Theta$. In this case, we find that the same correlator is equal to
\begin{calign}
    \tr\bigg[\rho_\beta
    \tilde\cT\Big( \hat A^{(2n)}(-t_{2n}-i\beta) \ldots \Big)
    \cT\Big( \hat A^{(1)}(-t_1)\ldots \Big)
    \tilde\cT\Big( \hat A^{(2)}(-t_{2})\ldots \Big)
    \ldots 
    \cT\Big( \hat A^{(2n-1)}(-t_{2n-1}) \ldots \Big)
    \bigg],
\end{calign}
up to an overall sign for the collative $\Theta$-eigenvalue of the operator insertions.
The $\Theta$-operation does two things: it flips the sign of time argument of each operator insertion and it also flips the order in which they are inserted. Note that the two operations are such that the time-ordered set of operators remain mutually time-ordered, but in flipped time arguments. This is now a genuine symmetry of the theory. In terms of fold labels, it reads that for generic operators $\cO$, we have
\begin{alignat}{2}
    \rmK_{(1)}\cO_k(t,\bx) 
    &= \eta_\Theta\cO_{2n-k}(-t+i\beta/2,\eta_\rmP\bx), \qquad 
    && (k\neq 2n), \nn\\
    \rmK_{(1)}\cO_{2n}(t,\bx) 
    &= \eta_\Theta\cO_{2n}(-t-i\beta/2,\eta_\rmP\bx).
\end{alignat}
Here $\eta_\Theta$ denotes the $\Theta$-eigenvalue of the operator. As we have defined in the main text, $\eta_\rmP = \pm 1$ depending on whether $\Theta$ involves a spatial parity transformation. In writing this, we have performed an additional time-translation with parameter $i\beta/2$ for convenience. One can check that $\rmK_{(1)}^2 = 1$ and thus generates a $\bbZ_2$ group. 

However, there is another non-trivial symmetry available to us for $n>1$. Let us go back to \cref{eq:KMS-one-circle} and take another set of operators across the density matrix. We find
\begin{calign}
    \tr\bigg[\rho_\beta
    \ldots
    \tilde\cT\Big( \hat A^{(2)}(t_{2})\ldots \Big)
    \cT\Big( \hat A^{(1)}(t_1)\ldots \Big)
    \tilde\cT\Big( \hat A^{(2n)}(t_{2n}+i\beta) \ldots \Big)
    \cT\Big( \hat A^{(2n-1)}(t_{2n-1}+i\beta) \ldots \Big)
    \bigg].
\end{calign}
This is already in the original form in \cref{eq:2n-corr}, without having to use $\Theta$. So, we can read off another symmetry
\begin{alignat}{2}
    \rmK_{(2)}\cO_k(t,\bx) 
    &= \cO_{k+2}(t-i\beta/n,\bx), \qquad 
    && (k\neq 2n,2n-1), \nn\\
    \rmK_{(2)}\cO_{k}(t,\bx) 
    &= \cO_{k-2n+2}(t+i\beta(n-1)/n,\bx), \qquad 
    && (k= 2n,2n-1).
\end{alignat}
Here we have performed an additional time translation with parameter $-i\beta/n$. With this choice, $\rmK_{(2)}$ forms a $\bbZ_n$ group, corresponding to taking two consecutive folds of the closed-time contour across the thermal circle. However, the two generators $\rmK_{(1)}$ and $\rmK_{(2)}$ do not generally commute. It is easy to see that 
\begin{align}
    \rmK_{(1)}^2 = 1, \qquad 
    \rmK_{(2)}^n = 1, \qquad 
    \rmK_{(1)}\rmK_{(2)}
    = \rmK_{(2)}^{n-1}\rmK_{(1)}.
\end{align}
This is precisely the Dihedral group $\rmD_{2n} = \bbZ_{n}\rtimes \bbZ_2$. Note that the two operations commute for $n=2$. On the other hand, for $n=1$, the generator $\rmK_{(2)}$ is just identity.

The above arguments only apply to correlation functions of operators whose equilibrium value remains unchanged when taken across the thermal circle. The reparametrisation fields $F_i$ discussed in the main text have equilibrium value $t/\beta$, which does shift when taken across the thermal circle. So we have to add another shift in the transformation rules:
\begin{alignat}{2}
    \rmK_{(1)}F_k(t) 
    &= \eta_\Theta \Big( F_{2n-k}(-t+i\beta/2,\eta_\rmP\bx) - i/2 \Big), \qquad 
    && (k\neq 2n), \nn\\
    \rmK_{(1)}F_{2n}(t) 
    &= \eta_\Theta \Big( F_{2n}(-t-i\beta/2,\eta_\rmP\bx) + i/2
    \Big),
\end{alignat}
and similarly 
\begin{alignat}{2}
    \rmK_{(2)}F_k(t,\bx) 
    &= F_{k+2}(t-i\beta/n,\bx) + i/n, \qquad 
    && (k\neq 2n,2n-1), \nn\\
    \rmK_{(2)}F_{k}(t,\bx) 
    &= F_{k-2n+2}(t+i\beta(n-1)/n,\bx) -i(n-1)/n, \qquad 
    && (k= 2n,2n-1),
\end{alignat}
where $\eta_\Theta = -1$ for reparametrisation fields. Note that these transformations preserve the thermal state $F_k(t,\bx) = t/\beta$.

For our EFT, where temperature is a dynamical quantity, this representation of the KMS symmetry is slightly unnatural as it makes reference to the particular equilibrium temperature $\beta$ in its definition. However, recall that our EFT is invariant under time translations on individual folds. Therefore, we can obtain a more robust version of the KMS symmetry by undoing the shifts of time coordinates with time translations in imaginary direction. This leads to the ``KMS shift'' symmetry:
\begin{alignat}{2}
    \rmK^{\rm shift}_{(1)}F_k(t) 
    &= \eta_\Theta \Big( F_{2n-k}(-t,\eta_\rmP\bx) - i/2 \Big), \qquad 
    && (k\neq 2n), \nn\\
    \rmK^{\rm shift}_{(1)}F_{2n}(t) 
    &= \eta_\Theta \Big( F_{2n}(-t,\eta_\rmP\bx) + i/2
    \Big),
\end{alignat}
and similarly 
\begin{alignat}{2}
    \rmK^{\rm shift}_{(2)}F_k(t,\bx) 
    &= F_{k+2}(t,\bx) + i/n, \qquad 
    && (k\neq 2n,2n-1), \nn\\
    \rmK^{\rm shift}_{(2)}F_{k}(t,\bx) 
    &= F_{k-2n+2}(t,\bx) -i(n-1)/n, \qquad 
    && (k= 2n,2n-1),
\end{alignat}
We have used the $n=2$ version of these transformations in the main text in \cref{eq:KMS-four}.

\subsection{KMS invariance of the factorised four-fold SK EFT}
\label{app:check-kms}

Under KMS$_{(1)}$, different components of the four-fold effective action in \cref{eq:4fold-action} transform as
\begin{alignat}{2}
    S\big[\rmK_{(1)}F_1,\rmK_{(1)}F_2\big]
    &= S\big[{-}\tilde F_3,-\tilde F_2\big]
    &&= -S\big[{-}\tilde F_2,-\tilde F_3\big]^* \nn\\
    - S\big[\rmK_{(1)}F_1,\rmK_{(1)}F_3\big]
    &= - S\big[{-}\tilde F_3,-\tilde F_1\big]
    &&= S\big[{-}\tilde F_1,-\tilde F_3\big]^* \nn\\
    S\big[\rmK_{(1)}F_1,\rmK_{(1)}F_4\big]
    &= S\big[{-}\tilde F_3,-\tilde F_4 - i\big]
    &&= S\big[F_3,F_4\big] \nn\\
    S\big[\rmK_{(1)}F_2,\rmK_{(1)}F_3\big]
    &= S\big[{-}\tilde F_2,-\tilde F_1\big]
    &&= -S\big[{-}\tilde F_1,-\tilde F_2\big]^* \nn\\
    - S\big[\rmK_{(1)}F_2,\rmK_{(1)}F_4\big]
    &= - S\big[{-}\tilde F_2,-\tilde F_4 - i\big]
    &&= - S\big[F_2,F_4\big]\nn\\
    S\big[\rmK_{(1)}F_3,\rmK_{(1)}F_4\big]
    &= S\big[{-}\tilde F_1,-\tilde F_4 - i\big] 
    &&= S\big[F_1,F_4\big] 
\end{alignat}
whereas under KMS$_{(2)}$, they transform as
\begin{alignat}{2}
    S\big[\rmK_{(2)}F_1,\rmK_{(2)}F_2\big]
    &= S\big[F_3,F_4\big] \nn\\
    - S\big[\rmK_{(2)}F_1,\rmK_{(2)}F_3\big]
    &= - S\big[F_3,F_1-i\big]
    &&= S\big[{-}\tilde F_1,-\tilde F_3\big]^* \nn\\
    S\big[\rmK_{(2)}F_1,\rmK_{(2)}F_4\big]
    &= S\big[F_3,F_2 - i\big]
    &&= -S\big[{-}\tilde F_2,-\tilde F_3\big]^* \nn\\
    S\big[\rmK_{(2)}F_2,\rmK_{(2)}F_3\big]
    &= S\big[F_4,F_1-i\big]
    &&= -S\big[{-}\tilde F_1,-\tilde F_4\big]^* \nn\\
    - S\big[\rmK_{(2)}F_2,\rmK_{(2)}F_4\big]
    &= - S\big[F_4,F_2 - i\big]
    &&= S\big[{-}\tilde F_2,-\tilde F_4\big]^* \nn\\
    S\big[\rmK_{(2)}F_3,\rmK_{(2)}F_4\big]
    &= S\big[F_1,F_2\big].
\end{alignat}
Here we have used the fact that the two-fold effective action is invariant under simultaneous constant shifts of fields.
It is then straightforward to see that the factorised form of the effective action is not generically invariant under either of the KMS symmetries. However, if we assume that the underlying theory features $\Theta$-odd dissipation, i.e. satisfies \cref{eq:cond-odd-dissipation}, we have that
\begin{align}\label{eq:consequence-of-odd-dissipation}
    S[F_1,F_2] + S[{-}\tilde F_1,{-}\tilde F_2]^*
    &= 2S_0[F_1] - 2S_0[F_2],
\end{align}
where $S_0[-\tilde F] = S_0[F]$. In this case, we can check that the residual terms perfectly cancel and the factorised form in \cref{eq:4fold-action} satisfies both components of the KMS symmetry. The discussion is more subtle for theories that do not respect $\Theta$-odd dissipation and requires appropriately modifying the four-fold effective action with non-factorised terms, even at cubic order in fluctuations. We plan to return to this point in future work.

\bibliographystyle{JHEP}
\setlength{\bibsep}{0.2em}
\bibliography{biblio}

\end{document}